# Variational Results and Solutions in Gauge Gravity and a Bifurcation Analysis of Black Hole Orbital Dynamics

Bruce Herold Dean

a Dissertation
Submitted to the College of Arts and Sciences
at West Virginia University
In Partial Fulfillment of the Requirements
for the degree of

**Doctor of Philosophy**
in
Physics

Larry Halliburton, Ph.D., Chair
James Crawford, Ph.D. (Physics Department / Penn State)
Marty Ferer, Ph.D.
Richard Treat, Ph.D.
H. Arthur Weldon, Ph.D.
Joseph Wilder, Ph.D. (WVU Department of Mathematics)

Department of Physics

Morgantown, West Virginia
1999

*Keywords*: general relativity, Palatini variational formalism, gauge gravity, conformal gravity, phase – plane and bifurcation analysis, relativistic orbital dynamics, black holes, Reissner - Nordström solution, periastron precession, perihelion precession

# *Abstract*

## Variational Results and Solutions in Gauge Gravity and a Bifurcation Analysis of Black Hole Orbital Dynamics

Bruce H. Dean

An analysis of all known spherically symmetric solutions to the field equations originating from the Riemann tensor quadratic curvature Lagrangian is presented. A new exact solution is found for the field equation originating from the "energy-momentum" equation of the gauge gravity theory. Imposing equivalence between the Palatini and standard variational field equations yields an algebraic condition that restricts the number spacetime solutions to gauge gravity. A class of spherically symmetric solutions to the conformally invariant theory of gravitation is shown to be shared by the gauge gravity field equations. An analysis of a spherically symmetric solution to the conformal gravity field equations is also presented.

Point particle orbital dynamics in both the Schwarzschild and Reissner-Nordström black hole spacetimes are analyzed as 2-d conservative bifurcation phenomena. The classification is based on a study of coalescing fixed points and the parameter values at which these bifurcations occur. Physically distinct behaviors are separated by bifurcation points while dynamically distinct cases are divided into various regions of the phase-plane by the separatrix. The Schwarzschild dynamics exhibit both saddle-center and transcritical bifurcation points and a calculation of periastron precession is presented that incorporates a phase-plane analysis of the relativistic equations of motion. Level curves of constant energy are illustrated for both timelike and null geodesics and a phase-plane analysis of dynamical invariance between the proper and coordinate time reference frames is discussed. The Reissner-Nordström dynamics exhibit saddle-center, transcritical, pseudo-transcritical, and additional bifurcations that combine all three previous bifurcations in various combinations. Periastron precession in the Reissner-Nordström spacetime is analyzed using the phase-plane and bifurcation techniques and extended to include a bifurcation point of the dynamics. A numerical solution at these parameter values illustrates that such orbits typically yield a much larger precession value compared to the standard value for timelike precession. The "acausal" geodesics considered by Brigman are also discussed and their precession value is calculated.

*for my Parents*

Sterl Herold Dean

and

Donna Kaye Kerr

# Contents

// Acknowledgements

Any work of this nature is a serious undertaking in both time and personal commitment which in many ways has been shared by several people. My wife Rebecca has shared the most and therefore I owe her my deepest appreciation. I thank my parents, Sterl H. Dean and late Donna Kerr Dean for their encouragement and opportunities they have provided for my education. Richard and Donna Shuttlesworth have also been very supportive and have contributed in many personal ways toward the development of this work. I am also grateful to my uncle Summers Dean for his many insights and characteristic way of probing deeply into problems, and I thank my brother Brian for his support and advice on computer issues.

During the development of this work I have benefited greatly from many helpful discussions with my research advisor, Dr. James Crawford. His guidance and perspective on many aspects of physics as well as sharing his research notes and pointing out difficulties in certain calculations have been invaluable. I also thank Dr. Richard Treat for helpful discussions on the technical content which have helped to clarify many of the calculations and also for his support and advice on many other topics during my graduate study. I thank Dr. Larry Halliburton for his support and for helping to setup my research with Dr. Crawford as well as my other committee members: Dr. Marty Ferer for financial support as a Research Assistant, Dr. Arthur Weldon for his participation and suggestions regarding the Reissner-Nordström calculation, and Dr. Joseph Wilder for his advice on many topics during my graduate study and also for financial support as a Research Assistant and Mathematics Workshop Instructor in the Department of Mathematics at WVU. Other members of the Physics Department staff have helped with many other details and have given personal support, particularly Sherry Puskar. I also thank Siobhan Byrne for helping with several details and Greg Puskar and Carl Rotter for their support of my work as a Teaching Assistant, and the rest of the "Gibbies" team for many enlightened discussions: Matt Balkey, Jack and Becky Littleton, John Kline, and Drew Milsom.

I am thankful for several helpful comments and pointers into the general relativity and nonlinear dynamics literature given to me by Abhay Ashtekar, Beverly Berger, Bryce DeWitt, Bill Hiscock, Charles Jaffe, Demos Kazanas, H. G. Loos, Ted Newman, Jorge Pullin, Carlo Rovelli, Lee Smolin, Steve Strogatz, Kip Thorne, and Clifford Will. The Goddard Space Flight Center staff have also been very supportive, particularly, Pam Davila, Pete Maymon, and Howard Herzig. I also thank Ray Boucarut, Laura Burns, Dave Content, John Hagopian, Doug Leviton, Sandra Laase, Cathy Marx, Eric Mentzell, Ken Stewart, Mark Wilson, and Eric Young as well as other members of team DCATT and the staff at JPL for their support during this past year.

Many other friends and family members have contributed in personal ways toward the development of this work. I thank especially: Joe and Stacy Brown, Gene and Leona Shuttlesworth, Anthony Dukes, Shirley Dean, Amber Hevener, Robin Dean, Jim and Katherine Simpson, Dottie Odell, Rob and Terri Nucilli, John, Donna, Nicholas, and Nicole Dorenzi, Jim Severino, Sam Severino, Renaud Stauber, Cat Dailey, and Peter Zurbuch.

# *List of Figures*

# Chapter 1 Introduction

Einstein's theory of general relativity [1] is a classical theory of the gravitational field. A very compelling aspect of the theory is that it is expressible within the mathematical framework of Riemannian geometry. As a result, the Newtonian concepts associated with gravitational forces are replaced by more abstract ideas developed from analysis on manifolds. The power in this identification is evidenced by the range of new and subtle physical phenomena that are predicted by Einstein's theory – e.g., light bending, periastron precession, red shift, and time delay of radar signals - phenomena that might otherwise have gone unnoticed or unexplained had it not been for the subsequent geometric analogies drawn between space, time, and curved manifolds. But the pattern in these developments is not altogether new - physicists have always benefited from more abstract representations of their concepts and theories. The hope is that by translating familiar ideas into a more abstract setting that new relations or physical consequences of a given theory will be made immediately obvious. Therefore, it comes as no surprise that general relativity itself has been analyzed and recast into a variety of different forms in attempts to generalize it. The earliest attempts along these lines were made by Einstein himself [2].

## Part I. Gauge Gravity

Of the many approaches that have been considered in attempts to generalize Einstein's theory, probably the simplest is to start from the alternative scalar invariants that may be used to base an action principle. Since Einstein's theory is already linear in the Ricci scalar - a natural choice would seem to be the quadratic curvature Lagrangians that may be formed by contracting the Riemann tensor onto itself. The combinations that may be considered are given by $R^2$, $R_{\rho\sigma}R^{\rho\sigma}$, and $R^{\rho}{}_{\sigma\mu\nu}R^{\sigma\cdot\mu\nu}_{\cdot\rho\cdot\cdot}$ ($R$ is the Ricci scalar, $R_{\mu\nu}$ is the Ricci tensor, and $R^{\rho}{}_{\sigma\mu\nu}$ is the Riemann tensor). As a result, there have been many investigations that begin with these quadratic curvature Lagrangians and various linear combinations of them as the starting point for alternative theories of gravitation. The original motivation for such work came from the desire to unify gravitation with electromagnetism, and also to find a more general algebraic starting point from which to base Einstein's theory. The earliest such attempts were made by Weyl [3], Pauli [4], Eddington [5], and Lanczos [6]. Eddington considered the Lagrangians: $R_{\rho\sigma} R^{\rho\sigma}$ and $R^{\rho}{}_{\sigma\mu\nu} R^{\sigma\cdot\mu\nu}_{\rho\cdot\cdot}$, while Weyl and Lanczos considered these and also $R^2$. Additional analysis was later given by Gregory [7], Buchdahl [8], Arnowitt [9], Stephenson [10], Kilmister [11], and Thompson [12]. The work of Lanczos was especially important in that he showed a certain linear combination of these Lagrangians gives an identity with respect to variations of the metric, i.e., the Gauss-Bonnet topological invariant. As a result, the number of independent quadratic curvature contractions is reduced from three to two.

A secondary motivation for considering the quadratic curvature Lagrangians came from the realization that the gravitational field should be regarded as an interaction in the same sense as the other fundamental interactions of nature. The hope was that by imitating the mathematical form of Yang-Mills gauge theory that gravitation could be derived in a framework that would suggest an obvious avenue towards unification with the other fundamental interactions. Therefore, by following a parallel with Yang-Mills gauge theory, it was believed that general relativity could be recast as a gauge theory based on the Lorentz group and that gravitation would arise as an interaction manifested through the local Lorentz invariance of the theory. This approach was first considered by Utiyama [13] based on the Lorentz group and then later extended by Kibble [14] to include the full inhomogeneous Lorentz group. But these investigations did not consider a quadratic curvature Lagrangian in their approach and even Yang [15] commented later that Utiyama's investigation was an unnatural interpretation of gauge theory. The most direct algebraic parallel with Yang-Mills gauge theory required the Lagrangian to be quadratic in the Riemann tensor.

Lichnerowicz [16] was the first to consider the field equations, $\nabla_{\rho} R^{\rho}{}_{\sigma\mu\nu} = 0$, by themselves, as the basis for a theory of gravitation. These field equations originate from the Riemann tensor quadratic Lagrangian under the assumption that the connection is in Christoffel form after a Palatini variational procedure is applied with respect to the connection. Kilmister and Newman [17] also considered this equation as the basis for a theory of gravitation and suggested that it is analogous to Maxwell's equations for electromagnetism. But the first real identification with Yang-Mills gauge theory was made by Loos-Treat [18] and Treat [19]. However, Loos and Treat did not consider it seriously as a generalization of Einstein's theory, but rather as a mathematical consequence of the dynamical equivalence between the Yang-Mills and gravitational fields. Subsequently, Yang [15] suggested this theory and it was also considered by Camenzind [20] and Shankar [21] (see also Ref's [22]). But in the subsequent literature, the equation $\nabla_\rho R^\rho{}_{\sigma\mu\nu} = 0$, taken by itself, became known as "Yang's gauge theory of gravity," or "gauge gravity" for short, although Yang was one of the last to suggest it.

But many problems surfaced in this approach. Particularly troublesome were the appearance of multiple nonphysical solutions and the fact that the field equations involve higher than second derivatives of the metric. The appearance of extraneous solutions was considered a nuisance by Pavelle [23], Thompson [24], Ni [25], and Fairchild [26], who cataloged several of the spherically symmetric cases and discussed their nonphysical nature. Hayashi [27] has considered the consequences of the theory under the assumption of a non-metric connection and concluded that the theory must very likely be metric to be consistent, although the field equations themselves originate by treating the metric and connection as independent quantities. Subsequent analysis on the linearized version of the theory was given by Aragone and Restuccia [28] and the PPN formalism (post parametrized Newtonian formalism [29]) was applied to the theory by Camenzind [30]. In summary, the general consensus was not in favor of a sound physical theory.

In subsequent work there were efforts to show that the gauge gravity theory might still be viable physically by finding a constraint to eliminate the nonphysical solutions. The possibility of finding such a constraint was initially suggested by Pavelle [31]. Thompson [32] also suggested this possibility but was more specific by hinting that the identity, $R_{\rho\sigma[\mu\nu}R^{\sigma}_{\lambda]} = 0$, may be helpful in eliminating all solutions that are not Einstein spaces. But no further analysis was given by Thompson although he did comment that the Petrov classification of spaces might be restricted by an unspecified condition related to this identity. In another attempt to eliminate all solutions that are not Einstein spaces, Fairchild [33] considered the field equations that originate by variation with respect to the metric, $R^\rho{}_{\sigma\mu\lambda} R^\sigma{}_{\rho\nu\cdot}{}^{\lambda} - \frac{1}{4} g_{\mu\nu} R^{\rho}{}_{\sigma\cdot\cdot}{}^{\lambda\kappa} R^\sigma{}_{\rho\lambda\kappa} = 0$, in addition to the gauge gravity equations, $\nabla_\rho R^\rho{}_{\sigma\mu\nu} = 0$, in an attempt to eliminate the extraneous solutions. But this approach failed as a result of the fact that both sets of field equations shared a class of non-physical solutions that are conformally flat and also have a vanishing Ricci scalar, i.e., they satisfy Nordström's field equations [34] (as discussed in detail in Chapter 3). In a more recent paper, Guilfoyle and Nolan [35] have reconsidered the identity suggested by Thompson, $R_{\rho\sigma[\mu\nu}R^{\sigma}_{\lambda]} = 0$, which they have relabeled as the "curvature orthogonality condition (COC)." But apparently, these authors have misinterpreted this condition as making a stronger statement on the classification of spacetime solutions of the gauge gravity field equations. Their analysis considers the various Petrov types of spacetimes allowed by the COC according to a Segré classification of the Ricci tensor. But in fact this condition says nothing more than, $0 = 0$, since the equation is an identity. This detail is discussed at greater length in Chapter 3.

Later on, some clarification with regard to the extraneous solutions was given by Havas [36] who showed that any set of field equations involving higher than second order derivatives of the metric must either give multiple spherically symmetric solutions or have a bad Newtonian limit. But even with these results, there was sufficient interest in the theory that Baekler and Yasskin [37] considered an analysis that gives a comparison of the spherically symmetric solutions to the field equations, $\nabla_{\rho} R^{\rho}{}_{\sigma\mu\nu} = 0$, to those of the quadratic curvature generalization based on Poincaré invariance proposed by Heyl, et. al. [38] (see also Ref's [39]). More recent work stemming from the original gauge gravity theory has been considered in attempts to quantize the gravitational field [40] and is still an active area of investigation even at the classical level (see [35] and the other investigations related to the gauge gravity theory in [41]).

In parting ways with the original gauge gravity approach, it was noted by Fairchild [42] that the analogies drawn between the Yang-Mills gauge theory and gravitation as a gauge theory are strictly kinematic. Therefore, in an attempt to bring gravitation into strictly Yang-Mills form,† Fairchild has considered a Lagrangian that is closely related to the Lagrangian considered by Leutwyler [43] in his investigation of the generally covariant Dirac equation. Additional analysis on the equivalence between solutions of this theory and Einstein's theory were given by Debney, Fairchild, and Siklos [44]. They showed that the only vacuum solutions of the theory are also those of Einstein's. A further investigation is given by Yasskin [45] who considers the Newtonian limit and reports that the spatially flat Friedmann-Robertson-Walker solution of Einstein's theory is an exact solution of Fairchild's theory.

---

† many authors have claimed that their Lagrangians or field equations are the most analogous to the Yang-Mills theory. See e.g., Gronwald and Heyl [89].

Although the original goals of the quadratic curvature generalizations have not been fully realized, the formalism that was developed from such investigations has proven general enough to suggest alternative approaches toward quantization of the gravitational field (although very few of these approaches are tied directly to the original gauge gravity theory). Indeed, there are many investigations underway toward quantization including the superstring-theoretic approach [46], the connection dynamics proposal [47], non-commutative geometries [48], generalized gauge-theoretic formulations [49], quantization of topologies [50], topological geons [51], gravity as an induced phenomenon [52], and so on (see also Burton and Mann [53] for a discussion of the recent literature). In each of these approaches a fundamental role has been given to the variational procedure - particularly to the Palatini approach in which the metric and connection are treated as independent dynamical variables. The advantages in this formalism were appreciated early on from the parallels made with gauge theory and quadratic curvature attempts at unification. As a result of the early work on quadratic curvature generalizations, the mathematical groundwork for the Palatini procedure has been developed and explored in considerable detail, but there are still details of the procedure that have not been fully investigated, at least insofar as the gauge gravity field equations are concerned.

Therefore, one goal of this thesis has been to reconsider earlier work on the gauge gravity theory to show that an additional algebraic constraint arises by imposing equivalence between the standard and Palatini variational procedures. A basic result is that this constraint gives additional insight into the classical gauge theoretic structure of gravitation and eliminates many (but not all) of the nonphysical solutions to the gauge gravity theory. The basis for the analysis is given by noting that the Einstein-Hilbert action is special in that its variation gives identical results using either the standard or Palatini variational procedures. However, this "symmetry" no longer applies when the action is taken in quadratic form. As a result of imposing this condition onto the gauge gravity field equations, an auxiliary algebraic condition is obtained that restricts the class of spacetime solutions to the field equations. This condition is shown to be similar to a condition that was proven earlier using the Segré classification of the Ricci tensor by Debney, Fairchild, and Siklos [44] which has been used by these authors to eliminate the "non-Einsteinian" solutions of Fairchild's theory [42] (as discussed above).

Background for the analysis is given in Chapters 2 and 3. In Chapter 2, the field equations that originate from the Einstein-Hilbert action are derived using the Palatini variational procedure from a slightly more general viewpoint than what is usually discussed in the literature. The analysis considers the possibility for both non-vanishing covariant derivative of the metric and also by assuming a non-symmetric connection, i.e., Schouten's $L_4$. The goal in this approach is to not only provide an introduction to the Palatini variational procedure used in Chapter 3, but also to make as many properties as possible follow from the variational procedure itself, while minimizing the number of additional constraints that are imposed on the variations. As a result, the relationship that exists between the metric and connection is determined by the secondary set of field equations in the Palatini variation with respect to the connection. For the Einstein-Hilbert action this relationship is expressed by a Lemma that is discussed in Chapter 2. A similar result has also been obtained by Papapetrou and Stachel [54] who have considered the torsion as an independent variational parameter. In the following two subsections of Chapter 2, the Schwarzschild and Reissner-Nordström solutions are then derived as the basis for the dynamical analysis presented in Chapters 5 and 6.

In the early sections of Chapter 3, the variational formulation of Yang-Mills gauge theory is reviewed to motivate the gauge theory kinematics based on the Lorentz group. In the following subsection, the gauge gravity field equations are derived and then the spherically symmetric solution structure of the field equations is presented in detail in the sub-subsection entitled "Solutions." An additional unreported solution to the free-field "energy momentum" equations is presented in addition to analysis on the conformally flat solutions that are shared by Nordström's theory [34]. Finally, the analysis for the additional algebraic constraints is presented in the final sub-subsection of Chapter 3.

In Chapter 4, an evaluation of the spherically symmetric solution structure for another attempted generalization of Einstein's theory is presented. The theory was proposed long ago by Weyl [55] as a generalization based on conformal invariance but has been more recently considered by Mannheim and Kazanas [56, 57] to explain the rotation curves of galaxies without the need for postulating the existence of dark matter. As a consequence of basing a theory on conformal invariance the action considered for the theory must also be quadratic in the curvature tensors, so there are features of the analysis that are shared by the quadratic curvature Lagrangians discussed in Chapter 3. Therefore, the purpose of this Chapter is to emphasize this similarity and to show that a class of conformally flat solutions to the gauge gravity field equations, namely those also satisfying Nordström's theory [34] are shared by conformal gravity.

## Part II. Dynamics

In the second part of the thesis (Chapters 5 and 6), the orbital dynamics of point particles in the Schwarzschild and Reissner-Nordström spacetimes are investigated using methods from dynamical systems theory. The motivation for considering this analysis is to gain additional insight into the physical and analytic structure of the solutions. It is a fact that the underlying mathematical structure for an arbitrary set of equations is often more clearly elucidated by applying techniques developed from the study of dynamical systems (for instance, Marsden [58]) – particularly in those cases where a direct physical interpretation is not always of central importance (e.g., Korteweg-deVries [59]; Hénon-Heiles [60], etc.). Therefore, even in dynamical systems that are considered to be well understood, additional qualitative information or obscure details of a given solution may sometimes be more clearly conveyed or classified by applying newer and more modern techniques of analysis. This is the goal in the second part of the thesis.

A great deal is known about the general relativistic orbital dynamics. Indeed, the earliest predictions of Einstein's theory concerned the bending of light and the perihelion precession of Mercury. Motivated by these and other solar system observations - the Schwarzschild, Reissner-Nordström, Kerr, and Kerr-Newman orbital dynamics and their analytic structure have been analyzed extensively in the literature (e.g., Darwin [61]; Kruskal [62]; Boyer and Lindquist [63]; Graves and Brill [64]; Carter [65]; Misner,

Thorne, and Wheeler [66]; Sharp [67], and Chandrasekhar [68]). The most common method of dynamical analysis has been based on perturbative techniques or numerical integration [69]. However, exact solutions have been discussed by many authors for circular orbits, light rays, and radial motion - see e.g., Chandrasekhar [68]. But surprisingly, given that the general relativistic equations of motion are integrable, an investigation of these dynamics that utilizes the phase-plane and bifurcation techniques in combination with a linear stability analysis has not appeared widely in the literature. However, earlier work by Szydlowski, et. al. [70] and Collins [71] considers an application of the phase-plane method to study the stability properties of cosmological solutions. A similar analysis has also been discussed by Khalatnikov [72]. A more recent phase-plane analysis of the Friedmann-Robertson-Walker cosmology in Brans-Dicke gravity is considered in Ref. [73]. Other recent work using methods from dynamical systems theory has been concerned with the detection and analysis of chaotic orbits [74]. In contrast to these studies the emphasis of the analysis considered here will be an application of the bifurcation and linear stability techniques to classify the stability properties of the orbits - particularly to the Reissner-Nordström solution. Specifically, the goal in this analysis is to solve the bifurcation problem for the Reissner-Nordström system. That is – to present a summary of the phase-plane topological structure based on a study of coalescing fixed points and identify the parameter values at which these bifurcations occur. The Schwarzschild dynamics have been analyzed with a pedagogical emphasis using the phase-plane and bifurcation techniques in [75] (thesis Chapter 5); the Reissner-Nordström periastron advance is considered in [76] (thesis Chapter 6).

Comparatively speaking, the Reissner-Nordström orbital dynamics have received relatively little attention in the literature compared to the corresponding Schwarzschild, Kerr, or Kerr-Newman dynamics. Probably the reason for this omission is due to the assumption that a black hole possessing a sizable charge would be rapidly neutralized by the inflow of oppositely charged particles. As a result, the solution has not often been considered seriously as a base model for astrophysical processes. But the work of Eddington [77] has shown that stars are positively charged and that electric fields play an essential role in stellar structure. Furthermore, the Reissner-Nordström solution could have added physical relevance if primordial black holes are proven to exist,[†] since then it is conceivable that a sizable charge could accumulate without neutralization by surrounding matter. But experimental evidence for such objects is questionable (see e.g. Ohanian and Ruffini [78], p. 487-489).

As an application of Eddington's model, Harrison [79] has considered order of magnitude calculations that place an upper bound on the size of such "gravity induced" charge and discusses the physical mechanism for its production within the stellar interior. The mechanism is based on the relative mass difference between the proton and electron. The reasoning is as follows: within a star, proton and electron gases contribute equally to the total pressure gradient. But gravity acts mainly on the relatively massive protons, and therefore an electric field mediates between the electron and proton gases. In effect, electrons have velocities greatly exceeding escape velocity, and therefore some electrons escape leaving the star positively charged with an electric field that retains the remaining electrons. However, from the cosmic censorship conjecture (discussed in more detail in Chapter 2), an upper bound is placed on the charge contribution to the total mass of the Reissner-Nordström black hole. Therefore, at certain parameter values the dynamics may be implausible physically, but are nevertheless dynamical consequences of the Reissner-Nordström solution, albeit only mathematical. The viewpoint adopted here is non-committal on the physical existence of such solutions – the full range of parameter values are simply explored as dynamical possibilities (Chapter 6), not necessarily physical ones.

The contents of Chapters 5 and 6 are organized as follows: in Chapter 5, the Schwarzschild orbital dynamics are first analyzed using the phase-plane and bifurcation techniques. The purpose of this Chapter is to setup the analysis that will be applied to the Reissner-Nordström dynamics in Chapter 6. Although the Schwarzschild orbital dynamics are well understood and have been analyzed using a variety methods in the literature, there are still qualitatively new results that have not been reported for the Schwarzschild dynamics. Essentially, these results correspond to the limitations that are placed upon the types of orbits that may exist before an unstable orbit is reached and the kinematic classification of the separatrices as distinct unstable orbits. In essence, the separatrix gives a graphic representation of the critical relationship that exists between energy and angular momentum at the unstable orbital radius. The separatrices themselves are therefore classified kinematically as unstable hyperbolic, parabolic, and elliptical orbits. Furthermore, the Schwarzschild orbital dynamics may be interpreted and analyzed as a conservative 2-d bifurcation phenomenon which summarizes the range of orbits that may occur as the energy and angular momentum of the system are varied. As a result, the bifurcation diagram divides the dynamics into physically distinct regions in contrast to the phase-plane that is divided into dynamically distinct regions.

---

[†] I thank H. A. Weldon for this comment.

In a later Section of Chapter 5, a phase-plane analysis of dynamical invariance between the coordinate and proper time reference frames is given. Although the dynamical structure (i.e. the effective potential) is demonstrated to be invariant between the two reference frames, the phase diagrams in each case are not identical. This is due to the existence of an additional phase-plane fixed point that appears in the coordinate reference frame at the event horizon. This fixed point is obviously coordinate dependent, but must exist to explain the apparent "slowing down" of objects (and redshift of signals) approaching the horizon boundary as seen by an observer in the coordinate reference frame. For comparison, the corresponding Newtonian phase-plane results are considered in an Appendix. Not only does this analysis complement the discussion of the Schwarzschild dynamics considered in Chapter 5, but it is of pedagogical value to show that an analysis of Newtonian orbits using time as an independent variable is just as instructive and no more complicated in principle than using the equatorial angle (however, the opposite is true when using the standard methods of analysis, e.g. [80]]). Finally, the phase-plane analysis is applied to the kinematics of light rays in the Schwarzschild black hole spacetime. The standard results are discussed and then compared with the phase-plane results. The added significance of the photon orbits in the phase-plane context is that the equilibrium points of the differential equations exhibit a transcritical bifurcation (i.e. an exchange in stability) at these parameter values.

As a result of the spherical symmetry of the Reissner-Nordström solution there are many similarities shared with the corresponding Schwarzschild analysis of Chapter 5. However, in the Reissner-Nordström case there are three fixed points compared to only two in the Schwarzschild example. As a result, the dynamics are considerably more complex since the parameter space is two dimensional compared to only one dimensional for the Schwarzschild case. These dynamics are presented in Chapter 6 and are organized as follows: the phase-plane equations and fixed points for the system are first derived. The Reissner-Nordström parameter space is then presented which is an important diagrammatic tool in the subsequent dynamical classification. The parameter space organizes and locates the bifurcations that are classified in the following sub-section entitled *Bifurcation Analysis*. The discussion on the stability properties of the orbits is augmented with exact phase-diagrams that illustrate the underlying "global" phase-plane structure of the system at selected parameter values. In the following sub-section the separatrix structure of the Reissner-Nordström phase diagrams is discussed as a straightforward generalization of the Schwarzschild case.

As a consequence of the additional dynamical complexity associated with the Reissner-Nordström system, there are multiple periodic solutions that yield orbits with finite precession values. An analysis and comparison of several such orbits is given in addition to the standard precession for timelike orbits. A linear stability phase-plane calculation of periastron advance is presented in the sub-section entitled *Periastron Precession* and then extended to (a) precession about a bifurcation point of the dynamics and (b) precession about a secondary center node fixed point that exists as a consequence of the black hole charge. The bifurcation point considered in (a) has not been reported in the literature and is given by parameter values that would presumably correspond to timelike orbits about a naked singularity, although it is not expected that any physical interpretation can be given for these cases. The "acausal" geodesics discussed by Brigman [81] are identified with (b), which is a center node fixed point positioned between the outer horizon, $r_+$, and the interior horizon, $r_-$. Finally, the photon orbits give a special case of the dynamics as a transcritical bifurcation point (as in the Schwarzschild case) and are presented in the final sub-section entitled *Light Rays*.

# Chapter 2 Field Equations and Variational Formalism

## Einstein-Hilbert

The goal in the first section of this Chapter is to not only provide background for the Palatini procedure used in Chapter 3, but to make as many properties as possible follow from the variational procedure, while minimizing the number of additional constraints that are imposed on the variations themselves. Therefore, we consider the field equations that result from a variational procedure applied to the Einstein-Hilbert action in which the connection is neither metric or symmetric, and no functional relation is assumed to exist between the metric and connection.

The fact that the spacetime manifold is defined with a connection that has no assumed symmetries gives the $L_4$ geometric manifold discussed by Schouten. But the addition of a symmetric metric tensor, $g_{\mu\nu}$, is necessary to define inner products between vector and tensor quantities as well as defining the length element (the addition of the metric into the $L_4$ is discussed by Heyl [82], i.e., the $(L_4, g)$).† Similar field equations have been considered by Papapetrou and Stachel [54] and Burton and Mann [54] who have assumed a symmetric connection and include all terms in the action that are at most quadratic in derivatives and/or connection variables. But here a simpler case is considered but with the added generality of a non-symmetric connection to explore the relationship that exists between the metric and connection, particularly in a spacetime originating from the Einstein-Hilbert action. In the Palatini style variation this relationship is determined by the secondary field equations that are obtained by variation with respect to the connection. A simpler calculation that is related to this approach was considered previously by Stephenson [10] for the case that the connection is symmetric and is discussed in greater detail in Chapter 3.

There are essentially two variational methods that may be used to calculate the field equations of general relativity. In the first (labeled as the "standard" variational procedure) the metric is the sole variational parameter of the theory and is implied by the condition that the connection be metric† (Schouten [83], p. 132):

$$\nabla_\rho g_{\mu\nu} = 0 \,. \tag{2.1}$$

---

† I thank Richard Treat for this clarifying comment on the $(L_4, g)$.

The solution to (2.1) is given by the Christoffel form:

$$\overset{o}{\Gamma}{}^{\sigma}_{\mu\nu} \equiv \tfrac{1}{2} g^{\sigma\lambda}(\partial_\mu g_{\lambda\nu} + \partial_\nu g_{\lambda\mu} - \partial_\lambda g_{\mu\nu}) \,, \tag{2.2}$$

when the connection is symmetric (i.e., $\Gamma^{\rho}_{[\mu\nu]} = 0$), and is derived as follows - expand the covariant derivative in (2.1) and then cyclically permute indices to obtain the linear combination:

$$\begin{aligned}\nabla_\rho g_{\mu\nu} + \nabla_\mu g_{\nu\rho} - \nabla_\nu g_{\rho\mu} = 0 = \partial_\rho g_{\mu\nu} - \Gamma^{\lambda}_{\mu\rho} g_{\lambda\nu} - \Gamma^{\lambda}_{\nu\rho} g_{\mu\lambda} \\ + \partial_\mu g_{\nu\rho} - \Gamma^{\lambda}_{\nu\mu} g_{\lambda\rho} - \Gamma^{\lambda}_{\rho\mu} g_{\nu\lambda} - \partial_\nu g_{\rho\mu} + \Gamma^{\lambda}_{\rho\nu} g_{\lambda\mu} + \Gamma^{\lambda}_{\mu\nu} g_{\rho\lambda} \,.\end{aligned} \tag{2.3}$$

Rewriting this result gives

$$2\Gamma^{\lambda}_{(\mu\rho)} g_{\lambda\nu} = \partial_\rho g_{\mu\nu} + \partial_\mu g_{\nu\rho} - \partial_\nu g_{\rho\mu} + T^{\lambda}_{\mu\nu} g_{\rho\lambda} + T^{\lambda}_{\rho\nu} g_{\lambda\mu} \,, \tag{2.4}$$

where the torsion tensor is defined: ‡

$$T^{\lambda}_{\mu\nu} = 2\Gamma^{\lambda}_{[\mu\nu]} = \Gamma^{\lambda}_{\mu\nu} - \Gamma^{\lambda}_{\nu\mu} \,. \tag{2.5}$$

If the connection is symmetric then (2.4) simplifies to the Christoffel form (2.2).

As a result of (2.2), the corresponding Einstein free-field equations are obtained from $\delta S_{EH} / \delta g^{\mu\nu} = 0$, using the standard variational procedure (Appendix A):

$$R_{\mu\nu} - \tfrac{1}{2} g_{\mu\nu} R = 0 \;\Rightarrow\; R_{\mu\nu} = 0 \,;\; \mu,\nu,\text{etc.} = 0,1,2,3 \,, \tag{2.6}$$

where $S_{EH}$ is the Einstein-Hilbert action (the constant factor is required for Newtonian correspondence which is discussed in the next section; $G$ is Newton's gravitational constant, $c$ is the speed of light):

$$S_{EH}[g] = \frac{1}{16\pi G c^{-4}} \int d^4q \; g^{\frac{1}{2}} g^{\rho\sigma} R_{\rho\sigma} = \frac{1}{16\pi G c^{-4}} \int d^4q \; g^{\frac{1}{2}} R \,. \tag{2.7}$$

---

† in the literature there are widely varying labels attached to the condition that $\nabla_\rho g_{\mu\nu} = 0$; e.g., *metricity*, *Riemann condition*, *metric compatibility*, *Einstein-Hilbert constraint*, etc., but we shall adopt the conventions and terminology of Schouten – that the connection is metric with respect to $g_{\mu\nu}$.

In (2.7), $q$ denote generalized coordinates, $g_{\mu\nu}$ is the metric, $g \equiv -\det g_{\mu\nu}$, and $R_{\rho\sigma}$ is the Ricci tensor:

$$R_{\sigma\nu} \equiv R^{\rho}{}_{\sigma\rho\nu} = \partial_{\rho}\Gamma^{\rho}_{\sigma\nu} - \partial_{\nu}\Gamma^{\rho}_{\sigma\rho} + \Gamma^{\rho}_{\lambda\rho}\Gamma^{\lambda}_{\sigma\nu} - \Gamma^{\rho}_{\lambda\nu}\Gamma^{\lambda}_{\sigma\rho}, \qquad (2.8)$$

which is a derived quantity from the transvection† (see Schouten [83], p. 8 and 14) of the Riemann curvature tensor:

$$R^{\rho}{}_{\sigma\mu\nu} = \partial_{\mu}\Gamma^{\rho}_{\sigma\nu} - \partial_{\nu}\Gamma^{\rho}_{\sigma\mu} + \Gamma^{\rho}_{\lambda\mu}\Gamma^{\lambda}_{\sigma\nu} - \Gamma^{\rho}_{\lambda\nu}\Gamma^{\lambda}_{\sigma\mu}, \qquad (2.9)$$

[*note*: the signature adopted here is $diag\ \eta_{\mu\nu} = (+1,-1,-1,-1)$].

The second variational procedure is given by the Palatini approach which assumes no *a-priori* relationship between the metric and connection. As a result, a more subtle role is played in the Palatini variation – for in addition to the field equations (2.6) obtained by variation with respect to $\delta g_{\mu\nu}$, the variation with respect to $\delta\Gamma^{\rho}_{\mu\nu}$ yields the metric condition (2.1) as a field equation extremizing the Einstein-Hilbert action. It is interesting to note that Einstein [84] was apparently the first to have treated the connection and metric as independent variational parameters and that Palatini never actually used this procedure (see Heyl, et. al [85] and Ref. [84]).

The fact that this seemingly independent approach confirms that the connection is metric has been viewed by many as a kind of "proof" of the validity of the Einstein-Hilbert action as a starting point for the correct theory of gravitation. But the Palatini procedure also suggests a natural starting point for generalizations of Einstein's theory by relaxing the constraint (2.1). In fact, it has been noted long ago by Schrödinger [86] and more recently by Hehl [87] that in a generalized theory of gravitation one expects (2.2) to be modified in some manner that is typically not obvious. Indeed, the Palatini approach provides the starting point for most generalized (quantum) theories of gravity mentioned in the Introduction and has also been considered in other classical generalizations based on the spin connection and local automorphism invariance (see Crawford [88]) and the "gauge gravity" theory that is considered in Chapter 3.

---

‡ Note: the conventions and notation used here will generally follow Schouten but differ in at least two respects, namely, the torsion tensor (2.5) differs by a factor of two, and the index ordering placed on the kernel symbol, $R$, of Riemann tensor (in (2.9)) is modified, i.e., Schouten uses $R_{\mu\nu\sigma}{}^{\rho}$ to denote (2.9).

† a technical point: *transvection* is the operation of multiplying by a Kronecker delta, $\delta^{\rho}_{\sigma}$, and has the operator form of an upper and lower index – in contradistinction to *contraction* - which involves the raising or lowering of an index using a metric, $g_{\rho\sigma}$.

Probably the most obvious attempt to generalize Einstein's theory based on a modification of (2.2) is to allow the connection to have a non-symmetric or torsion contribution. Although the theories stemming from this approach are not considered in this thesis, it is still worthwhile to make a few comments on these cases given that nonzero torsion will be utilized for calculational purposes in this Chapter and in Chapter 3. For example, torsion plays an essential role in Poincaré gravity which is the generalization based on the Poincaré group suggested by Hehl, et. al (e.g., [49, 82, 89]; otherwise known as the Einstein-Cartan $U_4$ theory). The physical motivation is given by noting that in general relativity the source of the gravitational field is mass, causing the curvature of spacetime, but no obvious geometric role has been given to spin. Furthermore, it has been demonstrated (see Heyl [82] and also Crawford [90]) that the spin angular momentum of a system determines the torsion of the space when spinor matter is present. Therefore, it is appealing to consider spaces with both curvature and torsion as a framework for constructing gravitational theories at the microscopic level. Macroscopically, these theories would presumably give back the usual version of general relativity since mass (being of monopole character) "adds up," while spin (being of dipole character) averages out in the large. In essence, this picture forms the basis of the view presented by Hehl et. al. [82].

Continuing with the variational calculation - the results may be summarized as follows: variation with respect to $\delta g_{\mu\nu}$ produces field equations that require the symmetric contribution to the Ricci tensor to vanish. But no constraint is placed on the antisymmetric part which may be expressed using Bianchi's second identity (Schouten [83], p. 150; see also Crawford [90]):

$$\begin{aligned} R^{\rho}{}_{\sigma\mu\nu} + R^{\rho}{}_{\mu\nu\sigma} + R^{\rho}{}_{\nu\sigma\mu} \equiv \nabla_{\sigma} T^{\rho}_{\mu\nu} + \nabla_{\mu} T^{\rho}_{\nu\sigma} + \nabla_{\nu} T^{\rho}_{\sigma\mu} \\ - T^{\rho}_{\kappa\sigma} T^{\kappa}_{\mu\nu} - T^{\rho}_{\kappa\mu} T^{\kappa}_{\nu\sigma} - T^{\rho}_{\kappa\varpi} T^{\kappa}{}_{\cdot\sigma\mu} \,. \end{aligned} \tag{2.10}$$

Thus, the transvection of $\sigma \to \rho$ gives (let $T_{\alpha} \equiv T^{\rho}_{\rho\alpha}$):

$$R_{\mu\nu} - R_{\nu\mu} = \nabla_{\mu} T_{\nu} - \nabla_{\nu} T_{\mu} - \nabla_{\rho} T^{\rho}_{\mu\nu} + T_{\rho} T^{\rho}_{\mu\nu} \,. \tag{2.11}$$

But the second set of field equations obtained from the variation, $\delta\Gamma^{\rho}_{\mu\nu}$, of the Einstein-Hilbert action imposes a relationship between the metric and connection – namely, that (2.1) is valid if and only if the connection is symmetric. The calculation is presented in detail below.

First consider the variation of (2.7) with respect to $\delta g_{\mu\nu}$:

$$\left.\frac{\delta S_{EH}[g,\Gamma]}{\delta g^{\mu\nu}(q')}\right|_{\Gamma} = 0 = \int d^4q \frac{\delta g^{1/2}}{\delta g^{\mu\nu}(q')} g^{\rho\sigma} R_{\rho\sigma} + \int d^4q\, g^{1/2} \frac{\delta g^{\rho\sigma}}{\delta g^{\mu\nu}(q')} R_{\rho\sigma}. \tag{2.12}$$

Using the variational results (see also Crawford [90]):

$$\frac{\delta g^{1/2}(q)}{\delta g^{\mu\nu}(q')} = -\tfrac{1}{2} g_{\mu\nu}\delta^4(q-q'); \quad \frac{\delta g^{\rho\sigma}(q)}{\delta g^{\mu\nu}(q')} = \tfrac{1}{2}\left(\delta^{\rho}_{\mu}\delta^{\sigma}_{\nu} + \delta^{\rho}_{\nu}\delta^{\sigma}_{\mu}\right)\delta^4(q-q')/g^{1/2}, \tag{2.13}$$

(2.12) becomes:

$$\begin{aligned}\left.\frac{\delta S_{EH}[g,\Gamma]}{\delta g^{\mu\nu}(q')}\right|_{\Gamma} &= \int d^4q\; g^{1/2}\; \tfrac{1}{2}\left(\delta^{\rho}_{\mu}\delta^{\sigma}_{\nu} + \delta^{\rho}_{\nu}\delta^{\sigma}_{\mu}\right)\delta^4(q-q')/g^{1/2} R_{\rho\sigma} \\ &\quad - \int d^4q\; \tfrac{1}{2} g_{\mu\nu}\delta^4(q-q')\, g^{\rho\sigma} R_{\rho\sigma}.\end{aligned} \tag{2.14}$$

Integrating (2.14) and then simplifying algebraically gives the field equations:

$$R_{(\mu\nu)} - \tfrac{1}{2} g_{\mu\nu} R = 0, \tag{2.15}$$

noting that $R_{\mu\nu}$ is not symmetric when torsion is present as noted above in (2.11). Therefore, the field equations resulting by variation with respect to $\delta g_{\mu\nu}$ require that the symmetric part of the Ricci tensor vanish:

$$\left.\frac{\delta S_{EH}[g,\Gamma]}{\delta g^{\mu\nu}}\right|_{\Gamma} : \quad R_{(\mu\nu)} = 0, \tag{2.16}$$

which is expected by variation with respect to a symmetric tensor, $\delta g^{\mu\nu}$. From (2.11) it is obvious that the free-field Einstein equations (2.6) are obtained as a special case of (2.16) when the torsion vanishes.

The secondary field equations are obtained by variation with respect to $\delta\Gamma^{\rho}_{\mu\nu}$:

$$\left.\frac{\delta S_{EH}[g,\Gamma]}{\delta\Gamma^{\alpha}_{\mu\nu}(q')}\right|_{g} = 0 = \int d^4q\; g^{1/2} g^{\rho\sigma} \frac{\delta R_{\rho\sigma}}{\delta\Gamma^{\alpha}_{\mu\nu}(q')}, \tag{2.17}$$

and then expanding the variation of $R_{\rho\sigma}$ using the definition of the Ricci tensor from (2.8) gives:

$$\left.\frac{\delta R_{\rho\sigma}}{\delta \Gamma^{\alpha}_{\mu\nu}}\right|_g g^{1/2} g^{\rho\sigma} = \partial_\lambda \left(\frac{\delta \Gamma^{\lambda}_{\rho\sigma}}{\delta \Gamma^{\alpha}_{\mu\nu}}\right) g^{1/2} g^{\rho\sigma} - \partial_\sigma \left(\frac{\delta \Gamma^{\lambda}_{\rho\lambda}}{\delta \Gamma^{\alpha}_{\mu\nu}}\right) g^{1/2} g^{\rho\sigma} + \frac{\delta}{\delta \Gamma^{\alpha}_{\mu\nu}} \left(\Gamma^{\lambda}_{\omega\lambda} \Gamma^{\omega}_{\rho\sigma} - \Gamma^{\lambda}_{\omega\sigma} \Gamma^{\omega}_{\rho\lambda}\right) g^{1/2} g^{\rho\sigma}. \tag{2.18}$$

The variational terms in (2.18) may be expressed:

$$\begin{aligned}
\partial_\lambda \left(\frac{\delta \Gamma^{\lambda}_{\rho\sigma}}{\delta \Gamma^{\alpha}_{\mu\nu}}\right) g^{1/2} g^{\rho\sigma} &= \partial_\lambda \left(\frac{\delta \Gamma^{\lambda}_{\rho\sigma}}{\delta \Gamma^{\alpha}_{\mu\nu}} g^{1/2} g^{\rho\sigma}\right) - \frac{\delta \Gamma^{\lambda}_{\rho\sigma}}{\delta \Gamma^{\alpha}_{\mu\nu}} \partial_\lambda \left(g^{1/2} g^{\rho\sigma}\right) \\
\partial_\sigma \left(\frac{\delta \Gamma^{\lambda}_{\rho\lambda}}{\delta \Gamma^{\alpha}_{\mu\nu}}\right) g^{1/2} g^{\rho\sigma} &= \partial_\sigma \left(\frac{\delta \Gamma^{\lambda}_{\rho\lambda}}{\delta \Gamma^{\alpha}_{\mu\nu}} g^{1/2} g^{\rho\sigma}\right) - \frac{\delta \Gamma^{\lambda}_{\rho\lambda}}{\delta \Gamma^{\alpha}_{\mu\nu}} \partial_\sigma \left(g^{1/2} g^{\rho\sigma}\right),
\end{aligned} \tag{2.19}$$

noting that the first terms on the RHS in each case are boundary terms and will be assumed to vanish. Substituting (2.19) into (2.18) and then expanding out the other variations gives the result:

$$\left.\frac{\delta R_{\rho\sigma}}{\delta \Gamma^{\alpha}_{\mu\nu}}\right|_g g^{1/2} g^{\rho\sigma} = -\frac{\delta \Gamma^{\lambda}_{\rho\sigma}}{\delta \Gamma^{\alpha}_{\mu\nu}} \partial_\lambda \left(g^{1/2} g^{\rho\sigma}\right) + \frac{\delta \Gamma^{\lambda}_{\rho\lambda}}{\delta \Gamma^{\alpha}_{\mu\nu}} \partial_\sigma \left(g^{1/2} g^{\rho\sigma}\right) + \left(\frac{\delta \Gamma^{\lambda}_{\omega\lambda}}{\delta \Gamma^{\alpha}_{\mu\nu}} \Gamma^{\omega}_{\rho\sigma} + \Gamma^{\lambda}_{\omega\lambda} \frac{\delta \Gamma^{\omega}_{\rho\sigma}}{\delta \Gamma^{\alpha}_{\mu\nu}} - \frac{\delta \Gamma^{\lambda}_{\omega\sigma}}{\delta \Gamma^{\alpha}_{\mu\nu}} \Gamma^{\omega}_{\rho\lambda} - \Gamma^{\lambda}_{\omega\sigma} \frac{\delta \Gamma^{\omega}_{\rho\lambda}}{\delta \Gamma^{\alpha}_{\mu\nu}}\right) g^{1/2} g^{\rho\sigma}. \tag{2.20}$$

Note that no symmetries have been assumed on the $\Gamma^{\rho}_{\mu\nu}$, therefore

$$\frac{\delta \Gamma^{\lambda}_{\rho\sigma}(q)}{\delta \Gamma^{\alpha}_{\mu\nu}(q')} = \delta^{\lambda}_{\alpha} \delta^{\mu}_{\rho} \delta^{\nu}_{\sigma} \frac{\delta^4(q-q')}{g^{1/2}}. \tag{2.21}$$

Substituting this result back into (2.20) then gives:

$$\frac{\delta R_{\rho\sigma}(q)}{\delta \Gamma^{\alpha}_{\mu\nu}(q')} g^{1/2} g^{\rho\sigma} = \Bigg[ \underbrace{-\partial_\alpha (g^{1/2} g^{\mu\nu}) - g^{1/2}\, \Gamma^{\mu}_{\rho\alpha}\, g^{\rho\nu} - g^{1/2}\, \Gamma^{\nu}_{\alpha\rho}\, g^{\mu\rho} + g^{1/2}\, \Gamma^{\lambda}_{\alpha\lambda}\, g^{\mu\nu}}_{t_1} + \underbrace{\delta^{\nu}_{\alpha} [\partial_\sigma (g^{1/2} g^{\mu\sigma}) + g^{1/2}\, \Gamma^{\mu}_{\rho\sigma}\, g^{\rho\sigma}]}_{t_2} \Bigg] \frac{\delta^4(q-q')}{g^{1/2}}. \tag{2.22}$$

The structure of (2.22) is further clarified by rewriting $t_1$ and $t_2$ in terms of covariant derivatives. Beginning with the first term of $t_1$, note that

$$\nabla_\alpha (g^{1/2} g^{\mu\nu}) = \nabla_\alpha (g^{1/2})\, g^{\mu\nu} + g^{1/2}\, \nabla_\alpha\, g^{\mu\nu} \,, \tag{2.23}$$

but to expand (2.23) further, an intermediate calculation of the covariant derivative of the tensor density $g^{1/2}$ is required (tensor densities are discussed in Schouten [83], p. 12). To obtain these results, consider the covariant derivative of a tensor density, $\Pi$, defined by (in 4 dimensions):

$$\Pi = \Pi^{\mu\nu\rho\sigma} \varepsilon_{\mu\nu\rho\sigma} \,, \tag{2.24}$$

where $\varepsilon_{\mu\nu\rho\sigma}$ is the totally antisymmetric tensor density of weight +1 and $\Pi^{\mu\nu\rho\sigma}$ is also totally antisymmetric. The covariant derivative is thus calculated:

$$\nabla_\lambda \Pi = \partial_\lambda \Pi \;+\; (\Gamma^{\mu}_{\omega\lambda} \Pi^{\omega\nu\rho\sigma} + \Gamma^{\nu}_{\omega\lambda} \Pi^{\mu\omega\rho\sigma} + \Gamma^{\rho}_{\omega\lambda} \Pi^{\mu\nu\omega\sigma} + \Gamma^{\sigma}_{\omega\lambda} \Pi^{\mu\nu\rho\omega}) \varepsilon_{\mu\nu\rho\sigma} \,. \tag{2.25}$$

But notice from (2.24) that:

$$\Pi^{\mu\nu\rho\sigma} = -\tfrac{1}{24} \Pi\, \varepsilon^{\mu\nu\rho\sigma} \,, \tag{2.26}$$

which gives

$$\nabla_\lambda \Pi = \partial_\lambda \Pi + \tfrac{1}{4} (\Gamma^{\mu}_{\omega\lambda} \delta^{\omega}_{\mu} + \Gamma^{\nu}_{\omega\lambda} \delta^{\omega}_{\nu} + \Gamma^{\rho}_{\omega\lambda} \delta^{\omega}_{\rho} + \Gamma^{\sigma}_{\omega\lambda} \delta^{\omega}_{\sigma}) \,. \tag{2.27}$$

Finally, (2.27) simplifies to

$$\nabla_\lambda \Pi = \partial_\lambda \Pi + \Gamma^{\mu}_{\mu\lambda} \Pi \,, \tag{2.28}$$

or similarly, for a tensor density of weight –1:

$$\nabla_\lambda \Pi = \partial_\lambda \Pi - \Gamma^{\mu}_{\mu\lambda} \Pi \,. \tag{2.29}$$

As an application of these results the covariant derivative of $g^{1/2}$ is thus obtained:

$$\nabla_\alpha\, g^{1/2} = \partial_\alpha\, g^{1/2} - g^{1/2}\, \Gamma^{\lambda}_{\lambda\alpha} \,. \tag{2.30}$$

Expanding the covariant derivatives on (2.23) and solving for the term, $\partial_\alpha (g^{1/2} g^{\mu\nu})$ gives finally

$$\partial_\alpha (g^{1/2} g^{\mu\nu}) = \nabla_\alpha (g^{1/2} g^{\mu\nu}) + g^{1/2} \Gamma^{\rho}_{\rho\alpha}\, g^{\mu\nu} - g^{1/2} \Gamma^{\mu}_{\rho\alpha}\, g^{\rho\nu} - g^{1/2} \Gamma^{\nu}_{\rho\alpha}\, g^{\mu\rho} \,. \tag{2.31}$$

As a result, $t_1$ may be expressed:

$$\begin{aligned} t_1 = &-\nabla_\alpha (g^{1/2} g^{\mu\nu}) - g^{1/2} \Gamma^{\rho}_{\rho\alpha}\, g^{\mu\nu} + g^{1/2} \Gamma^{\mu}_{\rho\alpha}\, g^{\rho\nu} - g^{1/2} \Gamma^{\nu}_{\rho\alpha} g^{\mu\rho} \\ &- g^{1/2} \Gamma^{\mu}_{\rho\alpha}\, g^{\rho\nu} - g^{1/2} \Gamma^{\nu}_{\alpha\rho}\, g^{\mu\rho} + g^{1/2} \Gamma^{\lambda}_{\alpha\lambda}\, g^{\mu\nu} . \end{aligned} \tag{2.32}$$

The third and fifth terms cancel leaving two torsion terms and a covariant derivative:

$$t_1 = -g^{1/2} [\nabla_\alpha (g^{1/2} g^{\mu\nu}) - g^{1/2} T^{\nu}_{\rho\alpha}\, g^{\mu\rho} + T_\alpha\, g^{\mu\nu}] \,. \tag{2.33}$$

Similarly, the $t_2$ term may be expressed:

$$t_2 = \nabla_\sigma(g^{1/2}g^{\mu\sigma})/g^{1/2} + T_\sigma\, g^{\mu\sigma}\,. \tag{2.34}$$

Combining $t_1$ and $t_2$ with (2.22) and performing the integration, the final field equations are obtained:

$$\left.\frac{\delta S_{EH}[g,\Gamma]}{\delta\Gamma^{\alpha}_{\mu\nu}}\right|_g :\quad 0 = \delta^{\nu}_{\alpha}\left[\nabla_\sigma(g^{1/2}g^{\mu\sigma})/g^{1/2} + T_\rho\, g^{\mu\rho}\right] - \left[\nabla_\alpha(g^{1/2}g^{\mu\nu})/g^{1/2} + T^{\nu}_{\ \alpha\sigma}\, g^{\mu\sigma} + T_\alpha\, g^{\mu\nu}\right]. \tag{2.35}$$

Considering various contractions of (2.35), in general it does not appear that that either $t_1$ or $t_2$ will vanish independently of the other (unless the torsion is zero as discussed below). Therefore, it is not sufficient to consider the vanishing of either term separately.

The field equations (2.35) determine the geometry of the Einstein-Hilbert spacetime when the connection is neither metric or symmetric – in the same sense that $\Gamma^{\sigma}_{\mu\nu} = \overset{o}{\Gamma}{}^{\sigma}_{\mu\nu}(g)$ determines the geometry of the space when $\nabla_\rho\, g_{\mu\nu} = 0$ is satisfied. As a result, (2.35) gives an implied relation between the metric and connection (or as expressed in (2.35) – between the metric and the torsion tensor). As a consequence of this relationship the following Lemma holds in a spacetime originating from the Einstein-Hilbert action:

$$\nabla_\kappa\, g^{\mu\nu} = 0 \Leftrightarrow T^{\nu}_{\ \kappa\lambda} = 0\,. \tag{2.36}$$

The Lemma is proven by first assuming that (2.1) holds. In this case (2.35) reduces to

$$0 = \left(\delta^{\nu}_{\kappa}\, T_\sigma - T^{\nu}_{\ \kappa\sigma}\right) g^{\mu\sigma} - T_\kappa g^{\mu\nu}\,, \tag{2.37}$$

and then solving for the torsion tensor gives the result:

$$T^{\nu}_{\ \kappa\lambda} = \delta^{\nu}_{\kappa}\, T_\lambda - \delta^{\nu}_{\lambda} T_\kappa\,. \tag{2.38}$$

Transvecting (2.38) shows that the torsion trace vanishes:

$$T_\lambda = 0\,, \tag{2.39}$$

and therefore, in turn the torsion must vanish from (2.38). The $\Leftarrow$ case is proven by first expanding the covariant derivatives in (2.35) using (2.30). The result simplifies to

$$0 = \delta^{\nu}_{\alpha}\left[\nabla_{\sigma} g^{\mu\sigma} + (\partial_{\sigma} g^{1/2}/g^{1/2} - \Gamma^{\lambda}_{\sigma\lambda})\, g^{\mu\rho}\right] \\ - \left[\nabla_{\alpha} g^{\mu\nu} + T^{\ \nu}_{\alpha\sigma}\, g^{\mu\sigma} + (\partial_{\alpha} g^{1/2}/g^{1/2} - \Gamma^{\lambda}_{\alpha\lambda})\, g^{\mu\nu}\right], \tag{2.40}$$

But as noted earlier below (2.35), neither term of (2.40) can be assumed to vanish independently of the other. However, if the torsion vanishes:

$$0 = \delta^{\nu}_{\alpha}\left[\nabla_{\sigma} g^{\mu\sigma} + (\partial_{\sigma} g^{1/2}/g^{1/2} - \Gamma^{\lambda}_{\sigma\lambda})\, g^{\mu\sigma}\right] \\ - \left[\nabla_{\alpha} g^{\mu\nu} + (\partial_{\alpha} g^{1/2}/g^{1/2} - \Gamma^{\lambda}_{\alpha\lambda})\, g^{\mu\nu}\right], \tag{2.41}$$

then the first term is a contraction of the second and therefore it is sufficient to consider only the second term:

$$\nabla_{\alpha} g^{\mu\nu} + (\partial_{\alpha} g^{1/2}/g^{1/2} - \Gamma^{\lambda}_{\lambda\alpha})\, g^{\mu\nu} = 0\,. \tag{2.42}$$

By inspection of (2.42) it is obvious that $\nabla_{\alpha} g^{\mu\nu} = 0$ implies that $\partial_{\alpha} g^{1/2}/g^{1/2} = \Gamma^{\lambda}_{\alpha\lambda}$ must be satisfied. But it is left to prove that $\partial_{\alpha} g^{1/2}/g^{1/2} = \Gamma^{\lambda}_{\alpha\lambda} \;\Rightarrow\; \nabla_{\alpha} g^{\mu\nu} = 0$. If so, this is equivalent to the statement that $T^{\ \nu}_{\alpha\sigma} = 0 \Rightarrow \nabla_{\alpha} g^{\mu\nu} = 0$. First expand the covariant derivative in (2.42) to obtain:

$$\partial_{\alpha} g^{\mu\nu} + \Gamma^{\mu}_{\beta\alpha}\, g^{\beta\nu} + \Gamma^{\nu}_{\beta\alpha}\, g^{\mu\beta} + g^{\mu\nu}\partial_{\alpha} g^{1/2}/g^{1/2} - g^{\mu\nu}\Gamma^{\lambda}_{\alpha\lambda} = 0\,, \tag{2.43}$$

and then multiplying both sides by $g_{\mu\nu}$ gives the result:

$$g_{\mu\nu}\partial_{\alpha} g^{\mu\nu} + 4\,\partial_{\alpha} g^{1/2}/g^{1/2} - 2\Gamma^{\lambda}_{\lambda\alpha} = 0\,. \tag{2.44}$$

Next expand the partial derivative, $\partial_{\alpha} g^{1/2}$, using the identity

$$\partial_{\alpha} g^{1/2} = \tfrac{1}{2} g^{1/2} g^{\rho\sigma} \partial_{\alpha} g_{\rho\sigma} = -\tfrac{1}{2} g^{1/2} g_{\rho\sigma} \partial_{\alpha} g^{\rho\sigma}\,, \tag{2.45}$$

which is obtained by differentiating the identity:

$$\det M = \exp[tr(\ln M)]\,, \tag{2.46}$$

which is true for an arbitrary (non-singular) matrix *M*. Substituting (2.45) into (2.44) then gives

$$\Gamma^{\lambda}_{\lambda\alpha} = -\tfrac{1}{2} g_{\rho\sigma} \partial_{\alpha} g^{\rho\sigma}\,, \tag{2.47}$$

and therefore (2.42) simplifies to

$$\nabla_{\alpha} g^{\mu\nu} = 0\,, \tag{2.48}$$

and completes the proof of the Lemma (2.36).

It is important to emphasize that neither side of (2.36) is obtained independently of the other. But if the $\Rightarrow$ case is assumed then as a consequence (albeit trivial) the antisymmetric contribution to the Ricci tensor must vanish as seen from the second contracted Bianchi identity (2.11).

## Derivation of the Schwarzschild Solution

The unique static spherically symmetric solution to Einstein's field equations describing the exterior gravitational field of a spherically symmetric distribution of matter was found by Schwarzschild [91]. The term static implies that the metric is both explicitly time independent (stationary) and invariant under $t \to -t$. The relativistic effects associated with a spinning distribution of matter (Kerr solution) are entirely negligible insofar as the classic solar system experimental tests are concerned.† As a result, the Schwarzschild solution has been the most important for testing the predictions of general relativity. The stability properties of the associated orbital dynamics are considered in detail in Chapter 5.

A derivation of the solution is obtained by assuming a spherically symmetric metric of the form:

$$\begin{aligned} ds^2 = g_{\mu\nu}\, dq^u dq^n = c^2 A(r)\, dt^2 - B(r)\, dr^2 - r^2 d\Omega^2 \\ d\Omega^2 = d\theta^2 + \sin^2\theta\, d\varphi^2. \end{aligned} \tag{2.49}$$

As discussed by many authors (e.g., Weinberg [92], pp. 176-179; Ohanian and Ruffini [78], pp. 391-396) the metric (2.49) is the most general starting form since cross terms involving $dt \cdot dr$ and scale factors preceding the $d\Omega^2$ term may be eliminated by suitable transformations of the time and radial coordinates. For the analysis considered here the "standard" form (see Weinberg, p. 177) shall be adopted but it is worth noting that the metric may also be expressed in isotropic form:

$$ds^2 = c^2 A(r)\, dt^2 - H(r)(dr^2 + r^2 d\Omega^2)\,, \tag{2.50}$$

by a suitable re-definition of $r$. The Christoffel symbols are thus calculated from (2.49) using the Christoffel formula (equation (2.2)):

---

† however, there are proposed experiments to measure the frame dragging effects of a spinning distribution of matter (gravity probe B).

$$\Gamma^0_{10} = \Gamma^0_{01} = A'/2A$$
$$\Gamma^1_{00} = A'/2B\ ;\ \Gamma^1_{11} = B'/2B\ ;\ \Gamma^1_{22} = -r/2B\ ;\ \Gamma^1_{33} = r\sin^2\theta/B \tag{2.51}$$
$$\Gamma^2_{12} = \Gamma^2_{21} = \Gamma^3_{13} = \Gamma^3_{31} = 1/r\ ;\ \Gamma^2_{33} = -\cos\theta\sin\theta\ ;\ \Gamma^3_{32} = \cot\theta,$$

and the free-field equations (2.6) become (Appendix A):

$$\begin{aligned} R_{tt} &= -\ [(2A''A - A'^2)B - A'B'A]/4AB^2 + A'/rB = 0 \\ R_{rr} &= +\ [(2A''A - A'^2)B - A'B'A]/4A^2B + B'/rB = 0 \\ R_{\theta\theta} &= (AB' - BA')r/2AB^2 + (B-1)/B = 0 \\ R_{\varphi\varphi} &= \sin^2\theta\ R_{\theta\theta}\ . \end{aligned} \tag{2.52}$$

For later reference, the Ricci scalar, $R = g^{\rho\sigma}R_{\rho\sigma}$, is also calculated:

$$\begin{aligned} R = [(ABA'' - AA'B' - BA'^2) + [2(A'B - B'A)/AB^2]/r \\ -2(1 - B^{-1})r^2]/2(AB)^2\ , \end{aligned} \tag{2.53}$$

which must vanish by contraction of (2.6).

The derivation of the Schwarzschild solution to equations (2.52) has been widely discussed in the literature (e.g., Ref's [93]) using a variety of methods. But probably the simplest procedure is to note the algebraic similarity of $R_{tt}$ and $R_{rr}$. Dividing these respective components of (2.52) by $A$ and $B$ and then adding together gives the following result

$$\frac{R_{tt}}{A} + \frac{R_{rr}}{B} = \left(\frac{A'B + B'A}{r\,AB^2}\right) = 0\ , \tag{2.54}$$

and therefore the following relationship must hold between $A$ and $B$:

$$A'B + B'A = 0 \Rightarrow e^{\ln A} = e^{-\ln B + const.} \Rightarrow A = \frac{const.}{B}\ . \tag{2.55}$$

The constant is determined from the Minkowski limit as $r \to \infty$ in which case, $A = B = 1$, giving

$$A = B^{-1}. \tag{2.56}$$

As a consequence of (2.56) the volume element of 4-space is well defined regardless of what coordinate singularities might exist for either $A$ or $B$.

Using (2.56) the field equations (2.52) simplify to (primes denote derivatives with respect to $r$):

$$
\begin{aligned}
R_{tt} &= -\frac{A}{2r}\frac{d}{dr}(1 - A - rA') = 0 \\
R_{rr} &= \frac{1}{2rA}\frac{d}{dr}(1 - A - rA') = 0 \\
R_{\theta\theta} &= 1 - A - rA' = 0 \\
R_{\varphi\varphi} &= \sin^2\theta \; R_{\theta\theta} \; .
\end{aligned}
\tag{2.57}
$$

Therefore, the field equations (2.57) imply that the following differential equation must be satisfied:

$$rA' + A - 1 = 0\,, \tag{2.58}$$

which has the solution:

$$A(r) = 1 - c_1/r\,, \tag{2.59}$$

where $c_1$ is an integration constant. Note also that the Ricci scalar simplifies in this case to

$$R = \frac{r^2 A'' + 4rA' + 2A - 2}{r^2} = \frac{d^2(r^2 A)/dr^2 - 2}{r^2}\,, \tag{2.60}$$

and therefore the vanishing of the Ricci scalar leads to a more general solution which gives (2.59) as a special case when $c_2 = 0$:

$$
\begin{aligned}
R = 0 &\Rightarrow d(r^2 A)/dr = 2r - c_1 \\
&\Rightarrow A = 1 - c_1/r + c_2/r^2 .
\end{aligned}
\tag{2.61}
$$

As illustrated in the next section the constant $c_2$ is due to a nonzero stress energy tensor and therefore (2.61) does not correspond to a vacuum solution of the field equations. Furthermore, it is worth noting that the differential equation resulting from the vanishing of the Ricci scalar is more general than simply (2.60). As a result there are other spherically symmetric solutions to $R = 0$ besides (2.61). In fact, Littlewood [94] has proposed a theory based on this condition which was later shown by Pirani [95] to have a nonphysical solution. These details are considered further in Chapter 3.

The Schwarzschild solution is thus given by:

$$
\begin{aligned}
ds^2 &= c^2 \Lambda\, dt^2 - \Lambda^{-1} dr^2 - r^2 d\Omega^2 \\
\Lambda &= 1 - r_s/r\;;\; d\Omega^2 = d\theta^2 + \sin^2\theta\, d\varphi^2 ,
\end{aligned}
\tag{2.62}
$$

and the integration constant is determined from Newtonian correspondence which is discussed below and in Chapter 5:

$$r_s = 2MG/c^2\,, \tag{2.63}$$

where $r_s$ defines the Schwarzschild radius. The Newtonian limit for a general spherically symmetric metric may be obtained in two ways – in the first by considering the geodesic motion of point mass, $m_0$. The second method considers the correspondence with Poisson's equation for the gravitational field. In summary, the former limit gives the value of the Schwarzschild radius as well as the interpretation of the total energy for the system (discussed in Chapter 5). In addition, the limiting form of $g_{00}$ is obtained from this method which is discussed below. The second limit provides the coefficient of $16\pi Gc^{-4}$ preceding the Einstein-Hilbert action (presented earlier in the previous section), but this derivation is not considered here (see for instance, Anderson [96]).

To consider the first limit note that the Euler-Lagrange equations of motion may be expressed in either of two forms - in the standard Lagrangian form (dots denote differentiation with respect to the proper time $\tau$):

$$\frac{d}{d\tau}\frac{\partial L}{\partial \dot{q}^{\mu}} - \frac{\partial L}{\partial q^{\mu}} = 0\,, \tag{2.64}$$

or in "geodesic" form

$$\ddot{q}^{\lambda} + \Gamma^{\lambda}_{\mu\nu}\,\dot{q}^{\mu}\dot{q}^{\nu} = 0\,. \tag{2.65}$$

Equation (2.64) is an alternative expression of (2.65) since the relativistic Lagrangian consists only of a "kinetic" term (i.e., $L = \frac{1}{2} g_{\mu\nu}\,\dot{q}^{\mu}\dot{q}^{\nu}$). As a result, the connection coefficients, $\Gamma^{\sigma}_{\mu\nu}$, are defined in terms of the gravitational field (the "metric") according to the Christoffel form, $\overset{o}{\Gamma}{}^{\sigma}_{\mu\nu} \equiv \frac{1}{2} g^{\sigma\lambda}(\partial_{\mu} g_{\lambda\nu} + \partial_{\nu} g_{\lambda\mu} - \partial_{\lambda} g_{\mu\nu})$. Expanding the summation on the second term in (2.65) gives (the Latin indices, $i, j$ denote spatial coordinates):

$$\Gamma^{\lambda}_{\mu\nu}\,\dot{q}^{\mu}\dot{q}^{\nu} = \Gamma^{\lambda}_{00}\,(\dot{q}^{0})^2 + 2\Gamma^{\lambda}_{0i}\,\dot{q}^{0}\dot{q}^{i} + \Gamma^{\lambda}_{00}\,(\dot{q}^{i})^2\,, \tag{2.66}$$

but in the Newtonian limit it is assumed that $\dot{q}^{i} << c$, and therefore,

$$\Gamma^{\lambda}_{\mu\nu}\,\dot{q}^{\mu}\dot{q}^{\nu} \approx \Gamma^{\lambda}_{00}\,(\dot{q}^{0})^2 = c^2\,\Gamma^{\lambda}_{00}\,(\dot{t})^2\,. \tag{2.67}$$

The $\Gamma^{\lambda}_{00}$ coefficient is calculated using the Christoffel formula:

$$\Gamma^{\lambda}_{00} = \tfrac{1}{2} g^{\lambda\rho}(2\partial_0\, g_{\rho 0} - \partial_{\rho}\, g_{00}) = -\tfrac{1}{2} g^{\lambda\rho}\,\partial_{\rho}\, g_{00}\,, \tag{2.68}$$

where the first term is zero since the metric is assumed static. Separating the spatial and time components gives

$$\Gamma^{\lambda}_{00} = -\tfrac{1}{2} g^{\lambda 0}\, \partial_0\, g_{00} - \tfrac{1}{2} g^{\lambda i}\, \partial_i\, g_{00} = -\tfrac{1}{2} g^{\lambda i}\, \partial_i\, g_{00}\,, \tag{2.69}$$

and then assuming the metric is diagonal - from (2.69) a further result is that

$$\Gamma^{0}_{00} \propto g^{0i} = 0\,, \tag{2.70}$$

and therefore

$$d^2 q^0 / d\tau^2 = c^2 (d^2 t / d\tau^2) = 0 \Rightarrow dt / d\tau = const. = 1\,. \tag{2.71}$$

The constant is obtained from the Minkowski line element when $v << c$. As a result the $\Gamma^{\lambda}_{\mu\nu}\, \dot{q}^{\mu} \dot{q}^{\nu}$ term is expressed finally:

$$\Gamma^{\lambda}_{\mu\nu}\, \dot{q}^{\mu} \dot{q}^{\nu} \;\rightarrow\; \Gamma^{i}_{00} (\dot{q}^0)^2 = c^2\, \Gamma^{i}_{00} = -\tfrac{1}{2} c^2 g^{ij}\, \partial_j\, g_{00}\,, \tag{2.72}$$

and then combining with (2.65) gives the equation of motion for the point mass, $m_0$:

$$m_0\, \ddot{q}^i = \tfrac{1}{2} m_0 c^2 g^{ij}\, \partial_j\, g_{00}\,. \tag{2.73}$$

In the Newtonian limit it is reasonable to assume that the metric differs by some small deviation, $h_{\mu\nu}$, from the Minkowski line element:

$$g_{\mu\nu} = \eta_{\mu\nu} + h_{\mu\nu}\,, \tag{2.74}$$

and therefore to first order in $h_{00}$ (noting that $\eta^{ij} = -\delta^{ij}$):

$$m_0\, \ddot{q}^i = -\tfrac{1}{2} m_0 c^2\, \delta^{ij}\, \partial_j\, h_{00}\,, \tag{2.75}$$

or in coordinate free notation:

$$m_0\, \ddot{\vec{q}} = -\tfrac{1}{2} m_0 c^2\, \vec{\nabla} h_{00}\,. \tag{2.76}$$

The corresponding Newtonian equation of motion for a particle in a gravitational field is given by

$$m_0\, \ddot{\vec{q}} = -\vec{\nabla}\Phi = -\vec{\nabla}\left(\frac{GMm_0}{r}\right), \tag{2.77}$$

and therefore the following correspondence is made:

$$h_{00} = \frac{2\Phi}{m_0\, c^2}\,. \tag{2.78}$$

As a result, the “00” component of the metric is expressed

$$g_{00} \approx 1 + \frac{2\Phi}{m_0 c^2} . \tag{2.79}$$

Comparing (2.79) with the Schwarzschild line element (2.62) shows that for proper Newtonian correspondence:

$$g_{00} = 1 - \frac{r_s}{r} \approx 1 + \frac{2\Phi}{m_0 c^2} \Rightarrow r_s = \frac{2MG}{c^2} , \tag{2.80}$$

(note: $r_s \approx 1\ cm$ for the Earth; $r_s \approx 3\ km$ for the Sun; $r_s \approx 3\ meters$ for Jupiter).

## Derivation of the Reissner-Nordström Solution

The static spherically symmetric solution to Einstein's field equations describing the exterior gravitational field of a spherically symmetric charged distribution of matter was found in 1916 by Reissner [97] and independently in 1918 by Nordström [98]. The Schwarzschild solution is a vacuum solution of Einstein's field equations, but in the Reissner-Nordström case the event geometry is coupled to the stress energy tensor of the electromagnetic field. The total action is thus defined (EH ≡ Einstein-Hilbert; EM ≡ *electromagnetic*):

$$S_{total}[g] = S_{EH} + S_{EM\ Field} , \tag{2.81}$$

and therefore the most general form of the Einstein field equations couple matter fields to the event geometry through the stress energy tensor of the source. The stress energy tensor is defined as the symmetric tensor:

$$\delta S_{EM}[g] / \delta g^{\mu\nu} = -\tfrac{1}{2}\Theta_{\mu\nu} , \tag{2.82}$$

and as result, the Einstein field equations including sources are given by (on the RHS $G$ is Newton's gravitational constant):

$$G_{\mu\nu} \equiv R_{\mu\nu} - \tfrac{1}{2} g_{\mu\nu} R = 8\pi G c^{-4}\, \Theta_{\mu\nu} , \tag{2.83}$$

where $G_{\mu\nu}$ is the Einstein tensor and the proportionality constant on the RHS of (2.83) gives the proper Newtonian limit with Poisson's equation. But since an electromagnetic field is present Maxwell's equations must also be satisfied, and therefore, the field equations are given by the coupled Einstein-Maxwell system:

$$G_{\mu\nu} = 8\pi G c^{-4}\,\Theta_{\mu\nu}$$
$$0 = \nabla_\mu F^{\mu\nu}. \tag{2.84}$$

First consider the $G_{\mu\nu} = 8\pi G c^{-4}\,\Theta_{\mu\nu}$ equations. The stress-energy tensor for the electromagnetic field is derived from the $S_{EM}$ action using (2.82) and the variational results, (2.13), presented earlier in the first section of this Chapter. The result of this calculation is given by ($k$ is determined by the system of units - see equation (2.95)):

$$\Theta_{\mu\nu} = \frac{1}{4\pi k}(F^{\lambda}{}_{\mu}F_{\nu\lambda} + \tfrac{1}{4}g_{\mu\nu}F_{\rho\sigma}F^{\rho\sigma}), \tag{2.85}$$

where

$$S_{EM}[g] = -\frac{1}{16\pi k}\int d^4q\, g^{1/2}F_{\rho\sigma}\,F^{\rho\sigma}. \tag{2.86}$$

The electromagnetic field is assumed to be static and spherically symmetric. As a result, the only nonzero components of the field strength tensor are in the radial direction:

$$F_{\mu\nu} = \begin{pmatrix} 0 & E(r) & 0 & 0 \\ -E(r) & 0 & 0 & 0 \\ 0 & 0 & 0 & 0 \\ 0 & 0 & 0 & 0 \end{pmatrix}, \tag{2.87}$$

where $E(r)$ is the electric field. Raising and lowering indices on the field strength tensor using the metric – the stress energy tensor (2.85) has the components:

$$\Theta_{\mu\nu} = \frac{1}{8\pi k}\begin{pmatrix} E^2/B & 0 & 0 & 0 \\ 0 & -E^2/A & 0 & 0 \\ 0 & 0 & r^2E^2/AB & 0 \\ 0 & 0 & 0 & r^2E^2\sin^2\theta/AB \end{pmatrix}, \tag{2.88}$$

which is traceless. As a result, from (2.83) the solution will have a vanishing Ricci scalar and therefore (2.84) is replaced by the simpler system of equations:

$$R_{\mu\nu} = 8\pi G c^{-4}\,\Theta_{\mu\nu}$$
$$0 = \nabla_\mu F^{\mu\nu}. \tag{2.89}$$

The first set of equations in (2.84) are obtained by combining (2.52) and (2.88):

$$\underbrace{-[(2A''A - A'^2)B - A'B'A]/4AB^2 + A'/rB = GE^2/kc^4B}_{00}$$

$$\underbrace{+[(2A''A - A'^2)B - A'B'A]/4A^2B + B'/rB = -GE^2/kc^4A}_{11} \tag{2.90}$$

$$\underbrace{(AB' - BA')r/2AB^2 + (B-1)/B = GE^2r^2/kc^4AB}_{22} \;,$$

and only the first three components are independent since the "33" component is "22" $\times\sin^2\theta$. From (2.90) add and subtract the following linear combination to obtain:

$$(\text{"}00\text{"}/A(r) \;+\; \text{"}11\text{"}/B(r)) \Rightarrow A'B + B'A = 0\,, \tag{2.91}$$

and therefore $A = B^{-1}$ as in the Schwarzschild case. Now consider the second field equation of (2.84):

$$\nabla_\mu F^{\mu\nu} = 0 = \begin{pmatrix} -\left[\frac{1}{2}E\left(\frac{A'}{A}+\frac{B'}{B}+\frac{4}{r}\right)+E'\right](AB)^{-1} \\ 0 \\ 0 \\ 0 \end{pmatrix}, \tag{2.92}$$

and after substituting the result, $B = A^{-1}$, the equation simplifies to

$$E' + 2E/r = 0\,. \tag{2.93}$$

The solution for $E(r)$ is thus given by

$$E(r) = \frac{c_3}{r^2}\,, \tag{2.94}$$

where $c_3$ is an integration constant corresponding to the charge of the system. In arbitrary units the electric field is given by

$$\frac{E(r)}{k^{1/2}} = \frac{k^{1/2}e}{r^2}\,. \tag{2.95}$$

where the constant $k$ is determined by the system of units (in Gaussian and electrostatic units $k$ = 1; in MKS units $k = (4\pi\varepsilon_0)^{-1}$; see Jackson [99] or Wangsness [100] for additional discussion on unit conventions). Using this result in combination with $B = A^{-1}$, the field equations (2.90) simplify to

$$\underbrace{\frac{A}{r^4}(rA' + A + Gke^2/c^4r^2 - 1) = 0}_{00}\ ;\ \underbrace{\frac{1}{Ar^2}(rA' + A + Gke^2/c^4r^2 - 1) = 0}_{11}$$
$$\underbrace{\frac{r}{2}\frac{d}{dr}(rA' + A + Gke^2/c^4r^2 - 1) = 0}_{22}\ . \tag{2.96}$$

By inspection the differential equation that must be satisfied is given by an inhomogeneous version of (2.58):

$$rA' + A - 1 = Gke^2/c^4r^2\,. \tag{2.97}$$

The solution to (2.97) is thus given by

$$A(r) = 1 - c_1/r + Gke^2/c^4r^2\,, \tag{2.98}$$

where $c_1$ is an integration constant which must reduce to (2.63) in the limit as $e \to 0$. For later reference it will be useful to express the Reissner-Nordström solution in the form (let $A \to \Lambda$):

$$\begin{aligned} ds^2 &= \Lambda\, dt^2 - \Lambda^{-1}dr^2 - r^2 d\Omega^2 \\ \Lambda &= 1 - x + x^2/2\lambda \\ d\Omega^2 &= d\theta^2 + \sin^2\theta\, d\varphi^2, \end{aligned} \tag{2.99}$$

by defining the dimensionless variable, $x = r_s/r$, and the dimensionless parameter:

$$\lambda = \frac{c^4}{2Gk}\left(\frac{r_s}{e}\right)^2 = \frac{2G}{k}\left(\frac{M}{e}\right)^2. \tag{2.100}$$

The most obvious difference between the Reissner-Nordström and Schwarzschild solutions is that the Reissner-Nordström spacetime now has two horizons located at

$$x_\pm = \lambda \pm \sqrt{\lambda(\lambda - 2)} = r_s/r_\mp\,, \tag{2.101}$$

where $r_+$ is the exterior horizon and $r_-$ is the interior horizon as illustrated in the phase diagrams of Chapter 6.

At the beginning of this chapter it was noted that the Reissner-Nordström solution corresponds physically to the exterior gravitational field of a charged spherically symmetric black hole. This case is given by the parameter value $\lambda > 2$. But there are other interpretations that may be given to (2.99) based on the ratio of $e$ and $M$. Specifically, the interpretation given to the case $\lambda = 2$ (i.e., $r_+ = r_-$) is to the spacetime exterior of a charged dust cloud in equilibrium between electrostatic repulsion and gravitational attraction. Finally when $\lambda < 2$, the singularity is exposed at $r = 0$. Although (2.99) was discovered in 1916 shortly after the Schwarzschild solution, the physical interpretation for the case, $\lambda = 2$, was not given until 1965 by Bonnor [101] (see Carter [102]).

The parameter value $\lambda < 2$ is widely considered to be nonphysical based on the cosmic censorship conjecture (Penrose [103]) which states essentially that naked singularities cannot exist in nature. Therefore, assuming that the cosmic censorship conjecture is valid, the maximum charge contribution to the total mass of the black hole is given by the parameter value $\lambda = 2$. This case gives the "extremal" Reissner-Nordström black hole for which $r_+ = r_-$ and corresponds to the maximum attainable charge that can accumulate based on the mechanism discussed by Eddington [77] and Harrison [79] (see the Introduction for a further discussion). Thus, the Reissner-Nordström line element (2.98) (letting $A(r) \to \Lambda(r)$) is expressed in terms of two fundamental length scales associated with the mass and charge ($r_s$ and $r_e$) respectively:

$$\Lambda(r) = 1 - \underbrace{\left(\frac{2MG}{c^2}\right)}_{r_s}\frac{1}{r} + \underbrace{\left(\frac{Gke^2}{c^4}\right)}_{r_e^2}\frac{1}{r^2}, \tag{2.102}$$

which according to the cosmic censorship conjecture must satisfy the relation:

$$\frac{r_e}{r_s} = \frac{1}{\sqrt{2\lambda}} \Rightarrow \frac{r_e}{r_s} \leq \frac{1}{2}. \tag{2.103}$$

From the definition of $\lambda$ the ratio of $e/M$ is thus calculated for the extremal case ($C \equiv$ Coulombs):

$$\frac{e_{\max}}{M} = \sqrt{\frac{G}{k}} \approx 8.61 \times 10^{-11}\ C/kg. \tag{2.104}$$

Using dimensionless units (where $\sqrt{G/k} = 1$) the charge to mass ratio of the electron is given by

$$\frac{e}{m_e} \approx 2 \times 10^{-21}, \tag{2.105}$$

and therefore

$$\frac{e_{\max}}{M} \approx 10^{-21}\left(\frac{e}{m_e}\right). \tag{2.106}$$

For an additional order of magnitude comparison the value of $\lambda$ given by (2.100) is thus calculated for both an electron and proton:

$$\lambda_e \approx 4.8\times 10^{-43};\ \lambda_p \approx 1.5\times 10^{-36}, \tag{2.107}$$

and therefore $\lambda << 2$ in these cases. For a star like the Sun the order of magnitude estimate given by Eddington [77] (see also Harrison [79]) gives approximately 100 *C* of charge per solar mass and therefore

$$\frac{e_{Sun}}{M} \approx 10^{-40}\left(\frac{e}{m_e}\right);\ \frac{1}{\lambda} \approx 1.7\times 10^{-37}. \tag{2.108}$$

But the cosmic censorship conjecture has not been proven, and recently a detailed examination of several gravitational collapse scenarios [104] has shown that naked singularities may develop in a variety of circumstances such as the collapse of radiation shells; spherically symmetric collapse of perfect fluid; collapse of a spherical inhomogeneous dust cloud [105], and the spherical collapse of a massless scalar field [106]. Additional numerical studies have shown that under special circumstances, e.g., highly elongated "cigar" shaped mass distributions, that the formation of naked singularities does occur. But the conclusions that may be drawn from these numerical studies are questionable given that the numerical solutions are not always well defined over all spacetime domains of interest (e.g., the Riemann curvature tensor may become singular; see also [78]).

Additional evidence that questions the plausibility of naked singularity solutions, specifically in the Reissner-Nordström case, was given some time ago by Boulware [107]. Boulware shows that a thin charged spherical shell will collapse to form a naked singularity if and only if the matter energy density of the shell is negative. But as Morris and Thorne [108] have emphasized in their investigation of wormhole solutions to the Einstein field equations, this does not necessarily prove that $\lambda < 2$ is nonphysical. In fact, they give physical examples where negative energy densities can be physically reasonable and are actually required in special circumstances. Therefore, it is not inconceivable that a Reissner-Nordström object with $\lambda < 2$ could exist, at least insofar as a negative energy density is concerned, although such objects must certainly be exotic as noted by Morris and Thorne. The viewpoint adopted here is non-committal on the physical existence of such solutions - these parameter values are simply explored as dynamical possibilities in Chapter 5 - not necessarily physical ones.

# Chapter 3 Gauge Gravity

As discussed in the Introduction several alternative quadratic curvature Lagrangians have been considered as possible generalizations of Einstein's theory. The motivation for considering such work has been to setup a framework for gravitation that closely resembles a Yang-Mills type gauge theory in the hopes of unifying gravitation with the other fundamental interactions. As a consequence the action must be quadratic in the Riemann curvature tensor - in contrast to the linear Einstein-Hilbert action discussed in Chapter 2 (for a similar approach based on the local gauge invariance of the Clifford algebra basis elements see Ref. [88]). Therefore, to trace this development, a brief review of the variational formulation of Yang-Mills gauge theory is discussed in the first section of this Chapter. Gauge Kinematics are then discussed in the following sub-section followed by Gauge Dynamics which includes an analysis of the spherically symmetric solutions. In the final section of this Chapter an algebraic constraint is derived by imposing equivalence between the standard and "gauge gravity" Palatini variational procedures.

## Yang-Mills Formalism

In 1954, Yang and Mills [109] introduced the idea of gauge fields through local isotopic spin transformations. If dynamical equations are to be defined over the internal isovectors on a background Riemannian manifold, then a gauge covariant derivative must be constructed ($\psi$ are assumed to be Lorentz scalars):

$$\nabla_\mu \psi^a = \partial_\mu \psi^a + \Gamma^a_{b\mu} \psi^b \qquad (a,b \equiv \text{isospace};\ \mu,\nu \equiv \text{event}), \tag{3.1}$$

to insure invariance under the local isospin transformations:

$$\psi^a \rightarrow \psi^{a'} = S^{a'}_b(x) \psi^b , \tag{3.2}$$

(or what is the same, to define equivalence of isovectors at neighboring events of the spacetime manifold (the kernel index method [83] is adopted here to denote the gauge transformation, $a \rightarrow a'$)). In (3.1), the connection (or "gauge potential") is expressed as a linear combination of the algebra basis, $\sigma_k$:

$$\Gamma^{a}_{\ b\mu} \equiv [\Gamma^{k}_{\ \mu}\sigma_k]^{a}_{\ b} = \Gamma^{k}_{\ \mu}\sigma^{a}_{\ bk} \qquad (3.3)$$
$$(i, j, k \equiv \text{isovector}),$$

generating the $su(2)$ Lie algebra:

$$[\sigma_i, \sigma_j] = \varepsilon^{k}_{\ ij}\sigma_k\,, \qquad (3.4)$$

with

$$S^{a}_{\ b} = [e^{\theta^k\sigma_k}]^{a}_{\ b} \in SU(2)\,, \qquad (3.5)$$

an element of the Lie group, $SU(2)$ † (see also Appendix D). Applying (3.1) to (3.2), and then requiring that

$$\nabla_\mu \psi^{a} \to \nabla_\mu \psi^{a'} = S^{a'}_{\ b}\nabla_\mu \psi^{b}\,, \qquad (3.6)$$

gives the required inhomogeneous transformation law of the gauge potentials:

$$\Gamma^{a'}_{\ b'\mu} = S^{a'}_{\ c}\Gamma^{c}_{\ d\mu}S_{b'}^{\ d} - (\partial_\mu S^{a'}_{\ c})S_{b'}^{\ c}\,, \qquad (3.7)$$

or using matrix notation

$$\Gamma'_\mu = S\,\Gamma_\mu S^{-1} - (\partial_\mu S)S^{-1}. \qquad (3.8)$$

Explicitly, we may consider (3.3) in the adjoint representation (i.e. the dimension of the carrier space = the dimension of the algebra):

$$\Gamma^{i}_{\ j\mu} = \begin{pmatrix} 0 & \Gamma^3_\mu & -\Gamma^2_\mu \\ -\Gamma^3_\mu & 0 & \Gamma^1_\mu \\ \Gamma^2_\mu & -\Gamma^1_\mu & 0 \end{pmatrix}, \qquad (3.9)$$

since in this case

$$[\sigma_k]^{i}_{\ j} = \varepsilon^{i}_{\ jk}\,, \qquad (3.10)$$

are the structure constants of $su(2)$. In the spinor representation (3.3) becomes

$$\Gamma^{A}_{\ B\mu} = \begin{pmatrix} \Gamma^3_\mu & \Gamma^1_\mu - \mathrm{i}\,\Gamma^2_\mu \\ \Gamma^1_\mu + \mathrm{i}\,\Gamma^2_\mu & -\Gamma^3_\mu \end{pmatrix}, \qquad (3.11)$$

since now the generators are the Pauli matrices, $\tau^{A}_{\ Bk}$:

$$[\sigma_k]^{A}_{\ B} = \tfrac{1}{2}\tau^{A}_{\ Bk}\,, \qquad (3.12)$$

with $k =$ isovector index; $A,B =$ isospinor index.

The field equations for $\Gamma^a_{b\mu}$ are derived from a variational principle. The free-field Lagrangian (see Uzes [110]) is taken in quadratic form:

$$L_{YM} = -\tfrac{1}{4}\, g^{\lambda\rho} g^{\omega\sigma} \varphi^{c}_{\cdot d\lambda\omega} \varphi^{d}_{\cdot c\rho\sigma} , \tag{3.13}$$

but for calculations let us use the first order form (for instance, Crawford [90]):

$$\begin{aligned} L_{YM} &= \tfrac{1}{4} g^{\lambda\rho} g^{\kappa\sigma} \varphi^{c}_{d\lambda\kappa} \varphi^{d}_{c\rho\sigma} \\ &\quad - \tfrac{1}{2} g^{\lambda\rho} g^{\kappa\sigma} \varphi^{c}_{d\lambda\kappa} \left( \partial_\rho \Gamma^d_{c\sigma} - \partial_\sigma \Gamma^d_{c\rho} + \Gamma^d_{f\rho} \Gamma^f_{c\sigma} - \Gamma^l_{f\sigma} \Gamma^f_{c\rho} \right) , \end{aligned} \tag{3.14}$$

given that this choice yields the defining relation of the gauge fields (or isospin curvature) in terms of the gauge potentials with respect to $\delta\varphi^{b\,\mu\nu}_{a\cdot\cdot}$ :

$$\delta\varphi^{b\,\mu\nu}_{a\cdot\cdot} : \qquad \varphi^a_{b\mu\nu} = \partial_\mu \Gamma^a_{b\nu} - \partial_\nu \Gamma^a_{b\mu} + \Gamma^a_{c\mu} \Gamma^c_{b\nu} - \Gamma^a_{c\nu} \Gamma^c_{b\mu} , \tag{3.15}$$

and in addition satisfies the Bianchi identity:

$$\nabla_\rho \varphi^a_{b\mu\nu} + \nabla_\mu \varphi^a_{b\nu\rho} + \nabla_\nu \varphi^a_{b\rho\mu} = 0 . \tag{3.16}$$

From the non-abelian structure in (3.15), the gauge field transforms as a rank 2 isotensor under the local gauge transformation, $\varphi^a_{b\mu\nu} \to \varphi^{a'}_{b'\mu\nu}$ :

$$\varphi^{a'}_{b'\mu\nu} = S^{a'}_{c} S^{d}_{b'} \varphi^{c}_{d\mu\nu} , \tag{3.17}$$

insuring that the action constructed from (3.13) is both a scalar and isoscalar. In the interest of generality, let us consider the field equations resulting from variation with respect to $\delta\Gamma^b_{a\mu}$, on a curved background spacetime. The action in this case must be

$$S_{YM} = \int d^4q \; g^{1/2} L_{YM} . \tag{3.18}$$

Performing the variation we get (*note*: this calculation is identical in structure to the calculation worked out in detail beginning at (3.53); see also Treat [19] - equation (59)):

$$\partial_\rho ( g^{1/2} \varphi^{a\,\rho\mu}_{b\cdot\cdot} ) + g^{1/2} ( \Gamma^a_{c\nu} \varphi^{c\,\rho\mu}_{b\cdot\cdot} - \Gamma^c_{b\nu} \varphi^{a\,\rho\mu}_{c\cdot\cdot} ) = 0 , \tag{3.19}$$

since $\Gamma^\sigma_{\mu\nu}$ is symmetric. We may express this result in terms of the covariant derivative noting from (2.30) that $\nabla_\mu g^{1/2} = 0 \Rightarrow \partial_\mu g^{1/2} = g^{1/2} \Gamma^\lambda_{\lambda\mu}$, and calculating:

$$\nabla_\rho \varphi^{a\,\rho\mu}_{b\cdot\cdot} = \partial_\rho ( \varphi^{a\,\rho\mu}_{b\cdot\cdot} ) + \Gamma^\rho_{\nu\rho} R^{a\,\nu\mu}_{b\cdot\cdot} + \underbrace{\Gamma^\mu_{\nu\rho} \varphi^{a\,\rho\nu}_{b\cdot\cdot}}_{zero} + \Gamma^a_{c\nu} R^{c\,\rho\mu}_{b\cdot\cdot} - \Gamma^c_{b\nu} R^{a\,\rho\mu}_{c\cdot\cdot} . \tag{3.20}$$

---

† let us note, however, that this construction may be generalized to other Lie groups.

Solving for $\partial_\rho \varphi^{a\ \rho\mu}_{\ b\cdot\cdot}$, gives the final result:

$$\delta\Gamma^{b}_{a\mu}: \qquad g^{\frac{1}{2}} \nabla_\rho \varphi^{a\ \rho\mu}_{\ b\cdot\cdot} = 0\,. \tag{3.21}$$

## Gauge Kinematics

Continuing with this pattern, the Yang-Mills formalism may be carried over to the general relativistic case beginning from the (global) invariance of special relativity, i.e. the Lorentz group, $SO(1,3)$ (see also Appendix D). But there is another viewpoint that can be taken in this respect (Fairchild [26]); that local invariance under general coordinate transformations may be accounted for by introducing $\Gamma^{\sigma}_{\mu\nu}$, i.e. the "gauge potential" of $Gl(4,\mathbb{R})$ (the linear group of general coordinate transformations in 4 dimensions). However, there are important differences to be noted in this approach compared to the former case that make it awkward to keep close analogy with the gauge theory formalism outlined in the previous section. Briefly, the idea is to note that if

$$dx^{\mu'} = A^{\mu'}_{\ \nu}\, dx^{\nu}\,, \tag{3.22}$$

denotes a general coordinate transformation on the spacetime manifold, then $A^{\mu'}_{\ \nu}$ is a non-singular $4\times 4$ matrix and therefore $A^{\mu'}_{\ \nu} \in Gl(4,\mathbb{R})$. Following the gauge theory pattern, next construct a covariant derivative with respect to $Gl(4,\mathbb{R})$ by introducing $\Gamma^{\sigma}_{\mu\nu}$ as a "gauge potential" to insure that $\nabla_\lambda b^\mu$ (for an arbitrary vector $b^\mu$) transforms as a tensor. The gauge field in this case is just the Riemann curvature tensor given by (2.9):

$$R^{\rho}_{\ \sigma\mu\nu} = \partial_\mu \Gamma^{\rho}_{\sigma\nu} - \partial_\nu \Gamma^{\rho}_{\sigma\mu} + \Gamma^{\rho}_{\lambda\mu}\Gamma^{\lambda}_{\sigma\nu} - \Gamma^{\rho}_{\lambda\nu}\Gamma^{\lambda}_{\sigma\mu}\,, \tag{3.23}$$

but here is where the analogy ends; the gauge potential, $\Gamma^{i}_{\ j\mu}$, of $SU(2)$ (for instance) has indices in both coordinate and internal spaces - so there can be no symmetries with respect to $j$ and $\mu$. However, as noted earlier in Chapter 2, $\Gamma^{\sigma}_{\mu\nu}$ is generally expressed as a sum of both symmetric and antisymmetric contributions which suggests that index symmetry on the lower indices should be considered fundamental in this case - in contrast to the gauge connection $\Gamma^{i}_{\ j\mu}$. Therefore, we will not proceed with the method based on $Gl(4,\mathbb{R})$, but instead carry through with calculations in a local Lorentz frame to follow closely (in formalism) the traditional Yang-Mills type gauge theory pattern with kinematics based on $SO(1,3)$. It is this observation which motivates the introduction of the tetrad basis as discussed below.

To begin, a brief review of local Lorentz invariance is given to set the notation (see also Appendix D on group theory). The discussion presented here is based on the earlier presentation by Crawford [90]. The Lorentz group is defined as the orthogonal group of transformations that leave invariant the Minkowski metric, $g_{ab} \equiv \eta_{ab} = (+1,-1,-1,-1)$,† where $a$, $b$ denote Lorentz indices. The transformations are defined such that

$$g_{a'b'} = \Lambda_{a'}^{\ a} \Lambda_{b'}^{\ b} g_{ab} , \tag{3.24}$$

with $g_{ab} = g_{a'b'} = (+1,-1,-1,-1)$ and $\Lambda_{a'}^{\ a}$ denotes the Lorentz transformation. A special transformation may be introduced that shifts the coordinate dependence of $g_{\mu\nu}$ to an auxiliary set of quantities labeled as *tetrads*, $e_{\mu}^{\ a}$ (note: additional discussion of the tetrad basis is given in the following section; these are equivalently referred to as vierbeins (or vierbein basis) by several authors, see e.g., Crawford [90]):

$$g_{\mu\nu}(q) = e_{\mu}^{\ a}(q) e_{\nu}^{\ b}(q)\, g_{ab} , \tag{3.25}$$

where $\mu,\nu$ denote the coordinate space indices. Note however that local Lorentz invariance may be considered in (3.24):

$$g_{a'b'} = \Lambda_{a'}^{\cdot a}(q)\, \Lambda_{b'}^{\cdot b}(q)\, g_{ab} , \tag{3.26}$$

still recovering the defining group relation, but $e_{\mu}^{\ a}$ in (3.25) is not a Lorentz transformation.

The covariant derivative in this case is defined by

$$\nabla_{\mu} V^{a} = \partial_{\mu} V^{a} + \Gamma^{a}_{b\mu} V^{b} , \tag{3.27}$$

---

† *note*: in the literature it is more common to denote the Minkowski metric using the kernel symbol $\eta$ rather than $g$. However, this notation is redundant when using the kernel index method since $g_{ab}$ is already $\eta_{ab}$.

which insures local Lorentz invariance of the derivative. Here $\Gamma^{a}_{b\mu}$ is the (spin) connection and $V^{a}$ denotes an arbitrary Lorentz vector. The connection is expressed as a linear combination of the algebra basis, $\sigma_{cd}$:

$$\Gamma^{a}_{b\mu} \equiv [\Gamma^{cd}{}_{\mu}\sigma_{cd}]^{a}{}_{\cdot b} = \Gamma^{cd}{}_{\mu}\sigma_{cd}{}^{a}{}_{b}, \tag{3.28}$$

generating the $so(1,3)$ Lie algebra:

$$[\sigma_{ab}, \sigma_{cd}] = g_{ac}\sigma_{bd} - g_{ad}\sigma_{bc} - g_{bc}\sigma_{ad} + g_{bd}\sigma_{ac}, \tag{3.29}$$

where

$$\Lambda^{a}{}_{b} = [e^{\frac{1}{2}\theta^{cd}\sigma_{cd}}]^{a}_{b} \in SO(1,3). \tag{3.30}$$

Again, by requiring that

$$\nabla_{\mu}V^{a} \rightarrow \nabla_{\mu}V^{a'} = \Lambda^{a'}_{b}\nabla_{\mu}V^{b}, \tag{3.31}$$

the transformation of $\Gamma^{a}_{b\mu}$ is determined by

$$\Gamma^{a'}_{b'\mu} = \Lambda^{a'}_{c}\Gamma^{c}_{d\mu}\Lambda^{d}_{b'} - (\partial_{\mu}\Lambda^{a'}_{d})\Lambda^{d}_{b'}. \tag{3.32}$$

## Tetrad Basis

A basic assumption of the gauge gravity theory is that the connection is in Christoffel form after the Palatini variation. As a result, the variational calculations are simplified considerably in this approach by calculating in the tetrad basis. In this section a review of several mathematical details of the tetrad basis are given and then adopted for calculations in the following section. To begin, consider the condition that the connection is metric as discussed in Chapter 2:

$$\nabla_{\mu}g_{\rho\sigma} = 0. \tag{2.1}$$

Assuming that (2.1) is valid, it follows from the definition of the covariant derivative that

$$\nabla_{\nu}e^{a}_{\mu} = 0 \Rightarrow \Gamma^{\sigma}_{\mu\nu} = e^{b}_{\mu}\Gamma^{a}_{b\nu}e^{\sigma}_{a} + (\partial_{\nu}e^{a}_{\mu})e^{\sigma}_{a}, \tag{3.33}$$

or conversely

$$\Gamma^{a}_{b\nu} = e^{a}_{\lambda}\Gamma^{\lambda}_{\mu\nu}e^{\mu}_{b} - (\partial_{\nu}e^{a}_{\mu})e^{\mu}_{b}. \tag{3.34}$$

However, it should be noted that $(b \rightarrow \mu)$ is not a coordinate transformation despite the formal similarity taken by (3.33) (or (3.34)) to the inhomogeneous transformations laws (3.7) (or (3.32)). The relations given in (3.33) and (3.34) will be considered basic for calculations in the tetrad basis; therefore, we will assume that the metric condition is valid throughout. In addition, we note that the index antisymmetry of the Riemann tensor (first index pair only):

$$R_{\rho\sigma\mu\nu} = -R_{\sigma\rho\mu\nu}, \tag{3.35}$$

is a consequence of $\Gamma^{\sigma}_{\mu\nu}$ being metric. To prove this consider

$$\left(\nabla_\mu \nabla_\nu - \nabla_\nu \nabla_\mu\right) g_{\rho\sigma}, \tag{3.36}$$

and then assuming that $\nabla_\mu g_{\rho\sigma} = 0$, we get (3.35) (see Misner, et. al. [66], p. 326 (and comment at bottom of p. 324). See also Schouten [83], p. 145. From (3.33), the torsion tensor (2.5) is expressed

$$T^{\sigma}_{\mu\nu} = -e_a^{\ \sigma}(\partial_\mu e^a_\nu - \partial_\nu e^a_\mu + \Gamma^a_{\ b\mu} e^b_\nu - \Gamma^a_{b\nu} e^b_{\ \mu}), \tag{3.37}$$

and for the case that $T^{\sigma}_{\mu\nu} = 0$, the spin connection and tetrads are no longer independent since $\Gamma^a_{\ b\mu}$ may be expressed in terms of the tetrads:

$$\Gamma^a_{\ b\mu} = \tfrac{1}{2} e^c_{\ \mu}(\Omega^a_{\ bc} + \Omega_{bc}^{\ \ a} - \Omega_{c\ b}^{\ a}), \tag{3.38}$$

where $\Omega^a_{\ bc} = e_b^{\ \mu} e_c^{\ \nu}(\partial_\mu e^a_\nu - \partial_\nu e^a_{\ \mu})$, is the object of anholonomity (Schouten [83], p. 100, equation (9.2) and p. 170, equation (9.9)).

In the tetrad basis the Einstein-Hilbert action is expressed ($e \equiv \det e^a_\mu$):

$$S_{EH}[e_\alpha^{\ \mu}] = \int d^4q\; e\, e^\rho_a\, e^{b\sigma} R^a_{\ b\rho\sigma}. \tag{3.39}$$

However, if we follow closely the gauge theory pattern, one would expect the correct Lagrangian to be given in quadratic form:

$$L = -\tfrac{1}{4}\, g^{\lambda\rho} g^{\omega\sigma} R^c_{\ d\lambda\omega} R^d_{\ c\rho\sigma}, \tag{3.40}$$

by analogy with (3.13); i.e. $R^a_{\ b\rho\sigma}$ plays the role of a field strength tensor for the gravitational field and $\Gamma^a_{\ b\mu}$ is the gauge potential. However, an important difference between the gravitational and Yang-Mills version should be noted: $R^a_{\ b\rho\sigma}$ and $\Gamma^a_{\ b\mu}$ are related to the event geometry, but this is not necessarily true for the corresponding Yang-Mills quantities. The relationship between the Yang-Mills field and the event geometry is discussed in detail in Ref. [111]. In the following section we give a discussion of the field equations resulting from (3.40) which provides the starting point for gauge gravity and a brief review of the other forms that have been considered.

## Gauge Dynamics

### Field Equations

In previous sections an outline has been given of the gauge theory formalism. A basic result is that if ones considers general relativity as a gauge kinematic theory based on $SO(1,3)$, then the action must be quadratic in the Riemann curvature. However, there are several quadratic invariants that can be formed using the Riemann tensor and calculations for these cases have already been examined in an early paper by Stephenson [10][†] discussing:

$$\begin{aligned} S_1 &= \int d^4q\ g^{\frac{1}{2}} R^2 \\ S_2 &= \int d^4q\ g^{\frac{1}{2}} R_{\rho\sigma} R^{\rho\sigma} \\ S_3 &= \int d^4q\ g^{\frac{1}{2}} R^{\rho}{}_{\sigma\mu\nu} R^{\sigma}{}_{\rho\cdot\cdot}{}^{\mu\nu}\ . \end{aligned} \tag{3.41}$$

The field equations resulting from (3.41) with respect to the Palatini variation (i.e. $\delta g_{\mu\nu}$ and $\delta\Gamma^{\sigma}_{\mu\nu}$ are assumed independent) are derived in [10] and are given by

$$\begin{aligned} &\delta S_1[g,\Gamma]: \begin{cases} \delta g^{\mu\nu}:\ 2R\left(R_{\mu\nu} - \tfrac{1}{4} g_{\mu\nu} R\right) = 0 \\ \delta\Gamma^{\sigma}_{\mu\nu}:\ \nabla_\sigma(g^{\frac{1}{2}} g^{\mu\nu} R) = 0 \end{cases} \\ &\delta S_2[g,\Gamma]: \begin{cases} \delta g^{\mu\nu}:\ R_{\mu}{}^{\lambda} R_{\nu\lambda} + R^{\lambda}{}_{\mu} R_{\lambda\nu} - \tfrac{1}{2} g_{\mu\nu} R_{\rho\sigma} R^{\rho\sigma} = 0 \\ \delta\Gamma^{\sigma}_{\mu\nu}:\ \nabla_\sigma(g^{\frac{1}{2}} R^{\mu\nu}) = 0 \end{cases} \\ &\delta S_3[g,\Gamma]: \begin{cases} \delta g^{\mu\nu}: -R_{\mu\cdot\cdot\cdot}{}^{\rho\sigma\lambda} R_{\nu\rho\sigma\lambda} + R^{\rho}{}_{\mu\cdot\cdot}{}^{\sigma\lambda} R^{\cdot}{}_{\rho\nu\sigma\lambda} + R^{\rho}{}_{\sigma\mu\lambda} R^{\sigma}{}_{\rho\nu\cdot}{}^{\lambda} - \tfrac{1}{4} g_{\mu\nu} R^{\rho}{}_{\cdot\sigma\cdot\cdot}{}^{\lambda\kappa} R^{\sigma}{}_{\rho\lambda\kappa} = 0 \\ \delta\Gamma^{\sigma}_{\mu\nu}:\ \nabla_\rho(g^{\frac{1}{2}} R^{\mu}{}_{\sigma\cdot\cdot}{}^{\rho\nu}) = 0\ , \end{cases} \end{aligned} \tag{3.42}$$

where Stephenson has assumed that $\Gamma^{\sigma}_{\mu\nu}$ is symmetric in $\mu\nu$, but no *a-priori* relationship is assumed to exist between the metric and connection (i.e., $\nabla_\rho g_{\mu\nu} \neq 0$). However, Stephenson and others have viewed (3.42) with little consequence and impose the

---

[†] Note: Stephenson did not have any intent along the lines of gauge theory. At the time of his paper, these formulations were of interest mainly for unifying the electromagnetic and gravitational fields and were also previously considered by Weyl, Pauli, Eddington, and Lanczos as discussed in the Introduction.

Christoffel form onto $\Gamma^{\sigma}_{\mu\nu}$ after the variation to obtain the following set of field equations:

$$\delta S_1[g,\Gamma]\Big|_{\Gamma=\overset{o}{\Gamma}}: \quad \begin{cases} \delta g_{\mu\nu}: \; R\left(R_{\mu\nu} - \tfrac{1}{4} g_{\mu\nu} R\right) = 0 \\ \delta \Gamma^{\sigma}_{\mu\nu}: \; g^{\mu\nu} \nabla_{\sigma} R = 0\,, \end{cases} \tag{3.43}$$

$$\delta S_2[g,\Gamma]\Big|_{\Gamma=\overset{o}{\Gamma}}: \quad \begin{cases} \delta g_{\mu\nu}: \; R_{\mu}^{\ \lambda} R_{\lambda\nu} - \tfrac{1}{4} g_{\mu\nu} R_{\rho\sigma} R^{\rho\sigma} = 0 \\ \delta \Gamma^{\sigma}_{\mu\nu}: \; \nabla_{\sigma} R^{\mu\nu} = 0\,, \end{cases} \tag{3.44}$$

$$\delta S_3[g,\Gamma]\Big|_{\Gamma=\overset{o}{\Gamma}}: \quad \begin{cases} \delta g_{\mu\nu}: \; R^{\rho}_{\ \cdot\sigma\mu\lambda} R^{\sigma}_{\ \rho\nu\cdot}{}^{\lambda} - \tfrac{1}{4} g_{\mu\nu} R^{\rho}_{\ \sigma\cdot\cdot}{}^{\lambda\kappa} R^{\sigma}_{\ \rho\lambda\kappa} = 0 \\ \delta \Gamma^{\sigma}_{\mu\nu}: \; \nabla_{\rho} R^{\rho}_{\ \sigma\mu\nu} = 0\,. \end{cases} \tag{3.45}$$

As discussed in Chapter 2, the Palatini variation of the Einstein-Hilbert action gives two field equations (for simplicity consider the free-field case): $\delta g_{\mu\nu}$ gives $R_{\mu\nu} = 0$ while $\delta\Gamma^{\sigma}_{\mu\nu}$ leads to the Christoffel form:

$$\overset{o}{\Gamma}{}^{\sigma}_{\mu\nu} = \tfrac{1}{2} g^{\sigma\lambda} (\partial_{\mu} g_{\lambda\nu} + \partial_{\nu} g_{\lambda\mu} - \partial_{\lambda} g_{\mu\nu})\,. \tag{3.46}$$

However, is not clear that imposing $\Gamma^{\sigma}_{\mu\nu} = \overset{o}{\Gamma}{}^{\sigma}_{\mu\nu}$ onto (3.42) should give anything useful since this relation originates from $\nabla_{\rho} g_{\mu\nu} = 0$ (and the Einstein-Hilbert action) while (3.42) is derived from (3.41). Furthermore, imposing this condition seems inconsistent with the Palatini approach - the consequences of each field equation for $\delta\Gamma^{\sigma}_{\mu\nu}$ in (3.42) should be worked out independently of (3.46) and therefore independently of the Einstein-Hilbert action. If this analysis leads to $\Gamma^{\sigma}_{\mu\nu} = \overset{o}{\Gamma}{}^{\sigma}_{\mu\nu}$ then fine. If not, then the secondary field equations of (3.42) should be considered over (3.45) as the starting point for a gauge theory of gravitation. In summary, the assumption of (3.46) after a Palatini style variation of $\delta S_3$ with respect to the connection constitutes the "gauge gravity" theory as discussed in the literature. At first glance this analysis would appear flawed and at least inconsistent with the Palatini variational procedure, but in the final section of this Chapter a constraint is obtained from this approach that places a restriction on the solutions satisfying (3.45). The condition is obtained by requiring equivalence between the field equations obtained from the standard variational procedure (Christoffel form is assumed at the outset) and the gauge gravity field equations that are derived by imposing the Christoffel form after the fact.

An important relation that will be used later in the analysis is discussed by Lanczos [6] who has shown that the field equations resulting from the variation of (3.41) are not all linearly independent since

$$\delta / \delta g_{\mu\nu} (S_1 - 4S_2 + S_3) \equiv 0 , \tag{3.47}$$

is an identity in a 4-dimensional spacetime when using the "standard" variational procedure. However, (3.47) is not an identity for the Palatini field equations obtained by imposing the Christoffel form of the connection after the variation. Therefore, (3.47) will have important consequences for the subsequent analysis. Furthermore, (3.47) may be expressed as the Euler topological invariant:

$$S_E[g] = \int d^4q \; g^{\frac{1}{2}} \varepsilon_{\rho\sigma\mu\nu} \varepsilon^{\lambda\kappa\omega\eta} R^{\rho\sigma}{}_{\cdot\,\lambda\kappa} R^{\mu\nu}{}_{\cdot\,\omega\eta} \, . \tag{3.48}$$

To see this, expand the product:

$$\varepsilon_{\rho\sigma\mu\nu} \varepsilon^{\lambda\kappa\omega\eta} = - \begin{vmatrix} \delta^\lambda_\rho & \delta^\lambda_\sigma & \delta^\lambda_\mu & \delta^\lambda_\nu \\ \delta^\kappa_\rho & \delta^\kappa_\sigma & \delta^\kappa_\mu & \delta^\kappa_\nu \\ \delta^\omega_\rho & \delta^\omega_\sigma & \delta^\omega_\mu & \delta^\omega_\nu \\ \delta^\eta_\rho & \delta^\eta_\sigma & \delta^\eta_\mu & \delta^\eta_\nu \end{vmatrix} , \tag{3.49}$$

and then (3.48) reduces to the *Gauss-Bonnet* form (for instance [112]; compare with (3.47) and (3.41)):

$$S_{GB}[g] = -\tfrac{1}{4} S_E[g] = \int d^4q \; g^{1/2} \left( R^2 - 4R_{\rho\sigma} R^{\rho\sigma} + R^{\rho}{}_{\sigma\mu\nu} R^{\sigma\cdot\,\mu\nu}_{\;\rho\cdot\cdot} \right) . \tag{3.50}$$

But since (3.48) is a topological invariant, inspection of (3.50) shows that (3.47) is true by identity.

The first set of equations (3.43) have been proposed by Littlewood [94] and then later considered by Pirani [95]. Pirani has shown that these equations have a solution leading to $1/6$ of the perihelion precession predicted by the Schwarzschild case, and in the opposite direction. The second set have been discussed by Misner and Wheeler [113] in their study of electromagnetism and gravitation as pure geometry. In their paper, the general relativistic field equations are coupled with electromagnetic sources (i.e. (2.85)) and then solved for the Maxwell field strength tensor, $F_{\rho\sigma}$, in terms of the Ricci tensor. Substituting this result into Maxwell's equations they obtain the coupled Einstein-Maxwell equations in purely geometric form (i.e. using only $R_{\rho\sigma}$). For the relevance of their work to $\delta S_2$ in (3.44), the algebraic condition that $F_{\rho\sigma}$ be expressed in terms of $R_{\rho\sigma}$ is equivalent to $R = 0$, and in addition that $R_{\mu}^{\ \lambda} R_{\lambda\nu} - \frac{1}{4} g_{\mu\nu} R_{\rho\sigma} R^{\rho\sigma} = 0$, is satisfied.

Returning to (3.41), the remaining discussion will focus on $S_3$ since this is the starting point for gauge gravity. This case has also been considered by Fairchild [33] given the minimal assumptions considered earlier in Chapter 2 that the connection is not assumed metric or symmetric. In the remaining part of this section the second $\delta S_3$ field equation in (3.45) will be derived by working in the tetrad basis. Therefore, the starting assumption is that the torsion is nonzero since $e_a^{\ \mu}$ and $\Gamma^b_{a\mu}$ are no longer dependent variables as may be seen from (3.37) and (3.38). Note however that the event connection is assumed to be metric to obtain (3.33)).

Since the "gauge gravity" theory consists of the field equations obtained with respect to the connection, we will consider in some detail a calculation of the secondary field equations for $\delta S_3$ in (3.45). To begin, the Lagrangian is taken in the first order form:

$$\begin{aligned} L_3[g,\Gamma] &= \tfrac{1}{4}\, g^{\rho\lambda} g^{\sigma\omega} R^{c}_{\ d\rho\sigma} R^{d}_{\ c\lambda\omega} \\ &- \tfrac{1}{2}\, g^{\rho\lambda} g^{\sigma\omega} R^{c}_{\ d\rho\sigma}\left(\partial_\rho \Gamma^{c}_{d\sigma} - \partial_\sigma \Gamma^{c}_{d\rho} + \Gamma^{c}_{f\rho}\Gamma^{f}_{d\sigma} - \Gamma^{c}_{f\sigma}\Gamma^{f}_{d\rho}\right), \end{aligned} \tag{3.51}$$

and the first set of field equations are expressed in the coordinate basis:

$$\delta g^{\mu\nu}: \quad H_{\mu\nu} \equiv R^{\rho}_{\ \sigma\mu\lambda} R^{\sigma\ \ \lambda}_{\ \rho\nu\cdot} - \tfrac{1}{4} g_{\mu\nu} R^{\rho\ \ \lambda\kappa}_{\ \sigma\cdot\cdot} R^{\sigma}_{\ \rho\lambda\kappa} = 0\,. \tag{3.52}$$

For simplicity we assume that no sources are present (Fairchild has considered the possibility for nonzero sources [26]). The field equations are thus calculated by variation with respect to the connection:

$$\left.\frac{\delta S_3[e,\Gamma]}{\delta \Gamma^{b}_{a\mu}}\right|_g = 0 = \int d^4q\, g^{\frac{1}{2}}\, \delta L_3 / \delta \Gamma^{b}_{a\mu}\,, \tag{3.53}$$

to obtain:

$$\left.\frac{\delta L_3[e,\Gamma]}{\delta\Gamma^{b}_{\ a\mu}}\right|_g e = -\tfrac{1}{2} R^{c\,\rho\sigma}_{\ d\cdot\cdot}\Big[\partial_\rho\Big(\delta\Gamma^{c}_{\ d\sigma}/\delta\Gamma^{b}_{\ a\mu}\Big) - \partial_\sigma\Big(\delta\Gamma^{c}_{\ d\rho}/\delta\Gamma^{b}_{\ a\mu}\Big)$$
$$+ (\delta\Gamma^{c}_{\ f\rho}/\delta\Gamma^{b}_{\ a\mu})\Gamma^{f}_{\ \cdot d\sigma} + \Gamma^{c}_{\ f\rho}(\delta\Gamma^{f}_{\ d\sigma}/\delta\Gamma^{b}_{\ a\mu})$$
$$- (\delta\Gamma^{c}_{\ f\sigma}/\delta\Gamma^{b}_{\ a\mu})\Gamma^{f}_{\ d\rho} - \Gamma^{c}_{\ f\sigma}(\delta\Gamma^{f}_{\ d\rho}/\delta\Gamma^{b}_{\ a\mu})\Big]\, e\,. \tag{3.54}$$

Noting that

$$R^{c\,\rho\sigma}_{\ d\cdot\cdot}\,\partial_\rho\Big(\delta\Gamma^{c}_{\ d\sigma}/\delta\Gamma^{b}_{\ a\mu}\Big)e = \partial_\rho\Big[R^{c\,\rho\sigma}_{\ d\cdot\cdot}\Big(\delta\Gamma^{c}_{\ d\sigma}/\delta\Gamma^{b}_{\ a\mu}\Big)\,e\Big]$$
$$- \Big(\delta\Gamma^{c}_{\ d\sigma}/\delta\Gamma^{b}_{\ a\mu}\Big)\,(\partial_\rho R^{c\,\rho\sigma}_{\ d\cdot\cdot})e - \Big(\delta\Gamma^{c}_{\ d\sigma}/\delta\Gamma^{b}_{\ a\mu}\Big)R^{c\,\rho\sigma}_{\ d\cdot\cdot}\,(\partial_\rho e)\,, \tag{3.55}$$

equation (3.54) becomes

$$0 = -\tfrac{1}{2}\Big\{\partial_\rho\Big[R^{c\,\rho\sigma}_{\ d\cdot\cdot}\Big(\delta\Gamma^{c}_{\ d\sigma}/\delta\Gamma^{b}_{\ a\mu}\Big)\,e\Big] - \Big(\delta\Gamma^{c}_{\ d\sigma}/\delta\Gamma^{b}_{\ a\mu}\Big)(\partial_\rho R^{c\,\rho\sigma}_{\ d\cdot\cdot})e$$
$$- \Big(\delta\Gamma^{c}_{\ d\sigma}/\delta\Gamma^{b}_{\ a\mu}\Big)R^{c\,\rho\sigma}_{\ d\cdot\cdot}(\partial_\rho e) - \partial_\sigma\Big[R^{c\rho\sigma}_{\ d\cdot\cdot}\Big(\delta\Gamma^{c}_{\ d\rho}/\delta\Gamma^{b}_{\ a\mu}\Big)\,e\Big]$$
$$+ \Big(\delta\Gamma^{c}_{\ d\rho}/\delta\Gamma^{b}_{\ a\mu}\Big)\,(\partial_\sigma R^{c\,\rho\sigma}_{\ d\cdot\cdot})e + \Big(\delta\Gamma^{c}_{\ d\rho}/\delta\Gamma^{b}_{\ a\mu}\Big)R^{c\,\rho\sigma}_{\ d\cdot\cdot}(\partial_\sigma e) \tag{3.56}$$
$$+ R^{c\,\rho\sigma}_{\ d\cdot\cdot}(\delta\Gamma^{c}_{\ f\rho}/\delta\Gamma^{b}_{\ a\mu})\Gamma^{f}_{\ \cdot d\sigma}\, e + R^{c\,\rho\sigma}_{\ d\cdot\cdot}\,\Gamma^{c}_{\ \cdot f\rho}(\delta\Gamma^{f}_{\ d\sigma}/\delta\Gamma^{b}_{\ a\mu})e$$
$$- R^{c\,\rho\sigma}_{\ d\cdot\cdot}(\delta\Gamma^{c}_{\ f\sigma}/\delta\Gamma^{b}_{\ a\mu})\Gamma^{f}_{\ d\rho} - R^{c\,\rho\sigma}_{\ d\cdot\cdot}\Gamma^{c}_{\ f\sigma}(\delta\Gamma^{f}_{\ d\rho}/\delta\Gamma^{b}_{\ a\mu})\Big]\, e\,,$$

and then setting the boundary terms:

$$\partial_\rho\Big[R^{c\,\rho\sigma}_{\ d\cdot\cdot}\Big(\delta\Gamma^{c}_{\ d\sigma}/\delta\Gamma^{b}_{\ a\mu}\Big)\,e\Big] \text{ and } \partial_\sigma\Big[R^{c\,\rho\sigma}_{\ d\cdot\cdot}\Big(\delta\Gamma^{c}_{\ d\rho}/\delta\Gamma^{b}_{\ a\mu}\Big)e\Big], \tag{3.57}$$

to zero this equation simplifies to

$$0 = -\tfrac{1}{2}\Big\{-\Big(\delta\Gamma^{c}_{\ d\sigma}/\delta\Gamma^{b}_{\ a\mu}\Big)\,(\partial_\rho R^{c\rho\sigma}_{\ d\cdot\cdot})e - \Big(\delta\Gamma^{c}_{\ d\sigma}/\delta\Gamma^{b}_{\ a\mu}\Big)R^{c\rho\sigma}_{\ d\cdot\cdot}(\partial_\rho e)$$
$$+ \Big(\delta\Gamma^{c}_{\ d\rho}/\delta\Gamma^{b}_{\ a\mu}\Big)(\partial_\sigma R^{c\,\rho\sigma}_{\ d\cdot\cdot})e - \Big(\delta\Gamma^{c}_{\ d\rho}/\delta\Gamma^{b}_{\ a\mu}\Big)R^{c\,\rho\sigma}_{\ d\cdot\cdot}\,(\partial_\sigma e)$$
$$+ R^{c\,\rho\sigma}_{\ d\cdot\cdot}(\delta\Gamma^{c}_{\ f\rho}/\delta\Gamma^{b}_{\ a\mu})\Gamma^{f}_{\ d\sigma}\, e + R^{c\,\rho\sigma}_{\ d\cdot\cdot}\Gamma^{c}_{\ f\rho}(\delta\Gamma^{f}_{\ d\sigma}/\delta\Gamma^{b}_{\ a\mu})e \tag{3.58}$$
$$- R^{c\,\rho\sigma}_{\ d\cdot\cdot}(\delta\Gamma^{c}_{\ f\sigma}/\delta\Gamma^{b}_{\ a\mu})\Gamma^{f}_{\ d\rho}\, e - R^{c\,\rho\sigma}_{\ d\cdot\cdot}\Gamma^{c}_{\ f\sigma}(\delta\Gamma^{f}_{\ d\rho}/\delta\Gamma^{b}_{\ a\mu})e\Big]\Big\}.$$

By using the identity

$$\delta\Gamma^{c}_{\ d\sigma}(q) = \int d^4q'\; e(q')\,\tfrac{1}{2}\Big(\delta^c_b\delta^a_d - g^{ca}g_{db}\Big)\delta^\mu_\sigma\;\delta\Gamma^{b}_{\ a\mu}\,\delta^4(q'-q)/e\,, \tag{3.59}$$

(3.58) expands to

$$
\begin{aligned}
0 = -\tfrac{1}{2}\Big\{ & -\tfrac{1}{2}\left(\delta^c_b\delta^a_d - g^{ca}g_{db}\right)\delta^\mu_\sigma(\partial_\rho R^{c\,\rho\sigma}_{d\cdot\cdot})e \\
& -\tfrac{1}{2}\left(\delta^c_b\delta^a_d - g^{ca}g_{db}\right)\delta^\mu_\sigma R^{c\,\rho\sigma}_{d\cdot\cdot}(\partial_\rho e) + \tfrac{1}{2}\left(\delta^c_b\delta^a_d - g^{ca}g_{db}\right)\delta^\mu_\rho(\partial_\sigma R^{c\,\rho\sigma}_{d\cdot\cdot})e \\
& -\tfrac{1}{2}\left(\delta^c_b\delta^a_d - g^{ca}g_{db}\right)\delta^\mu_\rho R^{c\,\rho\sigma}_{d\cdot\cdot}(\partial_\sigma e) + R^{c\,\rho\sigma}_{d\cdot\cdot}\tfrac{1}{2}\left(\delta^c_b\delta^a_f - g^{ca}g_{fb}\right)\delta^\mu_\rho\Gamma^f_{d\sigma}e \\
& + R^{c\,\rho\sigma}_{d\cdot\cdot}\Gamma^c_{f\rho}\tfrac{1}{2}\left(\delta^f_b\delta^a_d - g^{fa}g_{db}\right)\delta^\mu_\sigma e - R^{c\,\rho\sigma}_{d\cdot\cdot}\tfrac{1}{2}\left(\delta^c_b\delta^a_f - g^{ca}g_{fb}\right)\delta^\mu_\sigma\Gamma^f_{d\rho} \\
& - R^{c\,\rho\sigma}_{d\cdot\cdot}\Gamma^c_{f\sigma}\tfrac{1}{2}\left(\delta^f_b\delta^a_d - g^{fa}g_{db}\right)\delta^\mu_\rho\Big\} \times \delta^4(q'-q)/e.
\end{aligned}
\tag{3.60}
$$

Simplifying algebraically and then integrating this result, the following field equation is obtained (compare to (3.19)):

$$\partial_\rho(eR^{a\,\rho\mu}_{b\cdot\cdot}) + e\left(\Gamma^a_{cv}R^{c\,\rho\mu}_{b\cdot\cdot} - \Gamma^c_{bv}R^{a\,\rho\mu}_{c\cdot\cdot}\right) = 0. \tag{3.61}$$

Equation (3.61) may be expressed in the alternative form

$$\partial_\rho(R^{a\,\rho\mu}_{b\cdot\cdot}) + \Gamma^\rho_{v\rho}R^{a\,v\mu}_{b\cdot\cdot} + T_\rho R^{a\,\rho\mu}_{b\cdot\cdot} + \Gamma^a_{cv}R^{c\,\rho\mu}_{b\cdot\cdot} - \Gamma^c_{bv}R^{a\,\rho\mu}_{c\cdot\cdot} = 0, \tag{3.62}$$

after substituting (this equation may be derived from (2.30)):

$$\partial_\rho e = e\,(\Gamma^v_{\rho v} + T^v_{v\rho}) = e\,(\Gamma^v_{\rho v} + T_\rho), \tag{3.63}$$

into (3.61) and then switching dummy indices on the first term. Next use the definition of the covariant derivative:

$$\nabla_\rho R^{a\,\rho\mu}_{b\cdot\cdot} = \partial_\rho R^{a\,\rho\mu}_{b\cdot\cdot} + \Gamma^\rho_{v\rho}R^{a\,v\mu}_{b\cdot\cdot} - \tfrac{1}{2}T^{\ \mu}_{\rho v}R^{a\,\rho v}_{b\cdot\cdot} + \Gamma^a_{cv}R^{c\,\rho\mu}_{b\cdot\cdot} - \Gamma^c_{bv}R^{a\,\rho\mu}_{c\cdot\cdot},$$

and solve for $\partial_\rho R^{a\,\rho\mu}_{b\cdot\cdot}$ to obtain the generally covariant result:

$$\nabla_\rho R^{a\,\rho\mu}_{b\cdot\cdot} + T_\rho R^{a\,\rho\mu}_{b\cdot\cdot} + \tfrac{1}{2}T^{\ \mu}_{\rho v}R^{a\,\rho v}_{b\cdot\cdot} = 0. \tag{3.64}$$

The corresponding coordinate basis result is therefore:

$$\nabla_\rho R^{\lambda\,\rho\mu}_{\kappa\cdot\cdot} + T_\rho R^{\lambda\,\rho\mu}_{\kappa\cdot\cdot} + \tfrac{1}{2}T^{\ \mu}_{\rho v}R^{\lambda\,\rho v}_{\kappa\cdot\cdot} = 0. \tag{3.65}$$

If the torsion vanishes the gauge gravity field equations are thus obtained (lowering the $\lambda$ and $\mu$ indices, changing $\kappa \to v$, and noting symmetry on the first and last index pair):

$$Y_{\sigma\mu v} \equiv \nabla_\rho R^\rho_{\ \sigma\mu v} = 0. \tag{3.66}$$

The field equations (3.66) may be expressed in an equivalent form using the torsion-free Bianchi identity:

$$\nabla_\lambda R^\rho_{\ \sigma\mu v} + \nabla_\mu R^\rho_{\ \sigma v\lambda} + \nabla_v R^\rho_{\ \sigma\lambda\mu} \equiv 0. \tag{3.67}$$

Transvecting on $\lambda$ and $\rho$ (3.67) becomes

$$\nabla_\rho R^\rho_{\ \sigma\mu v} - \nabla_\mu R_{\sigma v} + \nabla_v R_{\sigma\mu} \equiv 0, \tag{3.68}$$

and therefore (3.66) is equivalent to

$$Y_{\sigma\mu\nu} \equiv \nabla_{[\mu} R_{\nu]\sigma} = 0 . \tag{3.69}$$

Contracting once again on (3.68) gives

$$\nabla_{\rho} R^{\rho}{}_{\mu} \equiv \tfrac{1}{2} \nabla_{\mu} R , \tag{3.70}$$

but then contracting (3.69) and combining with (3.70) shows that:

$$\nabla_{\mu} R = 0 \quad \Rightarrow \quad R = const.. \tag{3.71}$$

**Spherically Symmetric Solutions Part I:** (+)(+)

As a result of (3.71) solutions to (3.66) will be spaces of constant curvature although spaces of constant curvature need not be Einstein spaces (as noted by Schouten [83], p. 148; note: by definition - an Einstein space is one in which the Ricci tensor is a scalar multiple of the metric – see Petrov [114]). This statement is easily proven by counter-example which is provided by (3.73) and (3.74) discussed below. In addition, the fact that $R$ is constant does not necessarily imply that the metric satisfies (3.66) (although the reverse is certainly true as pointed out above). An example is given by (3.103) as discussed below.

An immediate result is that solutions to the Einstein field equations

$$R_{\mu\nu} = 0 , \tag{3.72}$$

satisfy (3.69). However, there have been objections to (3.69) on the grounds that in addition to (3.72), several other nonphysical solutions satisfy these equations. For example, Pavelle [31] and Thompson [32] have shown that

$$\begin{gathered} ds^2 = (1 + c_1/r)^{-2} dt^2 - (1 + c_1/r)^{-2} dr^2 - r^2 d\Omega^2 \\ d\Omega^2 = d\theta^2 + \sin^2\theta \, d\varphi^2 , \end{gathered} \tag{3.73}$$

is a solution to (3.69). But this is the solution already shown by Pirani [95] to give an erroneous value for the perihelion precession (see comment below (3.50)). Furthermore, the metric discussed by Thompson [32]:

$$ds^2 = dt^2 - (1 - c_1/r)^{-1} dr^2 - r^2 d\Omega^2 , \tag{3.74}$$

satisfies (3.69) but gives an incorrect Newtonian limit. Another class of solutions found by Ni [25]:

$$ds^2 = dt^2 - (1 + c_1/r + c_2 r^2)^{-1} dr^2 - r^2 d\Omega^2 , \tag{3.75}$$

satisfies (3.69) with $c_1$ and $c_2$ = constant, but gives an incorrect value for the redshift as noted by Ni.

The origin of these solutions may be clarified by summarizing the differential equations that follow from (3.52) and (3.70). To begin, assume a spherically symmetric solution in exponential form:

$$ds^2 = e^{2A(r)} dt^2 - e^{2B(r)} dr^2 - r^2 d\Omega^2 , \tag{3.76}$$

which greatly simplifies the structure of the differential equations that are considered in this Chapter compared to those obtained from the "standard" form, (2.49). The Christoffel symbols and Ricci tensor for this metric are thus given by (Appendix B):

$$\begin{gathered} \Gamma^t_{rt} = \Gamma^t_{tr} = A' \\ \Gamma^r_{tt} = e^{2(A-B)} ; \Gamma^r_{rr} = B' ; \Gamma^r_{\theta\theta} = e^{-2B} ; \Gamma^r_{\varphi\varphi} = -e^{-2B} r \sin^2\theta \\ \Gamma^\theta_{r\theta} = \Gamma^\theta_{\theta r} = \Gamma^\varphi_{r\varphi} = \Gamma^\varphi_{\varphi r} = 1/r ; \Gamma^\theta_{\varphi\varphi} = -\cos\theta \sin\theta ; \Gamma^\varphi_{\varphi\theta} = \cot\theta , \end{gathered} \tag{3.77}$$

and

$$\begin{aligned} R_{tt} &= e^{2(A-B)}[(A'' + A'^2 - A'B') + 2A'/r] = 0 \\ R_{rr} &= (A'' + A'^2 - A'B') - 2B'/r \\ R_{\theta\theta} &= -re^{-2B}[(A' - B') - (e^{2B} - 1)] \\ R_{\varphi\varphi} &= \sin^2\theta \; R_{\theta\theta} . \end{aligned} \tag{3.78}$$

respectively. The Ricci scalar simplifies to

$$R = 2e^{-2B}[(A'' + A'^2 - A'B') + 2(A' - B')/r - (e^{2B} - 1)/r^2] , \tag{3.79}$$

and for later reference the Riemann and Weyl tensors are also calculated in Appendix B.

Beginning first with the $H_{\mu\nu} = 0$ field equations (3.52) the nonzero components are given by:

$$\begin{aligned} H_{tt} &= e^{2(A-2B)}[(A'' + A'^2 - A'B')^2 + 2(A'^2 - B'^2)/r^2 - (e^{2B} - 1)^2/r^4] \\ H_{rr} &= -e^{-2B}[(A'' + A'^2 - A'B')^2 - 2(A'^2 - B'^2)/r^2 - (e^{2B} - 1)^2/r^4] \\ H_{\theta\theta} &= r^2 e^{-4B}[(A'' + A'^2 - A'B')^2 - (e^{2B} - 1)^2/r^4] \\ H_{\varphi\varphi} &= H_{\theta\theta} \sin^2\theta , \end{aligned} \tag{3.80}$$

and the "gauge gravity" equations (3.70) simplify to

$$Y_{trt} = e^{2(A-B)}[\tfrac{d}{dr}(A'' + A'^2 - A'B') - 2(B' - \tfrac{1}{r})(A'' + A'^2 - A'B') - 2A'/r^2]$$
$$Y_{\theta\theta r} = -re^{-2B}[(A'' + A'^2 - A'B') - (A'' - B'') + 2B'(A' - B') - (e^{2B} - 1)/r^2] \quad (3.81)$$
$$Y_{ttr} = -Y_{trt} \;;\; Y_{\theta r\theta} = -Y_{\theta\theta r} \;;\; Y_{\varphi\varphi r} = -Y_{\varphi r\varphi} = Y_{\theta\theta r}\sin^2\theta \;.$$

Considering the first set of equations (3.80), by inspection there are four obvious solution possibilities:

$$\begin{aligned} A'' + A'^2 - A'B' &= \pm\,(e^{2B} - 1)/r^2 \\ A' &= \pm B'. \end{aligned} \quad (3.82)$$

Considering (3.81) separately leads to the system:

$$\begin{aligned} &\tfrac{d}{dr}(A'' + A'^2 - A'B') - 2(B' - \tfrac{1}{r})(A'' + A'^2 - A'B') - 2A'/r^2 = 0 \\ &(A'' + A'^2 - A'B') - (A'' - B'') + 2B'(A' - B') - (e^{2B} - 1)^2/r^2 = 0 \;, \end{aligned} \quad (3.83)$$

and then by substituting the various possibilities from (3.82) into (3.83) (i.e., $(+)(+)$, $(+)(-)$, $(-)(+)$, $(-)(-)$) it is apparent that only two cases will give a simultaneous solution of both $H_{\mu\nu} = 0$ and $Y_{\sigma\mu\nu} = 0$. These correspond to the $(+)(+)$ and $(-)(-)$ equations (see Appendix B).

To begin, the $(+)(+)$ differential equations of (3.82), are given by

$$\begin{aligned} A'' + A'^2 - A'B' &= +\,(e^{2B} - 1)/r^2 \\ A' &= +B', \end{aligned} \quad (3.84)$$

and as noted above satisfy $Y_{\sigma\mu\nu} = 0$ identically. Combining the two equations in (3.84) gives a second order, nonlinear, inhomogeneous differential equation:

$$B'' = (e^{2B} - 1)/r^2 \,. \quad (3.85)$$

Equation (3.85) is expressed in an alternative form using the variable substitution:

$$B \to -\tfrac{1}{2}\ln\Lambda(r)\,, \quad (3.86)$$

to give

$$\begin{aligned} &\Lambda''/2\Lambda - \Lambda'^2/2\Lambda^2 + 1/r^2\Lambda = 1/r^2 \\ &\Rightarrow \frac{d}{dr}\left(\frac{\Lambda'}{2\Lambda} + \frac{1}{r}\right) + \frac{1}{r^2\Lambda} = 0 \;. \end{aligned} \quad (3.87)$$

A solution to this differential equation is given by the Pavelle-Thompson solution, $\Lambda = (1 + c_1/r)^2$, where $c_1$ is an integration constant. It is worthwhile to pause for a moment to summarize the forms of the metric that result from the second $\pm$ equation of (3.82) using the variable substitution in (3.86):

$$\begin{aligned}
&\underbrace{(+)(+)}_{A=+B} \Rightarrow ds^2 = \Lambda^{-1} dt^2 - \Lambda^{-1} dr^2 - r^2 d\Omega^2 \\
&\underbrace{(+)(-)}_{A=-B} \Rightarrow ds^2 = \Lambda\, dt^2 - \Lambda^{-1} dr^2 - r^2 d\Omega^2 \\
&\underbrace{(-)(+)}_{A=+B} \Rightarrow ds^2 = \Lambda^{-1} dt^2 - \Lambda^{-1} dr^2 - r^2 d\Omega^2 \\
&\underbrace{(-)(-)}_{A=-B} \Rightarrow ds^2 = \Lambda\, dt^2 - \Lambda^{-1} dr^2 - r^2 d\Omega^2,
\end{aligned} \tag{3.88}$$

noting that only two of these forms are distinct. The Pavelle-Thompson solution has the form of the first equation in (3.88) with $\Lambda$ given above.

## Note on The Vanishing Ricci Scalar and Weyl Tensor Solutions

The Pavelle-Thompson metric (3.73) also has a vanishing Ricci scalar and Weyl tensor which has consequences for the theory of conformal gravity discussed in Chapter 5 as well as for the gauge gravity field equations. To see this note that the Weyl tensor calculated from (3.76) has 24 nonzero components, but interestingly every component contains the following differential equation as a common factor:

$$C^{\rho}{}_{\sigma\mu\nu} \sim [(A'' + A'^2 - A'B') - (A' - B')/r - (e^{2B} - 1)/r^2]\,. \tag{3.89}$$

Comparing (3.89) with the Ricci scalar of (3.79), the following result is obtained:

$$\begin{matrix} A'' + A'^2 - A'B' = +\,(e^{2B} - 1)/r^2 \\ A = +B \end{matrix} \Rightarrow \begin{cases} R = 0 \\ C^{\rho}{}_{\sigma\mu\nu} = 0\ , \end{cases} \tag{3.90}$$

which is just the $(+)(+)$ case discussed above. Therefore, we make the observation that a vanishing Ricci scalar and Weyl tensor gives a class of simultaneous solutions to the gauge gravity field equations, i.e., both $H_{\mu\nu} = 0$ and $Y_{\sigma\mu\nu} = 0$ are satisfied. The Pavelle-Thompson solution is just one example.

The theory originating from the field equations (3.90) is Nordström's theory [34] and was noted previously by Ni [25] and later by Baekler and Yasskin [37] as a simultaneous solution to the gauge gravity field equations. But also note that other differential equations may be satisfied that give sufficient conditions for a vanishing Ricci scalar:

$$\begin{matrix} A'' + A'^2 - A'B' = 0 \\ A' = B' + (e^{2B} - 1)/2r \end{matrix} \Rightarrow \{ \; R = 0 \, , \tag{3.91}$$

$$A = const. \Rightarrow \left[ B' = -(e^{2B} - 1)/2r \right] \Rightarrow \{ \; R = 0 \, , \tag{3.92}$$

$$A = -B \Rightarrow \left[ r^2(B'' - 2B'^2) + 4r\,B' = (e^{2B} - 1) \right] \Rightarrow \{ \; R = 0 \, , \tag{3.93}$$

and the Weyl tensor is considered separately:

$$\begin{matrix} A'' + A'^2 - A'B' = 0 \\ A' = B' - (e^{2B} - 1)/r \end{matrix} \Rightarrow \{ \; C^{\rho}_{\ \sigma\mu\nu} = 0 \, , \tag{3.94}$$

$$A = const. \Rightarrow \left[ B' = (e^{2B} - 1)/r = 0 \right] \Rightarrow \{ \; C^{\rho}_{\ \sigma\mu\nu} = 0 \, , \tag{3.95}$$

$$A = -B \Rightarrow \left[ r^2(B'' - 2B'^2) - 2r\,B' = -(e^{2B} - 1) \right] \Rightarrow \{ \; C^{\rho}_{\ \sigma\mu\nu} = 0 \, , \tag{3.96}$$

which are not equivalent to the simultaneous condition of (3.90), i.e., there could be cases where the Ricci scalar vanishes (recall that this case is Littlewood's theory [94]), but the Weyl tensor does not and vice-versa.

Considering first the Ricci scalar equations (3.91), eliminate $A'$ using the second equation and then the first equation simplifies to:

$$r^2B'' + (3e^{2B} - 1)rB'/2 + (e^{2B} - 1)(e^{2B} - 3)/4 = 0 \, , \tag{3.97}$$

which may be expressed in the alternative form using $B = -\frac{1}{2}\ln\Lambda(r)$:

$$2r^2(\Lambda'^2 - \Lambda\,\Lambda'') + r(\Lambda - 3)\Lambda' + (3\Lambda - 1)(\Lambda - 1) = 0 \, . \tag{3.98}$$

Although solutions of (3.98) imply that the Ricci scalar vanishes, the converse is not necessarily true since $\Lambda = (1 + c_1/r^2)^2$ (of the Pavelle-Thompson solution) does not satisfy (3.98), but $\Lambda = (1 + c_1/r)^{-1}$ of the Thompson solution (3.74) does. The difference is given by the assumed relation between $A$ and $B$: the Pavelle-Thompson solution corresponds to $A' = B'$ from (3.90), while $A'$ and $B'$ of the Thompson solution satisfies (3.97) with $A' = const.$. However, (3.97) could have more general solutions other than the Thompson solution. The case, $A' = const.$, is just one special form and is discussed next.

The second differential equation resulting from (3.92), $A' = const.$, is given by

$$B' = -(e^{2B} - 1)/2r\,, \tag{3.99}$$

which may be expressed using the change of variable $B = -\frac{1}{2}\ln\Lambda(r)$:

$$r\Lambda' + \Lambda - 1 = 0\,. \tag{3.100}$$

The differential equation (3.100) is identical to the differential equation discussed earlier for the Schwarzschild case and therefore, $\Lambda = 1 + c_1/r$, is the solution. As a result, this metric gives a (trivially) more general form of Thompson's solution (3.74) (since the term preceding $dt^2$ may be an arbitrary constant). The last case of $R = 0$ is obtained by substituting $A = -B$ into (3.79) and is given by (3.93). The differential equation is thus:

$$r^2(B'' - 2B'^2) + 4rB' = (e^{2B} - 1)\,, \tag{3.101}$$

which may be expressed using the change of variable (3.86):

$$r^2\Lambda'' + 4r\Lambda' + 2(\Lambda - 1) = 0\,. \tag{3.102}$$

This case is satisfied by the Schwarzschild function, $\Lambda = 1 + c_1/r$, as may be verified by direct substitution, but is not satisfied by either the Pavelle-Thompson or Thompson solutions. In addition, $\Lambda = 1 + c_1/r^2$, also satisfies (3.102) and therefore an additional vanishing Ricci scalar solution is given by

$$ds^2 = (1 + c_1/r^2)\,dt^2 - (1 + c_1/r^2)^{-1}dr^2 - r^2d\Omega^2\,. \tag{3.103}$$

However, one may check by direct substitution that (3.103) does not satisfy the $H_{\mu\nu} = 0$ or $Y_{\sigma\mu\nu} = 0$ field equations, nor does it have a vanishing Weyl tensor (Appendix B). Interestingly, note also that (3.103) corresponds to the Reissner-Nordström solution of the Einstein field equations with zero mass.†

Next consider the vanishing Weyl tensor equations (3.94). The second equation is now used to eliminate $A'$ so the first equation simplifies to

$$r^2B'' - (3e^{2B} - 1)rB' + e^{2B}(e^{2B} - 1) = 0\,, \tag{3.104}$$

which may be expressed in the alternative form using (3.86):

$$r^2(\Lambda\Lambda'' - \Lambda'^2) + r(\Lambda - 3)\Lambda' + 2(\Lambda - 1) = 0\,. \tag{3.105}$$

Equation (3.105) is satisfied by

---

† I thank Jim Crawford for pointing out this detail.

$$\Lambda = (1 + c_1 r)^{-1} , \tag{3.106}$$

and also

$$\Lambda = (1 + c_1 r^2)^{-1} . \tag{3.107}$$

However, in either case, we would still have to solve for $A'$ from the differential equation: $A' = B' - (e^{2B} - 1)/r$, to obtain the full solution. Next consider the case when $A'$ is constant giving

$$B' = (e^{2B} - 1)/r = 0 . \tag{3.108}$$

Using (3.86) results in the following differential equation in $\Lambda$:

$$r\Lambda' - 2\Lambda + 2 = 0 , \tag{3.109}$$

which has the solution:

$$\Lambda = 1 + c_1 r^2 . \tag{3.110}$$

Equation (3.110) has a vanishing Weyl tensor, but does not satisfy (3.90). However, this equation also originates from the $(-)(-)$ equations of (3.82) as a special case of (3.119) (which is discussed at greater length below), and therefore satisfies both the $H_{\mu\nu} = 0$ and $Y_{\sigma\mu\nu} = 0$ equations. The $A = -B$ case is given by (3.96) and results in the differential equation:

$$r^2(B'' - 2B'^2) - 2rB' = -(e^{2B} - 1) , \tag{3.111}$$

which is expressed using (3.86):

$$r^2\Lambda'' - 2r\Lambda' - 2\Lambda - 2 = 0 . \tag{3.112}$$

A solution to (3.112) is given by

$$\Lambda = 1 + c_1 r + c_2 r^2 , \tag{3.113}$$

and therefore

$$ds^2 = (1 + c_1 r + c_2 r^2)\, dt^2 - (1 + c_1 r + c_2 r^2)^{-1} dr^2 - r^2 d\Omega^2 , \tag{3.114}$$

gives a solution to the $C^{\rho}_{\sigma\mu\nu} = 0$ field equations as well as the special cases: $c_1$ or $c_2 = 0$. However, (3.114) does not satisfy either gauge gravity field equation since the Ricci scalar is nonzero in this case, but the solution does have consequences for the conformally invariant theory of gravitation discussed in Chapter 5 (as does as the special case, $c_2 = 0$).

**Spherically Symmetric Solutions Part II:** $(+)(-)$**,** $(-)(+)$**,** $(-)(-)$

Continuing with the other solution possibilities of (3.82), it is important to note that the first two ($(+)(-)$ and $(-)(+)$) are not necessarily solutions of both gauge gravity field equations but rather only for (3.82). Beginning with the $(+)(-)$ equations, these simplify to the following differential equation:

$$r^2\Lambda'' + 2\Lambda - 2 = 0\,, \tag{3.115}$$

which may be integrated to give

$$\Lambda = 1 + r^{1/2}\left[c_1\cos(\ln r^{\sqrt{7}/2}) - c_2\sin(\ln r^{\sqrt{7}/2})\right], \tag{3.116}$$

with $c_1$ and $c_2$ integration constants. Equation (3.116) gives an unreported exact solution to the $H_{\mu\nu} = 0$ equations but does not satisfy $Y_{\sigma\mu\nu} = 0$. Nevertheless, the solution has some interest given that all other previously reported exact solutions of the gauge gravity equations, besides the Pavelle Thompson solution, satisfy only $Y_{\sigma\mu\nu} = 0$ and not $H_{\mu\nu} = 0$.

The structure of the $(-)(+)$ equations of (3.82) is similar in form to the $(+)(+)$ equation (3.87):

$$\begin{gathered}\Lambda''/2\Lambda - \Lambda'^2/2\Lambda^2 - 1/r^2\Lambda = -1/r^2 \\ \Rightarrow \frac{d}{dr}\left(\frac{\Lambda'}{2\Lambda} - \frac{1}{r}\right) - \frac{1}{r^2\Lambda} = 0\,.\end{gathered} \tag{3.117}$$

However, the Pavelle-Thompson solution does not work in this case. Furthermore, there are no other nontrivial solutions known to satisfy (3.117). Finally, the $(-)(-)$ equations give the second set of equations that satisfy $Y_{\sigma\mu\nu} = 0$ identically. In this case, (3.82) simplifies to:

$$r^2\Lambda'' - 2\Lambda + 2 = 0\,, \tag{3.118}$$

which may be integrated to give the solution:

$$\begin{gathered}ds^2 = \Lambda dt^2 - \Lambda^{-1}dr^2 - r^2 d\Omega^2 \\ \Lambda = (1 + c_1/r + c_2 r^2)\,.\end{gathered} \tag{3.119}$$

Equation (3.119) is the solution first obtained by Kottler [115] as a solution to Einstein's field equations with cosmological constant, $c_2$:

$$R_{\mu\nu} = 3c_2 \, g_{\mu\nu} \,. \tag{3.120}$$

The solution was later noted as a solution to $Y_{\sigma\mu\nu} = 0$ by Pavelle [23]† and also by Fairchild [33]. As a result, this metric has the important property that its Ricci tensor is a scalar multiple of the metric and is therefore an Einstein space.

There are other solutions to the gauge gravity equations that do not necessarily satisfy the $H_{\mu\nu} = 0$ equations. One case to consider is by setting $A'$ to zero in (3.76) which results in (3.118) and therefore (3.119). But there is an important difference in this case since the coefficient of $dt^2$ is constant. This particular form was noted by Ni and was listed earlier in (3.75):

$$ds^2 = dt^2 - (1 + c_1/r + c_2 r^2)^{-1} dr^2 - r^2 d\Omega^2 \,. \tag{3.121}$$

The extraneous solution (3.74) noted by Pavelle [31] and Fairchild [26] is simply a special case of Ni's solution with $c_2 = 0$:

$$ds^2 = dt^2 - (1 + c_1/r)^{-1} dr^2 - r^2 d\Omega^2 \,. \tag{3.122}$$

Another special case is obtained when $c_1 = 0$ which gives the Einstein universe solution:

$$ds^2 = dt^2 - (1 + c_2 r^2)^{-1} dr^2 - r^2 d\Omega^2 \,, \tag{3.123}$$

which was originally pointed out as a gauge gravity solution by Thompson [32]. The Einstein universe solution also has a vanishing Weyl tensor but the Ricci scalar is non vanishing ($R = 6c_2$) and is discussed in more detail below.

As a check one might wonder if setting $B'$ to zero in the $Y_{\sigma\mu\nu} = 0$ equations would lead to any other solutions. But the only solution in this case is given by $A = const.$ since every component of the resulting equations contains a factor of $A'$. As a result, this case is trivial. Another possibility to consider is setting $A'' + A'^2 - A'B' = 0$ in (3.81). In this case the equations simplify to

$$2A'/r^2 = 0$$
$$(A'' - B'') - 2B'(A' - B') = -(e^{2B} - 1)/r^2\,, \tag{3.124}$$

which are only consistent if $A = const.$. Therefore the second equation reduces to the differential equation:

$$B'' - 2B'^2 = (e^{2B} - 1)/r^2\,, \tag{3.125}$$

or in terms of $\Lambda$:

$$r^2\Lambda'' - 2\Lambda + 2 = 0\,, \tag{3.126}$$

which again results in the Ni solution but in a trivially more general form since then $A$ is an arbitrary constant.

---

† note, however, that Pavelle's listing of Kottler's solution is incorrect: $\Lambda = (-c_1/r + c_2 r^2)$, which does not satisfy the gauge gravity field equations.

**Auxiliary Algebraic Constraints**

Given these extraneous solutions to the gauge gravity field equations, Pavelle [31] and Thompson [32] have suggested that there should be an additional constraint placed on the solution set of (3.69) to eliminate the nonphysical solutions. Following this suggestion, Fairchild [33] has proposed that the constraint be given by the "second set" of field equations obtained from the first equation of $\delta S_3$ in (3.45) - by variation with respect to $\delta g_{\mu\nu}$. However, this approach turned out to be incorrect as later noted by Fairchild [116, 42†]. The purpose of this section is to discuss a condition that eliminates several (but not all) of the extraneous solutions to the gauge gravity equations. The condition is derived by first noting that the Einstein-Hilbert action is special in that its variation gives identical results using either the standard or Palatini variational procedures. Specifically, the field equations that are derived from the Einstein-Hilbert action by assuming the connection is in Christoffel form from the outset are equivalent to those obtained when the connection is taken as an independent variational parameter. However, this "symmetry" no longer applies when the action is taken in quadratic form and the gauge gravity "Palatini" procedure is applied – which should be distinguished from the true Palatini procedure where the connection and metric are completely independent (as considered in Chapter 2). As a result, imposing this condition onto the resulting field equations originating from the quadratic curvature Lagrangians leads to an auxiliary algebraic constraint that restricts the class of spacetime solutions satisfying the gauge gravity field equations. These constraints may be equivalently derived as a contraction of the integrability conditions for the Ricci tensor as discussed below.

† see footnote 27 of Fairchild's paper.

First calculate the metric variation of the Gauss-Bonnet action (3.50) using the "standard" variational procedure. The result of this calculation is given by (Appendix B - see also Parker; Christensen [117]):

$$\begin{aligned} 0 \equiv & - R\left(R_{\mu\nu} - \tfrac{1}{4} g_{\mu\nu} R\right) - \left(R_{\mu\lambda\rho\sigma} R_{\nu\cdot\cdot\cdot}^{\cdot\lambda\rho\sigma} - \tfrac{1}{4} g_{\mu\nu} R_{\rho\sigma\lambda\kappa} R^{\rho\sigma\lambda\kappa}\right) \\ & + 2(R^{\rho}_{\cdot\mu} R_{\rho\nu} - \tfrac{1}{2} g_{\mu\nu} R_{\rho\sigma} R^{\rho\sigma}) + 2R^{\sigma}_{\cdot\rho} R^{\rho}_{\cdot\mu\sigma\nu} \ , \end{aligned} \tag{3.127}$$

noting especially that (3.127) is an identity. It is worthwhile to pause for a moment and consider another useful identity that may be derived from (3.127) using the decomposition (see for example Carmeli [93], p. 72; or Weinberg [92], p. 145):

$$\begin{aligned} R_{\rho\sigma\mu\nu} = & \ C_{\rho\sigma\mu\nu} + \tfrac{1}{2}\left(g_{\rho\mu} R_{\sigma\nu} - g_{\rho\nu} R_{\sigma\mu} - g_{\sigma\mu} R_{\rho\nu} + g_{\sigma\nu} R_{\rho\mu}\right) \\ & + \tfrac{1}{6}\left(g_{\rho\nu}\, g_{\sigma\mu} - g_{\rho\mu}\, g_{\sigma\nu}\right) R \ , \end{aligned} \tag{3.128}$$

where $C_{\rho\sigma\mu\nu}$ is the Weyl tensor defined in (3.128) (this quantity is discussed in more detail in Chapter 4). Eliminating the Riemann tensor from (3.127) using (3.129) gives the result (the algebra is listed in Appendix B):

$$C_{\mu\lambda\rho\sigma} C_{\nu\cdot\cdot\cdot}^{\cdot\lambda\rho\sigma} - \tfrac{1}{4} g_{\mu\nu} C_{\rho\sigma\lambda\kappa} C^{\rho\sigma\lambda\kappa} \equiv 0 \ , \tag{3.129}$$

which is Pirani's conformal identity (Ref. [118], p. 317),† and may be summarized by the relation:

$$\frac{\delta S_{Gauss-Bonnet}}{\delta g^{\mu\nu}} \equiv 0 \ \Rightarrow \ C_{\mu\lambda\rho\sigma} C_{\nu\cdot\cdot\cdot}^{\cdot\lambda\rho\sigma} - \tfrac{1}{4} g_{\mu\nu} C_{\rho\sigma\lambda\kappa} C^{\rho\sigma\lambda\kappa} \equiv 0 \ . \tag{3.130}$$

Therefore, Pirani's identity is the "field equation" originating from the Euler-Gauss-Bonnet action using the standard variational procedure.

Continuing with (3.127), add and subtract $\tfrac{1}{2} g_{\mu\nu} R_{\rho\sigma} R^{\rho\sigma}$ to obtain the result:

---

† note: this identity was only suggested by Pirani as an exercise and is not necessarily referred to as "Pirani's" identity in the literature (the derivation above gives a solution to it; an alternative solution is derived by Parker and Christensen [117]). But since Pirani has suggested the proof of this equation as an exercise we will refer to it as Pirani's conformal identify for lack of a better label.

$$0 \equiv -R\left(R_{\mu\nu} - \tfrac{1}{4}g_{\mu\nu}R\right) + \left(R_{\mu\lambda\rho\sigma}R_{\nu\cdot\cdot\cdot}^{\cdot\lambda\rho\sigma} - \tfrac{1}{4}g_{\mu\nu}R_{\rho\sigma\lambda\kappa}R^{\rho\sigma\lambda\kappa}\right)$$
$$+2(R^{\rho}_{\cdot\mu}R_{\rho\nu} - \tfrac{1}{4}g_{\mu\nu}R_{\rho\sigma}R^{\rho\sigma}) + 2\left(R^{\sigma}_{\cdot\rho}R^{\rho}_{\cdot\mu\sigma\nu} - \tfrac{1}{4}g_{\mu\nu}R_{\rho\sigma}R^{\rho\sigma}\right), \quad (3.131)$$

recalling that this field equation is an identity obtained from the Lanczos linear combination:

$$\delta/\delta g_{\mu\nu}(S_1 - 4S_2 + S_3) \equiv 0, \quad (3.132)$$

using a standard variational procedure. However, by substituting the field equations that are obtained with respect to $\delta g^{\mu\nu}$ from (3.43), (3.44), and (3.45), (i.e., by assuming the connection is in Christoffel form after the variation) then (3.132) (or equivalently (3.131) ) is no longer an identity. A basic result is that (3.131) is satisfied (with the connection in Christoffel form after variation) only if the following condition is met:

$$R^{\sigma}_{\ \rho}R^{\rho}_{\ \mu\sigma\nu} = R^{\rho}_{\ \mu}R_{\rho\nu}, \quad (3.133)$$

which is not an identity as may by checked by substituting several of the nonphysical solutions discussed in the preceding section. It is emphasized that (3.43), (3.44), and (3.45) were derived using the Palatini style variation of the quadratic curvature Lagrangians, and then by imposing the Christoffel form afterward. As a result, (3.133) gives a necessary condition that the standard and "gauge gravity Palatini" procedures give identical results.

The condition is also obtained by requiring the field equations resulting from $S_3$ of (3.41) to satisfy the same requirement, but in a weaker form. To derive this result consider the field equations resulting from the gauge gravity Palatini procedure applied to $S_3$ of (3.41) which are re-listed below:

$$\delta S_3: \begin{cases} \delta g_{\mu\nu}: \ R^{\rho}_{\ \sigma\mu\lambda}R^{\sigma\ \ \lambda}_{\ \rho\nu\cdot} - \tfrac{1}{4}g_{\mu\nu}R^{\rho\ \lambda\kappa}_{\ \sigma\cdot\cdot}R^{\sigma}_{\ \rho\lambda\kappa} = 0 \\ \delta\Gamma^{\sigma}_{\mu\nu}: \ \nabla_{\rho}R^{\rho}_{\ \sigma\mu\nu} = 0\,. \end{cases} \quad (3.134)$$

The corresponding standard variational results are given by (Appendix B):

$$\delta g_{\mu\nu}: (R_{\mu\lambda\rho\sigma}R_{\nu\cdot\cdot\cdot}^{\cdot\lambda\rho\sigma} - \tfrac{1}{4}g_{\mu\nu}R_{\rho\sigma\lambda\kappa}R^{\rho\sigma\lambda\kappa}) + 2\nabla_{\sigma}\nabla_{\rho}R^{\rho\ \sigma\cdot}_{\ \mu\cdot\nu} = 0, \quad (3.135)$$

which is the field equation considered by Eddington (this was noted in Ref. [37]). By inspection it is clear that a sufficient condition for (3.135) to be satisfied is given by both field equations of (3.134). In fact, since the Pavelle-Thompson solution (3.73) satisfies both field equations of (3.134), the field equations resulting from the "standard" variational procedure also admit a nonphysical solution. In fact, there could be even more solutions to (3.135) than merely those of (3.134) alone. For example, there might be cases where $\nabla_\rho R^{\rho\ \sigma\ \cdot}_{\ \mu\cdot\nu} \neq 0$, but $\nabla_\sigma\nabla_\rho R^{\rho\ \sigma\ \cdot}_{\ \mu\cdot\nu} = 0$, is satisfied. However, if (3.135) is required to vanish using only the $\delta g^{\mu\nu}$ field equation of (3.134), then it is straightforward to show that the vanishing of the second term implies that the auxiliary condition (3.133) is satisfied for a restricted class of spacetimes. To see this first note that the second term of (3.135) may be written as (using the Bianchi identity (3.67)):

$$\nabla_\sigma(\nabla_\rho R^\rho_{\ \mu\kappa\nu}) \equiv \nabla_\sigma(\nabla_\kappa R_{\mu\nu} - \nabla_\nu R_{\mu\kappa}) . \tag{3.136}$$

Raising the $\kappa \to \sigma$ index gives

$$\nabla_\sigma(\nabla_\rho R^{\rho\ \sigma}_{\ \mu\cdot\nu}) \equiv \nabla_\sigma\nabla^\sigma R_{\mu\nu} - \nabla_\sigma\nabla_\nu R^\sigma_{\ \mu} , \tag{3.137}$$

and note also for later reference that (this is equation (3.70)):

$$g^{\mu\nu}\nabla_\rho R^\rho_{\ \mu\kappa\nu}(= \nabla_\nu R^\nu_{\ \kappa}) \equiv \nabla_\kappa R - \nabla_\nu R^\nu_{\ \kappa} \Rightarrow \nabla_\nu R^\nu_{\ \kappa} \equiv \tfrac{1}{2}\nabla_\kappa R . \tag{3.138}$$

But since

$$\begin{aligned}(\nabla_\sigma\nabla_\nu - \nabla_\nu\nabla_\sigma) R^\sigma_{\ \mu} &= R^\rho_{\ \mu} R^\sigma_{\ \rho\sigma\nu} - R^\sigma_{\ \rho} R^\rho_{\ \mu\sigma\nu} \\ &= R^\rho_{\ \mu} R_{\rho\nu} - R^\sigma_{\ \rho} R^\rho_{\ \mu\sigma\nu} ,\end{aligned} \tag{3.139}$$

the $\nabla_\sigma\nabla_\nu R^\sigma_{\ \mu}$ term on the RHS of (3.137) is expressed using (3.139):

$$\nabla_\sigma\nabla_\nu R^\sigma_{\ \mu} \equiv R^\rho_{\ \mu} R_{\rho\nu} - R^\sigma_{\ \rho} R^\rho_{\ \mu\sigma\nu} + \nabla_\nu\nabla_\sigma R^\sigma_{\ \mu} , \tag{3.140}$$

and after using (3.138) this term becomes

$$\nabla_\sigma\nabla_\nu R^\sigma_{\ \mu} \equiv R^\rho_{\ \mu} R_{\rho\nu} - R^\sigma_{\ \rho} R^\rho_{\ \mu\sigma\nu} + \tfrac{1}{2}\nabla_\mu\nabla_\nu R . \tag{3.141}$$

Finally, substituting (3.141) into (3.137), the final result is obtained:

$$\nabla_\sigma\nabla_\rho R^{\rho\ \sigma}_{\ \mu\cdot\nu} \equiv \nabla^2 R_{\mu\nu} - (R^\rho_{\ \mu} R_{\rho\nu} - R^\sigma_{\ \rho} R^\rho_{\ \mu\sigma\nu}) - \tfrac{1}{2}\nabla_\mu\nabla_\nu R . \tag{3.142}$$

Therefore, the field equation originating from the standard variational procedure may be expressed in the equivalent form:

$$\begin{aligned}(R_{\mu\lambda\rho\sigma} R_{\nu\cdot\cdot\cdot}^{\ \cdot\lambda\rho\sigma} - \tfrac{1}{4} g_{\mu\nu} R_{\rho\sigma\lambda\kappa} R^{\rho\sigma\lambda\kappa}) - 2(R^\rho_{\ \mu} R_{\rho\nu} - R^\sigma_{\ \rho} R^\rho_{\ \mu\sigma\nu}) \\ + 2\nabla^2 R_{\mu\nu} - \nabla_\mu\nabla_\nu R = 0 .\end{aligned} \tag{3.143}$$

It is worthwhile to pause for a moment to summarize the field equations obtained using the standard and "gauge gravity" Palatini variational methods. The quotation marks on "Palatini" in the following listing emphasizes that these field equations are not "true"

Palatini results, but follow by imposing the Christoffel form of the connection as discussed earlier (the standard variational results listed below are derived in Appendix B):

$$\delta S_1 : \begin{bmatrix} \text{"Palatini"} : \begin{cases} \delta g_{\mu\nu} : \; R\left(R_{\mu\nu} - \frac{1}{4} g_{\mu\nu} R\right) = 0 \\ \delta \Gamma^{\sigma}_{\mu\nu} : \; g^{\mu\nu} \nabla_{\sigma} R = 0 \end{cases} \\ \text{Standard:} \left\{ \delta g_{\mu\nu} : \; R\left(R_{\mu\nu} - \frac{1}{4} g_{\mu\nu} R\right) - (\nabla_\mu \nabla_\nu R - g_{\mu\nu} \nabla^2 R) = 0 \right. , \end{bmatrix} \tag{3.144}$$

$$\delta S_2 : \begin{bmatrix} \text{"Palatini"} : \begin{cases} \delta g_{\mu\nu} : \; R^{\lambda}_{\cdot\mu} R_{\lambda\nu} - \frac{1}{4} g_{\mu\nu} R_{\rho\sigma} R^{\rho\sigma} = 0 \\ \delta \Gamma^{\sigma}_{\mu\nu} : \nabla_{\sigma} R^{\mu\nu} = 0 \end{cases} \\ \text{Standard:} \begin{cases} \delta g_{\mu\nu} : \; 2(R^{\lambda}_{\cdot\mu} R_{\lambda\nu} - \frac{1}{4} g_{\mu\nu} R_{\rho\sigma} R^{\rho\sigma}) - 2\nabla_\rho \nabla_\nu R^{\rho}_{\cdot\mu} \\ \qquad + \nabla^2 R_{\mu\nu} + g_{\mu\nu} \nabla_\rho \nabla_\sigma R^{\rho\sigma} = 0 \\ \Rightarrow 2(R_{\rho\sigma} R_{\lambda\nu} - \frac{1}{4} g_{\mu\nu} R_{\rho\sigma} R^{\rho\sigma}) - \nabla_\mu \nabla_\nu R + \nabla^2 R_{\mu\nu} \\ \qquad + \frac{1}{2} g_{\mu\nu} \nabla^2 R = 0 \; , \end{cases} \end{bmatrix} \tag{3.145}$$

$$\delta S_3 : \begin{bmatrix} \text{"Palatini"} : \begin{cases} \delta g_{\mu\nu} : \; R^{\rho}_{\cdot\sigma\mu\lambda} R^{\sigma\ \ \lambda}_{\cdot\rho\nu\cdot} - \frac{1}{4} g_{\mu\nu} R^{\rho\ \ \lambda\kappa}_{\cdot\sigma\cdot\cdot} R^{\sigma}_{\cdot\rho\lambda\kappa} = 0 \\ \delta \Gamma^{\sigma}_{\mu\nu} : \; \nabla_\rho R^{\rho}_{\cdot\sigma\mu\nu} = 0 \end{cases} \\ \text{Standard:} \begin{cases} \delta g_{\mu\nu} : \; (R_{\mu\lambda\rho\sigma} R_{\nu\cdot\cdot\cdot}^{\cdot\lambda\rho\sigma} - \frac{1}{4} g_{\mu\nu} R_{\rho\sigma\lambda\kappa} R^{\rho\sigma\lambda\kappa}) \\ \qquad + 2\nabla_\sigma \nabla_\rho R^{\rho\cdot\sigma\cdot}_{\cdot\mu\cdot\nu} = 0 \\ \Rightarrow (R_{\mu\lambda\rho\sigma} R_{\nu\cdot\cdot\cdot}^{\cdot\lambda\rho\sigma} - \frac{1}{4} g_{\mu\nu} R_{\rho\sigma\lambda\kappa} R^{\rho\sigma\lambda\kappa}) \\ -2(R^{\rho}_{\cdot\mu} R_{\rho\nu} - R_{\rho\sigma} R^{\rho\cdot\sigma\cdot}_{\cdot\mu\cdot\nu}) - \nabla_\mu \nabla_\nu R + 2\nabla^2 R_{\mu\nu} = 0 \; . \end{cases} \end{bmatrix} \tag{3.146}$$

By inspection of the "standard" variation results for each case above, we see that the Lanczos linear combination (3.47), eliminates terms like:

$$\nabla_\mu \nabla_\nu R \; ; \; g_{\mu\nu} \nabla^2 R \; ; \; \nabla^2 R_{\mu\nu} \, . \tag{3.147}$$

Returning to the discussion of the auxiliary condition, the point is that by requiring only the $\delta g^{\mu\nu}$ field equation of (3.134) to vanish implies that (3.133) is satisfied for a restricted class of spacetimes. I.e., substituting the $\delta g^{\mu\nu}$ field equation of (3.134) into (3.143) gives the result that

$$\nabla^2 R_{\mu\nu} - (R^{\rho}_{\ \mu} R_{\rho\nu} - R^{\sigma}_{\ \rho} R^{\rho}_{\ \mu\sigma\nu}) - \tfrac{1}{2} \nabla_\mu \nabla_\nu R = 0 \, , \tag{3.148}$$

but this condition admits additional solutions since it is just the condition that $\nabla_\sigma \nabla_\rho R^{\rho\ \ \sigma}_{\ \mu \cdot \nu}$ vanish (from (3.142)). However, a subset of (3.148) may be picked out from the result obtained earlier using the Gauss-Bonnet variation that only the middle term of the above equation vanish (i.e., (3.133)), which does not by itself imply that (3.148) is satisfied (an example is given by the Pavelle-Thompson solution, (3.73), for which $\nabla_\sigma \nabla_\rho R^{\rho\ \ \sigma}_{\ \mu \cdot \nu} = 0$ is satisfied but $R^{\rho}_{\ \mu} R_{\rho\nu} - R^{\sigma}_{\ \rho} R^{\rho}_{\ \mu\sigma\nu} = 0$ is not).

It is of interest to substitute the non-physical solutions discussed in the previous section into the auxiliary condition. The results are summarized below by letting

$$X_{\mu\nu} \equiv R^{\rho}_{\ \mu} R_{\rho\nu} - R^{\rho\sigma} R_{\rho\mu\sigma\nu} \,. \tag{3.149}$$

The nonzero components in each case are thus given by

$$\text{Pavelle-Thompson:} \begin{cases} X_{00} = -\dfrac{4c_1^2(c_1+3r)}{r^6(c_1+r)} \;;\; X_{11} = -\dfrac{4c_1^2(3c_1+5r)}{r^6(c_1+r)} \\ X_{22} = \dfrac{4c_1^2(c_1+r)}{r^6} \;;\; X_{33} = X_{22}\sin^2\theta \end{cases} \tag{3.150}$$

$$\text{Thompson:} \begin{cases} X_{00} = 0 \;;\; X_{11} = \dfrac{3c_1^2}{2r^5(c_1+r)} \\ X_{22} = \dfrac{3c_1^2}{4r^4} \;;\; X_{33} = X_{22}\sin^2\theta \end{cases} \tag{3.151}$$

$$\text{Ni:} \begin{cases} X_{00} = 0 \;;\; X_{11} = -\dfrac{3c_1(c_1-2c_2r^3)}{2r^5(c_1+r+c_2r^3)} \\ X_{22} = -\dfrac{3c_1(c_1-2c_2r^3)}{4r^4} \;;\; X_{33} = X_{22}\sin^2\theta \end{cases} \tag{3.152}$$

By inspection of these equations it is apparent that the Pavelle-Thompson and Thompson solutions satisfy the $X_{\mu\nu} = 0$ equations only if $c_1 = 0$, which results in the Minkowski line element. However, requiring the Ni solution to satisfy $X_{\mu\nu} = 0$ implies that $c_1 = 0$ which is the Einstein universe solution (3.123) - and is not an Einstein space since the Ricci tensor for the Einstein universe metric (3.123) is given by

$$\begin{aligned} &R_{tt} = 0 \\ &R_{ij} = 2c_2\, g_{ij} \;;\; (i,j \equiv 1,2,3), \end{aligned} \tag{3.153}$$

and the Ricci scalar is $6c_2$. Therefore, the Einstein universe solution is not eliminated as a nonphysical solution to the gauge gravity field equations by the auxiliary condition (3.149) (i.e., solutions that are not Einstein spaces). In addition, the Weyl tensor for the Einstein universe solution vanishes and therefore the space is conformally flat which is discussed further in Chapter 5 for more general cases. The origin of the Einstein universe solution is clarified by noting that it provides one of three static cosmological models (the de Sitter and Minkowski line elements give the other two; see e.g., Tolman [119]) and is therefore a solution of the Einstein field equations with a nonzero cosmological constant, $\lambda$, in a universe filled with matter and radiation (i.e., the stress energy tensor is nonzero):

$$R_{\mu\nu} - \tfrac{1}{2} g_{\mu\nu} R + \lambda g_{\mu\nu} = 8\pi G c^{-4} \Theta_{\mu\nu} . \tag{3.154}$$

The point is that it is not a solution to the Einstein free-field equations. Some additional insight on why this metric is not eliminated by (3.149) is obtained by expressing (3.149) in terms of the Weyl tensor and the trace-free Ricci tensor, $P_{\mu\nu}$, which is defined by

$$R_{\mu\nu} = \tfrac{1}{4} g_{\mu\nu} R + P_{\mu\nu} . \tag{3.155}$$

Eliminating the Ricci tensor using (3.155) and the Riemann tensor using the decomposition given in (3.128), (3.149) may be expressed (Appendix B):

$$\begin{aligned} X_{\mu\nu} = {} & \tfrac{1}{3} R P_{\mu\nu} + 2\left(P_\mu^{\ \rho} P_{\rho\nu} - \tfrac{1}{4} P_{\rho\sigma} P^{\rho\sigma}\right) - P^{\rho\sigma} C_{\rho\mu\sigma\nu} \\ & - \tfrac{1}{2} P_\lambda^\lambda (P_{\mu\nu} + \tfrac{5}{12} g_{\mu\nu} R) + \tfrac{1}{4} R C^{\lambda}_{\ \mu\lambda\nu} , \end{aligned} \tag{3.156}$$

but the last two terms vanish since $P_{\mu\nu}$ and $C^{\rho}_{\ \mu\sigma\nu}$ are traceless giving:

$$X_{\mu\nu} = \tfrac{1}{3} R P_{\mu\nu} + 2\left(P_\mu^{\ \rho} P_{\rho\nu} - \tfrac{1}{4} P_{\rho\sigma} P^{\rho\sigma}\right) - P^{\rho\sigma} C_{\rho\mu\sigma\nu} . \tag{3.157}$$

As a result, $X_{\mu\nu} = 0$ does not imply that $P_{\mu\nu} = 0$, i.e., there are other solutions to $X_{\mu\nu} = 0$ besides Einstein spaces. The example is given by the Einstein universe solution where

$$\begin{aligned} P_{00} &= -\tfrac{3}{2} c_2 \ ; \ P_{11} = \tfrac{1}{2} c_2 \, g_{11} \\ P_{22} &= \tfrac{1}{2} c_2 \, g_{22} \ ; \ P_{33} = \tfrac{1}{2} c_2 \, g_{33} . \end{aligned} \tag{3.158}$$

For this solution the $P^{\rho\sigma} C_{\rho\mu\sigma\nu}$ term is zero as a result of the vanishing Weyl tensor. Neither of the remaining two terms in (3.157) is zero, but their linear combination results in $X_{\mu\nu} = 0$.

The term

$$P^{\rho\sigma}C_{\rho\mu\sigma\nu}\,, \tag{3.159}$$

is closely related to an identity discussed by Thompson

$$R_{\rho\sigma[\mu\nu}R^{\sigma}_{\lambda]}=0\,, \tag{3.160}$$

which has recently been discussed by Guilfoyle and Nolan [35]. Their analysis considers the various Petrov types of spacetimes allowed by (3.160) according to a Segré classification of the Ricci tensor (this classification was previously discussed by Petrov [114] and also by Debney, Fairchild, and Siklos [44]). But in fact, the condition (3.160) says nothing more than, $0=0$, since the equation is an identity. The similarity to (3.159) is obtained by expressing (3.160) in terms of $P_{\mu\nu}$ using (3.155) and then eliminating the Riemann tensor using the Weyl tensor using (3.128). The result is given by (Appendix B):

$$C_{\rho\sigma[\mu\nu}P^{\sigma}_{\lambda]}=0\,, \tag{3.161}$$

and not surprisingly, all of the solutions discussed above satisfy (3.161) since this expression is an identity, but not every solution satisfies $P^{\rho\sigma}C_{\rho\mu\sigma\nu}=0$. It is of further interest to note that the theory proposed by Fairchild [42] – which is given by the simultaneous solution of ($\zeta$ is a constant):

$$\begin{aligned}\zeta R_{\mu\nu}+R^{\rho\sigma}C_{\rho\mu\sigma\nu}&=0\\ Y_{\sigma\mu\nu}\equiv\nabla_{[\mu}R_{\nu]\sigma}&=0,\end{aligned} \tag{3.162}$$

may be expressed using a term like (3.159). Additional analysis on the equivalence between solutions of this theory and Einstein's theory were given by Debney, Fairchild, and Siklos [44] showing that the only vacuum solutions of the theory are also those of Einstein's.

Finally, it is of interest to investigate other solution possibilities for $X_{\mu\nu}=0$ by substituting the arbitrary spherically symmetric metric into the auxiliary condition. The differential equations that result in this case are given by (see Appendix B):

$$\begin{aligned}
X_{tt}: &\quad r^2(2A'+B')(A''+A'^2-A'B')^2+r\,A'(A'+B')+A'(e^{2B}-1)=0\\
X_{rr}: &\quad r^2(A'+2B')(A''+A'^2-A'B')^2-rB'(A'+B')+B'(e^{2B}-1)=0\\
X_{\theta\theta}: &\quad r^2(A'-B')(A''+A'^2-A'B')^2+r(A'+B')^2+(A'-B')(e^{2B}-1)=0\\
X_{\varphi\varphi}: &\quad X_{\theta\theta}\sin^2\theta=0\ .
\end{aligned} \tag{3.163}$$

Considering the linear combination $X_{tt}/A'+X_{rr}/B'$ gives the condition that

$$\frac{X_{tt}}{A'}+\frac{X_{rr}}{B'}=0=-\frac{r(A'+B')[r(A'-B')(A''+A'^2-A'B')-2A'B']}{A'B'}, \tag{3.164}$$

and therefore either

$$A'+B'=0\ , \tag{3.165}$$

or

$$r(A'-B')(A''+A'^2-A'B')-2A'B'=0\ . \tag{3.166}$$

Considering first the $A=-B$ case and substituting into (3.163) shows that the following differential equation must be satisfied for each component of the $X_{\mu\nu}$:

$$B''-2B'^2-(e^{2B}-1)/r^2=0\ . \tag{3.167}$$

The differential equation (3.167) was discussed earlier in the analysis as the $(-)(-)$ case (see (3.125)) and results in the solution given by Kottler (3.119):

$$ds^2 \;=\; (1+c_1/r+c_2r^2)dt^2-(1+c_1/r+c_2r^2)^{-1}dr^2-r^2d\Omega^2\ , \tag{3.168}$$

which is an Einstein space. Another obvious solution to (3.166) is given by $A=const.$. In this case, $X_{tt}=0$ and the $X_{rr}$ and $X_{\theta\theta}$ both give the following differential equation:

$$X_{rr}\text{ and }X_{\theta\theta}=0: \qquad B'-(e^{2B}-1)/r=0\ . \tag{3.169}$$

The differential equation is simplified using the substitution, $B=-\frac{1}{2}\ln\Lambda$, to give

$$r\Lambda'-2\Lambda+2=0\ , \tag{3.170}$$

which results in the Einstein universe solution:

$$ds^2 \;=\; dt^2-(1+c_2r^2)^{-1}dr^2-r^2d\Omega^2\ , \tag{3.171}$$

as discussed earlier. Finally, setting $A=B$ satisfies (3.163) but only when $A$ and $B$ are both constant.

# Chapter 4 Conformal Gravity

As discussed earlier the quadratic curvature Lagrangians:

$$\begin{aligned} S_1 &= \int d^4q \; g^{\frac{1}{2}} R^2 \\ S_2 &= \int d^4q \; g^{\frac{1}{2}} R_{\rho\sigma} R^{\rho\sigma} \\ S_3 &= \int d^4q \; g^{\frac{1}{2}} R^{\rho}_{\;\sigma\mu\nu} R^{\sigma\cdot\mu\nu}_{\;\rho\cdot\cdot}, \end{aligned} \tag{4.1}$$

and various linear combinations of them have been considered as possible generalizations of Einstein's theory based on local Lorentz invariance. But other investigations based on local scale invariance have been considered, i.e., by postulating invariance under the conformal transformation:

$$`g_{\mu\nu} = \lambda(q)^2 \; g_{\mu\nu}, \tag{4.2}$$

where $\lambda$ is a locally varying scale factor, and $q$ denotes general coordinates. From a more physical viewpoint, the motivation for considering a conformally invariant theory is postulated from the transformation law of a length element:

$$ds \rightarrow `ds = \lambda\, ds. \tag{4.3}$$

The idea is that physical phenomena should be independent of locally chosen units for mass, length, time, etc.; i.e. the group of conformal transformations should be a symmetry group of nature. In fact, Maxwell's theory of electromagnetism is conformally invariant - as demonstrated by Cunningham and Batemann [120] (see also Fulton, et. al. [121] and Wald [122], p. 448); although it should be obvious that every conformally invariant theory would not necessarily be physical.

In the recent literature there have appeared several papers describing a theory of gravitation based upon an action quadratic in the Weyl conformal tensor, $C^{\rho}_{\cdot\,\sigma\mu\nu}$, which is defined as the totally traceless contribution to the Riemann curvature tensor according to the decomposition listed earlier in Chapter 3:

$$\begin{aligned} R_{\rho\sigma\mu\nu} = C_{\rho\sigma\mu\nu} &+ \tfrac{1}{2}\left(g_{\rho\mu}R_{\sigma\nu} - g_{\rho\nu}R_{\sigma\mu} - g_{\sigma\mu}R_{\rho\nu} + g_{\sigma\nu}R_{\rho\mu}\right) \\ &+ \tfrac{1}{6}\left(g_{\rho\nu}\,g_{\sigma\mu} - g_{\rho\mu}\,g_{\sigma\nu}\right)R. \end{aligned} \tag{4.4}$$

Therefore, to base a theory on conformal invariance the following conformally invariant combination has been considered for the action:

$$S = \int d^4q \, g^{\frac{1}{2}} C^{\rho}{}_{\sigma\mu\nu} C^{\sigma}{}_{\rho\cdot\cdot}{}^{\mu\nu} . \tag{4.5}$$

The field equations originating from (4.5) by using a standard variational procedure have recently been discussed by Mannheim and Kazanas (MK) [56, 57] from a strictly classical viewpoint. A spherically symmetric solution to these field equations that was originally discussed by Riegert [123] has been reconsidered by MK who have suggested that an apparent observational inconsistency with galactic rotation curves could be accounted for by a theory originating from (4.5).

It is apparent from (4.5) (and (4.4)) that a consequence of basing a theory on conformal invariance is that the action considered for the theory must be quadratic in the curvature tensors. Therefore, several features of the conformal analysis are shared by the quadratic curvature Lagrangians discussed in the previous Chapter. The purpose of this Chapter is to emphasize this similarity and to show that a class of conformally flat solutions to the gauge gravity field equations, namely those also satisfying Nordström's theory [34] are also shared by conformal gravity. In addition, an investigation of a spherically symmetric solution structure to the theory is presented – which is the solution that has been recently advocated by Mannheim and Kazanas.

## Conformal Transformations of the Curvature Tensors

To help motivate the proposed conformally invariant generalization of Einstein's theory we illustrate in this Section that none of (4.1) are invariant under (4.2). But there does exist a linear combination of $S_1$, $S_2$, and $S_3$ that is invariant. To begin, consider the transformation of the connection under (4.2) (see also Eddington [5] - pp. 200-222; Schouten [83] - pp. 133, 304):

$$\begin{aligned} {}^{\backprime}\Gamma^{\sigma}_{\mu\nu} &= \tfrac{1}{2} {}^{\backprime}g^{\sigma\kappa}(\partial_{\mu}{}^{\backprime}g_{\nu\kappa} + \partial_{\nu}{}^{\backprime}g_{\kappa\mu} - \partial_{\kappa}{}^{\backprime}g_{\mu\nu}) \\ &= \Gamma^{\sigma}_{\mu\nu} + \lambda^{-1}(\delta^{\sigma}_{\nu}\partial_{\mu}\lambda + \delta^{\sigma}_{\mu}\partial_{\nu}\lambda - g^{\sigma\kappa}g_{\mu\nu}\partial_{\kappa}\lambda) . \end{aligned} \tag{4.6}$$

Using the definition of the Riemann tensor:

$$R^{\rho}{}_{\sigma\mu\nu} = \partial_{\mu}\Gamma^{\rho}_{\sigma\nu} - \partial_{\nu}\Gamma^{\rho}_{\sigma\mu} + \Gamma^{\rho}_{\lambda\mu}\Gamma^{\lambda}_{\sigma\nu} - \Gamma^{\rho}_{\lambda\nu}\Gamma^{\lambda}_{\sigma\mu} , \tag{4.7}$$

and then (4.6), the transformation rule of $R^{\rho}{}_{\sigma\mu\nu}$ is thus calculated in powers of $\lambda$ (see Appendix E):

$$
\begin{aligned}
{}'R^{\rho}{}_{\sigma\mu\nu} = R^{\rho}{}_{\sigma\mu\nu} &+ \lambda^{-1}[(\delta^{\rho}_{\nu}\nabla_{\mu} - \delta^{\rho}_{\mu}\nabla_{\nu})\nabla_{\sigma}\lambda + (g_{\sigma\mu}\nabla_{\nu} - g_{\sigma\nu}\nabla_{\mu})\nabla^{\rho}\lambda] \\
&+ \lambda^{-2}[(\delta^{\rho}_{\nu}\, g_{\sigma\mu} - \delta^{\rho}_{\mu}\, g_{\sigma\nu})(\nabla\lambda)^2 + 2(\delta^{\rho}_{\mu}\nabla_{\nu}\lambda - \delta^{\rho}_{\nu}\nabla_{\mu}\lambda)\nabla_{\sigma}\lambda \\
&+ 2(g_{\sigma\nu}\nabla_{\mu}\lambda - g_{\sigma\mu}\nabla_{\nu}\lambda)\nabla^{\rho}\lambda]\,.
\end{aligned} \tag{4.8}
$$

Therefore, from (4.8) the transformation:

$$
g^{\kappa\mu}g^{\lambda\nu}R^{\rho}{}_{\sigma\mu\nu}R^{\sigma}{}_{\rho\kappa\lambda} \rightarrow {}'g^{\kappa\mu}\,{}'g^{\lambda\nu}\,{}'R^{\rho}{}_{\sigma\mu\nu}\,{}'R^{\sigma}{}_{\rho\kappa\lambda}\,, \tag{4.9}
$$

of $L$ from $S_3$ in (4.1) is given by

$$
\begin{aligned}
{}'g^{\kappa\mu}\,{}'g^{\lambda\nu}\,{}'R^{\rho}{}_{\sigma\mu\nu}\,{}'R^{\sigma}{}_{\rho\kappa\lambda} &= \lambda^{-4}\left(g^{\kappa\mu}g^{\lambda\nu}R^{\rho}{}_{\sigma\mu\nu}R^{\sigma}{}_{\rho\kappa\lambda}\right) + 8\lambda^{-5}\left(R^{\rho\sigma}\nabla_{\rho}\lambda\nabla_{\sigma}\lambda\right) \\
&+ 4\lambda^{-6}\left[R(\nabla\lambda)^2 - 2(\nabla_{\rho}\nabla^{\sigma}\lambda)^2 - (\nabla^2\lambda)^2 - 4R^{\rho\sigma}\nabla_{\rho}\nabla_{\sigma}\lambda\right] \\
&+ 8\lambda^{-7}\left[4(\nabla^{\rho}\lambda)(\nabla^{\sigma}\lambda)\nabla_{\rho}\nabla_{\sigma}\lambda - (\nabla\lambda)^2(\nabla^2\lambda)\right] - 24\,\lambda^{-8}(\nabla\lambda)^4\,,
\end{aligned} \tag{4.10}
$$

illustrating that the action $S_3$ is not invariant under (4.2) unless $\lambda$ is a scalar constant.

Similarly, the Ricci tensor transforms as

$$
\begin{aligned}
R_{\mu\nu} \rightarrow {}'R_{\mu\nu} = R_{\mu\nu} &- \lambda^{-1}\left(g_{\mu\nu}\nabla^2\lambda + \nabla_{\mu}\nabla_{\nu}\lambda\right) \\
&+ \lambda^{-2}\left[4\nabla_{\mu}\lambda\nabla_{\nu}\lambda - g_{\mu\nu}(\nabla\lambda)^2\right],
\end{aligned} \tag{4.11}
$$

so that

$$
\begin{aligned}
& g^{\kappa\mu}g^{\lambda\nu}R_{\mu\nu}R_{\kappa\lambda} \rightarrow {}'g^{\kappa\mu}\,{}'g^{\lambda\nu}\,{}'R_{\mu\nu}\,{}'R_{\kappa\lambda} \\
&= \lambda^{-4}\left(g^{\kappa\mu}g^{\lambda\nu}R_{\mu\nu}R_{\kappa\lambda}\right) - 2\lambda^{-5}\left(2R^{\rho\sigma}\nabla_{\rho}\lambda\nabla_{\sigma}\lambda - R\nabla^2\lambda\right) \\
&+ 2\lambda^{-6}\left[4R^{\rho\sigma}(\nabla_{\rho}\lambda)(\nabla_{\sigma}\lambda) - R(\nabla\lambda)^2 + 2(\nabla^{\rho}\nabla^{\sigma}\lambda)(\nabla_{\rho}\nabla_{\sigma}\lambda) + 4(\vec{\nabla}^2\lambda)^2\right] \\
&+ 4\lambda^{-7}\left[(\nabla\lambda)^2(\nabla^2\lambda) - 4(\nabla^{\rho}\lambda)(\nabla^{\sigma}\lambda)\nabla_{\rho}\nabla_{\sigma}\lambda\right] + 12\,\lambda^{-8}(\nabla\lambda)^4\,.
\end{aligned} \tag{4.12}
$$

We also find for the Ricci Scalar:

$$
R \rightarrow {}'R = \lambda^{-2}R - 6\lambda^{-3}(\nabla^2\lambda)\,, \tag{4.13}
$$

and therefore

$$
R^2 \rightarrow {}'R^2 = \lambda^{-4}R^2 - 12\lambda^{-5}R(\nabla^2\lambda) + 36\lambda^{-6}(\nabla^2\lambda)^2\,. \tag{4.14}
$$

As a result, $S_1$ and $S_2$ are also eliminated as possible actions from which to base a conformally invariant theory.

However, by substituting the transformation rules derived above for $R^{\rho}{}_{\sigma\mu\nu}$, $R_{\rho\sigma}$, $R$ into (4.4) along with (4.2), the definition of the Weyl tensor is thus viewed as the linear combination that eliminates all terms involving derivatives on $\lambda$. Hence, to construct a theory based on conformal invariance, the Weyl tensor is taken as the starting point given that from $L$ of (4.5):

$$g^{\frac{1}{2}}g^{\kappa\mu}g^{\lambda\nu}C^{\rho}_{\ \sigma\mu\nu}C^{\sigma}_{\ \rho\kappa\lambda} \rightarrow {}^{‘}g^{\frac{1}{2}}\,{}^{‘}g^{\kappa\mu}\,{}^{‘}g^{\lambda\nu}\,{}^{‘}C^{\rho}_{\ \sigma\mu\nu}\,{}^{‘}C^{\sigma}_{\ \rho\kappa\lambda} = g^{\frac{1}{2}}g^{\kappa\mu}g^{\lambda\nu}C^{\rho}_{\ \sigma\mu\nu}C^{\sigma}_{\ \rho\kappa\lambda}, \quad (4.15)$$

Note also that the combination:

$$g^{\frac{1}{2}}g^{\kappa\mu}g^{\lambda\nu} \rightarrow {}^{‘}g^{\frac{1}{2}}\,{}^{‘}g^{\kappa\mu}\,{}^{‘}g^{\lambda\nu} = g^{\frac{1}{2}}g^{\kappa\mu}g^{\lambda\nu}, \quad (4.16)$$

is itself conformally invariant.

## Field Equations

The field equations considered in the literature on conformal gravity (originating from (4.5)) are derived with respect to a standard variational procedure. To consider this variation first express $C^{\rho}_{\ \sigma\mu\nu}C^{\sigma\ \mu\nu}_{\ \rho\cdot\cdot}$ in terms of the Ricci tensor and Ricci scalar by solving for the Weyl tensor from (4.4). The result is given by (Appendix E):

$$C^{\rho}_{\ \sigma\mu\nu}C^{\sigma\ \mu\nu}_{\ \rho\cdot\cdot} = \left(2R_{\rho\sigma}R^{\rho\sigma} - \tfrac{1}{3}R^2 - R^{\rho}_{\ \sigma\mu\nu}R^{\sigma\ \mu\nu}_{\ \rho\cdot\cdot}\right), \quad (4.17)$$

so that (4.5) becomes:

$$\frac{\delta S}{\delta g_{\mu\nu}} = \frac{\delta}{\delta g_{\mu\nu}}\int d^4q\ g^{\frac{1}{2}}\left(2R_{\rho\sigma}R^{\rho\sigma} - \tfrac{1}{3}R^2 - R^{\rho}_{\ \sigma\mu\nu}R^{\sigma\ \mu\nu}_{\ \rho\cdot\cdot}\right) = 0. \quad (4.18)$$

However, before taking this variation, the linear combination above can be simplified further using the Gauss-Bonnet linear combination to eliminate $R^{\rho}_{\ \sigma\mu\nu}R^{\sigma\ \mu\nu}_{\ \rho\cdot\cdot}$ given that:

$$\varepsilon_{\rho\sigma\mu\nu}\varepsilon^{\lambda\kappa\omega\eta}R^{\rho\sigma}_{\ \cdot\cdot\lambda\kappa}R^{\mu\nu}_{\ \cdot\cdot\omega\eta} = R^2 - 4R_{\rho\sigma}R^{\rho\sigma} + R^{\rho}_{\ \sigma\mu\nu}R^{\sigma\ \mu\nu}_{\ \rho\cdot\cdot}. \quad (4.19)$$

Substituting into (4.18) gives

$$\frac{\delta S}{\delta g_{\mu\nu}} = -\frac{2\delta}{\delta g_{\mu\nu}}\int d^4q\ g^{\frac{1}{2}}\left(R_{\rho\sigma}R^{\rho\sigma} - \tfrac{1}{3}R^2 + \tfrac{1}{2}\varepsilon_{\rho\sigma\mu\nu}\varepsilon^{\lambda\kappa\omega\eta}R^{\rho\sigma}_{\ \cdot\cdot\lambda\kappa}R^{\mu\nu}_{\ \cdot\cdot\omega\eta}\right) = 0, \quad (4.20)$$

but since the last term in (4.20) is the Euler topological invariant:

$$\frac{\delta}{\delta g_{\mu\nu}}\int d^4q\ g^{\frac{1}{2}}\varepsilon_{\rho\sigma\mu\nu}\varepsilon^{\lambda\kappa\omega\eta}R^{\rho\sigma}_{\ \cdot\lambda\kappa}R^{\mu\nu}_{\ \cdot\omega\eta} \equiv 0, \quad (4.21)$$

we may consider the following variation in place of (4.18):

$$\frac{\delta}{\delta g_{\mu\nu}}\int d^4q\ g^{\frac{1}{2}}C^{\rho}_{\ \sigma\mu\nu}C^{\sigma\ \mu\nu}_{\ \rho\cdot\cdot} = -\frac{2\delta}{\delta g_{\mu\nu}}\int d^4q\, g^{\frac{1}{2}}\left(R_{\rho\sigma}R^{\rho\sigma} - \tfrac{1}{3}R^2\right). \quad (4.22)$$

The field equations resulting from (4.22) using the standard variational procedure are given by (Appendix E):

$$\begin{aligned} \delta g^{\mu\nu} :\quad 0 &= \tfrac{2}{3} R \left( R_{\mu\nu} - \tfrac{1}{4} g_{\mu\nu} R \right) - 2 \left( R^{\sigma}_{\ \rho} R^{\rho}_{\ \mu\sigma\nu} - \tfrac{1}{4} g_{\mu\nu} R_{\rho\sigma} R^{\rho\sigma} \right) \\ &+ \nabla^2 \left( R_{\mu\nu} - \tfrac{1}{3} g_{\mu\nu} R \right) - 2 \nabla_{\rho} \nabla_{\nu} R^{\rho}_{\ \mu} - 2 \nabla_{\rho} \nabla_{\sigma} R^{\sigma\ \rho}_{\ \mu \cdot \nu} \\ &+ g_{\mu\nu} \nabla_{\rho} \nabla_{\sigma} R^{\rho\sigma} + \tfrac{1}{3} \nabla_{\mu} \nabla_{\nu} R \ . \end{aligned} \tag{4.23}$$

and then using the Bianchi identities (4.23) is equivalent to the simpler form (see also Tsantilis, et. al. [124] for identical results):

$$\begin{aligned} W_{\mu\nu} = 0 &= \tfrac{2}{3} R \left( R_{\mu\nu} - \tfrac{1}{4} g_{\mu\nu} R \right) - 2 \left( R^{\rho}_{\ \sigma} R^{\sigma}_{\ \mu\rho\nu} - \tfrac{1}{4} g_{\mu\nu} R_{\rho\sigma} R^{\rho\sigma} \right) \\ &- \nabla^2 \left( R_{\mu\nu} - \tfrac{1}{6} g_{\mu\nu} R \right) + \tfrac{1}{3} \nabla_{\mu} \nabla_{\nu} R \ . \end{aligned} \tag{4.24}$$

However, these field equations differ from the those presented by MK [56]:

$$\begin{aligned} 0 &= \tfrac{2}{3} R \left( R_{\mu\nu} - \tfrac{1}{4} g_{\mu\nu} R \right) - 2 \left( R^{\rho}_{\ \mu} R_{\rho\nu} - \tfrac{1}{4} g_{\mu\nu} R_{\rho\sigma} R^{\rho\sigma} \right) \\ &+ \nabla^2 \left( R_{\mu\nu} - \tfrac{1}{6} g_{\mu\nu} R \right) - \nabla_{\rho} \left( \nabla_{\mu} R^{\rho}_{\ \nu} + \nabla_{\nu} R^{\rho}_{\ \mu} \right) - \tfrac{2}{3} \nabla_{\mu} \nabla_{\nu} R \ . \end{aligned} \tag{4.25}$$

Removing covariant derivatives on the MK result for comparison to (4.24), (4.25) is equivalent to

$$\begin{aligned} 0 &= \tfrac{2}{3} R \left( R_{\mu\nu} - \tfrac{1}{4} g_{\mu\nu} R \right) - 2 \left( R^{\rho}_{\ \mu} R_{\rho\nu} - \tfrac{1}{4} g_{\mu\nu} R_{\rho\sigma} R^{\rho\sigma} \right) \\ &+ \nabla^2 \left( R_{\mu\nu} - \tfrac{1}{6} g_{\mu\nu} R \right) - 2 \left( R^{\rho}_{\ \mu} R_{\rho\nu} - R^{\rho}_{\ \sigma} R^{\sigma}_{\ \rho\mu\nu} \right) - \tfrac{1}{3} \nabla_{\mu} \nabla_{\nu} R \ . \end{aligned} \tag{4.26}$$

Although (4.26) is traceless, the field equations are incorrect. To gain insight on this problem note that (4.24) and (4.26) would be equivalent provided that the auxiliary condition (3.133) is satisfied if

$$R^{\sigma}_{\ \rho} R^{\rho}_{\ \mu\sigma\nu} = R^{\rho}_{\ \mu} R_{\rho\nu} \ .$$

We also note that the spherically symmetric solution discussed by MK (discussed in detail in the following section) satisfies (4.24) but does not satisfy their own field equations. However, we view this as a relatively minor problem which probably indicates that the field equations (4.26) were derived by MK using a "gauge gravity" type Palatini procedure. Furthermore, Pawlowski and Raczka [125] have published field equations originating from (4.5):

$$\begin{aligned} 0 &= \tfrac{2}{3} R \left( R_{\mu\nu} - \tfrac{1}{4} g_{\mu\nu} R \right) - 2 \left( R^{\rho}_{\ \sigma} R^{\sigma}_{\ \mu\rho\nu} - \tfrac{1}{4} g_{\mu\nu} R_{\rho\sigma} R^{\rho\sigma} \right) \\ &- \nabla^2 \left( R_{\mu\nu} - \tfrac{1}{3} g_{\mu\nu} R \right) + \tfrac{1}{3} \nabla_{\mu} \nabla_{\nu} R \ , \end{aligned} \tag{4.27}$$

that differ from (4.24) by the term, $\nabla^2(\frac{1}{3} g_{\mu\nu} R)$, which should be $\nabla^2(\frac{1}{6} g_{\mu\nu} R)$, as given in (4.24). The result of this incorrect factor of 2 is that (4.27) is no longer traceless, as must be required by conformal invariance.

## A Spherically Symmetric Solution

A spherically symmetric solution to the field equations (4.24) has been found by Riegert [123] and expressed by MK [56] in the form:

$$\begin{aligned} ds^2 &= B(r)\, dt^2 - B(r)^{-1} dr^2 - r^2\, d\Omega^2 \\ B(r) &= 1 - \frac{\beta(2-3\beta\gamma)}{r} - 3\beta\gamma + \gamma r - k r^2, \end{aligned} \tag{4.28}$$

where $\beta, \gamma, k$ are 3 independent integration constants. To simplify the analysis on (4.28) we write $B(r)$ in the form:

$$B(r) = 1 - \frac{k_1}{r} - k_2 + k_3 r - k_4 r^2, \tag{4.29}$$

where $k_1$, $k_2$, $k_3$, and $k_4$ are now four arbitrary (and assumed) independent constants. Substituting (4.29) into (4.24) gives the result (Appendix E):

$$\begin{aligned} W_{00} &= 2\left(k_2^2 - 2k_2 + 3k_1k_3\right)\left[k_1 + r\left(k_4 r^2 - k_3 r + k_2 - 1\right)\right] / 3r^5 \\ W_{11} &= -\left(k_2^2 - 2k_2 + 3k_1k_3\right) / 3r^3 \left[k_1 + r\left(k_4 r^2 - k_3 r + k_2 - 1\right)\right] \\ W_{22} &= -2\left(k_2^2 - 2k_2 + 3k_1k_3\right) / 3r^2 \\ W_{33} &= W_{22} \sin^2\theta, \end{aligned} \tag{4.30}$$

and therefore if the field equation (4.24) is satisfied, we obtain a quadratic equation in $k_1$, $k_2$, and $k_3$ ($k_4$ is therefore an arbitrary independent parameter):

$$k_2^2 - 2k_2 + 3k_1k_3 = 0. \tag{4.31}$$

Solving (4.31), we find that

$$k_1 = \frac{k_2\left(2 - k_2\right)}{3k_3}, \tag{4.32}$$

which must hold for (4.29) to satisfy the field equations (4.24). Using (4.32), (4.29) is expressed:

$$B(r) = 1 - \frac{k_2(2-k_2)}{3k_3 r} - k_2 + k_3 r - k_4 r^2 , \tag{4.33}$$

and then by comparison with (4.28) and (4.29) we identify:

$$k_2 = 3\beta\gamma \;\; ; \;\; k_3 = \gamma \;\; ; \;\; k_4 = k \, . \tag{4.34}$$

Substituting into (4.32) we then recover the form of the MK solution (4.28). Note that (4.31) could be used to eliminate $k_2$ or $k_3$ rather than $k_1$:

$$k_2 = 1 \pm \sqrt{1-3k_1k_3} \quad or \quad k_3 = \frac{k_2(2-k_2)}{3k_1} , \tag{4.35}$$

and these substitutions give back exactly (4.28) - providing no new forms of the solution.

By inspection of (4.32) and (4.35) there will be several special cases of (4.28) to consider (the labeling of these solutions will be explained in the following discussion):

Schwarzschild:

$$k_3 = k_4 = k_2 = 0 \Rightarrow B(r) = 1 - k_1/r , \tag{4.36}$$

"Pseudo - Signature" Solution 1:

$$k_3 = k_4 = 0; \; k_2 = 2 \Rightarrow B(r) = -(1 + k_1/r) , \tag{4.37}$$

No Label:

$$k_4 = k = 0 \Rightarrow B(r) = 1 - k_1/r - k_2 + k_3 r , \tag{4.38}$$

Conformally Flat 1:

$$k_1 = k_2 = 0 \; (k = 0) \Rightarrow B(r) = 1 + k_3 r , \tag{4.39}$$

where (4.36) and (4.37) originate from the two solutions for $k_2$ in (4.35), with $k = 0$. The fact that $k_4 = k$ is arbitrary should be apparent from the condition that (4.24) vanishes as given by (4.31) - irrespective of *k*. This freedom gives three additional forms corresponding to $k \neq 0$:

Schwarzschild - de Sitter:

$$k_3 = k_2 = 0; \; k_4 \neq 0 \Rightarrow B(r) = 1 - k_1/r - k_4 r^2 , \tag{4.40}$$

"Pseudo - Signature" Solution 2:

$$k_3 = 0; k_2 = 2; k_4 \neq 0 \Rightarrow B(r) = -(1 + k_1/r + k r^2) , \tag{4.41}$$

Conformally Flat 2:

$$k_1 = k_2 = 0 \Rightarrow B(r) = 1 + k_3 r - k_4 r^2 . \tag{4.42}$$

**Schwarzschild and "Pseudo - Signature" Solutions**

To understand these cases from a more general viewpoint begin with the Schwarzschild and "Pseudo - Signature" solutions. We see by inspection of (4.29) that the Schwarzschild solution follows when $k_2 = k_3 = k_4 \to 0$, and $k_1 = 2MG/c^2$. However, a more general result is obtained by simply allowing:

$$B(r) = 1 - k_1/r - k_2 . \tag{4.43}$$

Substituting (4.43) into (4.24) we get

$$\begin{aligned} W_{00} &= -2k_2\left(2-k_2\right)\left[k_1 + r\left(k_2 - 1\right)\right]/3r^5 \\ W_{11} &= 2k_2\left(2-k_2\right)/3r^3\left[k_1 + r\left(k_2 - 1\right)\right] \\ W_{22} &= 2k_2\left(2-k_2\right)\sin^2\theta/3r^2 \\ W_{33} &= W_{22}\sin^2\theta , \end{aligned} \tag{4.44}$$

which is satisfied if

$$k_2(2-k_2) = 0 \Rightarrow k_2 = 0 \;\; or \;\; k_2 = 2 , \tag{4.45}$$

in agreement with the first equation of (4.35) when $k_3 = 0$. Hence, the $k_2 = 0$ (and $k = 0$) case gives the Schwarzschild solution (4.36) with signature: $\{1,-1,-1,-1\}$, while the $k_2 = 2$ case gives

$$B(r) = -\left(1 + k_1/r\right), \tag{4.46}$$

corresponding to (4.37):

$$ds^2 = -\left(1 + k_1/r\right)dt^2 + \left(1 + k_1/r\right)^{-1}dr^2 - r^2 d\Omega^2 . \tag{4.47}$$

Equation (4.47) is not equivalent to the Schwarzschild case - noting the unusual signature combination of negative signs in front of two spatial and the time differentials, while only one spatial term has a positive coefficient. To see that there is no correspondence with the Schwarzschild solution note that (4.47) is not a solution to the Einstein-Hilbert field equations since:

$$R_{\mu\nu} \neq 0;\; R_{22} = 2,\, R_{33} = R_{22}\sin^2\theta;\text{ and } R = -4/r^2\,(\neq 0) . \tag{4.48}$$

Since $k$ is arbitrary we obtain (4.41) as the complementary "pseudo-signature" solution #2 with $k \neq 0$. The special case (4.38) follows from (4.28) when $k = 0$.

**Conformally Flat Solutions**

There are other nontrivial solutions to the conformal field equations (4.24). To see this note that the vanishing of the Weyl tensor implies that the metric may be expressed:

$$C^{\rho}{}_{\sigma\mu\nu} = 0 \;\Rightarrow\; g_{\mu\nu} = \lambda^2(q)\eta_{\mu\nu}\,, \tag{4.49}$$

where $\eta_{\mu\nu}$ is the Minkowski form of the metric. In such cases, the $W_{\mu\nu} = 0$ field equations are satisfied identically, which is not necessarily obvious. The reason is that although a function (e.g. the Weyl tensor) could be made to vanish, the same does not necessarily hold for its derivatives (i.e., the field equations $W_{\mu\nu}$). However, the important thing to realize is that the field equations are conformally invariant under the transformation of (4.2), i.e., $`g_{\mu\nu} = \lambda(q)^2\, g_{\mu\nu}$. Therefore, by considering, $g_{\mu\nu} = \lambda^2(q)\eta_{\mu\nu}$, as a transformation of this form, the scalar factor $\lambda^2(q)$ passes through the $W_{\mu\nu} = 0$ field equations leaving derivatives on $\eta_{\mu\nu}$. As a result, $C^{\rho}{}_{\sigma\mu\nu} = 0 \Rightarrow W_{\mu\nu} \equiv 0$, and therefore solutions satisfying Nordström's theory (recalling that $C^{\rho}{}_{\sigma\mu\nu} = 0;\, R = 0$) and those satisfying $C^{\rho}{}_{\sigma\mu\nu} = 0$ will be solutions to $W_{\mu\nu} = 0$. A consequence is that the Pavelle-Thompson solution:

$$ds^2 = \left(1 + k_1/r\right)^{-2} dt^2 - \left(1 + k_1/r\right)^{-2} dr^2 - r^2 d\Omega^2\,, \tag{4.50}$$

is a solution to both the gauge gravity and conformal gravity field equations since Nordström's theory is implied by the $(+)(+)$ case discussed earlier in Chapter 3. The origin of the conformally flat solutions "one" and "two" noted above stem from $C^{\rho}{}_{\sigma\mu\nu} = 0$. These cases were derived earlier in Chapter 3 (see (3.114)) but do not satisfy the gauge gravity equations. As a result, the conformally invariant theory of gravitation also has nonphysical solutions - recalling that (4.50) gives 1/6 of the observed value for perihelion precession and in the opposite direction (Pirani [95]). Another characteristic of such spaces is that there will be no deflection of light rays. To see this note that the null-cone is invariant under (4.2), i.e.

$$ds^2 = 0 \rightarrow {}^{\backprime}ds^2 = 0 \,. \tag{4.51}$$

Therefore, for the special case of a conformally flat space, the null geodesics of $g_{\mu\nu}$ will be identical to the null geodesics of $\eta_{\mu\nu}$ and no deflection of light rays will occur. However, the fact that (4.24) has nonphysical solutions (in addition to possibly physical ones) does not necessarily rule out this formulation as a nonviable physical theory. For example, in classical electromagnetism one obtains both advanced and retarded potential solutions. But advanced potential solutions are considered nonphysical and therefore discarded; the point is that nonphysical solutions to a theory do not necessarily mean that the theory is incorrect.

# Chapter 5 Schwarzschild Dynamics

In the second part of the thesis the orbital dynamics of point particles in the Schwarzschild and Reissner-Nordström spacetimes are investigated using techniques from dynamical systems theory. The motivation for considering this analysis is to gain additional insight into the physical and analytic structure of the solutions, particularly with regard to the stability properties of the orbits. From the viewpoint of a bifurcation analysis the Schwarzschild solution is singled out in one respect - the solution has the lowest dimensional parameter space of all other black hole solutions of Einstein's field equations and is therefore the simplest to analyze. Therefore, although there are some new qualitative results to report on the stability properties and classification of the Schwarzschild orbits, the analysis presented in this Chapter will serve mainly as an introduction to the Reissner-Nordström bifurcation problem which is considered in Chapter 6.

In the first sub-section entitled *Orbital Equations*, the equations of motion for test particle orbits in the Schwarzschild geometry are derived. The goal in this section is to not only outline and introduce the analysis that will be applied in later sections, but to "tailor" the derivation towards a discussion emphasizing the bifurcation results. In the following sub-section entitled *Fixed Points and Linear Stability Analysis*, the phase-plane analysis is developed and then applied to obtain the well-known value of periastron precession. In the sub-section entitled *Phase Diagrams*, the Schwarzschild orbital dynamics are analyzed based on the level curves of the dynamical state space. The standard results are discussed but also an alternative viewpoint for analyzing the dynamics is presented based upon the separatrix structure of the phase-plane. In this approach, the critical relationship that occurs between energy and angular momentum at the unstable orbital radius (i.e. the separatrix) summarizes the range of physically possible orbits, and then a saddle-center bifurcation is identified while a dimensionless parameter involving the angular momentum is varied.

Although the dynamical structure (i.e. the effective potential) is invariant between the coordinate and proper time reference frames, the phase diagrams in each reference frame are not identical. This is due to the existence of an additional phase-plane fixed point that appears in the coordinate reference frame at the event horizon. This fixed point is obviously coordinate dependent, but must exist to explain the apparent "slowing down" of objects (and redshift of signals) approaching the horizon boundary as seen by an observer in the coordinate reference frame.

For comparison to the relativistic case, the corresponding Newtonian phase-plane results are discussed in Appendix F. Finally, the phase-plane analysis is applied to the kinematics of light rays in the Schwarzschild spacetime. The standard results are discussed and then compared with the timelike phase-plane results. The added significance of the photon orbits (in the phase-plane context) is that the equilibrium points of the differential equations exhibit a transcritical bifurcation, i.e. a an exchange of stability occurs at these parameter values.

## Orbital Equations

The equations of motion for a point mass with rest mass, $m_0$, orbiting a Schwarzschild black hole with mass, $M$ (assuming for simplicity that $m_0 << M$) originate from the line element (derived in Chapter 2):

$$\begin{gathered} ds^2 = c^2 \Lambda\, dt^2 - \Lambda^{-1} dr^2 - r^2 d\Omega^2 \\ \Lambda = 1 - r_s / r \\ d\Omega^2 = d\theta^2 + \sin^2\theta\, d\varphi^2 . \end{gathered} \tag{5.1}$$

Equation (5.1) is expressed using spherical coordinates and $r_s$ is the Schwarzschild radius obtained earlier in (2.80):

$$r_s = 2MG / c^2 . \tag{5.2}$$

The Lagrangian is a constant of the motion:

$$L = \tfrac{1}{2} m_0 \left( ds / d\tau \right)^2 = \tfrac{1}{2} m_0 c^2 \ ; \ \tau \equiv proper\ time , \tag{5.3}$$

and if the orbit is confined to the equatorial plane ($\theta = \pi / 2$), $\hat{L}$ takes the explicit form ($\dot{t} = dt / d\tau$, etc.):

$$2\hat{L} = 1 = \Lambda \dot{t}^2 - \Lambda^{-1} \dot{r}^2 - r^2 \dot{\varphi}^2 . \tag{5.4}$$

From the Euler-Lagrange equations there are two additional constants of motion ((5.4) is trivial; (5.5) and (5.6) are first integrals):

$$\hat{E} = \Lambda \dot{t} \; ; \; \hat{E} \equiv E / m_0 c^2 \, , \tag{5.5}$$

$$\partial L / \partial \varphi = 0 \Rightarrow \partial L / \partial \dot{\varphi} = J = m_0 r^2 \dot{\varphi} ; \; \hat{J} = J / m_0 c^2 \, . \tag{5.6}$$

Physically, $\hat{E}$ is the energy required for an observer at infinity to place $m_0$ in orbit about $M$ (total energy per unit rest energy). $\hat{J}$ is the angular momentum of the system (per unit rest energy) and since this is constant, there will be no precession of the equatorial plane.

The physical interpretation of $\hat{E}$ and the value of $r_s$ given by (5.2) are checked in the Newtonian limit which was also considered in Chapter 2. A slightly simpler approach for obtaining these limits is given by assuming only radial motion to obtain:

$$ds^2 = c^2 d\tau^2 = c^2 \Lambda dt^2 - \Lambda^{-1} dr^2 \, , \tag{5.7}$$

and then factoring $c^2 dt^2$ leads to

$$d\tau^2 = dt^2 (\Lambda - \Lambda^{-1} v^2 / c^2) \, . \tag{5.8}$$

Solving this equation for $dt / d\tau = \dot{t}$, and then assuming the mass is at rest (i.e., $v \to 0$):

$$\dot{t} \;=\; \Lambda^{-\frac{1}{2}} = (1 - r_s / r)^{-\frac{1}{2}} \, . \tag{5.9}$$

Finally, substituting for $\dot{t}$ into (5.5) (in the limit of small $r_s$) gives the result

$$\hat{E} = (1 - r_s / r)^{\frac{1}{2}} \;\approx\; 1 - r_s / 2r \, . \tag{5.10}$$

Comparing (5.10) with the Newtonian potential energy, the interpretation of $\hat{E}$ is in agreement with the Newtonian result - provided that $r_s$ is given by (5.2).

Continuing with the equations of motion, use (5.5) and (5.6) to eliminate $\dot{t}$ and $\dot{\varphi}$ from (5.4) and then re-arranging algebraically gives the result:

$$\dot{r}^2 / c^2 = \left( dr / ds \right)^2 = \hat{E}^2 - (1 + c^2 \hat{J}^2 / r^2) \Lambda \, . \tag{5.11}$$

Noting the functional dependence of $r$ on the equatorial angle (i.e. $r = r(\varphi) \Rightarrow \dot{r} = (dr / d\varphi) \dot{\varphi}$), (5.11) is further expressed in terms of the constant $\hat{J}$. Furthermore, the degree of this equation (in $r$) is reduced by making the change of variable: $x \equiv r_s / r$. Simplifying algebraically then gives the result:

$$\left( dx / d\varphi \right)^2 = 2\sigma (\hat{E}^2 - \hat{V}_{eff}^2) \, , \tag{5.12}$$

where $\sigma$ defines the dimensionless parameter:

$$\sigma = \tfrac{1}{2}\left(r_s / c\hat{J}\right)^2 , \tag{5.13}$$

and the effective potential is given by:

$$\hat{V}_{eff}^2 = (1 + x^2 / 2\sigma)\,\Lambda . \tag{5.14}$$

Differentiating (5.12) with respect to $\varphi$ then gives the standard second order equation in dimensionless form:

$$d^2x / d\varphi^2 + x = \sigma + \tfrac{3}{2}x^2 . \tag{5.15}$$

## Fixed Points and Linear Stability Analysis

A phase-plane and bifurcation analysis of (5.15) is considered by first converting this second order, nonlinear, inhomogeneous, differential equation into two first order equations by first introducing a new variable, $y = dx / d\varphi\ (\equiv x')$. For a general spherically symmetric metric of the form given by (5.1) (with $\Lambda$ unspecified) the phase-plane equations are given by

$$\begin{aligned} x' &= f(x,y) = y \\ y' &= g(x,y) = -\,[x\Lambda + (\sigma + \tfrac{1}{2}x^2)(d\Lambda / dx)] , \end{aligned} \tag{5.16}$$

which shows that if $\Lambda$ is a polynomial of degree *n*, there will be $n+1$ fixed points, i.e. equilibrium solutions of (5.17) obtained from $x' = y' = 0$ for $x$ and $y$. For the Schwarzschild solution, (5.15) reduces to

$$\begin{aligned} x' &= y \\ y' &= \tfrac{3}{2}x^2 - x + \sigma . \end{aligned} \tag{5.17}$$

and the fixed points, $\vec{x}^*$, are given by (denoting $\vec{x} = (x,y)$):

$$\vec{x}_1^* = \left(\frac{1+\sqrt{1-6\sigma}}{3}\ , 0\right);\ \vec{x}_2^* = \left(\frac{1-\sqrt{1-6\sigma}}{3}\ , 0\right). \tag{5.18}$$

Alternatively, by expressing *y* in terms of *x* using (5.12):

$$x' = y = \pm\,[2\sigma\hat{E}^2 - (2\sigma + x^2)(1-x)]^{1/2} = 0 , \tag{5.19}$$

and then solving simultaneously: $x' = y' = 0$, for $\hat{E}^2$ and $x$ rather than $x$ and $y$, the corresponding energies at each fixed point are expressed solely in terms of $\sigma$:

$$\hat{E}_1^2 - 1 = \frac{2\sigma[1 - 4\sigma - (1-6\sigma)^{1/2}]}{[(1-6\sigma)^{1/2} - 1]^3}\ ;\ \hat{E}_2^2 - 1 = \frac{2\sigma[-1 + 4\sigma - (1-6\sigma)^{1/2}]}{[(1-6\sigma)^{1/2} - 1]^3} , \tag{5.20}$$

respectively. Therefore, solving simultaneously for $\hat{E}^2$ and $x$ gives additional information on the dynamics compared to the standard technique of obtaining only $x$ and $y$ (e.g., Strogatz [126]; Tabor [127]). Furthermore, the phase-plane equations analogous to (5.17) that result from the proper and coordinate time analysis considered in a later section (and also in the Newtonian case) give non-physical roots when solving only for $x$ and $y$ (i.e. do not correspond to the effective potential extrema). However, these additional roots are eliminated by solving for $\hat{E}^2$ and $x$ as illustrated above and as discussed below.

A general classification of the fixed points (5.18) is obtained from a linear stability analysis (e.g., [128]). Essentially, this amounts to series expanding (5.17) about an arbitrary fixed point in the small parameters: $\delta x = x - x^*$ and $\delta y = y - y^*$. Dropping second order terms, the resulting first order linear equations are expressed in matrix form:

$$\begin{pmatrix} \delta x' \\ \delta y' \end{pmatrix} \approx \begin{pmatrix} \partial_x f & \partial_y f \\ \partial_x g & \partial_y g \end{pmatrix}\Bigg|_{\vec{x} = \vec{x}^*} \begin{pmatrix} \delta x \\ \delta y \end{pmatrix} = \begin{pmatrix} 0 & 1 \\ 3x - 1 & 0 \end{pmatrix}\Bigg|_{\vec{x} = \vec{x}^*} \begin{pmatrix} \delta x \\ \delta y \end{pmatrix} \equiv A\big|_{\vec{x} = \vec{x}^*} \delta\vec{x} \,. \qquad (5.21)$$

The general solution of (5.21) is therefore an exponential whose stability at each fixed point is analyzed by classifying the eigenvalues of the matrix $\boldsymbol{A}$. Solving the eigenvalue problem, we find roots to

$$|A - \lambda I| = 0 \,, \qquad (5.22)$$

but since $\boldsymbol{A}$ is $2 \times 2$, the characteristic polynomial may be expressed:

$$\lambda^2 - \tau\lambda + \Delta = 0 \,, \qquad (5.23)$$

where $\tau \equiv \text{trace}\, A$, and $\Delta \equiv \text{determinant}\ A$. The eigenvalues are roots to (5.23):

$$\lambda = \tfrac{1}{2}(\tau \pm \sqrt{\tau^2 - 4\Delta}) \,, \qquad (5.24)$$

and accordingly, the exponential solutions to (5.21) are classified by the various regions of Figure 1 (dots mark the location of the fixed points given in (5.18)).

Briefly, region *I* corresponds to a "saddle-node" fixed point, whose stable and unstable manifolds (corresponding to positive and negative (real) eigenvalues, respectively) are given by the eigenvectors of (5.21). Region *II* represents an "unstable node," i.e. 2 positive real eigenvalues with $\tau^2 - 4\Delta > 0$; region *III* gives solutions having one (positive) real and one complex eigenvalue ("unstable spirals"), while regions $II'$ and $III'$ are the complimentary stable solutions of region's $II$, and $III$, respectively. The "boundary" cases are given by $\tau^2 = 4\Delta$ (degenerate nodes and lines of fixed points) and $\tau = 0$, $\Delta > 0$ are "centers" giving periodic orbits in the phase-plane. A complete discussion will not be given on each case since it is only region $I$ and the boundary separating regions $II$ and $II'$ that are relevant for the analysis considered here (see Strogatz [126] for additional discussion and examples).

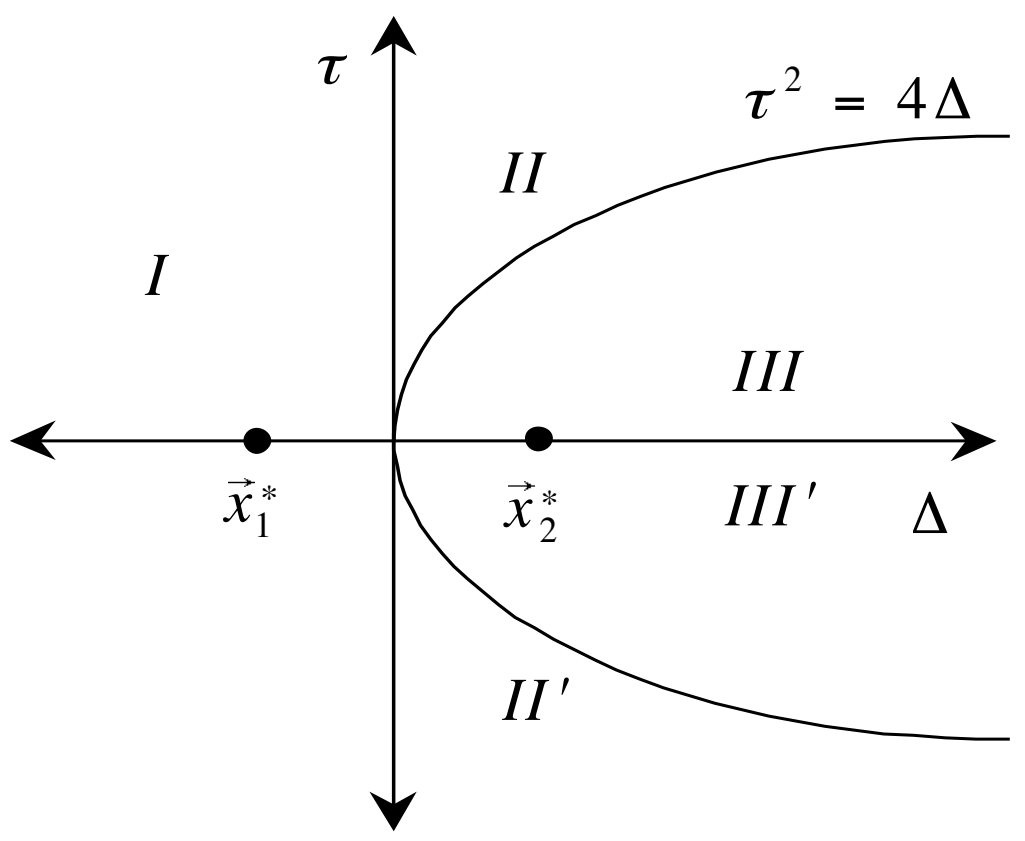

Figure 1 Eigenvalue Classification

By evaluating the matrix $\boldsymbol{A}$ in (5.21) at each fixed point of (5.18), the following classifications are obtained:

$$\begin{aligned} A\Big|_{\vec{x}_1^*} &= \begin{pmatrix} 0 & 1 \\ \sqrt{1-6\sigma} & 0 \end{pmatrix} \Rightarrow \underbrace{\tau = 0;\ \Delta = -\sqrt{1-6\sigma}}_{\text{"Saddle Node"}} \\ A\Big|_{\vec{x}_2^*} &= \begin{pmatrix} 0 & 1 \\ -\sqrt{1-6\sigma} & 0 \end{pmatrix} \Rightarrow \underbrace{\tau = 0;\ \Delta = +\sqrt{1-6\sigma}}_{\text{"Center"}}, \end{aligned} \tag{5.25}$$

corresponding to a "saddle" and "center-node" fixed point, respectively (see Figure 1 for the placement of these points). As previously discussed, the linear stability analysis gives an exponential solution about each fixed point with the phase-plane trajectories shown in Figure 2 (the directions follow from (5.17) and are indicated by arrows).

Physically, trajectories about the center-node fixed point correspond to precessing elliptical orbits (and will be used to obtain the value for periastron precession). However, the saddle-node that appears is not predicted by Newtonian theory, but is due to an unstable orbital radius originating from the $r^{-3}$ term of the effective potential (5.14). As a result of this instability, there are orbital effects not present in the Newtonian theory which have been summarized in the literature (see e.g., MTW [93], p. 637). An interesting consequence of the phase-plane approach to this analysis is that this result comes out very quickly in the analysis as a secondary fixed point.

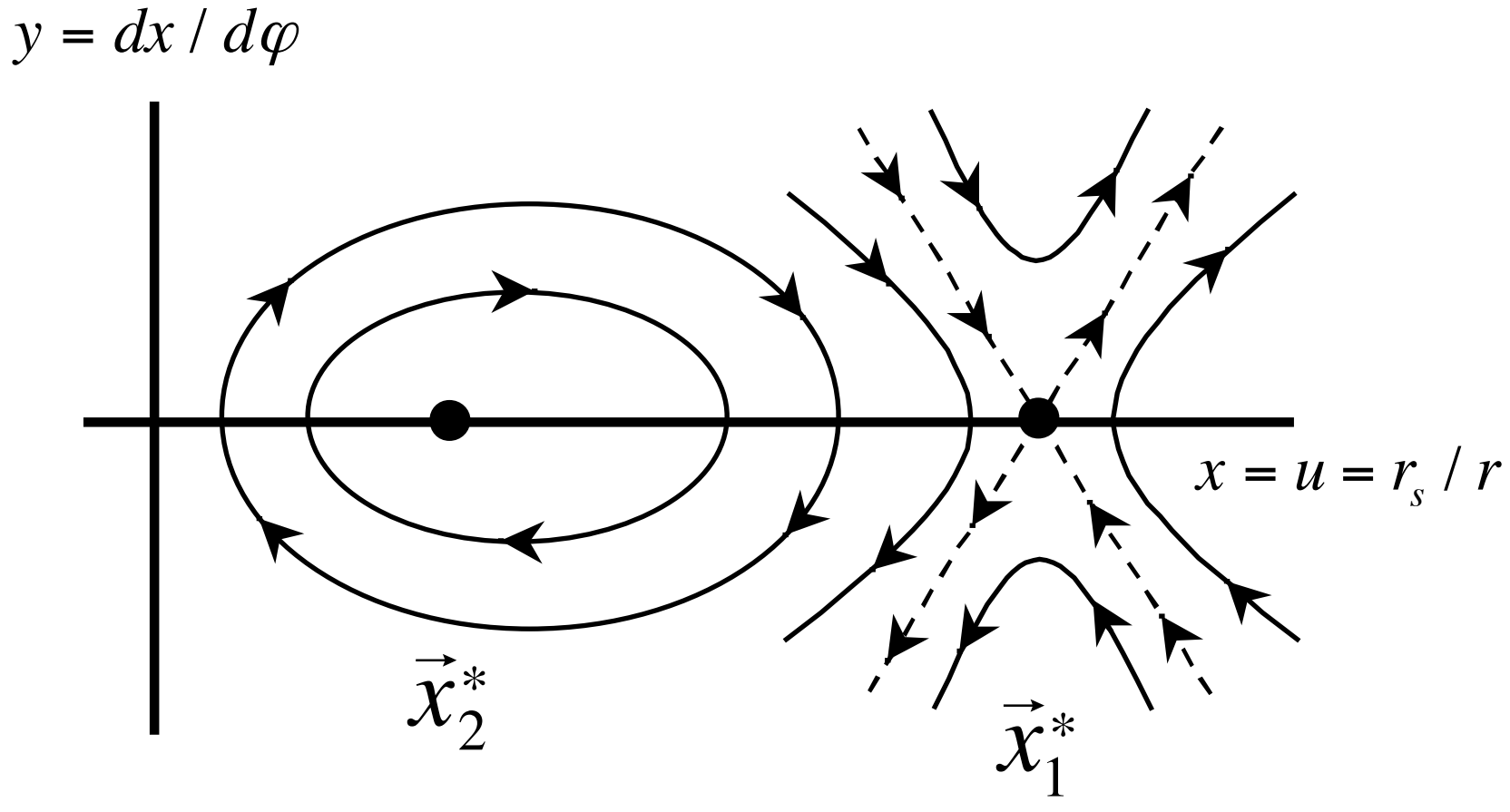

Figure 2 Linear Stability Phase-Plane

## Periastron Precession

The periastron shift, $\Delta\varphi$, of planetary orbits has provided one of the earliest and most important experimental test of Einstein's theory (see Figure 3). The lowest-order relativistic contribution to the periastron precession was first calculated by Einstein [129] to explain the anomalous perihelion shift of Mercury (the modern experimental value is approximately $\approx 43''$ per century). In binary systems the periastron shift is due to both Newtonian and relativistic contributions. The most important Newtonian contributions are due to gravitational interactions with other bodies and the quadruple moment induced by rotation and tides. For example, the planet Mercury has an observed total periastron shift of approximately 574 arc seconds per century but $532''$ are due to other Newtonian contributions. For a detailed discussion of Newtonian effects on the periastron shift see [130].

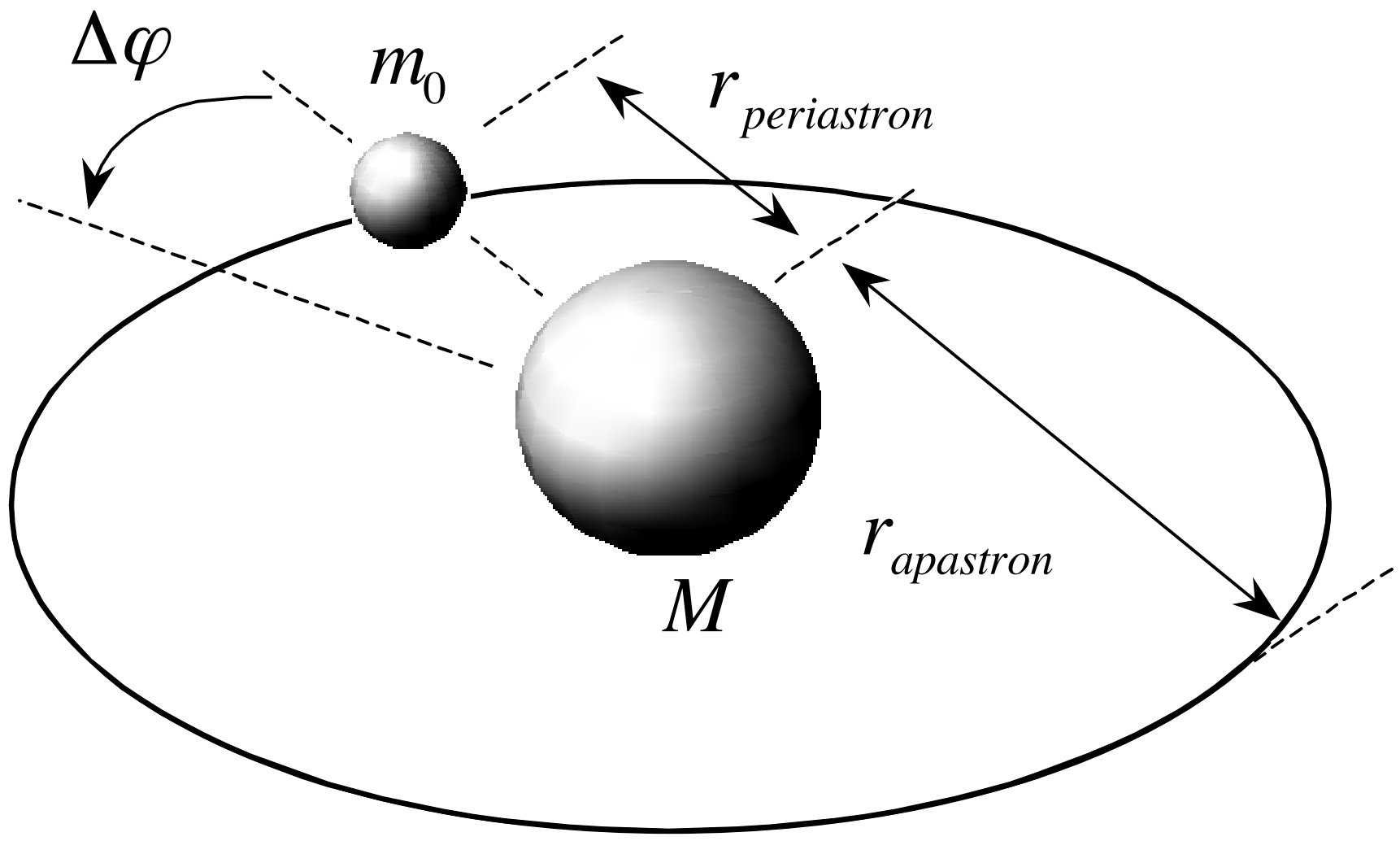

Figure 3 Schematic of Periastron Precession

More recent interest in periastron calculations is motivated by observations of relativistic binary systems (e.g., [131]) and has remained an active area of investigation. For example, Esteban and Diaz [132] have discussed higher order relativistic contributions to the periastron shift due to spin-orbit and spin-spin interactions. Damour and Schafer [133] have considered applications of these higher order contributions to binary pulsar systems. Such calculations are further motivated by suggestions [134] that neutron-star–black-hole and radio-pulsar–black-hole binaries could be detected in the near future, and therefore observational evidence might soon be available to cross check these astrophysical models.

The purpose of this section is to illustrate the phase-plane method as a calculational tool to obtain the standard periastron precession value. These results are then later generalized to a study of periastron precession in the Reissner-Nordström spacetime in Chapter 6. As a brief review, in the standard textbook presentation of this calculation [93] there are essentially two approaches taken to calculate its value from the nonlinear equations of motion:

(A) approximate an elliptic integral,

(B) a perturbative solution to the general relativistic equations.

Although (A) and (B) are the most common methods appearing in the literature, other approximation methods do exist, e.g. Wald [93] considers small oscillations about an elliptical orbit; Misner, Thorne, and Wheeler (MTW) [93] consider nearly circular orbits and then later using the PPN formalism.

Here we simply solve (5.21) about $\vec{x}_2^*$. The system is rewritten here as:

$$\delta x' = \delta y \ , \delta y' = -\omega^2 \delta x \ ; \omega = (1-6\sigma)^{1/4} . \tag{5.26}$$

The solutions are centers corresponding to precessing elliptical orbits (see Figure 2):

$$\begin{aligned} \delta x(\varphi) &= A\cos\omega\varphi + B\sin\omega\varphi \\ \delta y(\varphi) &= -\ \omega A\sin\omega\varphi + \omega B\cos\omega\varphi , \end{aligned} \tag{5.27}$$

with $A$ and $B$ arbitrary constants. Choosing initial conditions at the position of periastron:

$$\delta x(0) = u_0 \ ; \delta y(0) = du(0)/d\varphi = 0 , \tag{5.28}$$

(5.27) becomes

$$\begin{aligned} \delta x(\varphi) &= u(\varphi) = u_0 \cos\omega\varphi \\ \delta y(\varphi) &= u'(\varphi) = -\omega\, u_0 \sin\omega\varphi , \end{aligned} \tag{5.29}$$

giving a typical "center" solution about the fixed point $\vec{x}_2^*$.

As previously discussed, in "physical" space the orbit of $m_0$ about $M$ does not close. However, the phase-plane trajectory given by (5.29) must close after a single orbit since the system is conservative (ignoring radiative effects). Therefore, the period of a single orbit, $\Phi$, is defined from the period of the phase space trajectory given by (5.29):

$$\omega\Phi = 2\pi . \tag{5.30}$$

Solving for $\Phi$, and then substituting for $\omega$ in the limit of small $\sigma$ gives the result:

$$\Phi = 2\pi\omega^{-1} \approx 2\pi + 3\pi\sigma . \tag{5.31}$$

The Newtonian calculation gives only the first term, $\Phi = 2\pi$, as expected. However, as seen from (5.31), the Schwarzschild solution gives the correction:

$$\Delta\varphi = 3\pi\sigma = 6\pi\left(GM/\hat{J}\right)^2 , \tag{5.32}$$

which is the standard value expressed in the "geometrized" system of units (i.e., $G = c = 1$, $r_s = 2M$; see e.g., Shutz [1], p. 198) with $m_0$ taken as unity (see (5.36) and (5.37) for order of magnitude estimates of $\sigma$).

## Phase Diagrams

In the preceding Section a linear stability analysis has been considered about each fixed point. However, this procedure gives only "local" information on the general relativistic orbits, and is in fact one shortcoming of the linear stability analysis. Therefore, no correspondence can be made with parabolic, hyperbolic, or orbits near the black hole event horizon using Figure 2 alone. However, since the equations of motion are integrable as a result of the constants of motion that exist (i.e., (5.4), (5.5), and (5.6)), a complete phase diagram may be constructed and then several "global" features of these orbits may be deduced as a result. In addition, other qualitative features of the Schwarzschild orbital dynamics may be derived from this diagram (Figure 4) as discussed below. (Note: since the equations are integrable no chaos exists here. However, if additional degrees of freedom are allowed the possibility for chaos exists; see e.g. [135] for a discussion of chaos in relativistic orbital dynamics).

To obtain the complete phase-plane diagram, consider the "level curves" found by taking the ratio of $x'$ and $y'$ from (5.17), and then integrating to get a conserved quantity:

$$dy / dx = (\tfrac{3}{2}x^2 - x + \sigma) / y \Rightarrow y^2 = \beta + x^3 - x^2 + 2\sigma x . \qquad (5.33)$$

The value of the constant $\beta$ is easily found by comparison with (5.12):

$$\beta = 2\sigma\left(\hat{E}^2 - 1\right) , \qquad (5.34)$$

so that (5.33) may be alternatively expressed:

$$\hat{E}^2 - 1 = \left(y^2 + x^2 - x^3\right) / 2\sigma - x . \qquad (5.35)$$

In Figure 4, the level curves corresponding to different values of $\hat{E}$ in (5.35) are shown with the effective potential (5.14) (with $\sigma = \frac{1}{9}$). These curves correspond to solutions of (5.17) for various energies and initial conditions, and should be compared with the approximate solutions given by the linear stability analysis of Figure 2. The vertical dotted line at $r_s / r = 1$, labels the black hole event horizon.

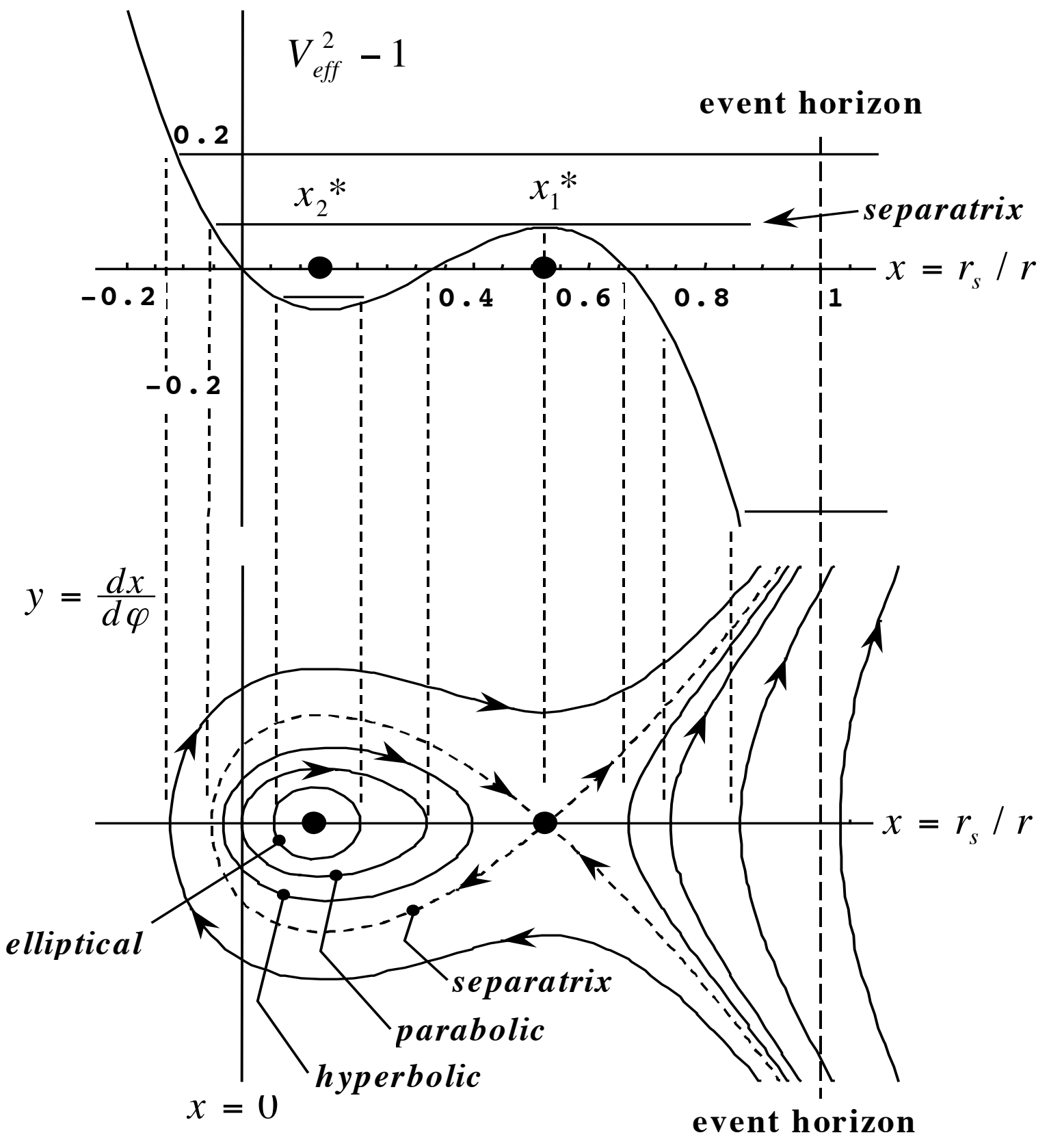

Figure 4 Complete Phase-Plane for $\sigma = 1/9$

The value of $\sigma$ used in (5.35) has been greatly exaggerated to better illustrate the qualitative features of the exact phase-plane. For a more realistic value of $\sigma$ consider Mercury's orbit - taking the value of $\Delta\varphi$ over a single orbit and then using (5.32):

$$3\pi\sigma \approx 0.104'' \Rightarrow \sigma \approx 5.3 \times 10^{-8}, \tag{5.36}$$

or for the binary pulsar system discovered by Hulse and Taylor [131]:

$$3\pi\sigma \approx 4° \Rightarrow \sigma \approx 7.4 \times 10^{-3}. \tag{5.37}$$

To check that $\sigma = \frac{1}{9}$ is a reasonable value in Figure 4, an upper bound may be placed on $\sigma$ for the existence of stable or unstable orbits from either phase-plane fixed point. By inspection of (5.18), if $\sigma > \frac{1}{6}$ then no (real) fixed points exist for a given value of energy and angular momentum. To trace the physical origin of this value and to understand the topological structure of Figure 4 from a more general viewpoint, note that when $y = 0$ in (5.35) the effective potential is obtained:

$$\hat{V}_{eff}^{2} - 1 = (x^2 - x^3)/2\sigma - x\,. \qquad (5.38)$$

The locations of the stable and unstable orbits are found as usual by solving: $\partial_x \hat{V}_{eff} = 0$ for $x$, which gives identically (5.18). From (5.18) no extrema exist for $\sigma > \frac{1}{6}$, establishing an upper bound on $\sigma$ for stable or unstable orbits. For $\sigma = \frac{1}{6}$, stable orbits (smallest value of) and unstable orbits (largest value of) coincide at

$$r_{1,2} = 3r_s\,, \qquad (5.39)$$

providing an inflection point in the plot of $\hat{V}_{eff}^{2} - 1$ *vs* $x$ as shown in Figure 5 for several values of $\sigma$ † (note: the standard presentations of this diagram are commonly displayed as: $\hat{V}_{eff}^{2} - 1$ *vs* 1/*x*; see e.g. Wald, Ohanian and Ruffini, or MTW [93] for an alternative parametrization using $r_s$ and $\hat{J}$). But another critical value of $\sigma$ occurs when $\hat{E}^2 = 1$. To see this, solve $\hat{E}_1^2 - 1 = 0$ using (5.20) to obtain $\sigma = \frac{1}{8}$, as displayed in Figure 5. Therefore, qualitatively distinct orbits exist based upon the following values of $\sigma$:

$$0 < \sigma < \tfrac{1}{8}\ ;\ \sigma = \tfrac{1}{8}\ ;\ \tfrac{1}{8} < \sigma < \tfrac{1}{6}\ ;\ \sigma = \tfrac{1}{6}\ ;\ \sigma > \tfrac{1}{6}\,. \qquad (5.40)$$

For $\sigma > \frac{1}{6}$ there is insufficient angular momentum for $m_0$ to sustain an orbit, therefore the mass simply falls into $M$ and correspondingly, $\hat{V}_{eff}$ has no extrema. The physical significance of $\sigma = \frac{1}{6}$ is discussed above (5.39). The physical meaning of the other values in (5.40) are understood by analyzing the separatrix‡ structure of (5.35). Essentially, this corresponds to a limitation placed upon the types of orbits that may exist before an unstable orbit is reached, and the kinematic classification of the separatrices as distinct unstable orbits.

---

† using $x$ rather than $1/x$ for the horizontal axis pushes the singularity at $r = 0$ to infinity. As a result, the relative locations of fixed points are more easily scaled and plotted in the phase plane with this choice of variables.

‡ also termed *homoclinic* orbit in the literature on nonlinear analysis.

In essence, the separatrix gives a graphic representation of the critical relationship that occurs between energy and angular momentum at the unstable orbital radius (see Figure 6). For a given angular momentum ($\sigma$), the critical energy of the unstable orbit is calculated from $\hat{E}_1$ of (5.20). For the values of $\sigma$ plotted in Figure 5, these energies are computed and marked with horizontal lines. Substituting these values of $\hat{E}^2 - 1$ into (5.35), the separatrices corresponding to (5.40) are plotted in Figure 6.

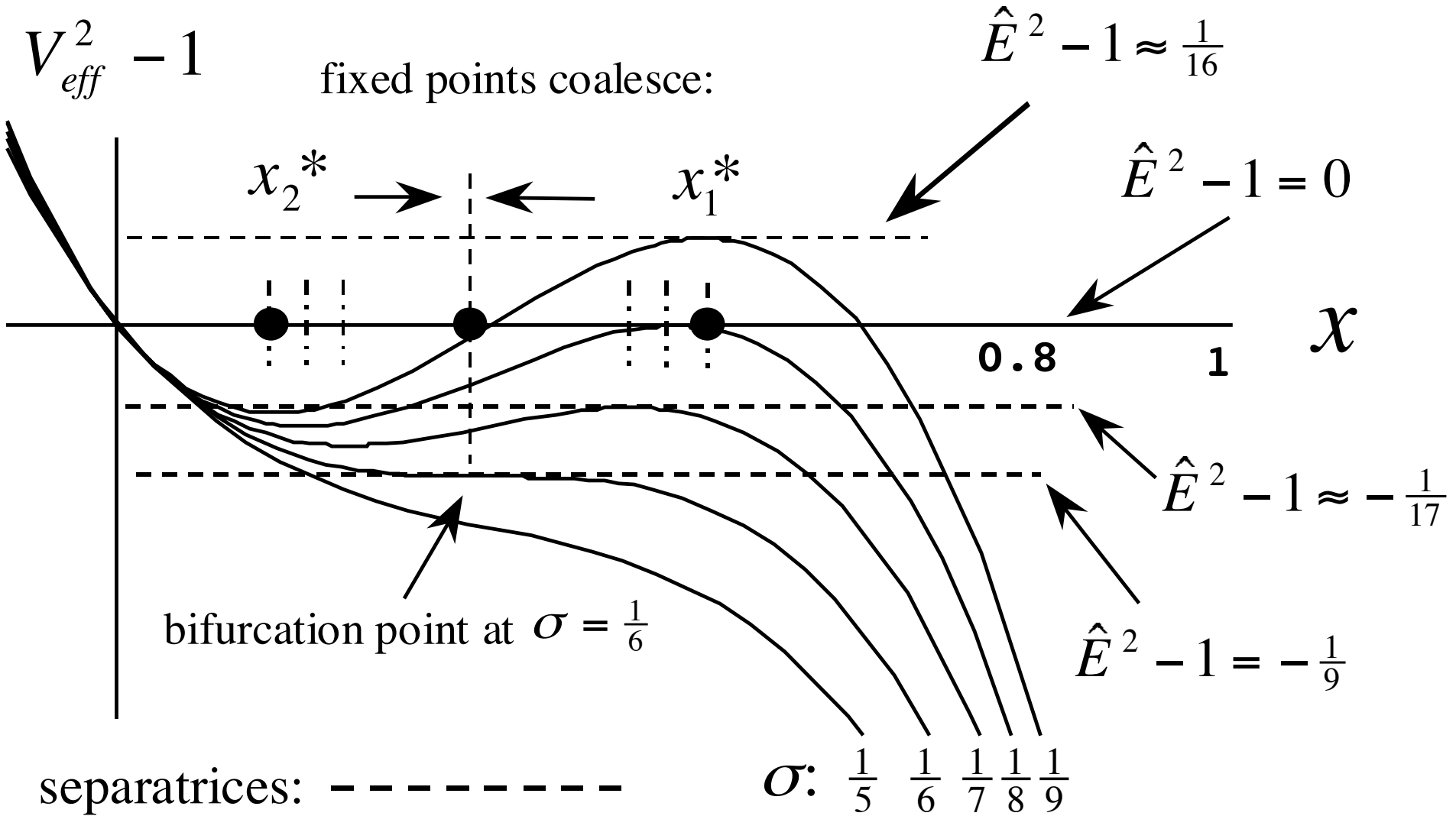

Figure 5 Schwarzschild Effective Potential

These distinct separatrices divide the phase-plane into four regions of motion for $0 < \sigma < \frac{1}{6}$ ($\sigma = \frac{1}{9}$ is just one special case in Figure 4). To begin, consider Figure 4 in the region surrounding the stable fixed point $\vec{x}_2^*$. The oval trajectory in this region corresponds to an elliptical orbit and was used earlier to find the value for periastron precession. A unique parabolic orbit occurs as the phase-plane trajectory just touches the $y$-axis and separates the hyperbolic and elliptic orbits. The hyperbolic orbit† is characterized by a trajectory approaching $M$ from infinity, but then returning to infinity with constant $dr/d\varphi$. Therefore, the separatrix of Figure 4 (typical for $\sigma < \frac{1}{8}$) corresponds to a critically unstable hyperbolic orbit that separates trajectories spiraling into $M$ (above the separatrix) or escaping to infinity (below the separatrix).

† these are actually *precessing* hyperbolic orbits if one allows negative $r$ values. Although nonphysical, these are shown to the left of the $y$-axis in Figure 4.

Similarly, as illustrated in Figure 6 these separatrices are summarized according to the following values of $\sigma$ as distinct unstable orbits:

$$\begin{aligned} 0 < \sigma < \tfrac{1}{8} &\Rightarrow \text{unstable hyperbolic}, \\ \sigma = \tfrac{1}{8} &\Rightarrow \text{unstable parabolic}, \\ \tfrac{1}{8} < \sigma < \tfrac{1}{6} &\Rightarrow \text{unstable elliptic}. \end{aligned} \tag{5.41}$$

It is obvious from Figure 6 that for $\sigma$ in the range: $\frac{1}{8} \le \sigma < \frac{1}{6}$, only elliptical orbits are possible (about $\vec{x}_2^*$) before the unstable orbit is reached, while the case $0 < \sigma < \frac{1}{8}$ allows all three: hyperbolic, parabolic and elliptic as discussed above.

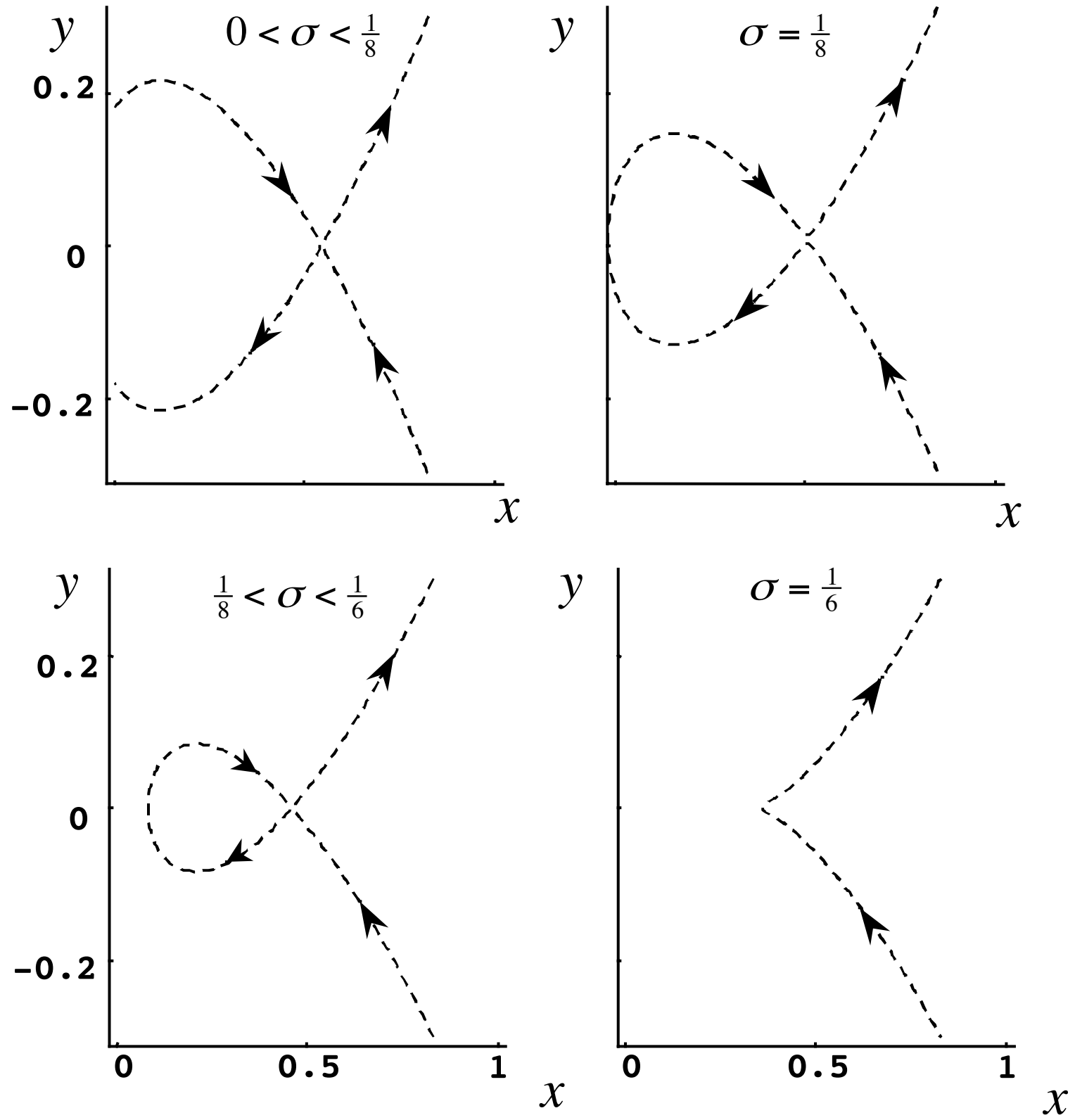

Figure 6 Separatrices for Selected Values of $\sigma$

But these results are consistent with the orbital motion obtained from inspection of the effective potential for different values of $\sigma$ in Figure 5. However, these qualitative differences over the range of unstable orbits have not been pointed out in the literature.

It should be noted that a physical orbit corresponding to the separatrix can never be achieved in finite proper time. To do so would imply that the phase-plane trajectories change direction at $\vec{x}_1^*$, which is not possible in a deterministic system. To see this, consider the proper time equivalent of (5.35) (this is (5.11) after rewriting the equation using the definition of $\sigma$ in (5.13) and again using $x = r_s / r$):

$$\left(dr / ds\right)^2 = \hat{E}^2 - 1 + (x^3 - x^2) / 2\sigma + x \,. \tag{5.42}$$

Separating variables gives an elliptic integral:

$$c\tau \;=\; \pm \int dr \Big/ \sqrt{\left(\hat{E}^2 - 1\right) + x - x^2 / 2\sigma + x^3 / 2\sigma} \,, \tag{5.43}$$

which diverges to $\pm\infty$ as $r$ approaches the unstable orbital radius $r_1$ of (5.18) (and (5.20) is substituted for $\hat{E}^2 - 1$), i.e. for a particle approaching the saddle-point along the separatrix.

From the separatrix analysis it is apparent that a bifurcation occurs at the critical value $\sigma = 1/6$, i.e. the topological structure of the phase-plane changes as the two fixed points move together, coalesce into a single fixed point, and then disappear from the phase-plane as $\sigma$ is further increased above the critical value 1/6. Therefore, the Schwarzschild orbital dynamics may be interpreted and analyzed as a conservative 2-d bifurcation phenomena. Specifically, this bifurcation is a saddle-center bifurcation [136] (see Figure 7), and summarizes the range of physically possible orbits that may occur as the energy and angular momentum are varied for $\sigma > 0$. But from a more general viewpoint one should also consider negative values of $\sigma$ (although it is clear that $\sigma < 0$ has no physical interpretation since $\sigma$ must be positive definite according to (5.13); note also that $\sigma = 0$ in (5.17) gives the phase-plane equations for light rays - see (5.49) below). For $\sigma < 0$ the two fixed points (equation (5.18)) exchange stability at $\sigma = 0$ as shown in Figure 7. Therefore, another (transcritical) bifurcation occurs at $\sigma = 0$ (see e.g. Strogatz [126], pp. 50-52), followed by the saddle-center bifurcation at $\sigma = 1/6$.

Finally, an interpretation of the phase-plane trajectories to the right of the separatrix should be given, namely those trajectories leaving and then returning through the event horizon. These trajectories are clearly nonphysical since it is impossible for any classical particle or light ray to escape from within the black hole horizon. The origin of these trajectories may be understood as a consequence of the symmetry of (5.17) under the interchange: $\varphi \to -\varphi$ ; $y \to -y$, where $\varphi \to -\varphi$, is due to the time-reversal symmetry of the Schwarzschild dynamics. As a consequence, this system is classified as *reversible* and gives the symmetry of Figure 4 (and Figure 6) about the $x$-axis, but with the vector field below the $x$-axis reversing direction.†

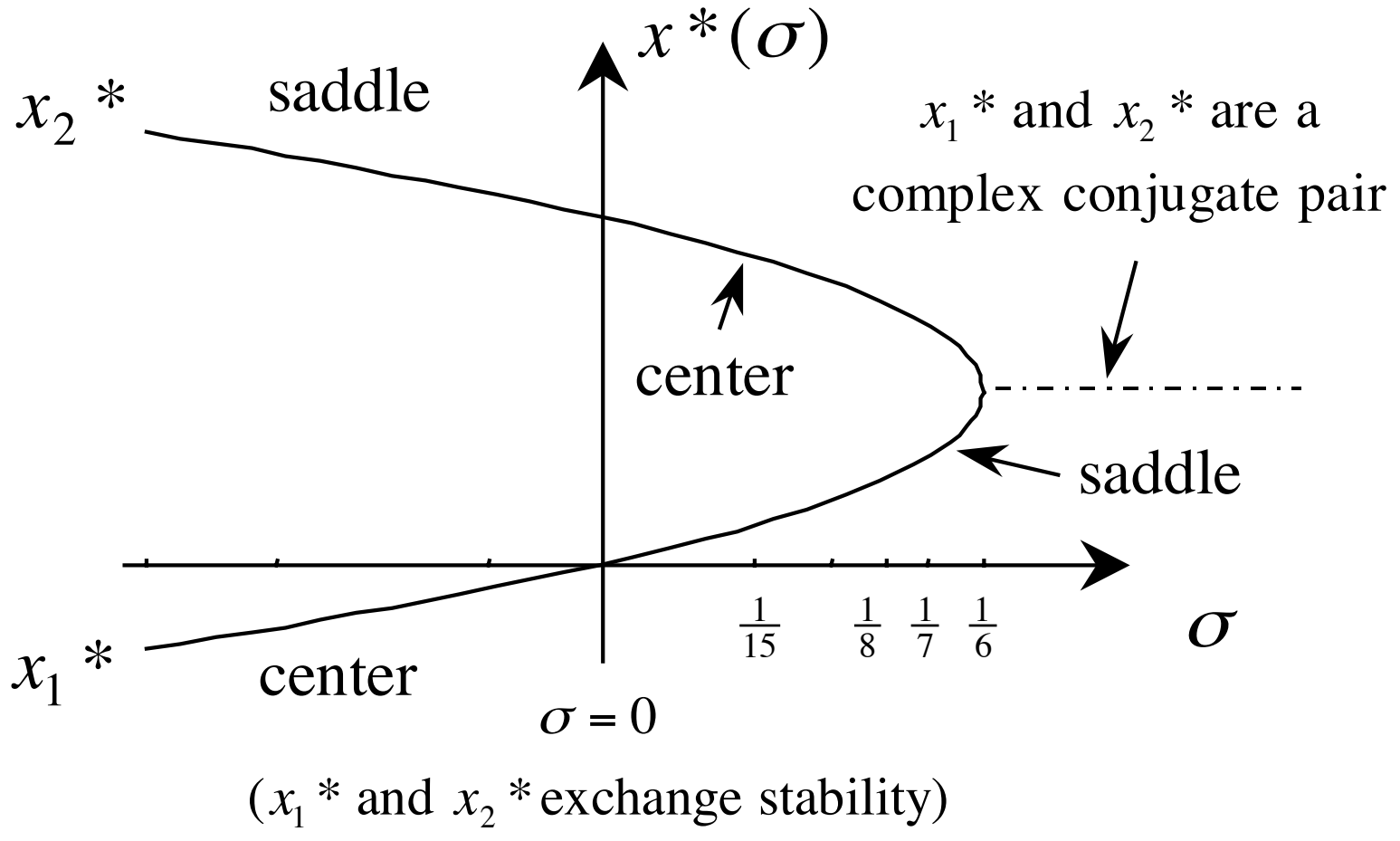

Figure 7 Schwarzschild Bifurcation Diagram

## Proper and Coordinate Time Analysis

In the standard analysis on relativistic orbital dynamics the proper time parameter is replaced by the equatorial angle as the independent variable. One advantage in this replacement is to simplify the algebra of a perturbative analysis, and is a carry over from the corresponding Newtonian analysis (see Appendix F). However, as far as the phase-plane analysis is concerned there are no essential difficulties analyzing the dynamics using the proper time (or coordinate time) as independent variables. To demonstrate the invariance of the effective potential between the proper and coordinate time reference frames start with (5.11) to obtain the proper time result (note: using $\varphi$ rather than $\tau$ eliminates the $x^4$ leading term appearing below):

$$(r_s / c)^2 \dot{x}^2 = x^4\left[\hat{E}^2 - \left(1 + x^2 / 2\sigma\right)\Lambda\right]. \tag{5.44}$$

† incidentally, $\vec{x}_2^*$ of Fig. 4 (and Fig. 3) is classified as a *nonlinear* center, see Strogatz [126], p. 164.

The corresponding coordinate time expression is obtained using: $\dot{x} = (dx / dt)\dot{t}$ in combination with (5.5) which gives

$$(r_s / c)^2 (dx / dt)^2 = x^4\left(\Lambda / \hat{E}\right)^2\left[\hat{E}^2 - \left(1 + x^2 / 2\sigma\right)\Lambda\right]. \tag{5.45}$$

Solving (5.45) for $\hat{E}^2$ gives the coordinate frame expression for the total energy:

$$\hat{E}^2 = \frac{x^4\left(1 + x^2 / 2\sigma\right)\Lambda^3}{x^4\Lambda^2 - (r_s / c)^2 (dx / dt)^2}. \tag{5.46}$$

By inspection of (5.46), as $dx / dt \to 0$, the effective potential (5.14) is recovered, i.e. $\hat{V}_{eff}^2 \to \hat{V}_{eff}'^{\,2} = \hat{V}_{eff}^2$, is invariant between the proper and coordinate time reference frames. Therefore, the dynamical structure is invariant, or alternatively stated, the extrema of $\hat{V}_{eff}^2$ are identical in either reference frame. However, the phase diagrams in each case are not identical due to the existence of an additional "frame-dependent" fixed point that appears in the coordinate reference frame at the event horizon (see Figure 8).

To summarize these results, the corresponding phase-plane equations analogous to (5.17) in both the proper and coordinate time reference frames are derived by differentiating (5.44) and (5.45), respectively. In each case the results are given by:

$$\begin{aligned} dx / d\tau &= y = \pm x^2 [\hat{E}^2 - (1 + x^2 / 2\sigma)\Lambda]^{1/2} \\ dy / d\tau &= x^3 [7x^3 - 6x^2 + 10x\sigma + 8\sigma(\hat{E}^2 - 1)] / 4\sigma, \end{aligned} \tag{5.47}$$

and

$$\begin{aligned} dx / dt &= y = \pm x^2\Lambda^{1/2}[\hat{E}^2 - (1 + x^2 / 2\sigma)\Lambda]^{1/2} / \hat{E} \\ dy / dt &= x^3\Lambda[9x^4 - 15x^3 + 2x^2(3 + 7\sigma) + 2x\sigma(6\hat{E}^2 - 11) - 8\sigma(\hat{E}^2 - 1)] / 4\sigma. \end{aligned} \tag{5.48}$$

Although (5.47) and (5.48) are more complicated algebraically than (5.17), the simultaneous solution of $\dot{x} = \dot{y} = 0$ for $E^2$ and $x$ in each case reduces to (5.18) and (5.20) identically, but with another fixed point, $x = 0$, at infinity and at $x = 1$ in the case of (5.48). However, the fixed point at infinity exists for the Newtonian case as well, and is considered in Appendix F. The fixed point at the event horizon is obviously coordinate dependent and does not correspond to any extrema of the effective potential. Nevertheless, this fixed point has physical consequences for observers in the coordinate reference frame - explaining the slowing down of objects and redshift of signals from objects approaching the event horizon.

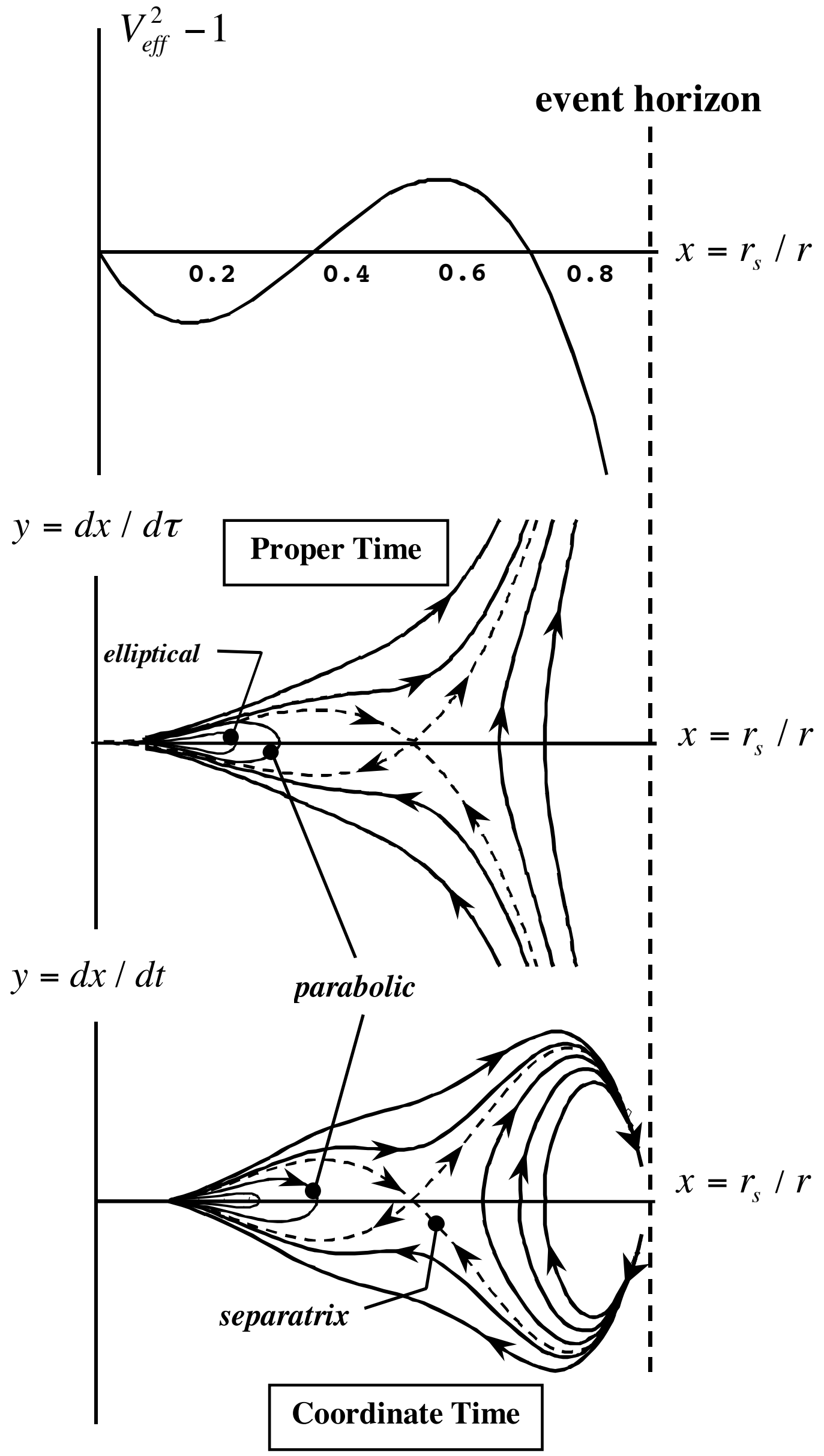

Figure 8 Proper and Coordinate Time Phase Diagrams

As discussed below (5.20), there are additional non-physical roots obtained when solving $\dot{x} = \dot{y} = 0$ only for $x$ and $y$. The non-physical nature of these fixed points is due to the fact that there must be a constraint placed upon $\hat{E}$ when $\tau$ or $t$ is used as the independent variable. Solving simultaneously the expressions for $y$ given in (5.47) or (5.48) gives the proper constraint on $\hat{E}$, and as a result forces these fixed points to coincide with the extrema of the effective potential. This is also a feature of the Newtonian dynamics when using $t$ as the independent variable (Appendix F).

## Light Rays

The analysis of photon orbits in the Schwarzschild spacetime is a straightforward application of the techniques discussed for timelike orbits. For light rays, $d\tau = 0$, which in turn implies that both $\hat{E}$ and $\hat{J}$ are divergent from (5.5) and (5.6), although their ratio remains finite. As a result, $\sigma \to 0$, and the phase-plane equations for light rays follow as a special case of (5.17):

$$\begin{aligned} x' &= y = \pm\, [1/b^2 - x^2(1-x)]^{1/2} \\ y' &= \tfrac{3}{2}x^2 - x\,, \end{aligned} \tag{5.49}$$

where $1/b^2 \equiv 2\sigma\hat{E}^2$, is a constant expressing the dimensionless impact parameter, $b$, as the finite ratio of $\hat{E}$, $\hat{J}$, and $r_s$.

The simultaneous solution of $x' = y' = 0$ for $1/b^2$ and $x$ results in two fixed points and the corresponding values of the impact parameter:

$$\{x_1 = \tfrac{2}{3}\,;\, 1/b^2 = \tfrac{4}{27}\} \text{ and } \{x_2 = 0\,;\, 1/b^2 = 0\}\,, \tag{5.50}$$

giving the standard results for the unstable orbital radius, $x_1$, and the impact parameter at which this instability occurs. The fixed point, $x_2$, is a center-node (at infinity) about which the hyperbolic orbits "precess" and gives the standard result on light bending. Therefore, the periastron precession of timelike orbits and light bending are actually special cases of one another: in the timelike case this center node fixed point is at finite $r$ and allows "real" circular orbits; but for light rays this fixed point moves to infinity and gives the precessing hyperbolic orbits pointed out above. However, a phase-plane calculation of light bending analogous to that discussed for the Schwarzschild case does not work here. This is due to the fact that a linear stability analysis (5.21) “kills” the necessary terms; namely, the impact parameter $b$ disappears from the matrix $\boldsymbol{A}$ (a similar result occurs when calculating the period of a simple pendulum for large angles using this technique).

The phase-plane level curves for light rays in Figure 9 correspond to different values of $1/b^2$. These are shown together with the locations of the fixed points and photon effective potential: $x^2(1-x)$. The most striking difference between the photon and timelike dynamics (comparing Figures 9 and 4) is that the center node fixed point moves to the origin as $\sigma \to 0$ (as discussed above). As a result, circular photon orbits do not exist in any dynamical sense, but become circular in geometry as the orbits approach the separatrix. To see this use the definition of $y$ in the first equation of (5.49), and then separating variables shows that $\varphi \to \infty$ as $x \to x_1$ and $1/b^2 \to 4/27$ (this result is analogous to the proper time divergence pointed out in (5.43)). Therefore, the separatrix corresponds to the unstable “photon sphere” that is commonly discussed in the literature (see e.g. Ohanian and Ruffini [93], p. 410).

The physical interpretation of the various phase-plane regions of Figure 9 is similar to that of Figure 4, but there are important differences. For light rays with impact parameter $1/b^2 = 0$, these orbits just graze the event horizon from the inside and simultaneously (in an unrelated trajectory) reach the center node fixed point of the effective potential (see Figure 9). For $1/b^2 < 0$, $b$ loses its interpretation as an impact parameter since the trajectories in this case originate from the singularity at $r = 0$ and lie within the horizon. For $1/b^2 < 4/27$, the trajectories are confined to within the separatrix and correspond to the light rays arriving from infinity, reaching a turning point (given by the appropriate root of the first equation in (5.49)), and then return to infinity as discussed below (5.50). For $1/b^2 > \frac{4}{27}$, a photon arrives from infinity (above the separatrix) and then falls through the event horizon. The corresponding time reversed trajectories are given below the separatrix. The trajectories to the right of the unstable orbital radius of Figure 9 are also interpreted as time reversed paths that reach a maximum distance from the event horizon and then return to the singularity.

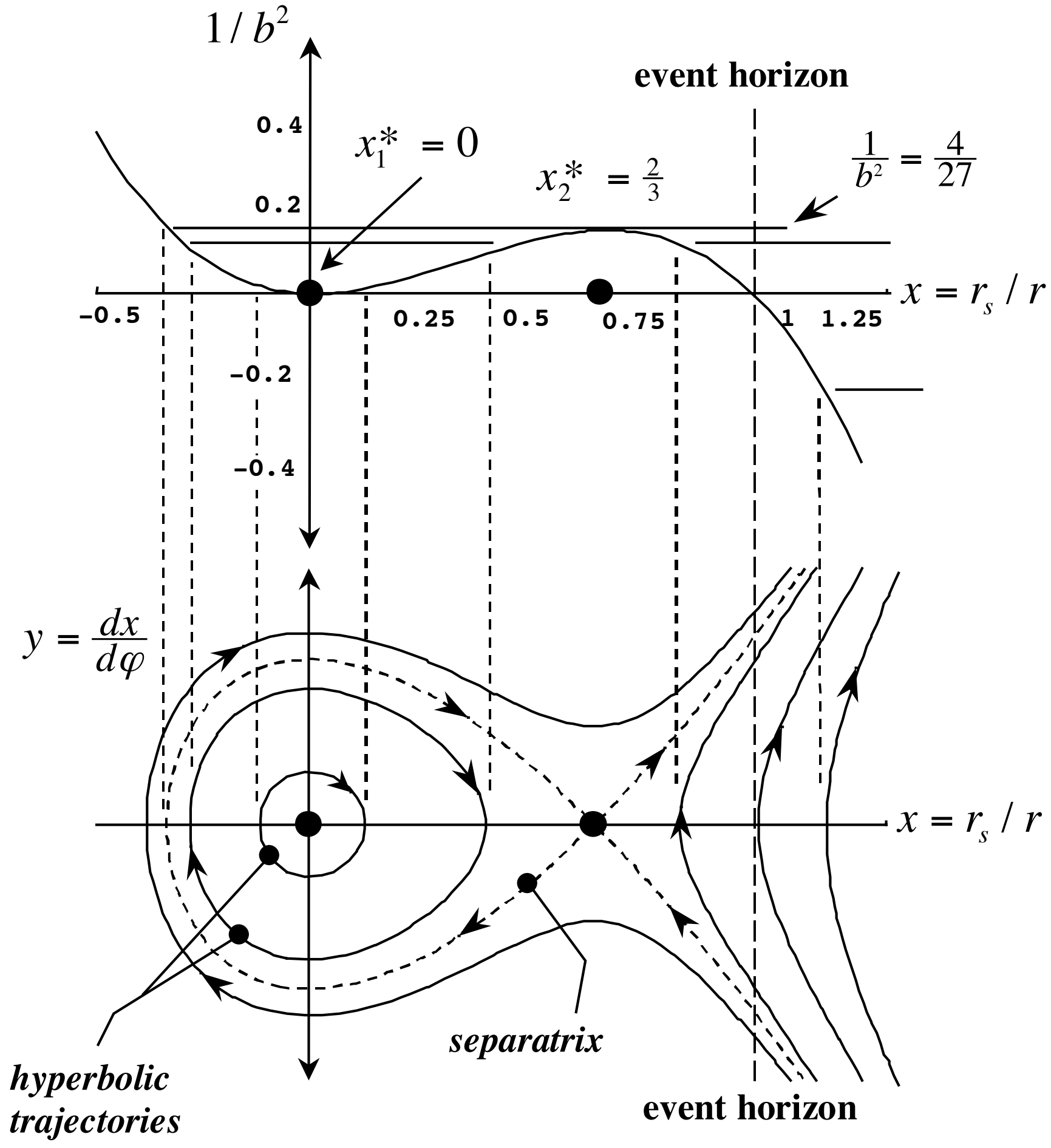

Figure 9 Null Geodesics Phase-Plane

# Chapter 6 Reissner-Nordström Dynamics

As discussed earlier in Chapter 2 the Reissner-Nordström solution is the exterior solution of a spherically symmetric charged distribution of matter that has collapsed to form a black hole. The line element was derived earlier and listed below in Schwarzschild coordinates:

$$\begin{aligned} ds^2 &= \Lambda\, dt^2 - \Lambda^{-1} dr^2 - r^2 d\Omega^2 \\ \Lambda &= 1 - x + x^2/2\lambda; \quad x = r_s/r, \end{aligned} \tag{6.1}$$

where $d\Omega^2 \equiv d\theta^2 + \sin^2\theta\, d\varphi^2$; $\lambda$ and $\sigma$ are the dimensionless parameters defined earlier in Chapter 2 (equation (2.100)):

$$\lambda = \frac{c^4}{2Gk}\left(\frac{r_s}{e}\right)^2 = \frac{2G}{k}\left(\frac{M}{e}\right)^2. \tag{6.2}$$

From a strictly dynamical viewpoint, the Reissner-Nordström solution exhibits a much richer orbital structure than the Schwarzschild case - yielding three fixed points whose relative positions and stability properties in state-space are determined by the values of black hole charge and orbital angular momentum. The purpose of this Chapter is to solve the bifurcation problem for the Reissner-Nordström system, i.e., to present a summary of the phase-plane topological structure based on a study of coalescing fixed points and to find the parameter values at which these bifurcations occur. The separatrix plays an important role in the analysis by distinguishing the dynamical regions of the phase-plane at given values of the black hole charge and orbital angular momentum. The main difference as far as the separatrix is concerned (compared to the Schwarzschild case) is that the Reissner-Nordström separatrix is dependent upon the black hole charge, but the structure of the separatrix resembles closely the Schwarzschild separatrix and therefore can be used in a similar way to base a qualitative classification of the orbital dynamics. The main results of this Chapter will be a discussion and graphical analysis of orbital stability which is presented using the bifurcation diagrams as discussed below.

As discussed in a later section of this Chapter, the presence of charge reduces the periastron precession compared to the standard Schwarzschild result. Therefore, an obvious experimental measurement of the gravity induced solar charge considered by Eddington [77] and Harrison [79] could be based on observations of perihelion advance. In fact, Burman has considered this application to Mercury's orbit using estimates of the solar charge given by Bailey [137], but concludes that the correction made to the Schwarzschild precession value is negligible. Subsequently, Treder, et. al. [138] have discussed perihelion advance in the Reissner-Nordström spacetime as a means for estimating the solar charge based on accurate perihelion data for planetary bodies. Following Burman, Teli and Palaskar [139] also consider the effect of a net solar charge on the perihelion advance of Mercury's orbit (as well as to the orbits of Venus and Icarus). But the difficulty in such measurements is that the Schwarzschild precession value is very small to begin with (at least in the solar system; see equation (5.36)).

## Orbital Equations and Fixed Points

The Reissner-Nordström and Schwarzschild solutions are both spherically symmetric and static and therefore the constants of motion and orbital equations obtained for both solutions have the same general form. The essential difference arises only from the form of $\Lambda$ . As a result, the phase-plane equations for the Reissner-Nordström system are obtained from (5.16) and (6.1) ($x \equiv r_s / r$ ; $x' \equiv y \equiv dx/d\varphi$ ):

$$\begin{aligned} x' &= y \\ y' &= -x^3/\lambda + 3x^2/2 - \left(1+\sigma/\lambda\right)x + \sigma , \end{aligned} \tag{6.3}$$

defining the equatorial motion of a non-charged unit point mass orbiting a charged spherically symmetric black hole. For comparison and later reference the dimensionless parameters $\sigma$ and $\lambda$ are listed together (defined in equations (5.13) and (2.100), respectively):

$$\sigma = \tfrac{1}{2}(c^{-2})(r_s/\hat{J})^2;\ \lambda = \tfrac{1}{2}(c^4/Gk)\left(r_s/e\right)^2 . \tag{6.4}$$

The Reissner-Nordström fixed points are the roots of a cubic equation while the Schwarzschild fixed points result from a quadratic. As a result, the Reissner-Nordström fixed points (and resulting dynamics) are considerably more complex algebraically. The fixed points, $\vec{x}^*$, are obtained from $x' = y' = 0$ , and expressed in the form (denoting $\vec{x}^* = (x_i^*, y^*)$ ; note that $y^* = 0$ ):

$$x_1^* = \lambda/2 + a \cdot g^{-\frac{1}{3}} + g^{\frac{1}{3}}/(2^{4/3} \cdot 3)$$
$$x_2^* = \lambda/2 + (1 + \mathrm{i}\sqrt{3})/(6g^{\frac{1}{3}}) + \mathrm{i}(1 + \mathrm{i}\sqrt{3})\, g^{\frac{1}{3}}/(2^{7/3} \cdot 3) \quad (6.5)$$
$$x_3^* = \lambda/2 + (1 + \mathrm{i}\sqrt{3})/(6g^{\frac{1}{3}}) - \mathrm{i}(1 + \mathrm{i}\sqrt{3})\, g^{\frac{1}{3}}/(2^{7/3} \cdot 3)\ ,$$

where i is a complex factor resulting from the cubic equation, $y' = 0$. For brevity, the following substitutions have been defined in (6.5):

$$g = 2b + \sqrt{4d^3 + 2916c^2}$$
$$a = 3\lambda^2 - 4(\lambda + \sigma);\ b = 27\lambda^3 - 54(\lambda^2 - \lambda\sigma) \quad (6.6)$$
$$c = \lambda^2(\lambda^2 - 2\lambda + 2\sigma);\ d = -9\lambda^2 + 12(\lambda + \sigma)\ .$$

## Parameter Space

The bifurcation problem for the Reissner-Nordström system consists of determining all possible parameter values of $\lambda$ and $\sigma$ that correspond to coalescing fixed points. A solution to the problem is thus given by basing an analysis on the descriminant of the cubic equation $y' = 0$. For reference, the descriminant $D$ of a general cubic equation:

$$x^3 + a_2 x^2 + a_1 x + a_0 = 0\ , \quad (6.7)$$

is given by (for instance, Abramowitz and Stegun [140]):

$$D = q^3 + p^2\ , \quad (6.8)$$

where

$$q = \tfrac{1}{3}a_1 - \tfrac{1}{9}a_2^2;\ p = \tfrac{1}{6}(a_1 a_2 - 3a_0) - \tfrac{1}{27}a_2^3\ . \quad (6.9)$$

and therefore, the Reissner-Nordström descriminant simplifies to

$$D_{\lambda\sigma} = \left[\sigma^3 - \tfrac{9}{16}\lambda\left(\lambda - \tfrac{16}{3}\right)\sigma^2 + \tfrac{3}{8}\lambda^2\left(9\lambda^2 - 21\lambda + 8\right)\sigma - \tfrac{9}{16}\lambda^3\left(\lambda - \tfrac{16}{9}\right)\right]/27\ . \quad (6.10)$$

The roots of (6.7) are thus classified according the sign of the descriminant:

$D > 0$: 1 real and 2 complex conjugate roots
$D = 0$: 3 real roots, but at most 2 distinct (a bifurcation point) (6.11)
$D < 0$: 3 real distinct roots .

By an application of (6.11) to (6.5) the bifurcations of the Reissner-Nordström fixed points are identified according to the numerical values of $D_{\lambda\sigma}$. But note that a bifurcation analysis of the Reissner-Nordström system based on (6.11) is complicated by the dual parameter dependence of $D_{\lambda\sigma}$. The solution to this difficulty is given by eliminating $\sigma$ using the descriminant of (6.10) (i.e., the descriminant of the descriminant). This secondary descriminant (of (6.10)), $D_\lambda$, simplifies to

$$D_\lambda = \left(45/32\right)^3 \lambda^8 (\lambda - 2)(\lambda - \tfrac{8}{5})^3 / 2\,, \tag{6.12}$$

and therefore the $\sigma_i$ roots of (6.10), corresponding to the case, $D_{\lambda\sigma} = 0$, are expressed in the form:

$$\begin{aligned} \sigma_1 &= \lambda(3\lambda/16 - 1) - 3\alpha/16\beta^{\frac{1}{3}} + 3\lambda^{\frac{4}{3}}\beta^{\frac{1}{3}} \\ \sigma_2 &= \lambda(3\lambda/16 - 1) - 3(1 + \mathrm{i}\sqrt{3})\alpha/32\beta^{\frac{1}{3}} + 3\mathrm{i}(1 + \mathrm{i}\sqrt{3})\lambda^{\frac{4}{3}}\beta^{\frac{1}{3}} \\ \sigma_3 &= \lambda(3\lambda/16 - 1) - 3(1 - \mathrm{i}\sqrt{3})\alpha/32\beta^{\frac{1}{3}} - 3\mathrm{i}(1 - \mathrm{i}\sqrt{3})\lambda^{\frac{4}{3}}\beta^{\frac{1}{3}}\ , \end{aligned} \tag{6.13}$$

where $\alpha$ and $\beta$ are functions of $\lambda$:

$$\begin{aligned} \alpha &= \lambda^{\frac{5}{3}}(31\lambda - 64) \\ \beta &= -47\lambda^2 + 352\lambda - 512 + 16(\lambda - 2)^{\frac{1}{2}}(5\lambda - 8)^{\frac{3}{2}}. \end{aligned} \tag{6.14}$$

Solving $D_\lambda = 0$ implies that the $\sigma_i$ are all real but with only two distinct at the parameter values:

$$\lambda = \{0, 8/5, 2\}. \tag{6.15}$$

In Figure 10, $\mathrm{Re}(\sigma_i)$ *vs.* $\lambda$ is plotted to demonstrate the classification of (6.13) based on (6.11) (dashed lines indicate the location of complex roots).

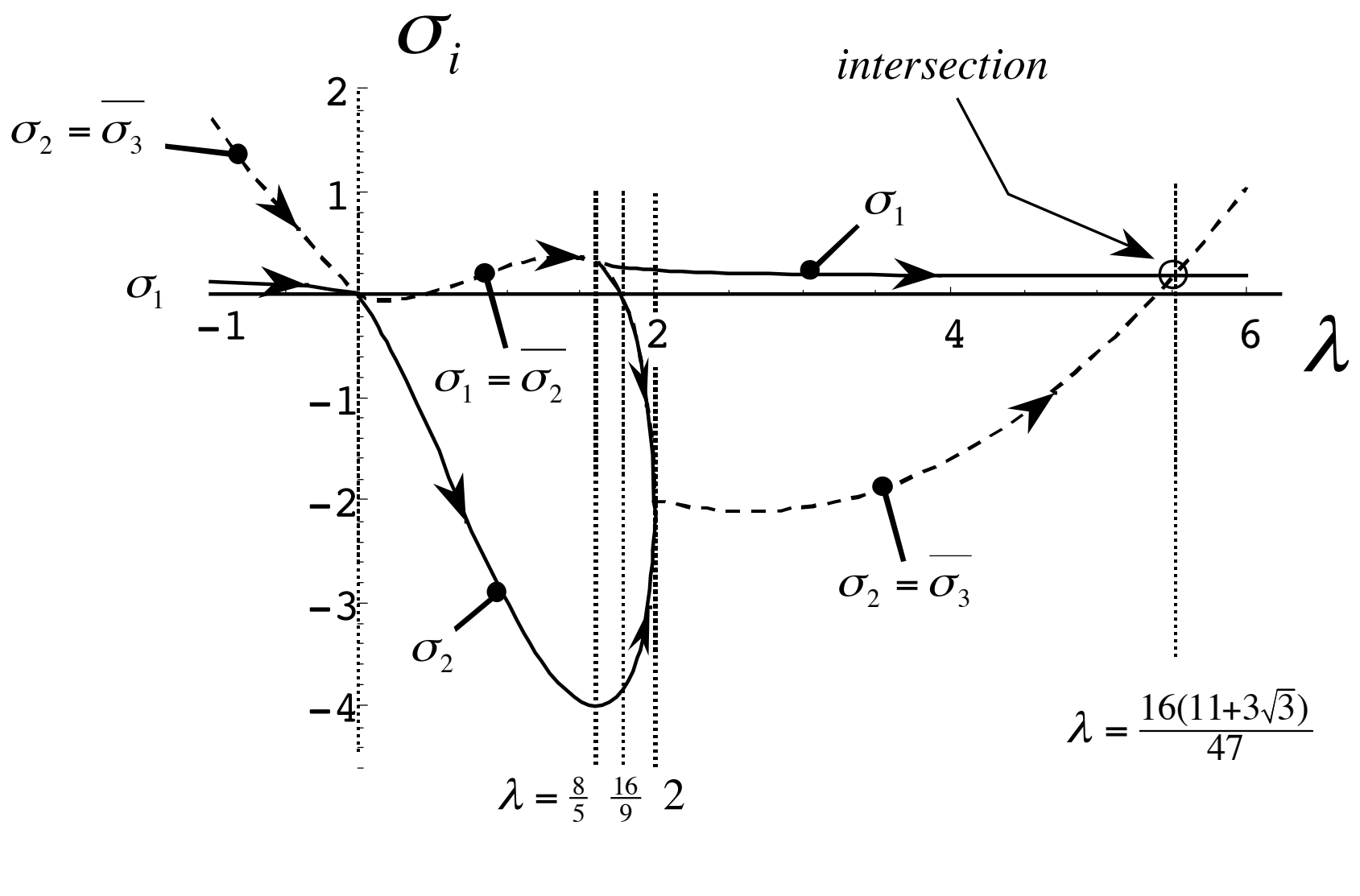

Figure 10 $\mathrm{Re}(\sigma_i)$ *vs.* $\lambda$

An additional parameter value of interest is given by $\lambda = 16/9$. At this parameter value (6.10) becomes quadratic in $\sigma$ when $D_{\lambda\sigma} = 0$ (but $D_\lambda < 0$ which implies that the $\sigma_i$ are all real and distinct). As a result, the numerical values of $\lambda$:

$$\lambda_i = \{0, \tfrac{8}{5}, \tfrac{16}{9}, 2\}, \tag{6.16}$$

when combined with (6.13) give the following values for $\sigma_i$:

$$\begin{aligned}
&\lambda_1 = 0 \Rightarrow \sigma_1 = \sigma_2 = \sigma_3 = 0 \\
&\lambda_2 = 8/5 \Rightarrow \sigma_1 = \sigma_3 = 8/25;\ \sigma_2 = -4 \\
&\lambda_3 = 16/9 \Rightarrow \sigma_1 \approx 0.275;\ \sigma_2 \approx -3.8;\ \sigma_3 = 0 \\
&\lambda_4 = 2 \Rightarrow \sigma_1 = 1/4;\ \sigma_2 = \sigma_3 = -2\ .
\end{aligned} \tag{6.17}$$

The point of intersection between $\mathrm{Re}(\sigma_i)$ (toward the right of Figure 10) is obtained by solving

$$\begin{aligned}
&\mathrm{Re}(\sigma_1) = \mathrm{Re}(\sigma_{2,3}) \Rightarrow (\beta - 31)\lambda + 64 = 0 \\
&\qquad \Rightarrow \lambda = \left\{ 0,\ 16(11 + 3\sqrt{3})/47 \approx 5.5 \right\},
\end{aligned} \tag{6.18}$$

where $\beta$ is given by (6.14). The second root of (6.18) gives the desired result but has no special significance for the dynamics since $D_{\lambda\sigma} > 0$ implies that $\sigma_2$ and $\sigma_3$ are complex conjugate (this is region I of Figure 11).

These results are further clarified by Figure 11 which defines the Reissner-Nordström parameter space. The bold line is given by the implicit solution of $D_{\lambda\sigma} = 0$ and divides the space into the seven physically distinct regions labeled in the Figure. Each segment of this line corresponds to the appropriate real root of $\sigma_i$ as labeled in Figure 10. The bifurcation points (6.17) are given by the intersections of this curve with the $\lambda_i$ of (6.16) and marked in the diagram with small circles. To summarize Figure 11, different regions of the Reissner-Nordström parameter space correspond to the separate cases of (6.11), which in turn distinguish the stability properties of the fixed points [note: the numerical values $\lambda = (-1, +1, \lambda > 2)$ in Figure 11 have no special significance other than as generic parameter values used to illustrate the dynamics in the indicated regions].

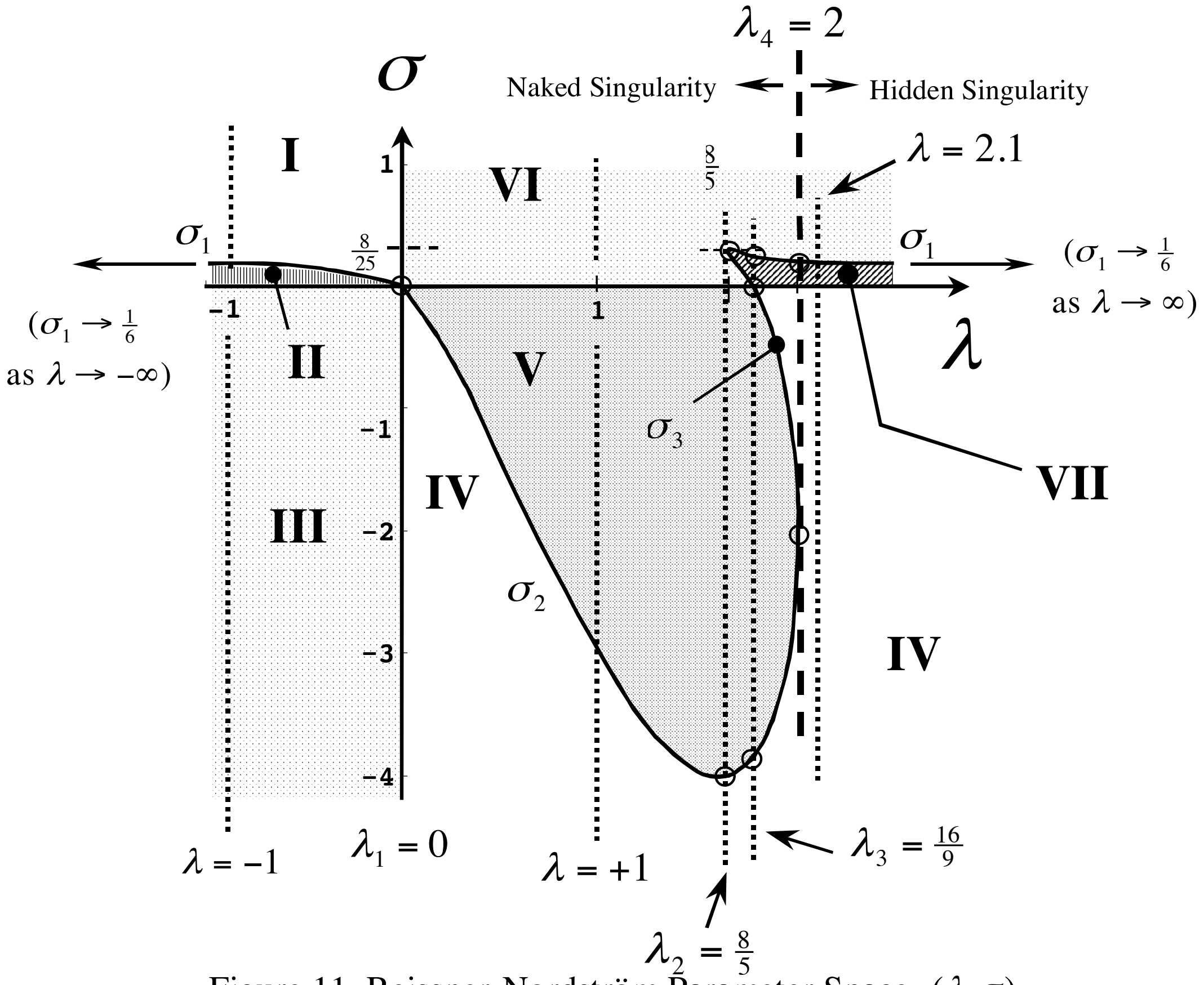

Figure 11 Reissner-Nordström Parameter Space, $(\lambda,\sigma)$.

It should be noted that the origin of the parameter space in Figure 11 is alternatively expressed in terms of Catastrophe Theory [141] which views the parameter space as a 2-d projection of a 3-d object whose surface is defined by the critical points of the effective potential. The folds in the surface correspond to the set of degenerate critical points (termed the singularity set in Catastrophe Theory). The significance is that the degenerate critical points mark a change of stability in the potential which is why the regions of Figure 11 are dynamically distinct to begin with.

The interpretation of the regions of Figure 11 are simple to deduce from the parameter values (6.4). The only physical possibilities are given by $\lambda$ and $\sigma$ both positive (regions VI and VII). Therefore, regions I - V are clearly nonphysical (the charge and/or angular momentum are complex valued), but as discussed in the next section, the underlying dynamical structure of regions VI and VII is more easily identified by considering the analysis for all regions of the parameter space. For reference, the first integral, $\hat{E}$, is identical in form to the Schwarzschild equation (5.12):

$$\left(dx/d\varphi\right)^2 = 2\sigma(\hat{E}^2 - \hat{V}_{eff}^2), \tag{6.19}$$

and the effective potential has the same general form:

$$\hat{V}_{eff}^2 = (1 + x^2/2\sigma)\Lambda. \tag{6.20}$$

Equivalently, the effective potential is obtained from the expression for Reissner-Nordström level curves (listed below) as $y \to 0$:

$$\hat{E}^2 - 1 = \underbrace{(y^2 + x^2 - x^3)/2\sigma - x}_{\text{Schwarzschild}} + (x^4/2\sigma + x^2)/2\lambda, \tag{6.21}$$

## Bifurcation Analysis

As discussed earlier in the Schwarzschild case, the stability properties of the fixed points are obtained by linearizing (6.3) and then classifying the resulting eigenvalues. But for the Reissner-Nordström system this approach is formidable algebraically and not very instructive. A more direct approach is to consider a graphical analysis that is based on the level curves of (6.21) in combination with the vector field of (6.3) and the effective potential at specific parameter values. The main results from the linear stability analysis are summarized by the bifurcation diagrams obtained by plotting $\vec{x}^*$ for fixed values of $\lambda$, while varying the orbital angular momentum. By including the limiting behavior of these dynamics (i.e., as $(\lambda,\sigma) \to \pm\infty$), all regions of the parameter space are covered in the analysis. These results are summarized in Figures 14 – 28 and in the discussion below. Samples of the *Mathematica* code (Wolfram [142]) used to obtain these results are listed in Appendix G.

$\lambda < 0$

---

First consider a cross section of the parameter space at $\lambda = -1$ (Figure 11). This parameter value is arbitrarily chosen to easily display the (graphical) locations of fixed points and generic behavior of the dynamics across regions *I*, *II*, and *III*. The bifurcation diagram is given in Figure 16 by plotting $\mathrm{Re}(\vec{x}^*)$ *vs.* $\sigma$. In summary for $\lambda < 0$, $\sigma_1$ is real while $\sigma_2$ and $\sigma_3$ are complex conjugate ($D_\lambda > 0$). Therefore, $\sigma_1$ separates regions *I* and *II* as illustrated in Figure 11. Referring to Figure 11, begin first at $\sigma < 0$ along the line $\lambda = -1$ and then proceed upward from region *III* to *II* and then finally to region *I*.

Region *III* is characterized by $D_{\lambda\sigma} < 0$ and therefore, the fixed points are real and distinct with the following stability properties: at $\lambda = -1$, $x_1^*$ is an unstable saddle node (Figure 12) converging to the numerical value, $x_1^* \to -1$ as $\sigma \to -\infty$, while $x_2^*$ and $x_3^*$ are center nodes diverging to $\pm\infty$, respectively.

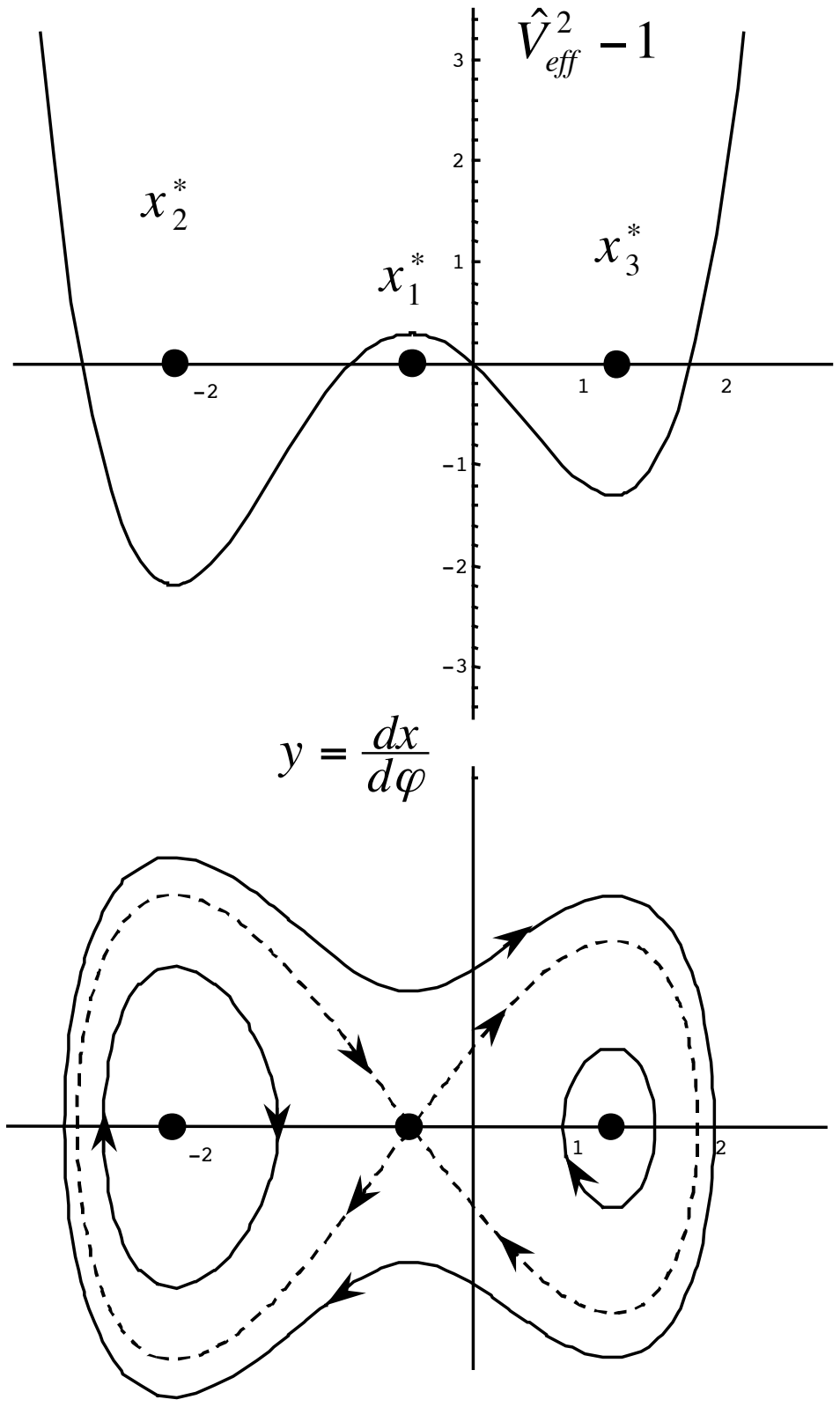

Figure 12 Phase Diagram for $\lambda = -1$; $\sigma < 0$ (region III).

As both $\sigma \to -\infty$ and $\lambda \to -\infty$, $x_1^*$ is an unstable saddle node diverging to $-\infty$ while $x_2^*$ and $x_3^*$ remain center nodes diverging to $\pm\infty$, respectively.

The parameter value, $\sigma = 0$, gives a transcritical bifurcation point dividing regions II and III and marks an exchange of stability between $x_1^*$, $x_2^*$, and $x_3^*$. As in region III, $D_{\lambda\sigma}$ is also negative in region II and therefore the fixed points are real and distinct but now $x_1^*$ is a center node while $x_2^*$ and $x_3^*$ become saddles in region II (Figure 13).

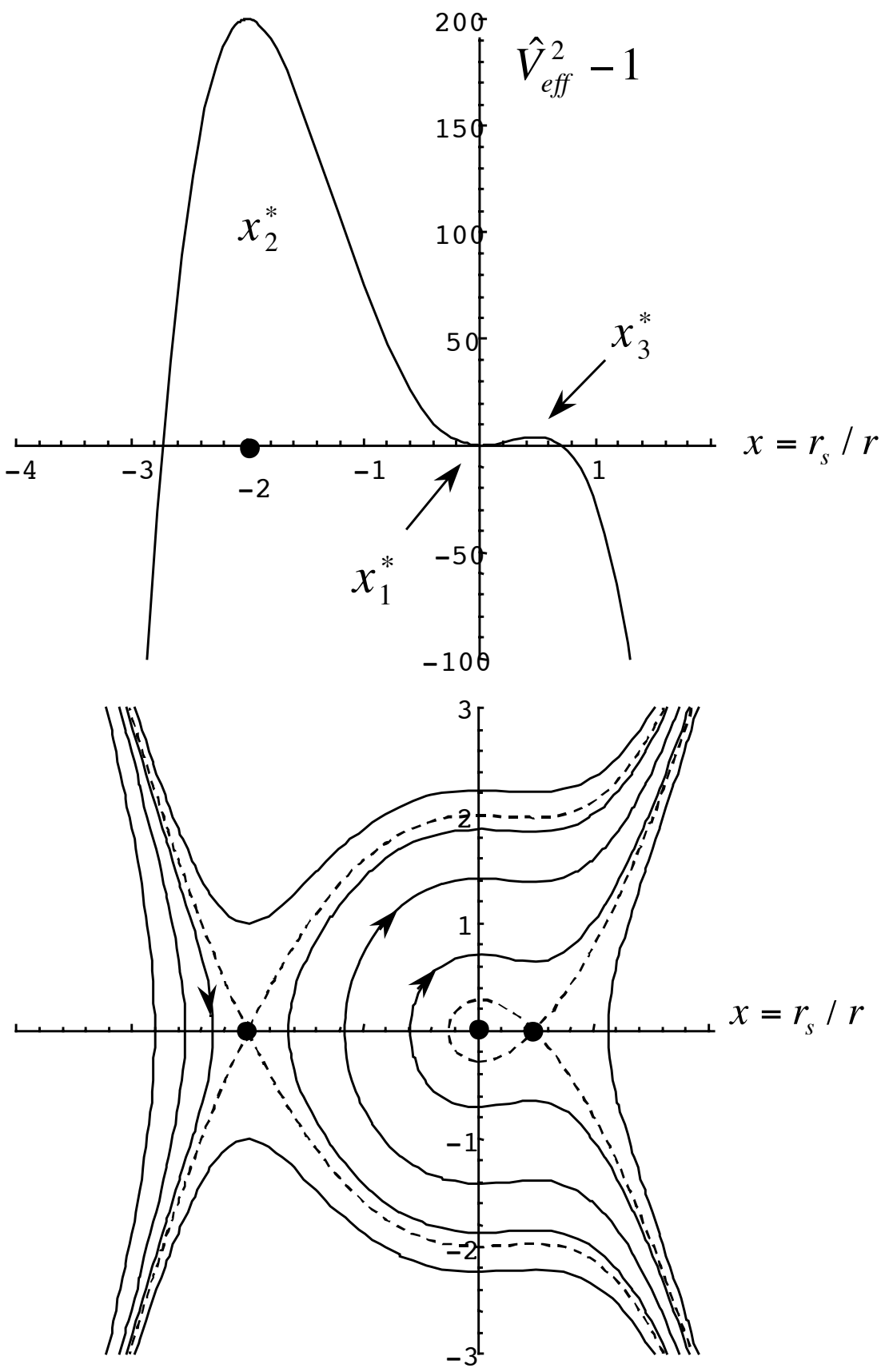

Figure 13 Phase Diagram for $\lambda = -1; 0 < \sigma < \sigma_1$ (region II).

At $\sigma = \sigma_1$, $D_{\lambda\sigma} = 0$, and a saddle-center bifurcation [136] occurs as the $x_1^*$ saddle node and $x_3^*$ center node merge together into an unstable degenerate inflection point as illustrated in Figure 14, however, $x_2^*$ retains its stability as a saddle node.

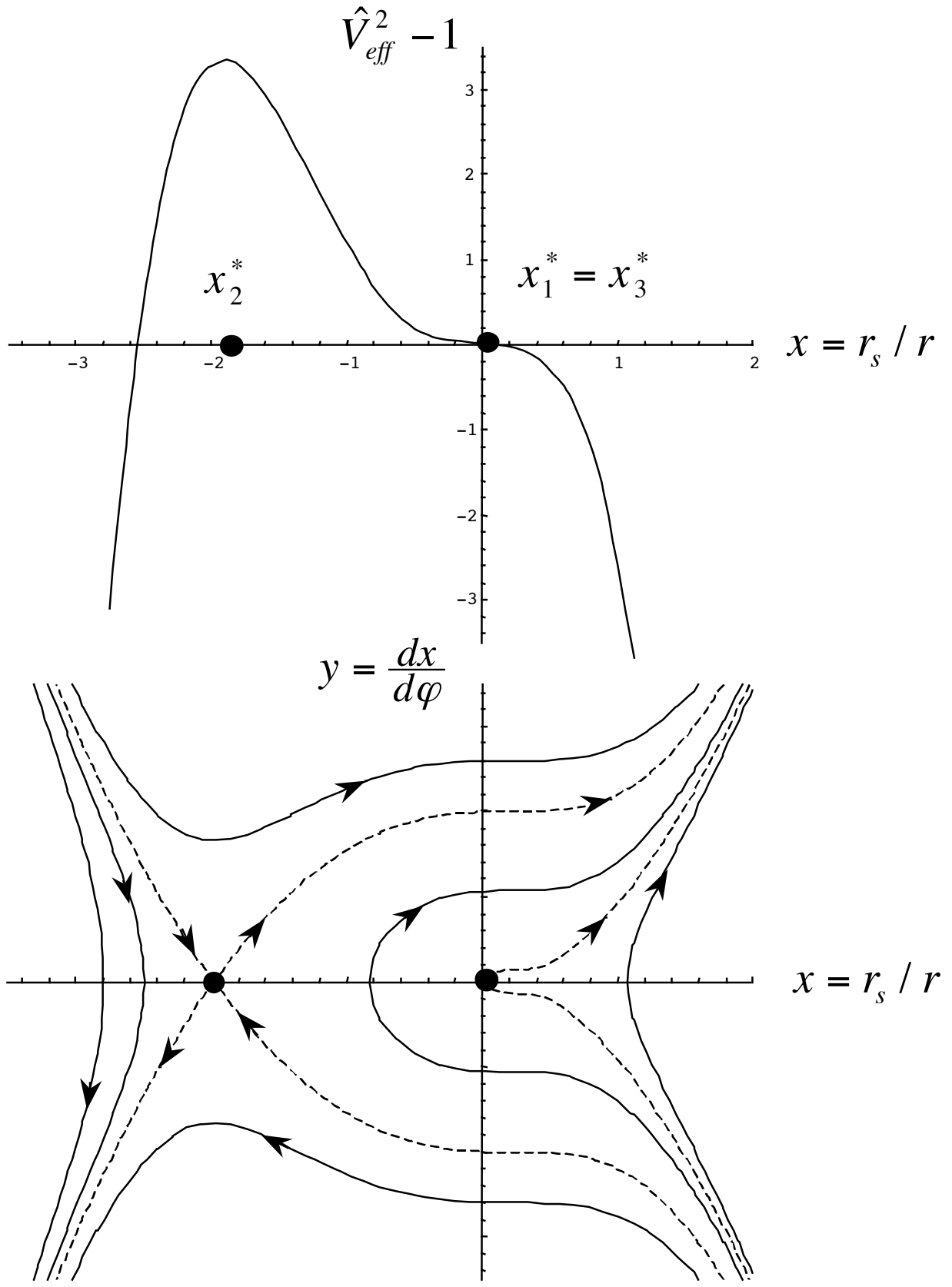

Figure 14 Phase Diagram for $\lambda = -1$; $\sigma = \sigma_1$.

In region I (Figure 15), $\sigma > \sigma_1$, $D_{\lambda\sigma} > 0$, and therefore, according to (6.11), only one real saddle node fixed point remains and is given by $x_3^*$. But interestingly, a "*pseudo*" transcritical bifurcation occurs at $\sigma = 7/4$ as $x_1^*$ and $x_2^*$ become complex conjugate while $x_2^*$ and $x_3^*$ exchange roles ($x_3^*$ becomes real at this parameter value). [*Note*: the term *pseudo* is used to indicate that an exchange of roles occurs - not an exchange of stability. In this case, it is the switching of real and complex conjugate roots that takes place while the dynamical properties of the system remain invariant].

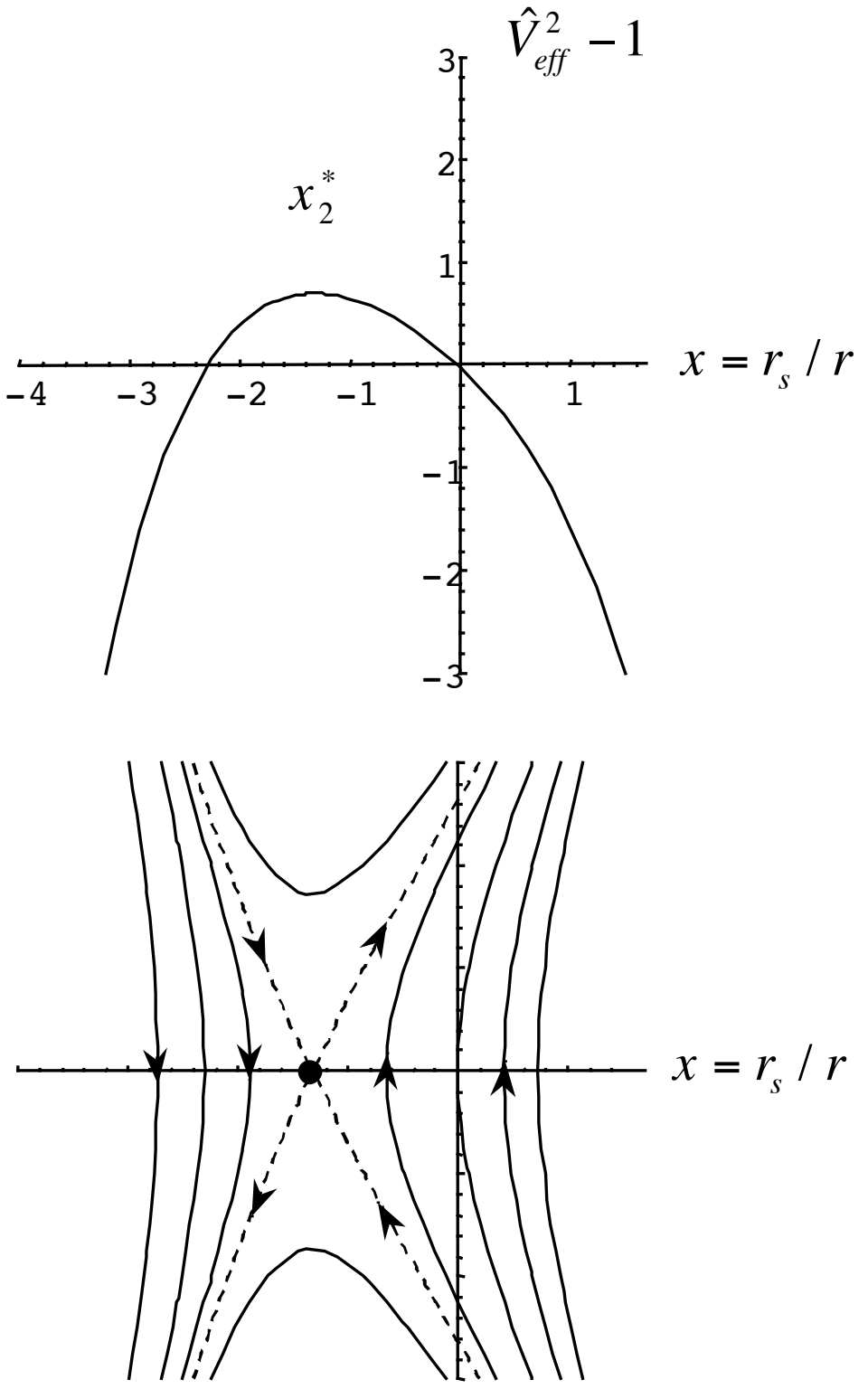

Figure 15 Phase Diagram for $\lambda = -1$; $\sigma > \sigma_1$ (region I).

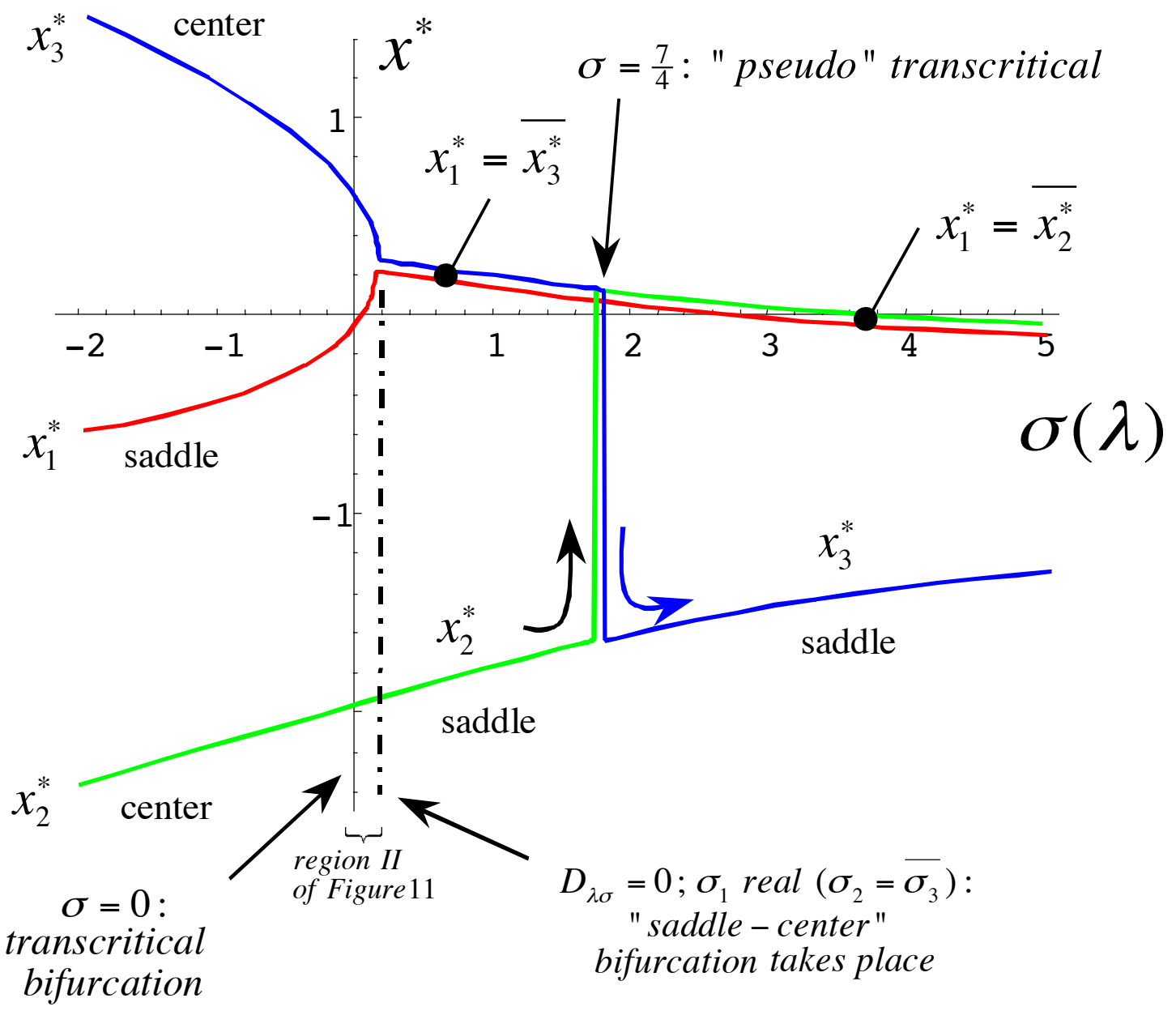

Figure 16 Bifurcation Diagram: $\lambda = -1$.

As $\sigma$ is further increased to $+\infty$, $x_3^* \to -1$ is the only real fixed point while $x_1^*$ and $x_2^*$ remain complex conjugate with $\mathrm{Re}(x_1^*) = \mathrm{Re}(x_2^*)$ converging to $-1/4$. A complementary limit of the Schwarzschild bifurcation diagram (Figure 7) is thus obtained as $\lambda \to -\infty$ and $x_2^*$ diverges to $-\infty$ as a center node, while $x_1^*$ ($<0$) and $x_3^*$ form saddle and a center nodes, respectively. For comparison, the true Schwarzschild limit is obtained as $\lambda \to +\infty$. In this case, $x_1^*$ is saddle node ($>0$) while $x_2^*$ is a center and $x_3^*$ diverges to $+\infty$ as a center node (this case is discussed in more detail for $\lambda > 2$).

**A "Pseudo" Transcritical Bifurcation Point**

Before discussing the bifurcation diagrams corresponding to positive values of $\lambda$, the structure of these dynamics will be more clearly illustrated by following the location of the pseudo-transcritical point as $\lambda$ is varied. The location is found analytically by solving for $\sigma$ from the equation:

$$\mathrm{Re}(x_2) = \mathrm{Re}(x_3) \Rightarrow A - \left(B + 6\sqrt{3C}\right)^{2/3} = 0 \qquad (6.22)$$

where

$$\begin{gathered} A = -9\lambda^2 + 12\lambda + 12\sigma;\ B = 27\lambda^3 - 54\lambda^2 + 54\sigma\lambda \\ C = 9(6\sigma - 1)\lambda^4 + (16 - 126\sigma)\lambda^3 + 3\sigma(16 - 3\sigma)\lambda^2 + 48\sigma^2\lambda + 16\sigma^3. \end{gathered} \qquad (6.23)$$

There are two solutions:

$$\sigma_I = \tfrac{3}{4}\lambda(\lambda - \tfrac{4}{3}),\ \sigma_{II} = -\tfrac{1}{2}\lambda(\lambda - 2), \qquad (6.24)$$

which are plotted in Figure 17. The $\sigma_I$ solution locates the pseudo-transcritical point which has a minimum value at $\sigma = -1/3$, while $\sigma_{II}$ locates the crossing point between $\mathrm{Re}(x_1) = \mathrm{Re}(\overline{x_2})$ and $x_3^*$ (the maximum is at $\sigma = 1/2$). This crossing point and the pseudo-transcritical point at $\sigma = -1/4$ are marked in Figure 17. The points of intersection between $\sigma_I$ and $\sigma_{II}$ determine the two totally degenerate bifurcation points of the Reissner-Nordström system given by

$$(\sigma = 8/25\ ;\ \lambda = 8/5) \Rightarrow x_1^* = x_2^* = x_2^* = 4/5, \qquad (6.25)$$

and

$$(\sigma = 0\ ;\ \lambda = 0) \Rightarrow x_1^* = x_2^* = x_2^* = 0. \qquad (6.26)$$

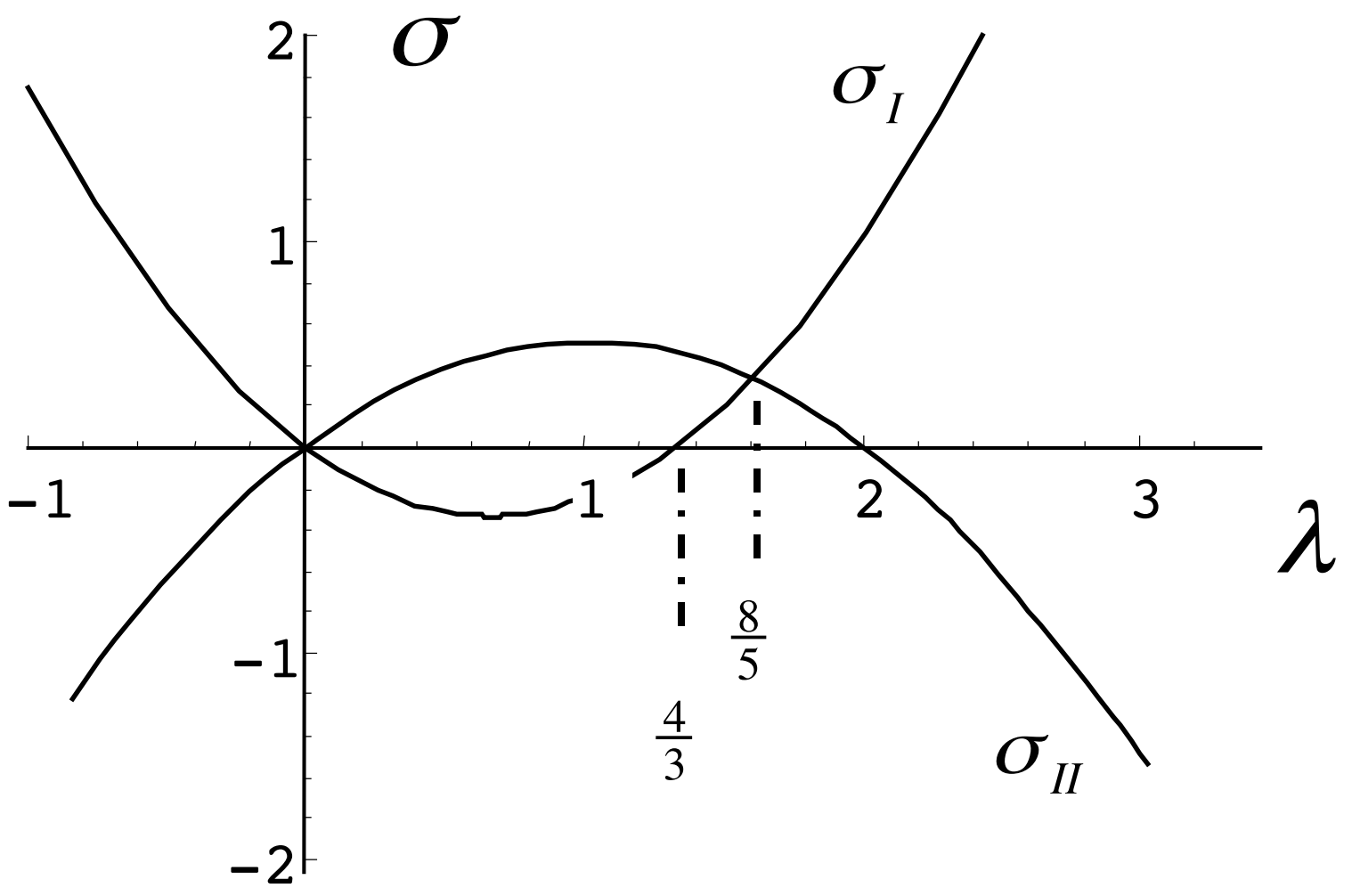

Figure 17 Positions of the Pseudo-Transcritical and Intersection Points of $\mathrm{Re}(x_{2,3})$ .

$\lambda = 0$

The interpretation of the bifurcation point (6.25) corresponds to timelike orbits about a singularity with no horizons ($\lambda < 2$). The parameter value (6.26) is a limiting case of the dynamics that is interpreted as corresponding to photon orbits ($\sigma = 0$) about a singularity possessing an infinite charge ($\lambda = 0$). However, it should be clear from the earlier discussion on the cosmic censorship conjecture that the labeling used for either bifurcation point is based solely on the parameter values and is not expected to correspond physically to any "real" dynamics. As $\lambda \to 0$ from the left, the $\sigma_1$ root separating regions I and II, and the complex conjugate exchange point between $x_2^*$ and $x_3^*$ both converge to zero. Therefore, the transcritical, saddle-center, and pseudo-transcritical bifurcation points merge together to form a stable center-node bifurcation point at $\lambda = \sigma = 0$ . This case is discussed in greater detail in the later section describing the Reissner-Nordström null geodesics.

$0 < \lambda \le 8/5$

For $0 < \lambda < 8/5$, the pseudo-transcritical and saddle-center bifurcation points separate from the transcritical point and are given by $\sigma_I$ of (6.24) and $\sigma_2$ of (6.13), respectively, at negative values of $\sigma$. Figure 21 summarizes these dynamics across regions IV, V, and VI of Figure 11. The structure of Figure 21 is similar to the bifurcation diagram of Figure 16, but the dynamics at these parameter values are quite different. For example, $D_{\lambda\sigma}$ is negative in both regions III and IV, and therefore, the fixed points are real and distinct. However, the stability is reversed in region IV (Figure 18) as $\lambda$ changes sign: $x_1^*$ is now a center node converging to $x_1^* \to 1$ as $\sigma \to -\infty$, while $x_2^*$ and $x_3^*$ are saddles nodes diverging to $\pm\infty$, respectively.

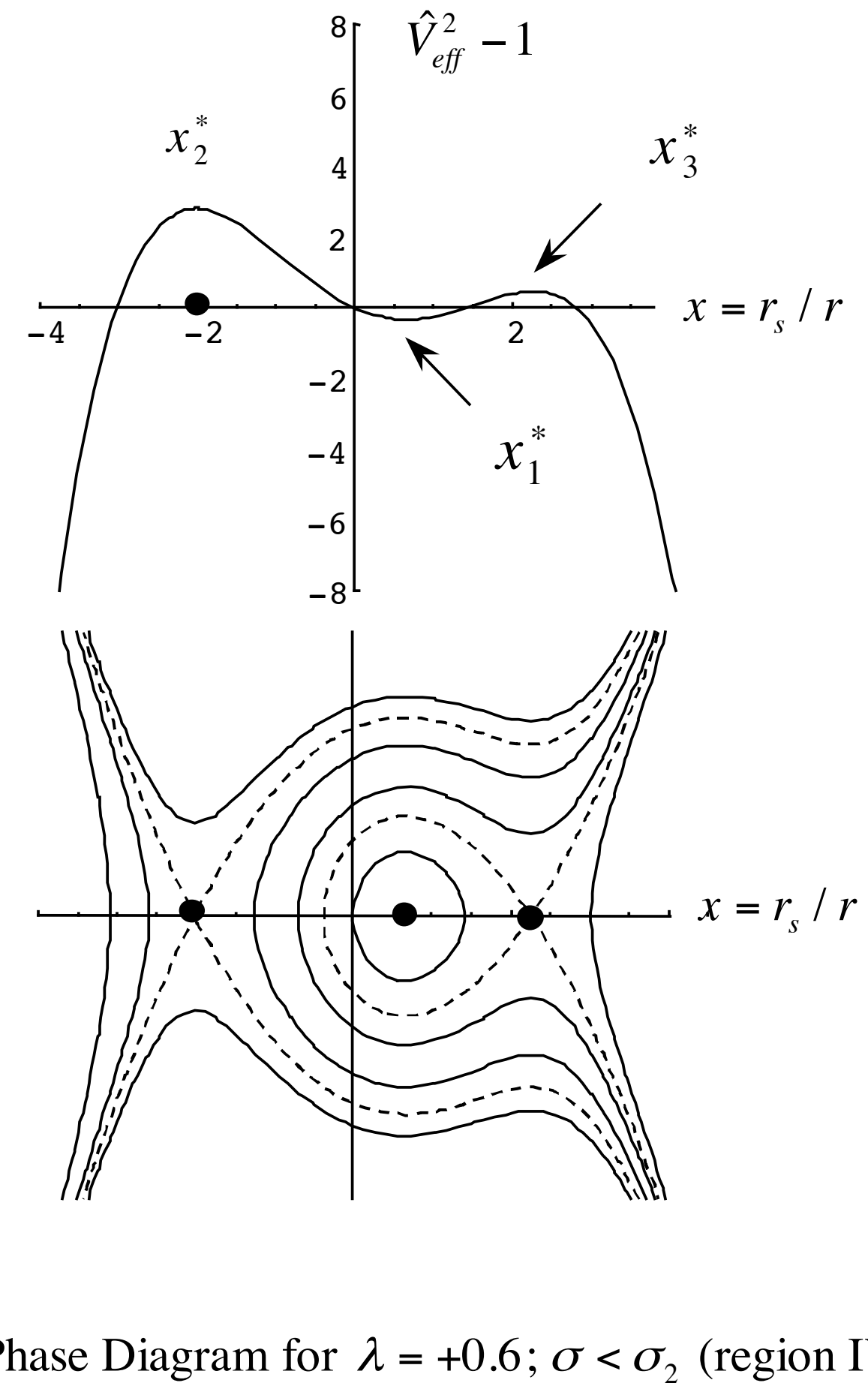

Figure 18 Phase Diagram for $\lambda = +0.6$; $\sigma < \sigma_2$ (region IV).

A saddle-center bifurcation occurs crossing into region V at $\sigma_2$, but the stability of $x_1^*$ and $x_3^*$ is reversed compared to Figure 16. The phase diagram at these parameter values is illustrated in Figure 19

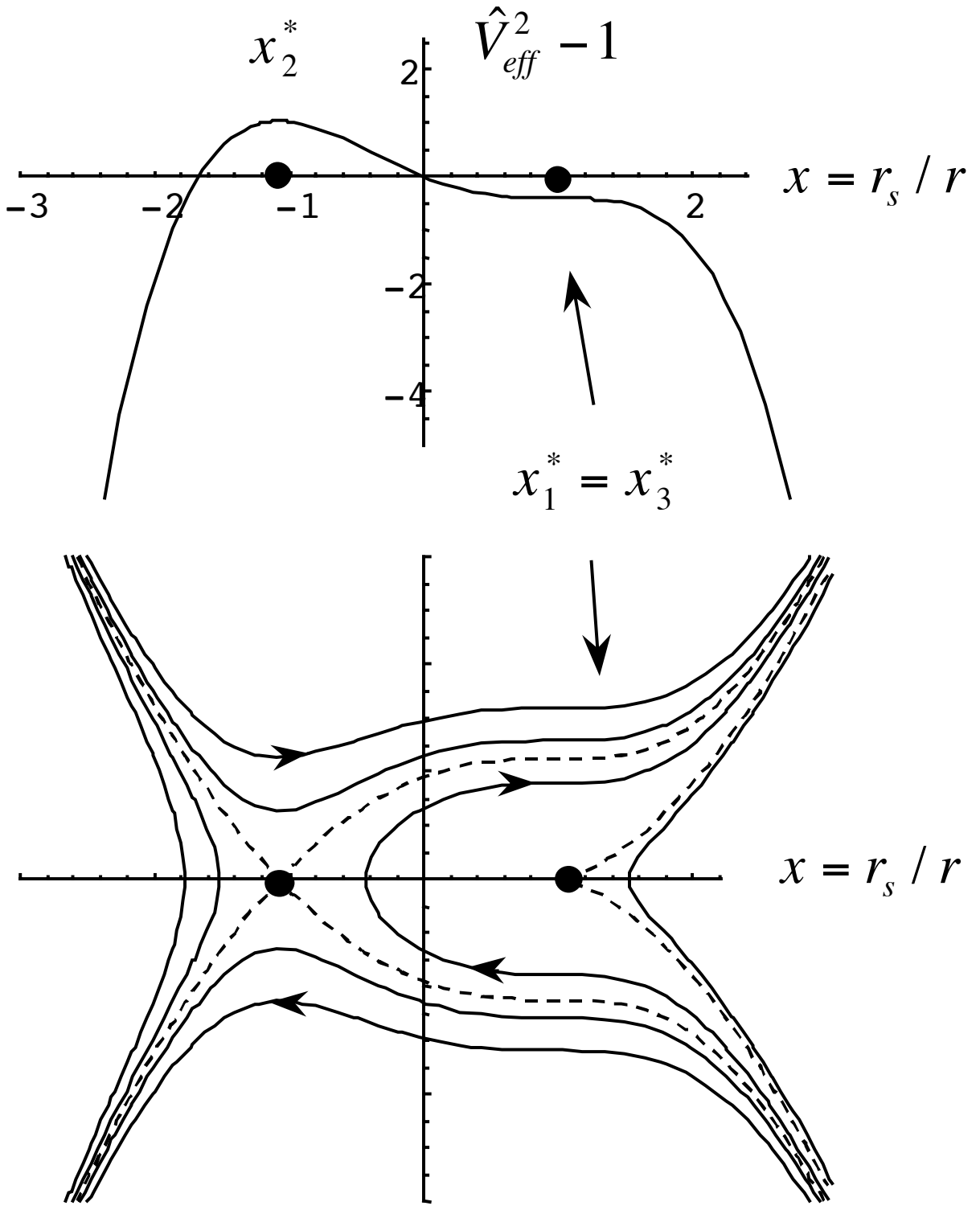

Figure 19 Phase Diagram for $\lambda = +0.6$; $\sigma(<0) = \sigma_2$.

In region V, $x_2^*$ is the only real fixed point (a saddle node; the phase diagram in this region is similar to Figure 15) while $x_1^*$ and $x_3^*$ are complex conjugate until the pseudo-transcritical bifurcation occurs at $\sigma_I < 0$. Once again, $x_1^*$ and $x_2^*$ become complex conjugate while $x_2^*$ and $x_3^*$ exchange roles ($x_3^*$ is real at this parameter value and is now a saddle node). As a result, the transcritical bifurcation point at $\sigma = 0$ (dividing regions V and VI) marks an exchange of stability for $x_3^*$ to a center node converging to +1, while $x_1^*$ and $x_2^*$ remain complex conjugate with real parts converging to $1/4$ as $\sigma \to +\infty$. Compared to Figure 16, $\sigma_{II}$ marks the crossing point of the real components of $x_1 = \overline{x_2}$ and $x_3^*$, which does not occur for $\lambda < 0$. In region VI, $D_{\lambda\sigma} > 0$, and therefore, $x_3^*$ is the only real fixed point that remains. Based on the parameter values of region VI, solutions about this fixed point are interpreted dynamically as giving periodic timelike orbits about a singularity with no horizons.

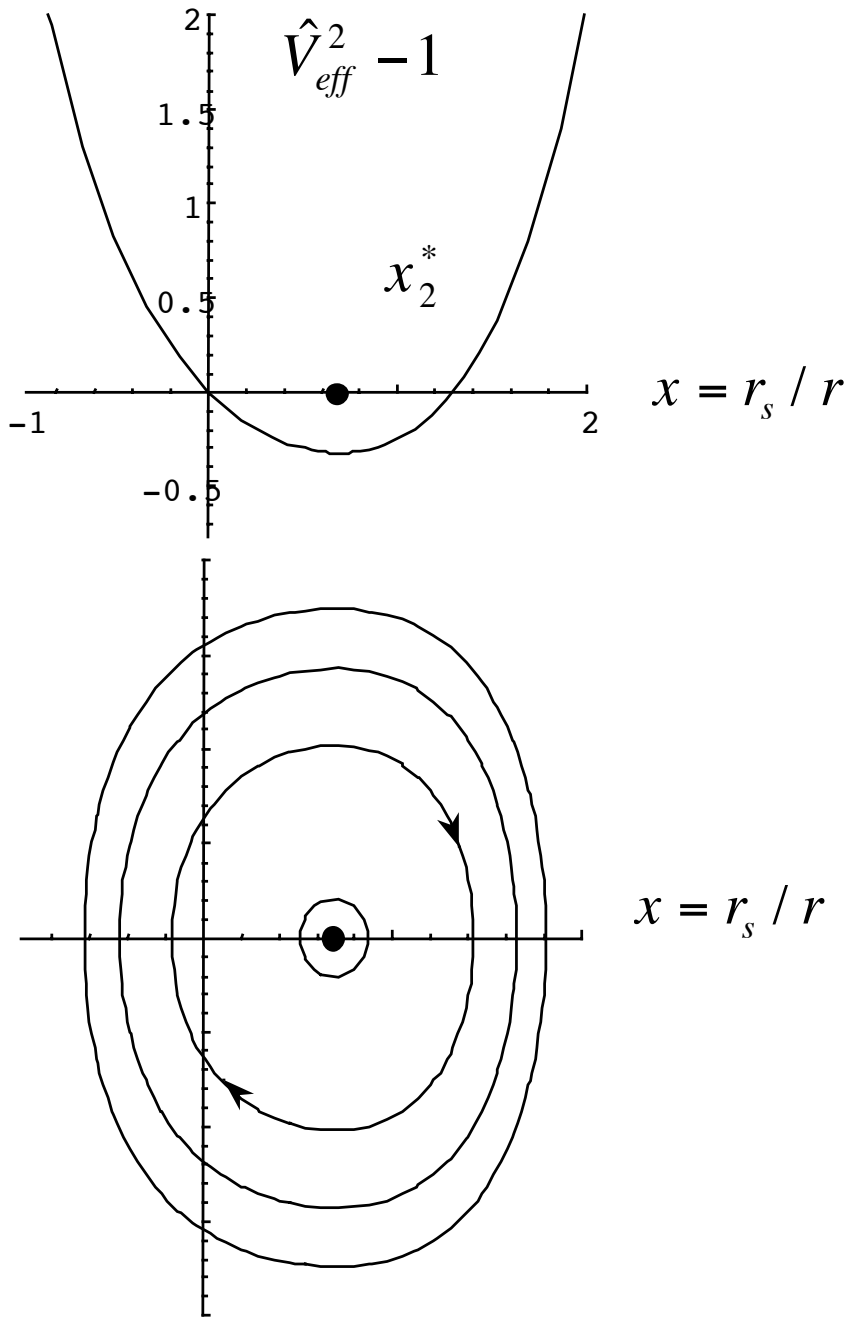

Figure 20 Phase Diagram for $\lambda = 1$; $\sigma = 1$ (region VI).

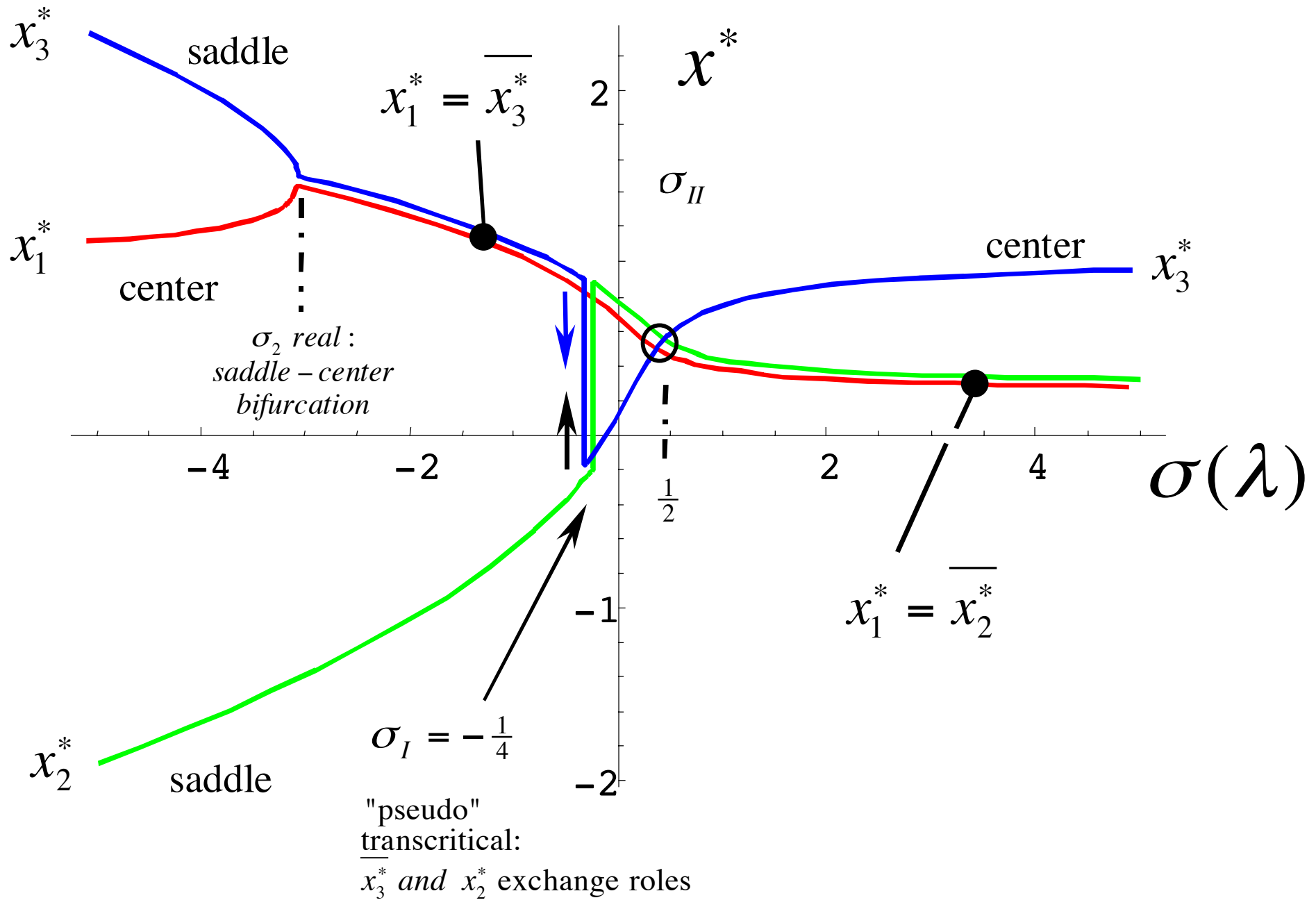

Figure 21 Bifurcation Diagram: $\lambda = +1$.

The parameter value $\lambda = 8/5$ gives a totally degenerate bifurcation point (equation (6.25)). The significance of this parameter value for the dynamics is seen by following the evolution of Figure 21 as $\lambda$ is increased to 8/5. As pointed out earlier, the $\sigma_I$ pseudo-transcritical point reaches a minimum value ($\sigma_I = -1/3$) at $\lambda = 2/3$, reverses direction, and then reaches zero at $\lambda = 4/3$ (see Figure 17). As $\lambda$ is further increased, $\sigma_I$ is positive and approaches $\sigma_{II}$. To better illustrate this behavior, the bifurcation diagram for $\lambda$ slightly less than 8/5 (= 3/2) is given in Figure 22. Finally, at $\lambda = 8/5$, the pseudo-transcritical point, $\sigma_I$, and the crossing point of the real and complex conjugate roots, $\sigma_{II}$, merge together to give the bifurcation diagram of Figure 23. At this parameter value the $\sigma_i$ are real (equation (6.13)) but with only two distinct ($\sigma_1 = \sigma_3 = 8/25$), and form the cusp point in Figure 11, while $\sigma_2$ has a minimum value at $-4$. For $\sigma > 8/25$, the stability properties of the fixed points are identical to those discussed for region V. But in this case, the real parts of $x_1^*$ and $x_2^*$ converge to 2/5 as $\sigma \to +\infty$, while $x_3^*$ remains a center node converging to 8/5. The dynamics and periastron precession at this cusp point are discussed in the section of this chapter entitled *Orbital Precession About a Bifurcation Point*.

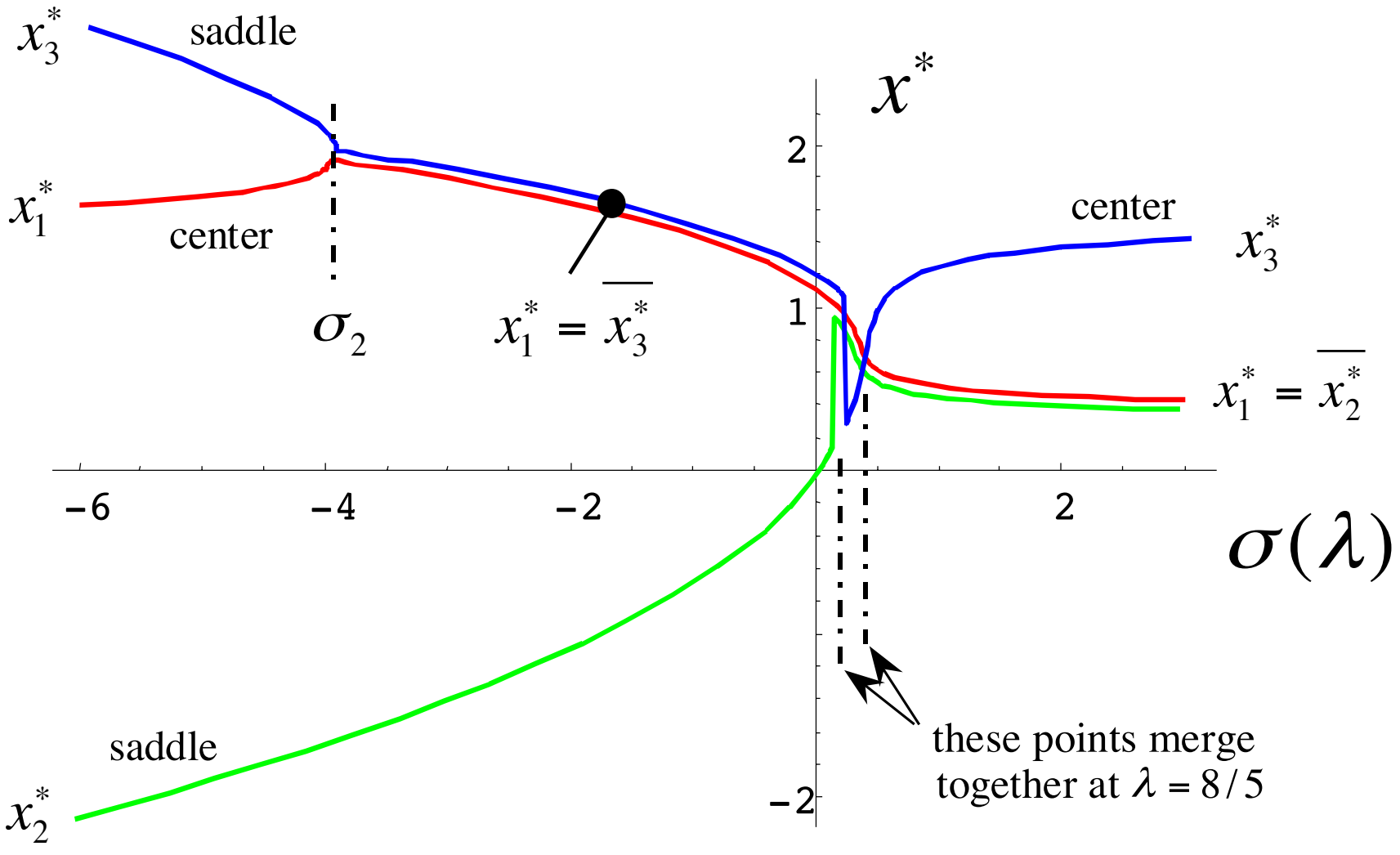

Figure 22 Bifurcation Diagram: $\lambda = 3/2$.

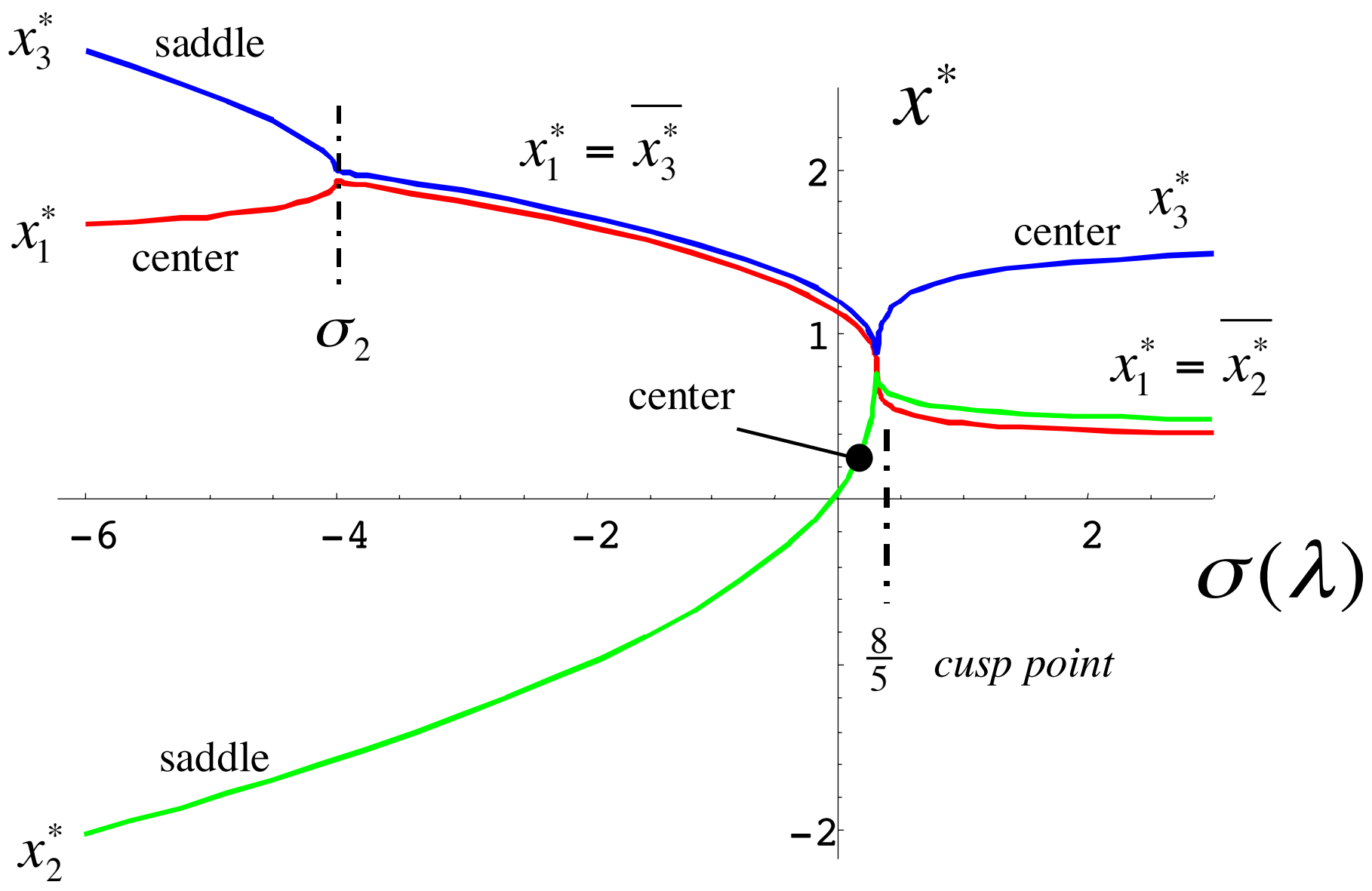

Figure 23 Bifurcation Diagram: $\lambda = 8/5$.

$8/5 < \lambda \le 16/9$

For $8/5 < \lambda < 16/9$, the $\sigma_i$ are all real and distinct ($D_\lambda < 0$) as the cusp bifurcation point separates into two additional saddle-center bifurcations – the first given at $\sigma_3$ – tending toward zero while the second is given by $\sigma_1$ (see Figure 11). As illustrated in Figure 11 and Figure 24, region VII corresponds to the interval: $0 < \sigma < \sigma_1$ and consists of three real fixed points: a saddle ($x_1^*$), center ($x_2^*$), and a saddle ($x_3^*$). At $\lambda = 16/9$, the $\sigma_3$ saddle-center moves to the origin and combines with the transcritical bifurcation at $\sigma = 0$ as shown in Figure 24 to form an unstable degenerate inflection point at $x_1^* = x_3^* = 4/3$. The phase diagram for these parameter values is illustrated in Figure 31. Therefore, two transcritical and three saddle center bifurcation points exist at $\lambda = 16/9$. Asymptotically, as $\sigma \to -\infty$, $x_1^* \to 16/9$ while $x_2^*$ and $x_3^*$ are saddles nodes diverging to $\pm\infty$, respectively. At $\sigma \to +\infty$ the real parts of $x_1^*$ and $x_2^*$ converge to 4/9 while $x_3^* = 16/9$.

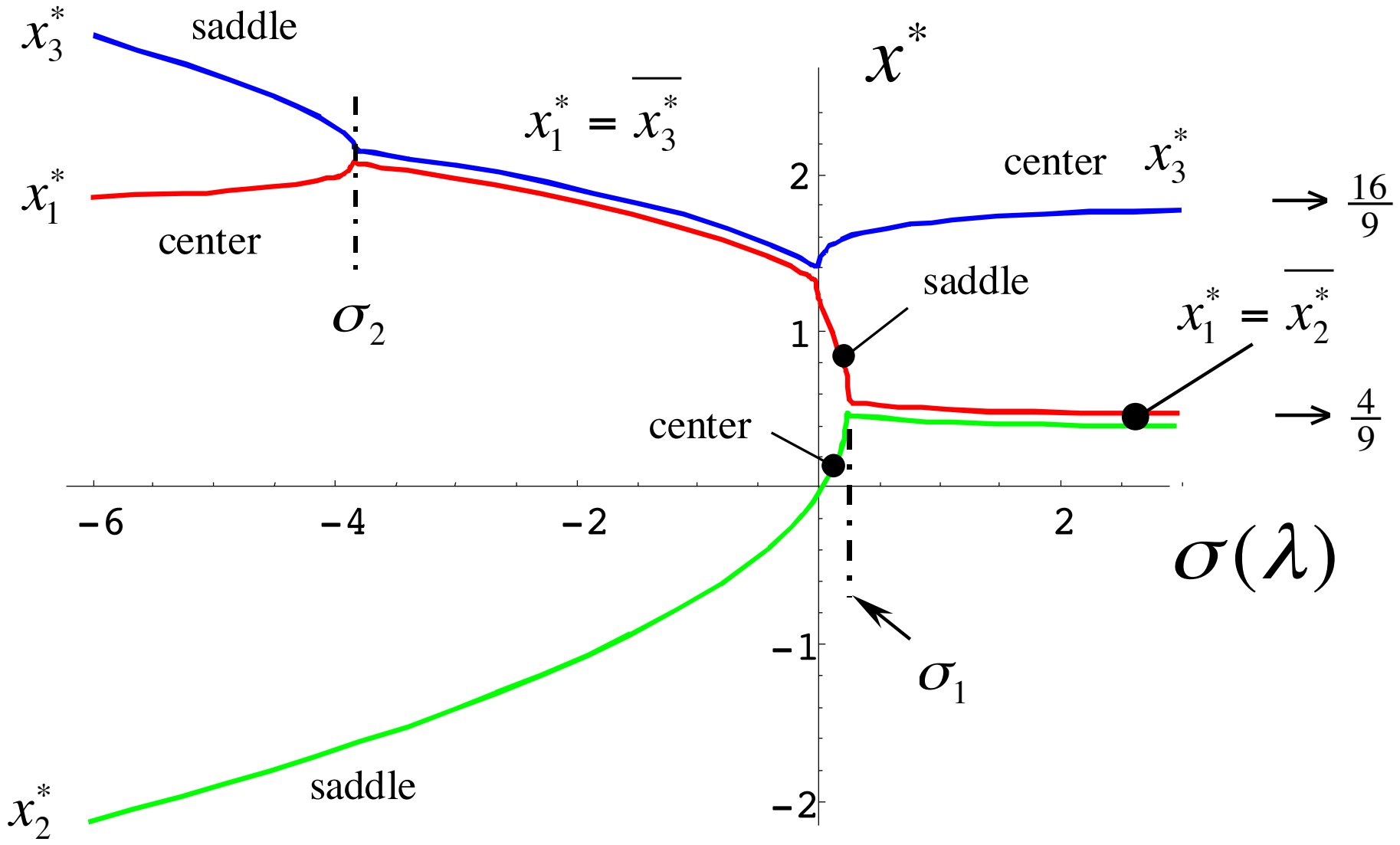

Figure 24 Bifurcation Diagram: $\lambda = 16/9$.

## $\lambda \geq 2$

The bifurcation point at $\lambda = 2$ (Figure 26) is more clearly illustrated by an intermediate bifurcation diagram at $\lambda = 1.99$ shown in Figure 25. For comparison, at $16/9 < \lambda < 2$, the $\sigma_3$ saddle-center bifurcation is decreasing, while simultaneously, the $\sigma_2$ saddle-center is increasing toward $\sigma_3$. In the intermediate region between $\sigma_2$ and $\sigma_3$ of Figure 25, $x_1^*$ and $x_3^*$ are complex conjugate while $x_2^*$ is a saddle node (this is region V of Figure 11). At $\lambda = 2$, the $\sigma_2$ and $\sigma_3$ roots merge together into an usual "saddle-center-saddle-center" bifurcation point of Figure 26 at $\sigma_2 = \sigma_3 = -2$; $\sigma_1 = 1/4$. For $\lambda > 2$ (Figure 27), the $x_1^*$ and $x_3^*$ fixed points separate leaving a transcritical (at $\sigma = 0$) and a single saddle-center bifurcation point at $\sigma = \sigma_1$.

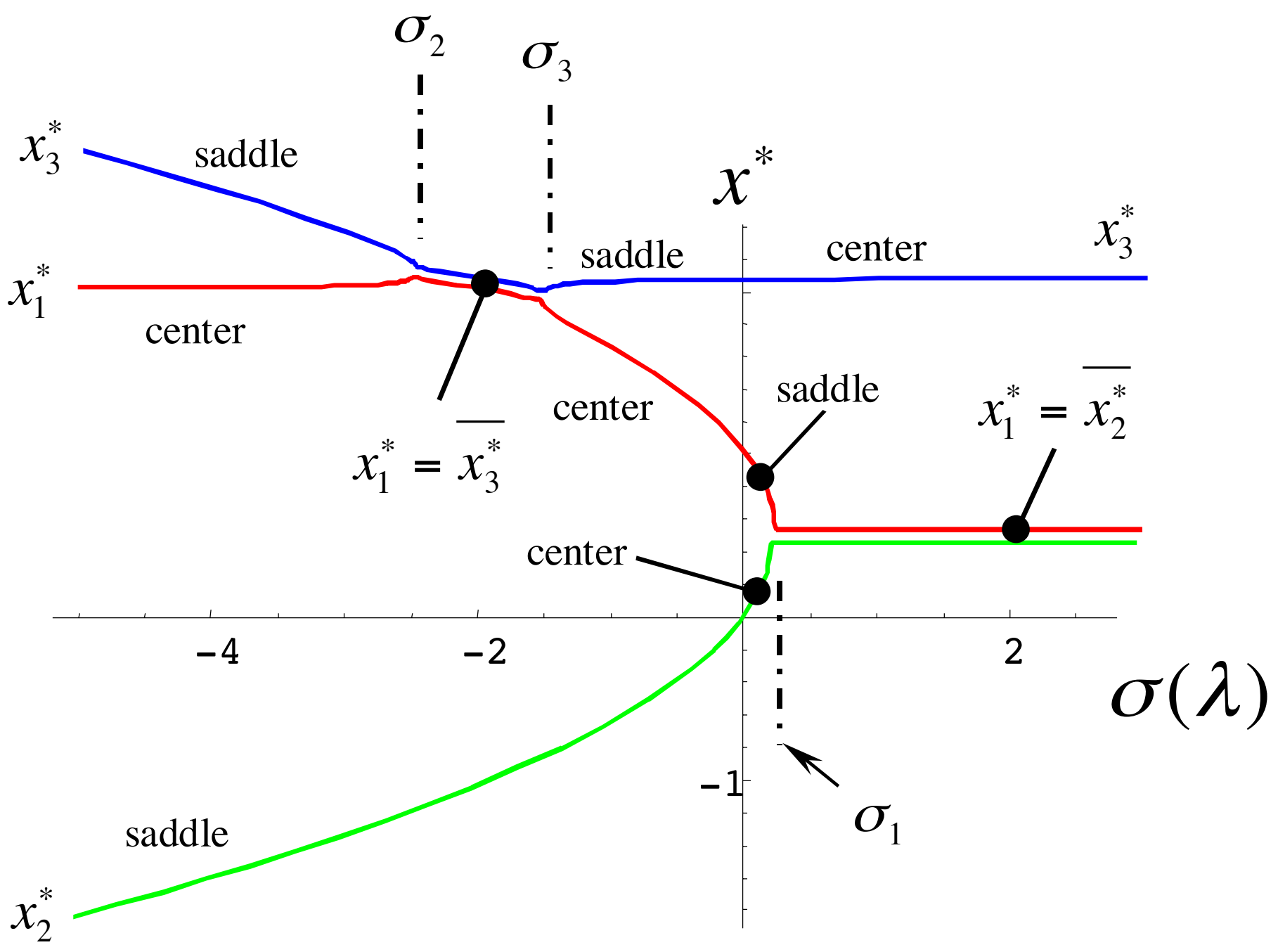

Figure 25 Bifurcation Diagram: $\lambda = 1.99$.

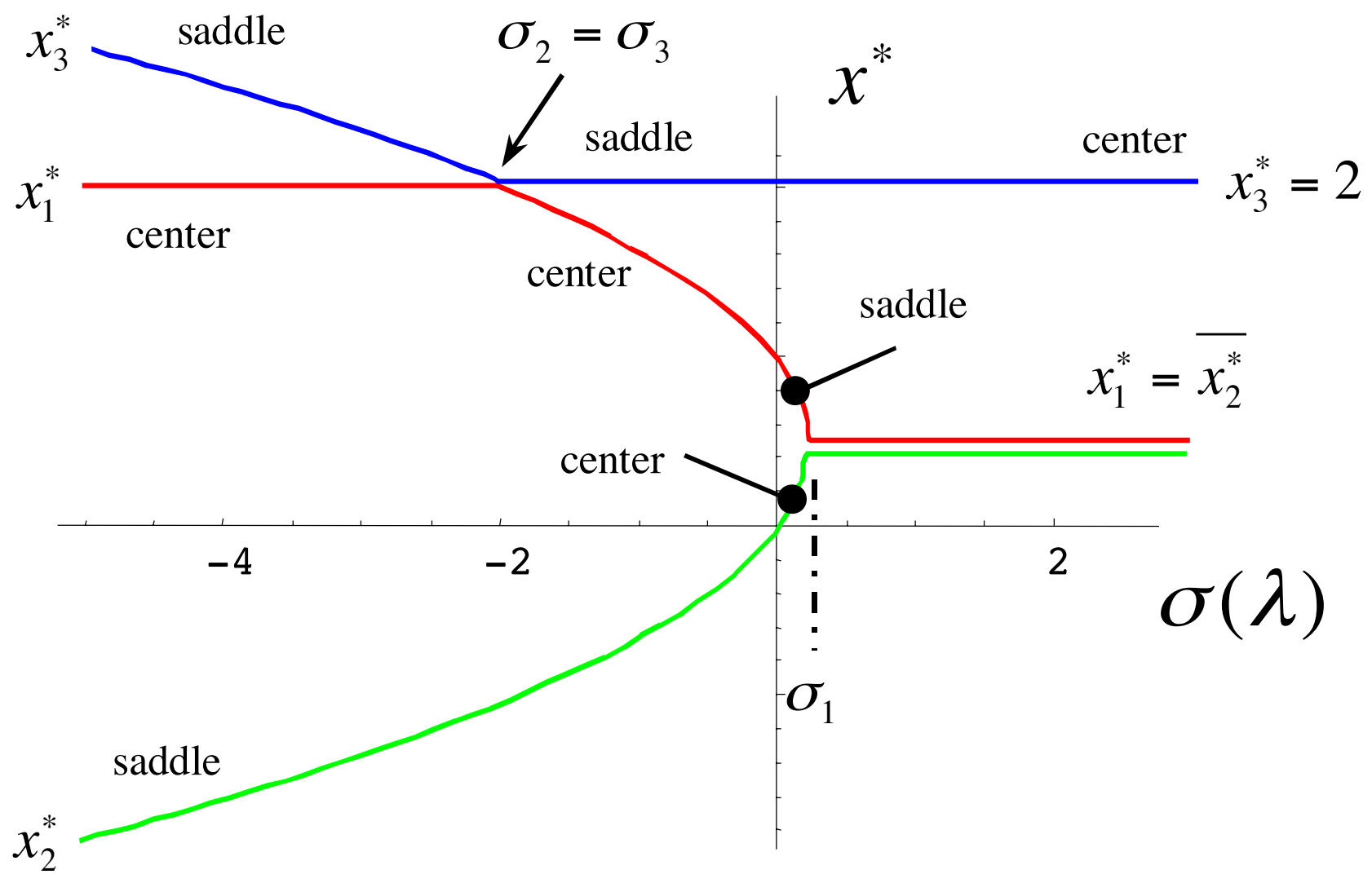

Figure 26 Bifurcation Diagram: $\lambda = 2$.

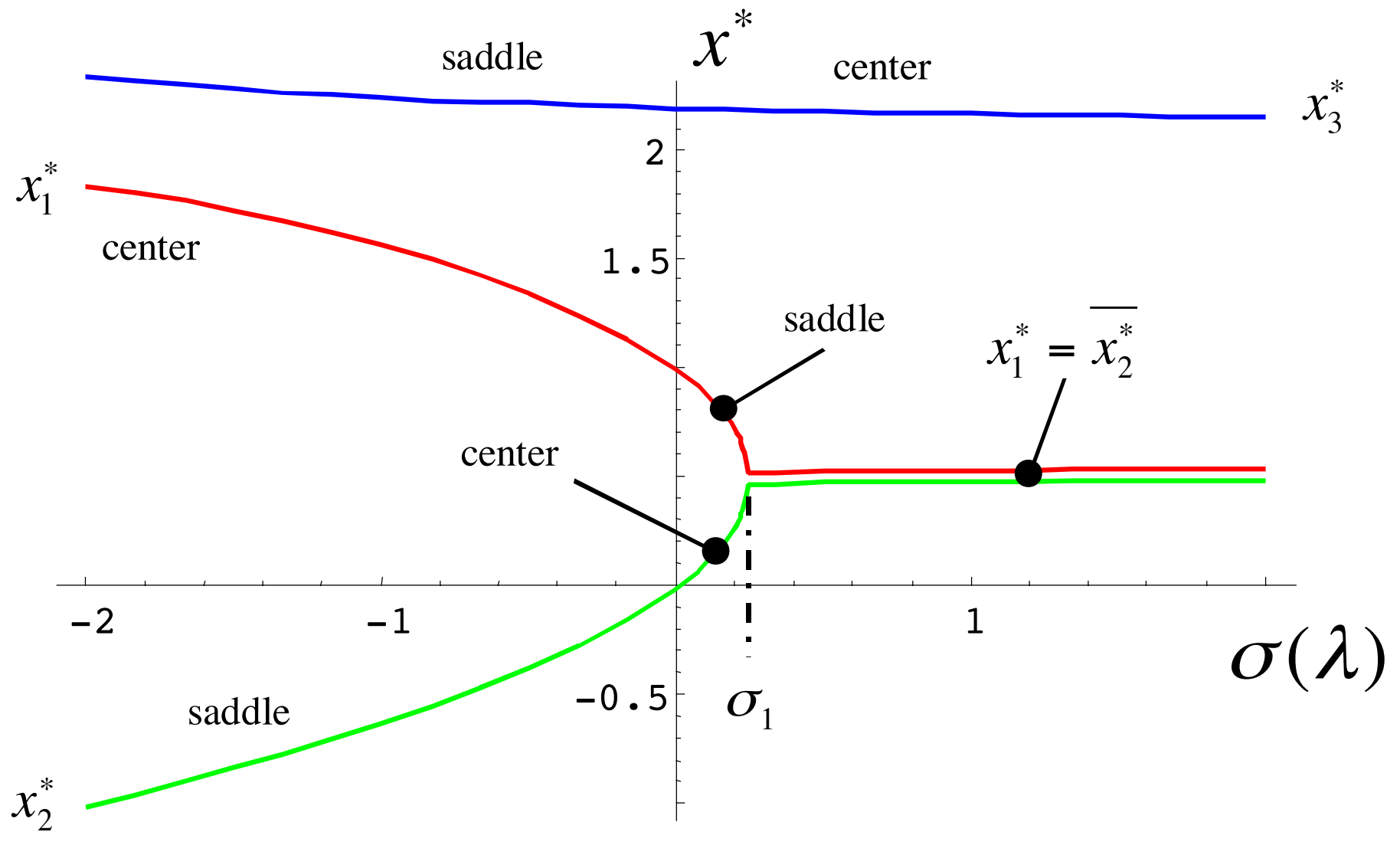

Figure 27 Bifurcation Diagram: $\lambda > 2$ $(= 2.1)$.

The Schwarzschild limit is obtained as $\lambda \to \infty$. In this case $x_3^*$ diverges to $+\infty$ giving the bifurcation diagram of Figure 7. For large $\lambda$ (and $\sigma << 1$) the fixed points (6.5) are series expanded to first order in $\lambda^{-1}$ giving (Appendix G):

$$\begin{aligned} x_1^* &\approx \frac{2}{3} - \sigma + \frac{2(4+3\sigma)}{27\lambda} + O(\lambda^{-2}) \\ x_2^* &\approx \sigma + \left(\frac{3}{2} - \frac{1}{\lambda}\right)\sigma^2 + O(\lambda^{-2}) \\ x_3^* &\approx -\frac{2}{3} + \frac{3}{2}\lambda - \frac{2(4+3\sigma)}{27\lambda} + O(\lambda^{-2}), \end{aligned} \tag{6.27}$$

which is consistent with the $\sigma << 1$ expansion of the Schwarzschild fixed points $x_1^* \approx 2/3 - \sigma$ and $x_2^* \approx \sigma$. For comparison with the Schwarzschild case the Reissner-Nordström phase diagram for a typical parameter value $\lambda > 2$ is given in Figure 28. Considering the fixed point $x_1^*$, the result in (6.27) indicates that the effect of adding a small charge to the black hole is to decrease the radius of the unstable orbit. Similarly, the $\lambda^{-1}$ term in the expression for $x_2^*$ gives a subtractive contribution and therefore the radii of circular orbits will increase (see also the expansion for $x_2^*$ given by (6.34)).

The locations of the horizons (given by (2.101): $x_{\pm} = \lambda \pm \sqrt{\lambda(\lambda - 2)} = r_s / r_{\mp}$) are also illustrated in the Figure noting that $x_+ = x_- = 2$ at $\lambda = 2$ which coincides with the fixed point $x_3^* = 2$ at these parameter values. The causal and dynamical interpretation of the orbits both in and about $r_+$ and $r_-$ have been previously discussed in the literature (for instance, Chandrasekhar [68], pp. 209-217) and are not included in this investigation. Here we have considered only the Reissner-Nordström bifurcation problem and a linear stability analysis about the orbital fixed points.

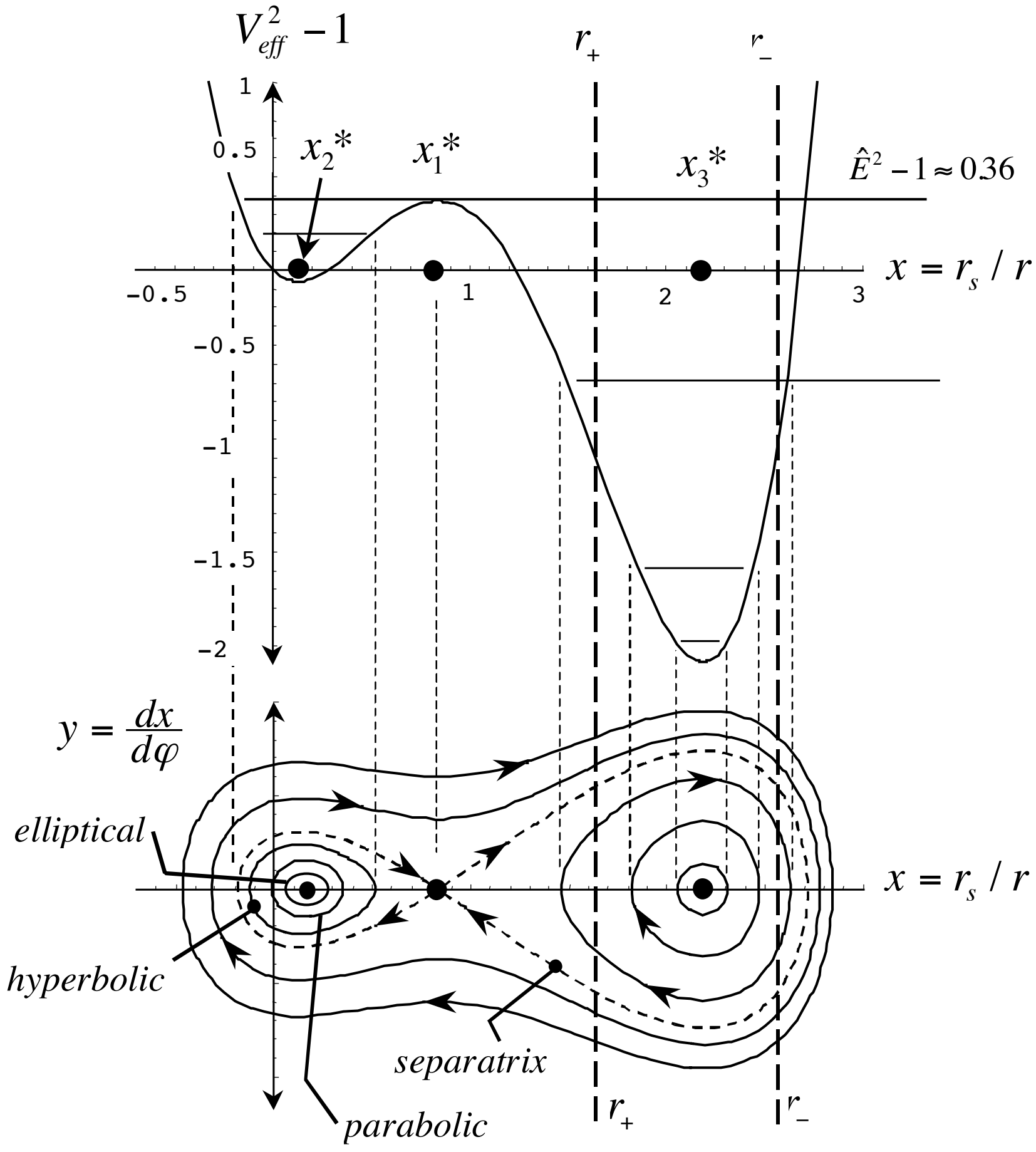

Figure 28 Phase Diagram: $\lambda > 2$ (= 2.1); $\sigma = 1/9$.

## Separatrix

The separatrix structure of the Reissner-Nordström dynamics is similar to that discussed earlier for the Schwarzschild case. As pointed out earlier in Chapter 5 - a classification of the dynamics that is based on the distinct separatrices that occur corresponds to the limitations placed upon the types of orbits that may exist before an unstable orbit is reached, and the kinematic classification of the separatrices as distinct unstable orbits. The separatrices themselves are therefore classified kinematically as unstable hyperbolic, parabolic, and elliptical orbits. The main difference between the Reissner-Nordström and Schwarzschild separatrices is given by noting that in the Reissner-Nordström case $\sigma = \sigma(\lambda)$ as given by (6.13). For comparison, the earlier classification of qualitatively distinct orbits from Chapter 5 (equation (5.40)) based on the values of $\sigma$ is listed below:

$$0 < \sigma < \tfrac{1}{8} \ ; \ \sigma = \tfrac{1}{8} \ ; \ \tfrac{1}{8} < \sigma < \tfrac{1}{6} \ ; \ \sigma = \tfrac{1}{6} \ ; \ \sigma > \tfrac{1}{6} . \tag{6.28}$$

Although a generalization (5.40) may be considered for all of the $\lambda$ values considered earlier for the bifurcation analysis, the case $\lambda > 2$ is the parameter value which may be compared to the Schwarzschild case. Therefore, the bifurcation diagram of Figure 27 applies to give a classification of the fixed point stability properties. The appropriate generalization of the bifurcation point is thus given by $\sigma_1(\lambda)$ with the limiting value displayed above, $\sigma_1(\lambda = \infty) \to 1/6$, as the charge goes to zero. The generalization for the value of 1/8 (the binding energy of the orbit is equal to the rest mass energy of $m_0$ at this parameter value) is more complicated to obtain algebraically, but in principle could be obtained by solving for $\sigma = \sigma(\lambda)$ from:

$$\hat{V}_{eff}^2 - 1 = \ 0 \ = (1 + x^2/2\sigma)(1 - x + x^2/2\lambda) - 1 , \tag{6.29}$$

at the unstable fixed point, $x = x_1^*$. Denoting this value of $\sigma$ by $\sigma_{up}(\lambda)$ (for 'unstable parabolic'), the appropriate generalization of (5.40) has the form:

$$\begin{gathered} 0 < \sigma < \sigma_{up}(\lambda) \ ; \ \sigma = \sigma_{up}(\lambda) \ ; \ \sigma_{up}(\lambda) < \sigma < \sigma_1(\lambda) \\ \sigma = \sigma_1(\lambda) \ ; \ \sigma > \sigma_1(\lambda) , \end{gathered} \tag{6.30}$$

and therefore the dynamics could be similarly classified as discussed earlier for the Schwarzschild case.

## Periastron Precession

Periastron precession in the Reissner-Nordström spacetime has received a surprisingly scanty treatment in the previous literature. Bronstein [143] was the first to discuss the effects of charge on the periastron advance, but as pointed out by Kudar [144] Bronstein's result was based essentially on a special relativistic calculation and shown later to be incorrect. Subsequently, an analysis of charged particle motion incorporating the approximation method given by Einstein, Infeld, and Hoffman [145] was discussed by Bertotti [146]. But a calculation of perihelion advance in the Reissner-Nordström spacetime was not given until 1969 by Burman [147]. Burman later considers an application of these results to Mercury's orbit using estimates of the solar charge given by Bailey [137] but concludes that the correction made to the Schwarzschild (perihelion) precession value is negligible. Burman [148] subsequently analyzes the perihelion advance of Icarus but considers a non-relativistic mechanism to explain the precession.

In a paper discussing the stability properties of circular orbits in the Reissner-Nordström spacetime, Armenti [149] mentions briefly that the presence of charge should have a subtractive effect on the periastron advance, although no calculation is given. Barker and O'Connell [150] have generalized earlier results to the case of the charged 2-body problem but give little discussion or interpretation of their results. Subsequently, Treder, et. al. [138] have discussed perihelion advance in the Reissner-Nordström spacetime as a means for estimating the solar charge based on accurate perihelion data for planetary bodies. Following Burman, Teli and Palaskar [139] also consider the effect of a net solar charge on the perihelion advance of Mercury's orbit (as well as to the orbits of Venus and Icarus). Finally, Rathod and Karade [151] discuss an alternative procedure for calculating periastron advance using the Hamilton-Jacobi formalism - in contrast to earlier calculations based solely on perturbative techniques.

In this Section the phase-plane calculation of periastron precession given earlier for the Schwarzschild solution is considered for the Reissner-Nordström case. An order of magnitude estimate is also given for the charge contribution to the periastron shift in the limit that the charge contribution to the total mass is small. As in the Schwarzschild example the calculation is based on the observation that the phase-plane trajectory about the center node $\vec{x}_2^*$, must close after a single period in phase space since the system is conservative. Therefore, the periastron advance is found by calculating the period of a phase-plane orbit, $\Phi$:

$$\Phi = 2\pi\omega^{-1}, \tag{6.31}$$

where $\omega$ is determined from the linear stability solution about $\vec{x}_2^*$ as illustrated below. For the Reissner-Nordström system the linear stability equations are identical to (5.26):

$$\delta x' = \delta y \ , \delta y' = -\omega^2 \delta x , \tag{6.32}$$

however, $\omega$ is given by

$$\omega^2 = [3x^{*^2} + (1-3x^*)\lambda + \sigma]/\lambda \Big|_{x^*=x_2^*} . \tag{6.33}$$

An expression for $\omega$ is obtained using a perturbative correction for $\vec{x}_2^*$ by writing to first order in $\lambda^{-1}$ (a brute force substitution of $\vec{x}_2^*$ from (6.5) into (6.33) is best performed using a computer algebra system; see Appendix G for a Mathematica code to generate (6.36) below):

$$x_2^{*(1)} = x_2^{*(0)} + \alpha/\lambda , \tag{6.34}$$

where $x_2^{*(0)}$ is given by (5.18). Substituting (6.34) into (6.33) and then solving for $\alpha$ subject to the condition that the first order correction vanishes (i.e. agrees with the Schwarzschild result) gives the following expression for $\alpha$:

$$\alpha = \left[4(1-\beta^{-1}) + 3(1+3\beta^{-1})\sigma\right]/27 \ ; \ \beta = \sqrt{1-6\sigma} . \tag{6.35}$$

The remaining calculation is straightforward algebra; substituting (6.35) into (6.34) and then finally into (6.33) gives the result:

$$\omega = (1 - 3\sigma + \sigma/\lambda)^{1/2} , \tag{6.36}$$

where

$$\frac{\sigma}{\lambda} = \frac{kG}{c^6}\left(\frac{e}{\hat{J}}\right)^2 . \tag{6.37}$$

Therefore, in the Reissner-Nordström spacetime the periastron advance is reduced compared to the Schwarzschild value (compare also with Burman [147]):

$$\Phi = 2\pi\omega^{-1} = 2\pi + \Delta\varphi \approx 2\pi + 3\pi\sigma - \pi\sigma/\lambda , \tag{6.38}$$

where $3\pi\sigma$ is the Schwarzschild result given in (5.32). However, from the order of magnitude estimates considered earlier for the Sun (equation (2.108)) the $1/\lambda$ factor is considerably smaller than the $\sigma$ contribution alone.

## Orbital Precession About a Bifurcation Point

As discussed earlier, a degenerate bifurcation point of the dynamics occurs at the parameter values:

$$\sigma = 8/25 \,;\, \lambda = 8/5 \Rightarrow x_1^* = x_2^* = x_3^* = 4/5\,, \tag{6.39}$$

and is classified as a center-node fixed point (Figure 29). The dynamical interpretation (which is not necessarily a physical interpretation) at this bifurcation point follows from the values of $\sigma$ and $\lambda$: timelike orbits (for $\sigma > 0$) about a naked singularity (for $\lambda < 2$). Similarly, another bifurcation point is given by

$$\sigma = 0 \,;\, \lambda = 16/9 \Rightarrow x_1 = x_3 = 4/3;\, x_2 = 0\,, \tag{6.40}$$

and corresponds to photon orbits ($\sigma = 0$) about a naked singularity. The bifurcation diagram for this case is given in Figure 23 along the y-axis (i.e., $\sigma = 0$) and is discussed further in the subsection entitled *Light Rays*.

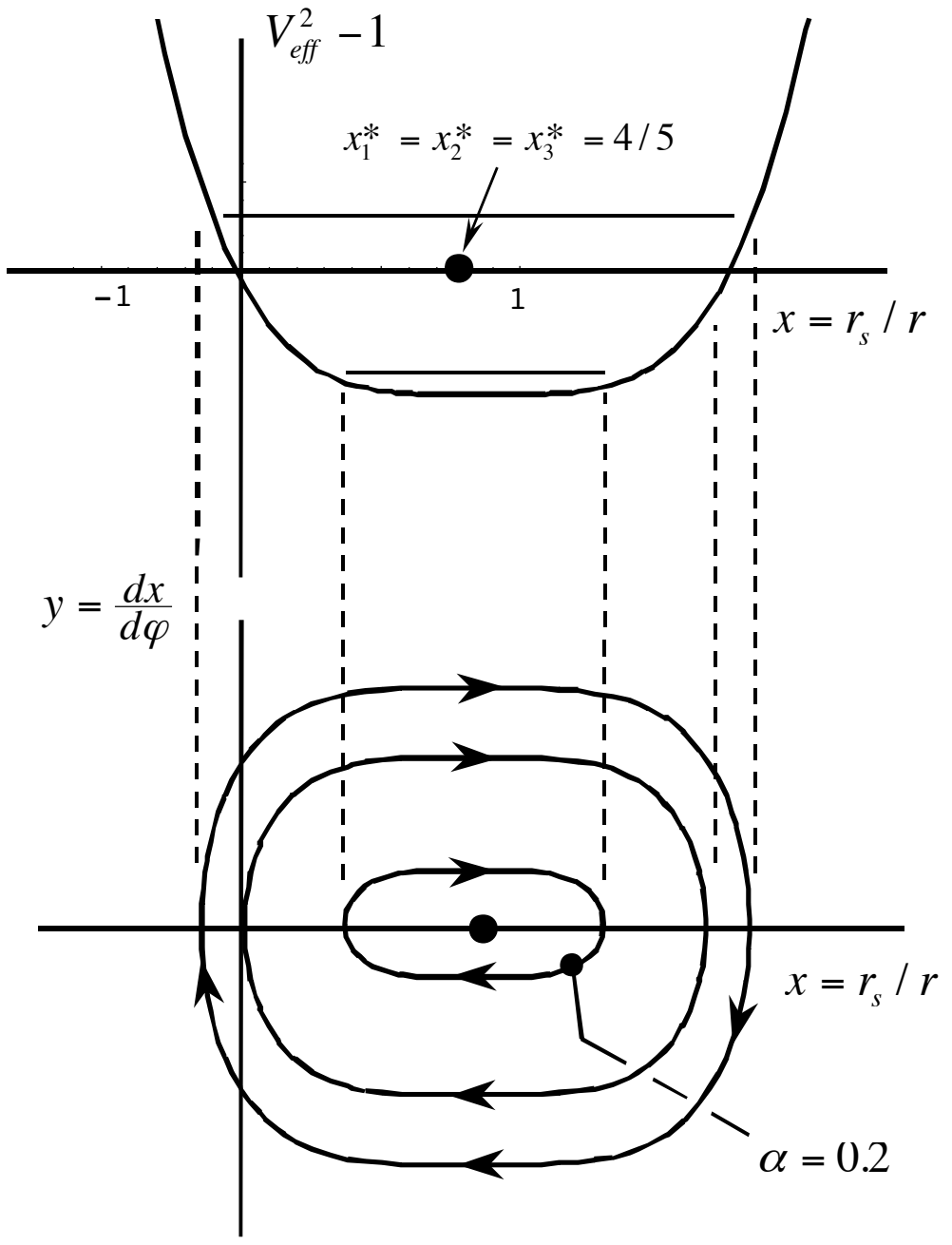

Figure 29 Phase-Diagram at the Bifurcation Point: $\sigma = \frac{8}{25}$; $\lambda = \frac{8}{5}$.

For comparison with the precession given by (6.38), it is of interest to calculate the precession value at the bifurcation point (6.25) which is given by the parameter values: $\sigma = 8/25$ ; $\lambda = 8/5$. But in this case the first and second order terms vanish identically. Therefore, the expansion must be carried out to third order giving the system:

$$\delta x' = \delta y \; ; \; \delta y' = -\tfrac{5}{8}\delta x^3 . \tag{6.41}$$

A first integral follows immediately from (6.41):

$$\tfrac{1}{2}(\delta y)^2 + \tfrac{5}{32}(\delta x)^4 = \gamma ; \; \gamma \equiv \text{constant} , \tag{6.42}$$

and therefore the level curves of (5.46) about $x^* = 4/5$ are given by the center-node trajectories of Figure 29. The period of a phase-plane orbit is thus calculated by integrating along a center node trajectory ( $= 4 \times \tfrac{1}{4}$ of a period):

$$\Phi = \tfrac{16}{\sqrt{5}} \int_0^{\alpha} d(\delta x) / \sqrt{\alpha^4 - (\delta x)^4} , \tag{6.43}$$

where $\alpha$ is the turning point defined by solving $\delta y = 0$ in (6.42) for $\delta x$. Equation (5.2) is solved using the Beta function (for instance, Abramowitz and Stegun [140]), $B(z,w) = B(\tfrac{1}{4},\tfrac{1}{2})$, expressed as the ratio of two Euler gamma functions, $\Gamma(z)$, to give the final result:

$$\Phi = 2\pi\omega^{-1} = \frac{4}{\alpha\sqrt{5}} B(\tfrac{1}{4},\tfrac{1}{2}) = \frac{4}{\alpha}\sqrt{\frac{\pi}{5}}\,\frac{\Gamma(\frac{1}{4})}{\Gamma(\frac{3}{4})} \approx \frac{9.4}{\alpha} . \tag{6.44}$$

By comparison with (6.38), the periastron advance at this bifurcation point will be larger than for the case $\lambda \geq 2$. A typical numerical example is illustrated using $\alpha = 0.2$ (see Figure 29). Integrating (6.41) numerically over the phase-plane period given by (6.43), and then plotting the orbital trajectory in the equatorial plane (Figure 30) illustrates that $m_0$ undergoes multiple revolutions in a radially oscillating orbit before finally completing a single period in phase-space to give $\Phi \approx 15\pi$ .

The "acausal" geodesics discussed by Brigman [81] are identified as the center node trajectories about $x_3^*$ in Figure 28. The distinguishing feature of such solutions is their proximity to the event horizon $r_+$ and interior Cauchy horizon $r_-$. As a result, these orbits are not periodic in the usual sense, but rather, alternate between the various regions of an extended Kruskal (or Penrose conformal) diagram to other asymptotically flat universes (see the discussion in Chandrasekhar [68]). Nevertheless, it is of interest to complete this calculation for comparison to the other cases considered earlier, although it should be clear that the term periastron has no meaning in this context.

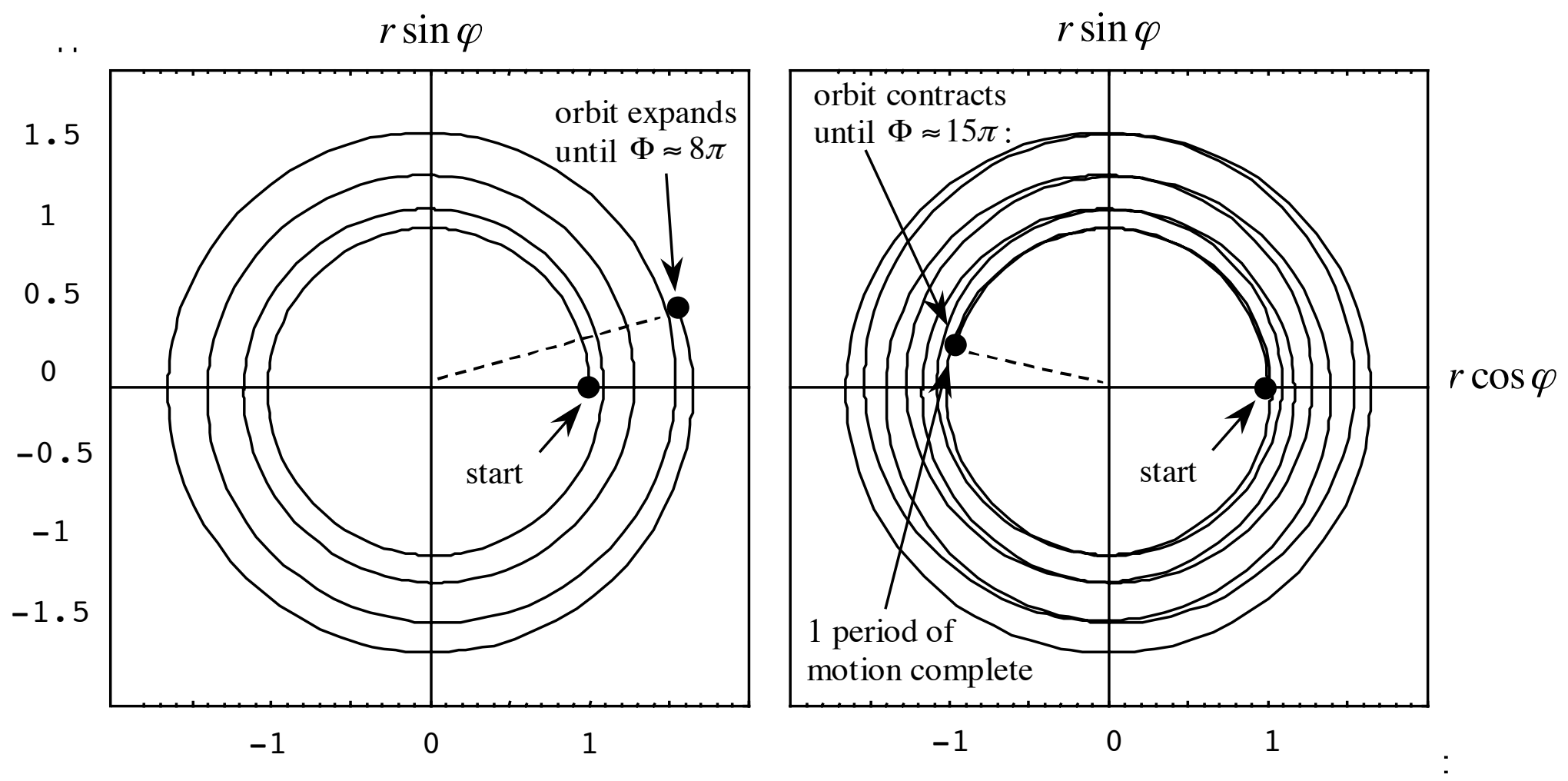

Figure 30 Equatorial Motion at the Bifurcation Point: $\sigma = \frac{8}{25}$; $\lambda = \frac{8}{5}$, ($\alpha = 0.2$).

The calculation is completed by substituting $x^* = x_3^*$ into the expression derived earlier for $\omega^2$ and then series expanding in $\lambda^{-1}$ to give (for large $\lambda$):

$$\begin{aligned} \omega^2 &= [3x^{*2} + (1-3x^*)\lambda + \sigma]/\lambda\Big|_{x^*=x_3^*} \\ &= 9\lambda/4 - 3 - (4+\sigma)/9\lambda. \end{aligned} \tag{6.45}$$

Therefore, the period of the orbit is given approximately by

$$\Phi \approx \frac{4\pi}{3\sqrt{\lambda}}, \tag{6.46}$$

and tending to zero in the Schwarzschild limit.

## Light Rays

Light ray kinematics in the Reissner-Nordström spacetime are similar to those discussed earlier for the Schwarzschild case but there are important differences. As in the Schwarzschild example, the phase-plane equations for the Reissner-Nordström photons are obtained as a special case of the timelike orbits as $\sigma \to 0$:

$$x' = \pm\sqrt{1/b^2 - x^2\Lambda}$$
$$y' = -\tfrac{1}{\lambda}x^3 + \tfrac{3}{2}x^2 - x, \tag{6.47}$$

where $1/b^2 \equiv 2\sigma\hat{E}^2$ is a constant expressing the dimensionless impact parameter, $b$, as the finite ratio of $\hat{E}$, $\hat{J}$, and $r_s$. However, the simultaneous solution of $x' = y' = 0$ for $1/b^2$ and $x$ results in three fixed points:

$$x_1^* = \tfrac{3}{4}\left[\lambda - \sqrt{\lambda(\lambda - \tfrac{16}{9})}\right]$$
$$1/b_1^2 = -\tfrac{3}{64}\lambda\left[9\lambda^2 - 24\lambda + \tfrac{32}{3} + 16\sqrt{\lambda(\lambda - \tfrac{16}{9})} - 9\sqrt{\lambda^3(\lambda - \tfrac{16}{9})}\right], \tag{6.48}$$

$$x_2^* = 0;\ 1/b_2^2 = 0, \tag{6.49}$$

$$x_3^* = \tfrac{3}{4}\left[\lambda + \sqrt{\lambda(\lambda - \tfrac{16}{9})}\right]$$
$$1/b_3^2 = -\tfrac{3}{64}\lambda\left[9\lambda^2 - 24\lambda + \tfrac{32}{3} - 16\sqrt{\lambda(\lambda - \tfrac{16}{9})} + 9\sqrt{\lambda^3(\lambda - \tfrac{16}{9})}\right], \tag{6.50}$$

compared to only two for the Schwarzschild equations which are listed below for comparison:

$$\textit{Schwarzschild}: \begin{cases} x_1^* = 2/3\ ;\ 1/b_1^2 = 4/27 \\ x_2^* = 0\ ;\quad 1/b_2^2 = 0\ . \end{cases} \tag{6.51}$$

The fixed point, $x_2^*$, is identical in both the Schwarzschild and Reissner-Nordström spacetimes and remains a center node at infinity for all parameter values. This fixed point gives the "precessing" hyperbolic orbits and is responsible for light bending. In addition, it is not surprising based on the earlier analysis of timelike orbits that a third fixed point $x_3^*$ appears.

For a comparison with the Schwarzschild fixed points $x_1^*$ and $x_3^*$ are series expanded to give for large $\lambda$:

$$x_1^* = \ \ 2/3 + 8/27\lambda\ ; \qquad 1/b_1^2 = \ \ 4/27 + 8/81\lambda$$
$$x_3^* = -2/3 - 8/27\lambda + 3\lambda/2\ ;\ 1/b_3^2 = -4/27 - 8/81\lambda - \lambda\ , \tag{6.52}$$

which is consistent with the Schwarzschild expressions for $x_1^*$ and $1/b^2$ as the charge goes to zero and noting also that $\{x_3^*, 1/b^2\}$ are divergent (or equivalently, $r \to 0$). Furthermore, considering the fixed point $x_1^*$, the result in (6.52) indicates the perturbative influence of adding a small charge to the black hole – the effect is to decrease the radius of the unstable "photon-sphere" while simultaneously reducing the impact parameter $b$.

By inspection of (6.50), a bifurcation occurs at the parameter value, $\lambda = 16/9$, which gives a degenerate unstable inflection point as $x_1^*$ and $x_3^*$ merge together at:

$$\lambda = 16/9: \quad x_1^* = x_3^* = 4/3\,; \; 1/b_1^2 = 1/b_3^2 = 8/27\,. \tag{6.53}$$

The dynamics at these parameter values are summarized by the phase-diagram and effective potential in Figure 31.

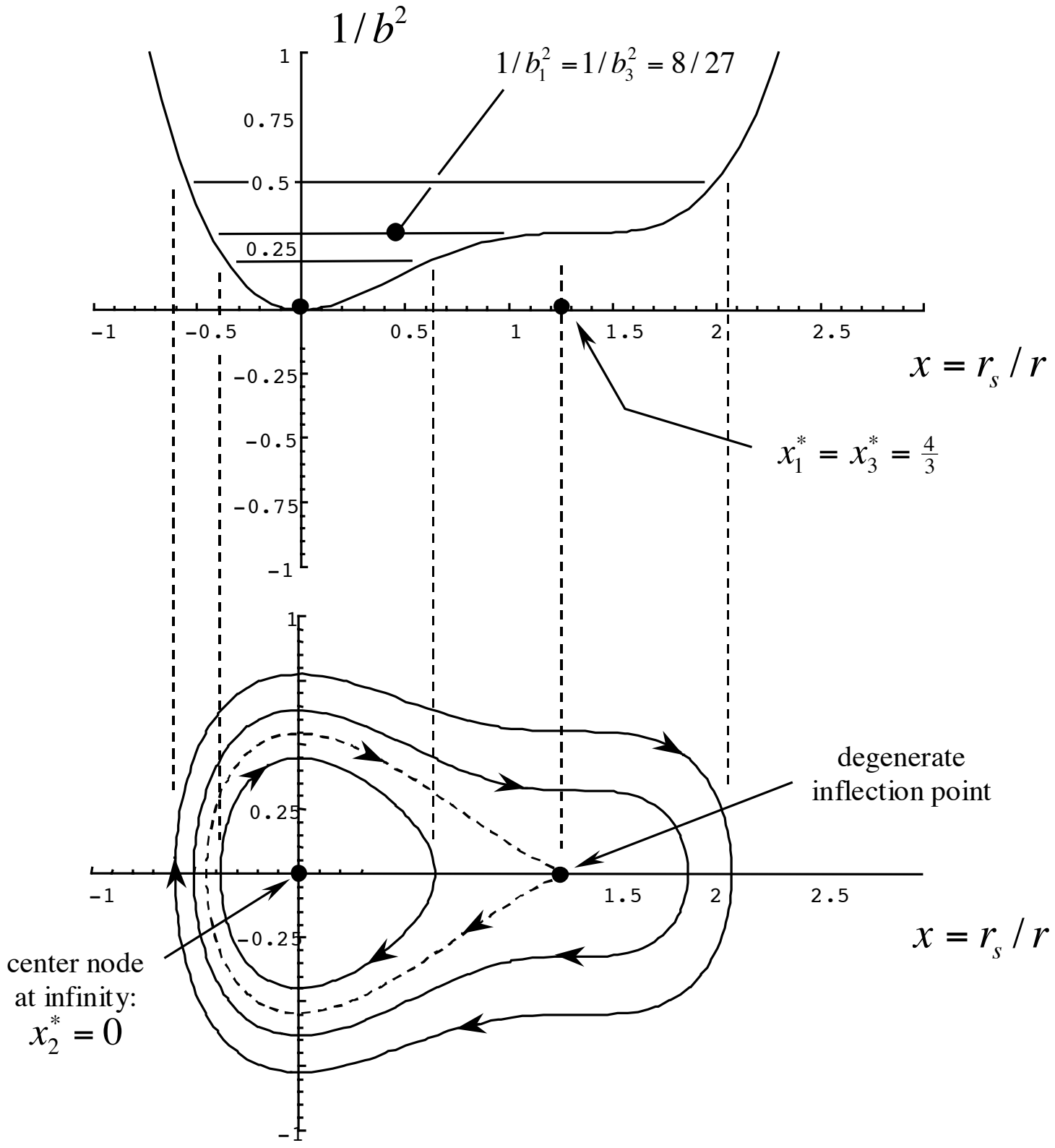

Figure 31 Phase Diagram for $\sigma = 0$; $\lambda = 16/9$.

But more generally, the significance of this bifurcation point and the stability properties of the fixed points at other parameter values are summarized in the bifurcation diagram of Figure 32, as the black hole charge is varied. The relevance of (6.53) from the perspective of the bifurcation analysis is that $x_1^* = x_3^* = 4/3$ is the first real fixed point for $\lambda > 0$. For $0 < \lambda < 16/9$, the fixed points are complex conjugate. As displayed in Figure 32, the numerical values of the real and imaginary components of the fixed points are equal at $\lambda = 8/9$, which is obtained by equating derivatives of $x_1^*$ and $x_3^*$ which simplifies to

$$\frac{9\lambda - 8}{6\sqrt{\lambda(\lambda - \frac{16}{9})}} = 0 \Rightarrow (x_1 = 2/3 - 2/3\mathrm{i}\ ;\ x_2 = 2/3 + 2/3\mathrm{i})\,. \tag{6.54}$$

However, there are no dynamical consequences for the parameter value $\lambda = 8/9$ since the fixed points are complex.

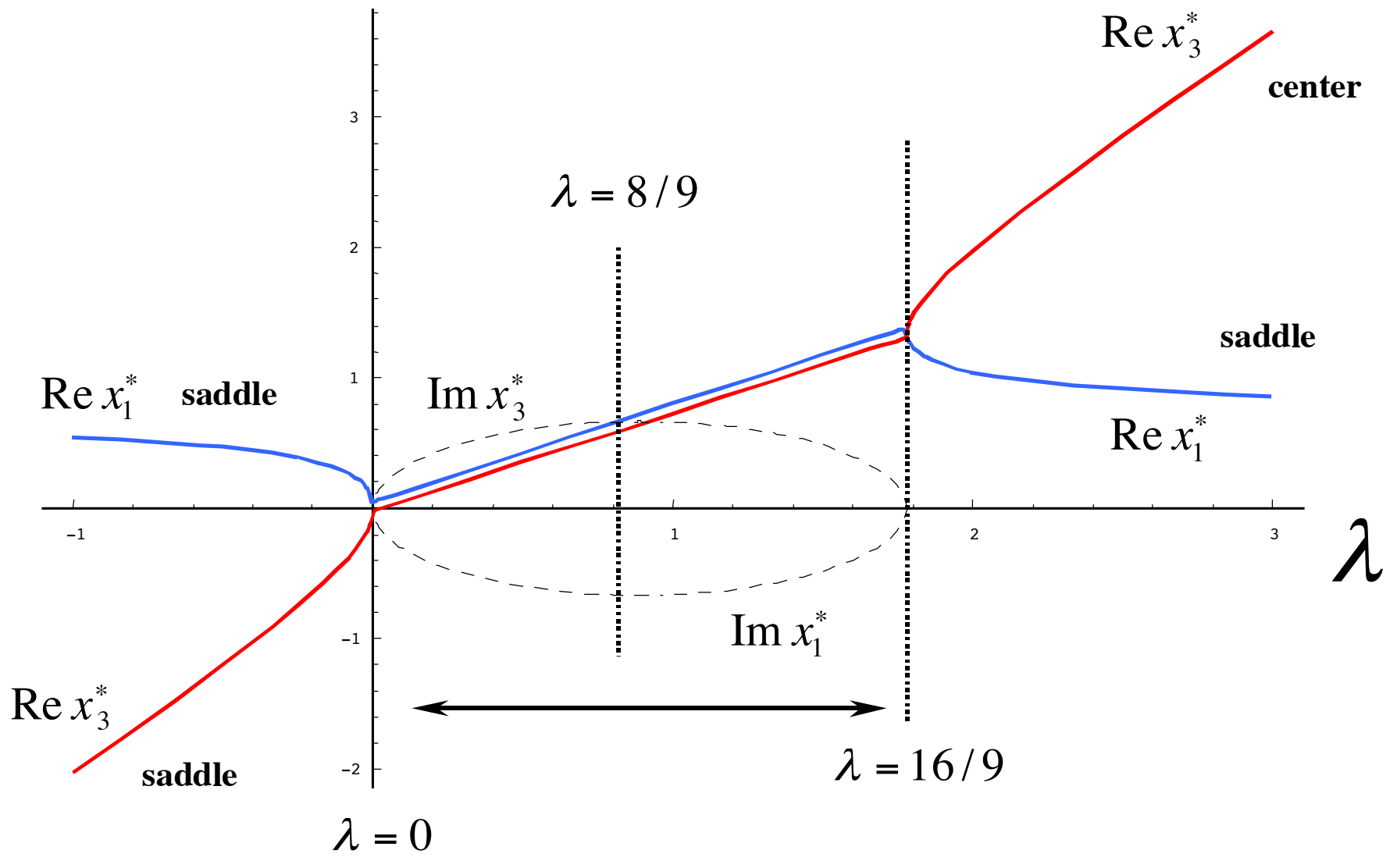

Figure 32 Reissner-Nordström Light Ray Bifurcation Diagram

The significance of the impact parameters is clarified by inspection of the phase-diagram of Figure 33. In each case, the value of the impact parameter corresponding to a given fixed point provides a maximum (or minimum) value for the capture of photons into one of several possible orbits. For example, as illustrated in Figure 33, for $1/b_3^2 < 1/b^2 < 0$, the only dynamics that are possible are given by the center node orbits about $x_3^*$. For $0 < 1/b^2 < 1/b_1^2$, the dynamics are shared by the center nodes at infinity and at $x_3^*$. The impact parameter, $1/b^2 = 1/b_1^2$, corresponds to the separatrix and is analogous to the unstable "photon sphere" that was discussed earlier for the Schwarzschild case. For $1/b^2 > 1/b_1^2$, the dynamics are librational - completing the outer loop orbits shown in Figure 33.

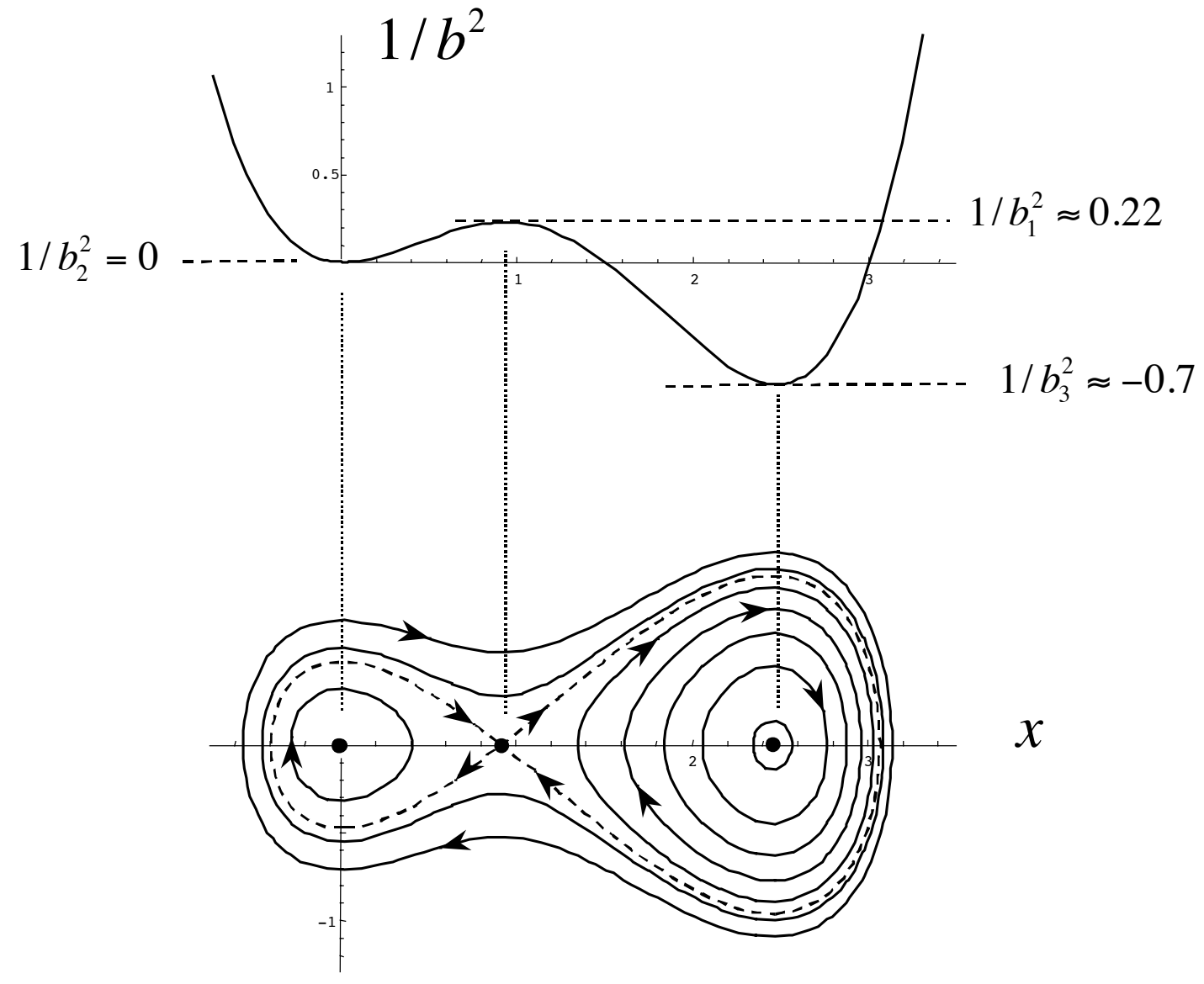

Figure 33 Phase Diagram for $\sigma = 0$; $\lambda = 7/3$.

The critical case, $\lambda = 2$, is significant dynamically in that the impact parameters $1/b_2^2$ and $1/b_3^2$ are equal and therefore the phase diagram and level curves are totally symmetric about $r_+ = r_- = 1$ as shown in Figure 34.

As a result of the spherical symmetry shared by the Reissner-Nordström and Schwarzschild solutions, the qualitative discussion given earlier for the light rays in the Schwarzschild spacetime is similar to those for the Reissner-Nordström case. The main difference in the Reissner-Nordström case is due to the appearance of a third fixed point and the fact that the impact parameters for orbits about $x_2^*$ and $x_3^*$ are dependent upon the black hole charge. Therefore, the physical interpretation is similar to that discussed in the Schwarzschild case (for $\lambda > 2$). The main qualitative features and interpretation of the impact parameters are discussed below (6.54).

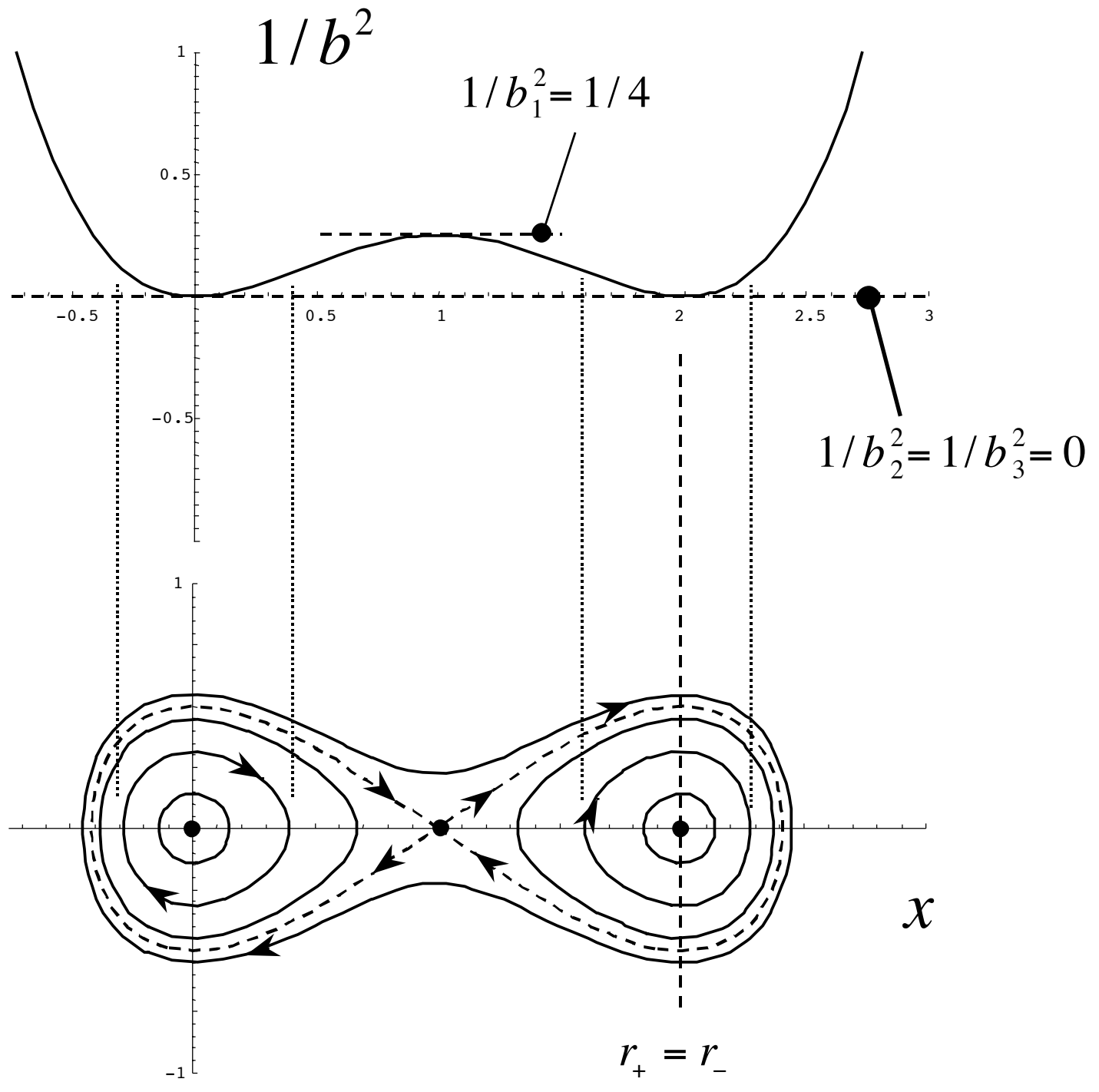

Figure 34 Phase Diagram for $\sigma = 0$; $\lambda = 2$.

# Chapter 7 Discussion

The field equations originating from the Riemann tensor quadratic curvature Lagrangian have been discussed and the analysis and classification of all known spherically symmetric solutions to the "gauge gravity" theory has been presented. In addition, a new exact solution has been found for the field equations originating from the "energy-momentum" equation of the gauge gravity theory. However, the analysis considered here does not preclude the existence of other spherically symmetric solutions since no solutions have been found for the (–)(+) differential equations that were discussed in Chapter 3:

$$\frac{d}{dr}\left(\frac{\Lambda'}{2\Lambda}-\frac{1}{r}\right)-\frac{1}{r^2\Lambda}=0. \tag{7.1}$$

Furthermore, Thompson [24] has found an axially symmetric solution that has been interpreted as a "dumbbell" consisting of two unsupported point masses. This indicates that a classification based on spherical solutions – although useful for investigating the gauge gravity theory – is not completely general and that other nonphysical solutions of the field equations should be expected based on the work of Havas [36].

The field equations that are derived from the Einstein-Hilbert action by using the standard variational procedure are equivalent to those obtained when the connection is taken as an independent variational parameter. As discussed in Chapter 3 this "symmetry" no longer applies when the action is taken in quadratic form and the gauge gravity "Palatini" procedure is applied – which should be distinguished from the true Palatini procedure where the connection and metric are completely independent. As a result, imposing this condition onto the resulting field equations leads to an auxiliary algebraic constraint that restricts the class of spacetime solutions satisfying the gauge gravity field equations. The constraints are equivalently derived as a contraction of the integrability conditions for the Ricci tensor.

The auxiliary condition gives an interesting result since it originates from the requirement that the quadratic curvature action behave analogously to the Einstein-Hilbert action with respect to both variational procedures. As a result, several (but not all) of the nonphysical solutions are eliminated from the gauge gravity equations. However, a general mathematical classification of such spaces has not been found in this analysis. A more restrictive condition must be met to eliminate all spaces that are not Einstein spaces since it is obvious from the analysis in Chapter 3 that

$$(R^{\rho}{}_{\mu} R_{\rho\nu} = R^{\rho\sigma} R_{\rho\mu\sigma\nu}) \Leftrightarrow (R_{\mu\nu} = \tfrac{1}{4} g_{\mu\nu} R), \tag{7.2}$$

is not valid in general. The counter-example is given by the Einstein universe metric, although it should be noted that the Einstein universe metric is not a solution to the Einstein free-field equations. In addition, the LHS of (7.2) eliminates all known solutions that are not solutions of the Einstein free-field equations. But it would be premature to claim any general results from this analysis without a general mathematical classification of the spacetimes implied by $R^{\rho}{}_{\mu} R_{\rho\nu} = R^{\rho\sigma} R_{\rho\mu\sigma\nu}$ (since only the spherically symmetric solutions were investigated as counter examples in this study). Although some progress has been made toward the goal of a general mathematical classification (from the study of the spherically symmetric solutions as considered in the latter part of Chapter 3) it is difficult to deduce general results from this analysis.

In the second part of the thesis the stability properties of point particle orbital dynamics in both the Schwarzschild and Reissner-Nordström black hole spacetimes have been analyzed using the phase-plane and bifurcation techniques. In general, the phase-plane and bifurcation techniques originate from dynamical systems theory as a means for classifying the solution structure and fixed points of nonlinear differential equations. The phase-plane analysis gives a method for classifying the local stability properties of fixed points, and when the equations are integrable, is useful for constructing phase-plane level curves that illustrate the global phase plane structure for a given system of nonlinear equations. The bifurcation analysis is a complementary method for identifying coalescing fixed points and the parameter values at which these "bifurcations" occur. When used in combination, a visual and quantitative method is available to analyze the relativistic orbital dynamics. As a result, a more intuitive approach can be taken based on the "energy method" to classify and summarize the various dynamics that may occur – particularly with regard to the stability properties of the orbits.

For additional applications it would be interesting to analyze solutions other than the Schwarzschild case, e.g., the Kerr solution (a rotating black hole), or the Kerr-Newman solution (a charged, rotating black hole). Further applications would include an analysis of cosmological solutions (although some work has already appeared on this

topic - see the Introduction for a discussion of the literature) and also non-conservative orbital dynamics (i.e. systems emitting gravitational radiation) and analysis of solutions stemming from alternative theories of general relativity, e.g., those considered in Chapters 3 and 4. The analysis on non-conservative dynamics is of special interest given that a considerable experimental effort has been recently organized to detect gravity waves (e.g., LIGO), and therefore, qualitatively simpler models that illustrate the dynamics of gravity wave production would certainly be of interest. Additional techniques of analysis that have been developed from the study of limit cycles and attractors in non-conservative systems would then be available and may prove useful in the classification and visualization of these dynamical processes.

Essentially, the utility of analyzing the relativistic dynamics using bifurcation analysis is due to the fact that the relevant physical parameters are isolated and provide a central role in the analysis. For example, the Schwarzschild orbital dynamics are characterized by a dimensionless parameter, $\sigma$, involving the angular momentum which is related to the energy of the system at the unstable orbital radius through the separatrix structure of the phase-plane. Identifying the bifurcations that occur as $\sigma$ is varied then provides a summary of the stability properties and dynamics over the complete range of values taken by the energy and angular momentum. As a result, seemingly unrelated orbital dynamics at different parameter values may be interpreted simply as different stages of a given bifurcation. These considerations become especially important when interpreting the dynamics associated with more complex physical situations such as the Reissner-Nordström, Kerr, and Kerr-Newman solutions. The Reissner-Nordström and Kerr solutions are similar in that one additional physical parameter (charge and spin angular momentum, respectively) beyond the orbital angular momentum of the system is required to characterize the dynamics. Therefore, a 2-d parameter space is required in both cases to classify the dynamics. A thorough analysis of the Kerr dynamics using traditional methods has been given earlier by Bardeen et. al. [152] and also Chandrasekhar [68] (see also the references given in the Introduction).

Although the Kerr solution has not been considered in this study, it is worthwhile to point out some difficulties encountered in the analysis when considering a direct application of the phase-plane and bifurcation techniques to the Kerr dynamics in the equatorial plane. To illustrate, the Lagrangian for the Kerr solution (for the simplifying case that $\theta = \pi/2$; $G = c = 1$) is expressed:

$$2\hat{L} = 1 = \Lambda \dot{t}^2 - \Delta^{-1}\dot{r}^2 - \Sigma\dot{\varphi}^2 + 2ax\dot{t}\dot{\varphi}\,, \tag{7.3}$$

where

$$\Lambda = (1-x)\;;\;\; \Delta = (1 - x + a^2x^2/r_s^2)\;;\;\; \Sigma = [r_s^2/x^2 + a^2(1+x)]\;;\;\; x = r_s/r\,. \tag{7.4}$$

The parameter $a$ corresponds to the spin angular momentum of the system per unit black hole mass. As in the Schwarzschild and Reissner-Nordström analysis the total energy and orbital angular momentum are constants of the motion:

$$\begin{aligned} \hat{E} &= \Lambda\dot{t} + ax\dot{\varphi} \\ \hat{J} &= -ax\dot{t} + \Sigma\dot{\varphi}, \end{aligned} \tag{7.5}$$

and therefore $\dot{t}$ and $\dot{\varphi}$ are expressed:

$$\begin{aligned} \dot{t} &= (\Sigma\hat{E} - ax\hat{J})(\Lambda\Sigma + a^2x^2)^{-1} \\ \dot{\varphi} &= (ax\hat{E} + \Lambda\hat{J})(\Lambda\Sigma + a^2x^2)^{-1}. \end{aligned} \tag{7.6}$$

The corresponding expression for $dx/d\varphi$ is thus obtained:

$$\left(dx/d\varphi\right)^2 = x^4\Delta\,[\Lambda\dot{t}^2 - \Sigma\dot{\varphi}^2 + 2ax\dot{t}\dot{\varphi} - 1]/r_s^2\,\dot{\varphi}\,, \tag{7.7}$$

and then substituting (7.6) into (7.7) gives the result:

$$\left(\frac{dx}{d\varphi}\right)^2 = \frac{x^4\Delta(\Lambda\Sigma + a^2x^2)}{r_s^2(ax\hat{E} + c^2\Lambda\hat{J})}\left[\Sigma(\Lambda - \hat{E}^2)\dot{t}^2 + a^2x^2 + \Lambda\hat{J}^2 + 2ax\hat{E}\hat{J}\right]. \tag{7.8}$$

To obtain the effective potential set $dx/d\varphi = 0$, solve for $\hat{E}$, and then substitute into $\hat{E}^2 - 1$ to obtain the following two cases (letting $\hat{J} \to r_s/\sqrt{2\sigma}$ ):

$$\begin{aligned} \hat{V}^2_{eff(\pm)} - 1 = \Lambda(1 + r_s^2/2\sigma\Sigma) - 1 \;\pm\; \frac{a\,r_s}{x\Delta^{1/2}\Sigma^2}[2x^4\Delta(a^2x^2 + \Lambda\Sigma)(\Sigma + r_s^2/2\sigma)/\sigma] \\ + a^2x^2(1 + r_s^2/\sigma\Sigma)/\Sigma\,, \end{aligned} \tag{7.9}$$

where the (–) case corresponds to a particle rotating in same sense as the Kerr black hole, and the (+) case is the effective potential for a particle that is orbiting in the opposite direction of rotation. To check the correspondence with the Schwarzschild result substitute for $\Delta$, $\Sigma$, $\Lambda$ and series expand to first order in $a$ to obtain:

$$\hat{V}^2_{eff(\pm)} - 1 = \underbrace{(x^2 - x^3)/2\sigma - x}_{Schwarzschild} \;\pm\; ax^3\sqrt{(1-x)(x^2 + 2\sigma)/\sigma^2}\,, \tag{7.10}$$

which reduces to the Schwarzschild effective potential as $a \to 0$. The main difficulty when applying the phase-plane analysis to the Kerr solution is that the algebraic equation that must be solved for the fixed points: $d\hat{V}_{eff(\pm)}/dx = 0$, is non-polynomial in *x*. Therefore, no closed form solution can be obtained in this case - at least for the choice of variables and coordinate system used above.

Finally, the pedagogical value of the phase-plane and bifurcation techniques to general relativity calculations should not be under emphasized. The literature on the subject is vast and alternative methods of presentation that make general relativity conceptually simpler for beginning students is an active area of investigation (e.g., [153]). The conceptual advantage of the phase-plane approach is based on the fact that the phase-plane method is a pictorial method closely related to the "energy-method" diagrams taught in introductory mechanics courses. As a result, a qualitative approach is emphasized that makes the underlying physical concepts easier to grasp for beginning students. For example, the Schwarzschild periastron calculation is tied directly to the existence of the center node fixed point and is a relatively straightforward calculation that eliminates much of the unnecessary algebra that appears in other presentations. The analysis presented in Section VI demonstrates that important topics such as dynamical invariance is also easily handled using the phase-plane approach. Such topics provide nontrivial and physically interesting examples which normally are difficult conceptually for beginning students.

In summary, constructing an exact phase-plane for an arbitrary solution will only be possible if the fixed point algebraic equation, $x' = y' = 0$, is of fourth order or less (and in addition that a sufficient number of first integrals exists). Otherwise, finding roots will be difficult if not impossible. However, a numerical approach could always be taken, and would be motivated by the interesting pictures that result from combining the fixed point structure of general relativity state-space into a diagram that includes the event horizon.

# Appendix A: Mathematica Calculations - Chapter 2

*Note*: The computer algebra system calculations in this appendix (and Appendices B, E, and G) were performed using *Mathematica* version 3.0 – a software package developed by Wolfram Research [142] and running under Microsoft Windows NT version 4.0 and Windows 98. Specialized tensor calculations were performed using the Mathematica add-on package *MathTensor* version 2.2, developed by Parker and Christensen [117].

Initialize MathTensor

```
<< Mathtens.m
Loading MathTensor for DOS/Windows . . .



===================================================
MathTensor (TM) 2.2 (DOS/Windows(R)) (Jun 1, 1994)
by Leonard Parker and Steven M. Christensen
Copyright (c) 1991-1994 MathSolutions, Inc.
Runs with Mathematica (R) Versions 2.X.
Licensed to one machine only, copying prohibited.
==================================================
No unit system is chosen. If you want one,
you must edit the file called Conventions.m,
or enter a command to interactively set units.
Units: {}
Sign conventions: Rmsign = 1 Rcsign = 1
MetricgSign = -1 DetgSign = -1
TensorForm turned on,
ShowTime turned off,
MetricgFlag = True.
=========================================
Dimension = 4;
SetDirectory["C:\Mathematica\Files\Dissertation\
   Tensor Calculations"];
```

## Einstein-Hilbert Standard Variation:

*Perform the variation ( $\delta h$ corresponds to $\delta g$ , semicolons denote covariant derivatives):*

```
Variation[Sqrt[Detg] ScalarR, Metricg]
```

$$\sqrt{g}\,(h_{pq;}{}^{pq}) - \sqrt{g}\,(h_p{}^p{}_{;q}{}^q) + \frac{1}{2}\sqrt{g}\,R\,(g^{pq})\,(h_{pq}) - \sqrt{g}\,(R_{pq})\,(h^{pq})$$

*Remove derivatives from $\delta g$ and factor det g (i.e., partial integration):*

```
Expand[PIntegrate[%, Metricg] / Sqrt[Detg]]
```

$$\frac{1}{2} R (g^{pq}) (h_{pq}) - (R_{pq}) (h^{pq})$$

*This is the result:*

```
VariationalDerivative[-% == 0, Metricg, ua, ub]
```

$$-\frac{1}{2} R (g_{ab}) + R_{ab} == 0$$

Schwarzschild and Reissner-Nordstrom Solutions:

*Load the File:*

```
<< spherical.m
MetricgFlag has been turned off.
```

*Display the Metric ("4" Labels the time coordinate):*

```
Table[Metricg[-i, -j], {i, 4}, {j, 4}]
```

$$\begin{pmatrix} -B(r) & 0 & 0 & 0 \\ 0 & -r^2 & 0 & 0 \\ 0 & 0 & -r^2 \sin^2(\text{theta}) & 0 \\ 0 & 0 & 0 & A(r) \end{pmatrix}$$

*Here are the non-zero Christoffel Symbols :*

```
connection = {};
Do[
  If[
    AffineG[i, -j, -k] =!= 0,
     connection = {connection,
        SubscriptBox[
      SubscriptBox[SuperscriptBox[Γ, i], j], k] ==
     AffineG[i, -j, -k] // DisplayForm},
    Continue[]],

  {i, 4}, {j, 4}, {k, 4}]
Flatten[connection]
```

$$\left\{\Gamma^1{}_{11} == \frac{B'[r]}{2 B[r]}, \Gamma^1{}_{22} == -\frac{r}{B[r]},\right.$$

$$\Gamma^1{}_{33} == -\frac{r \operatorname{Sin}[\text{theta}]^2}{B[r]}, \Gamma^1{}_{44} == \frac{A'[r]}{2 B[r]}, \Gamma^2{}_{12} == \frac{1}{r},$$

$$\Gamma^2{}_{21} == \frac{1}{r}, \Gamma^2{}_{33} == -\operatorname{Cos}[\text{theta}] \operatorname{Sin}[\text{theta}], \Gamma^3{}_{13} == \frac{1}{r},$$

$$\Gamma^3{}_{23} == \operatorname{Cot}[\text{theta}], \Gamma^3{}_{31} == \frac{1}{r}, \Gamma^3{}_{32} == \operatorname{Cot}[\text{theta}],$$

$$\left.\Gamma^4{}_{14} == \frac{A'[r]}{2 A[r]}, \Gamma^4{}_{41} == \frac{A'[r]}{2 A[r]}\right\}$$

*Ricci Tensor :*

```
Table[Simplify[RicciR[-i, -j]], {i, 4}, {j, 4}]
```

$$\left\{\left\{\frac{A[r]\,(4A[r]+r\,A'[r])\,B'[r]+r\,B[r]\,(A'[r]^2-2A[r]\,A''[}{4\,r\,A[r]^2\,B[r]}\right.\right.$$

$$\left.0,0,0\right\},\left\{0,\frac{1}{2}\left(2-\frac{2+\frac{r A'[r]}{A[r]}}{B[r]}+\frac{r\,B'[r]}{B[r]^2}\right),0,0\right\},$$

$$\left\{0,0,\frac{1}{2\,A[r]\,B[r]^2}\,(\mathrm{Sin}[theta]^2\right.$$
$$(-r\,B[r]\,A'[r]+A[r]\,(-2\,B[r]+2\,B[r]^2+r\,B'[r]))),0$$

$$\left\{0,0,0,\frac{1}{4\,r\,A[r]\,B[r]^2}\,(-r\,B[r]\,A'[r]^2+\right.$$
$$\left.\left.A[r]\,(-r\,A'[r]\,B'[r]+2\,B[r]\,(2\,A'[r]+r\,A''[r])))\right\}\right\}$$

## Schwarzschild Solution :

```
Collect[Expand[Table[RicciR[-i, -j] /. {B[r] -> 1/A[r
     B'[r] -> D[1/A[r], r]}, {i, 4}, {j, 4}]],
  r, Factor]
```

$$\left\{\left\{-\frac{A'[r]}{r\,A[r]}-\frac{A''[r]}{2\,A[r]},0,0,0\right\},\{0,1-A[r]-r\,A'[r],0,0\right.$$
$$\{0,0,-(-1+A[r])\,\mathrm{Sin}[theta]^2-r\,\mathrm{Sin}[theta]^2\,A'[r],0\}$$
$$\left.\left\{0,0,0,\frac{A[r]\,A'[r]}{r}+\frac{1}{2}\,A[r]\,A''[r]\right\}\right\}$$

```
Flatten[ExpandAll[DSolve[1 - A[r] - r A'[r] == 0, A[r], r
```

$$\left\{A[r]\to 1+\frac{C[1]}{r}\right\}$$

*Display the Ricci Scalar:*

```
ScalarR
```

$$\frac{1}{2\,r^2\,A[r]^2\,B[r]^2}\,(4\,A[r]^2\,B[r]-4\,A[r]^2\,B[r]^2+$$
$$4\,r\,A[r]\,B[r]\,A'[r]-r^2\,B[r]\,A'[r]^2-4\,r\,A[r]^2\,B'[r]-$$
$$r^2\,A[r]\,A'[r]\,B'[r]+2\,r^2\,A[r]\,B[r]\,A''[r])$$

```
Simplify[
 ScalarR /. {B[r] -> 1/A[r], B'[r] -> D[1/A[r], r]}]
```

$$\frac{-2+2\,A[r]+4\,r\,A'[r]+r^2\,A''[r]}{r^2}$$

## Reissner-Nordstrom Solution:

*Maxwell's Homogeneous Field Equations:*

```
f1[ub_] = CD[MaxwellF[ua, ub], la]
```

$F^{ab}{}_{;a}$

*Expand out the covariant derivatives*
*($G^{a}_{bc}$ is used to denote the connection, $\Gamma^{a}_{bc}$):*

```
f2[ub_] = CDtoOD[f1[ub]]
```

$-(G^{b}{}_{pq})\,(F^{pq})+(G^{p}{}_{pq})\,(F^{qb})+F^{pb}{}_{,p}$

*Finally, this defines source-free Maxwell:*

```
Vf[ub_] = f2[ub]
```

$-(G^{b}{}_{pq})\,(F^{pq})+(G^{p}{}_{pq})\,(F^{qb})+F^{pb}{}_{,p}$

*Define Components:*

```
Table[MaxwellF[i, j] = 0, {i, 4}, {j, 4}];
MaxwellF[4, 1] = H[r]; MaxwellF[1, 4] = -H[r];
```

```
Table[MaxwellF[i, j], {i, 4}, {j, 4}]
{{0, 0, 0, -H[r]}, {0, 0, 0, 0}, {0, 0, 0, 0},
 {H[r], 0, 0, 0}}
```

*Load and Display the Metric ("4" Labels the time coordinate):*

```
<< spherical.m
MetricgFlag has been turned off.
Table[Metricg[-i, -j], {i, 4}, {j, 4}]
```

$$\{\{-B[r], 0, 0, 0\}, \{0, -r^2, 0, 0\}, \{0, 0, -r^2\,\mathrm{Sin}[theta]^2, 0\}, \{0, 0, 0, A[r]\}\}$$

*Evaluate Ordinary Derivatives:*

```
On[EvaluateODFlag]
```

*Lower indices on the Field Strength tensor:*

```
Table[MaxwellF[-k, -l] = Sum[Metricg[-k, -i]
    Metricg[-l, -j] MaxwellF[i, j], {i, 4}, {j, 4}],
 {k, 4}, {l, 4}]
{{0, 0, 0, A[r] B[r] H[r]}, {0, 0, 0, 0}, {0, 0, 0, 0},
 {-A[r] B[r] H[r], 0, 0, 0}}
```

*Raise an Index:*

```
Table[MaxwellF[-k, j] =
  Sum[Metricg[-k, -i] MaxwellF[i, j], {i, 4}], {k, 4},
 {j, 4}]
{{0, 0, 0, B[r] H[r]}, {0, 0, 0, 0}, {0, 0, 0, 0},
 {A[r] H[r], 0, 0, 0}}
```

*Define the Energy Momentum Tensor:*

```
Θ[la_, lb_] =
 Factor[ApplyRules[MaxwellT[la, lb], MaxwellTtoFrule]
```

$$\frac{c^4\,(4\,(F_{pa})\,(F_b{}^p) + (F_{pq})\,(F^{pq})\,(g_{ab}))}{16\,(k_1)\,\pi}$$

```
Table[Θ[-i, -j] = MakeSum[Θ[-i, -j]], {i, 4}, {j, 4}]
```

$$\left\{\left\{-\frac{c^4 A[r]\,B[r]^2\,H[r]^2}{8\,(k_1)\,\pi}, 0, 0, 0\right\}, \left\{0, \frac{c^4 r^2 A[r]\,B[r]\,H[r]^2}{8\,(k_1)\,\pi}, 0, 0\right\}, \left\{0, 0, \frac{c^4 r^2 A[r]\,B[r]\,H[r]^2\,\mathrm{Sin}[theta]^2}{8\,(k_1)\,\pi}, 0\right\}, \left\{0, 0, 0, \frac{c^4 A[r]^2\,B[r]\,H[r]^2}{8\,(k_1)\,\pi}\right\}\right\}$$

*Traceless:*

```
Sum[Metricg[i, j] Θ[-i, -j], {i, 4}, {j, 4}]
0
```

*These are the field equations:*

```
fe = Simplify[
  Table[RicciR[-i, -j] - 8 π G c^-4 Θ[-i, -j], {i, 4}, {j,
```

$$\left\{\left\{\frac{G\,c^4\,A[r]\,B[r]^2\,H[r]^2}{c^4\,(k_1)} + \frac{A[r]\,(4A[r]+r\,A'[r])\,B'[r]+r\,B[r]\,(A'[r]^2-2\,A[r]\,A''}{4\,r\,A[r]^2\,B[r]}\right.\right.$$

$$,\ 0,\ 0,\ 0\Big\},$$

$$\left\{0,\ \frac{1}{2}\left(2-\frac{2\,G\,c^4\,r^2\,A[r]\,B[r]\,H[r]^2}{c^4\,(k_1)}-\frac{2+\frac{r\,A'[r]}{A[r]}}{B[r]}+\frac{r\,B'[r]}{B[r]^2}\right.\right.$$

$$0,\ 0\Big\},\ \{0,\ 0,\ (\mathrm{Sin}[\mathrm{theta}]^2\ (-2\,G\,c^4\,r^2\,A[r]^2\,B[r]^3\,H[r]^2-c^4\,(k_1)\,r\,B[r]\,A'[r]+c^4\,(k_1)\,A[r]\,(-2\,B[r]+2\,B[r]^2+r\,B'[r])))\,/\,(2\,c^4\,(k_1)\,A[r]\,B[r]^2),\ 0\},$$

$$\left\{0,\ 0,\ 0,\ -\frac{G\,c^4\,A[r]^2\,B[r]\,H[r]^2}{c^4\,(k_1)}-\frac{A'[r]^2}{4\,A[r]\,B[r]}+\frac{-r\,A'[r]\,B'[r]+2\,B[r]\,(2\,A'[r]+r\,A''[r])}{4\,r\,B[r]^2}\right\}\Big\}$$

```
Expand[r B[r] Expand[fe[[4, 4]] / A[r] + fe[[1, 1]] / B[r]
```

$$\frac{A'[r]}{A[r]}+\frac{B'[r]}{B[r]}$$

*Maxwell's homogeneous equations:*

```
Simplify[Table[MakeSum[∇f[i]], {i, 4}]]
```

$$\left\{0,\ 0,\ 0,\ H[r]\left(-\frac{2}{r}-\frac{A'[r]}{2\,A[r]}-\frac{B'[r]}{2\,B[r]}\right)-H'[r]\right\}$$

```
Factor[% /. {B[r] -> 1 / A[r], B'[r] -> D[1 / A[r], r]}]
```

$$\left\{0,\ 0,\ 0,\ -\frac{2\,H[r]+r\,H'[r]}{r}\right\}$$

```
Flatten[DSolve[2 H[r] + r H'[r] == 0, H[r], r]]
```

$$\left\{H[r]\to\frac{C[1]}{r^2}\right\}$$

*Substitute back into the field equations:*

```
Simplify[
 fe /. {B[r] -> 1 / A[r], B'[r] -> D[1 / A[r], r], H[r] -> -
```

$$\left\{\left\{-\frac{-\frac{2\,e^2\,G\,c^4}{c^4\,(k_1)\,r^4}+\frac{2\,A'[r]}{r}+A''[r]}{2\,A[r]},\ 0,\ 0,\ 0\right\},\right.$$

$$\left\{0,\ 1-\frac{e^2\,G\,c^4}{c^4\,(k_1)\,r^2}-A[r]-r\,A'[r],\ 0,\ 0\right\},$$

$$\left\{0,\ 0,\ \mathrm{Sin}[\mathrm{theta}]^2\left(1-\frac{e^2\,G\,c^4}{c^4\,(k_1)\,r^2}-A[r]-r\,A'[r]\right),\ 0\right\},$$

$$\left.\left\{0,\ 0,\ 0,\ \frac{1}{2}\,A[r]\left(-\frac{2\,e^2\,G\,c^4}{c^4\,(k_1)\,r^4}+\frac{2\,A'[r]}{r}+A''[r]\right)\right\}\right\}$$

*The Reissner-Nordstrom Solution:*

```
Flatten[
 ExpandAll[DSolve[1 - A[r] - r A'[r] == e^2 G / (k1 r^2), A[r], r]]]
```

$$\left\{A[r]\to 1+\frac{C[1]}{r}+\frac{e^2\,G}{r^2\,k_1}\right\}$$

# Schwarzschild Christoffel Symbols:

```
Simplify[Flatten[connection] /.
   {B[r] -> 1/A[r] /. {A[r] -> 1 - r_s/r}, B'[r] ->
     D[1/A[r] /. {A[r] -> 1 - r_s/r}, r], A[r] -> {1 - r_s/
    A'[r] -> D[A[r] /. {A[r] -> 1 - r_s/r}, r]} /.
  Subscript^(1,0)[r, s] -> 0]
```

$$\left\{\Gamma^1{}_{11} == \frac{r_s}{-2\,r^2 + 2\,r\,r_s},\ \Gamma^1{}_{22} == -r + r_s,\right.$$

$$\Gamma^1{}_{33} == -\text{Sin[theta]}^2\,(r - r_s),\ \Gamma^1{}_{44} == \frac{(r - r_s)\,r_s}{2\,r^3},\ \Gamma^2{}_{12} ==$$

$$\Gamma^2{}_{21} == \frac{1}{r},\ \Gamma^2{}_{33} == -\text{Cos[theta] Sin[theta]},\ \Gamma^3{}_{13} == \frac{1}{r},$$

$$\Gamma^3{}_{23} == \text{Cot[theta]},\ \Gamma^3{}_{31} == \frac{1}{r},\ \Gamma^3{}_{32} == \text{Cot[theta]},$$

$$\left.\Gamma^4{}_{14} == \left\{\frac{r_s}{2\,r^2 - 2\,r\,r_s}\right\},\ \Gamma^4{}_{41} == \left\{\frac{r_s}{2\,r^2 - 2\,r\,r_s}\right\}\right\}$$

# Reissner-Nordstrom Christoffel Symbols:

```
Simplify[Flatten[connection] /.
   {B[r] -> 1/A[r] /. {A[r] -> 1 + r_s/r + (e^2 G)/(r^2 k_1)},
    B'[r] -> D[1/A[r] /. {A[r] -> 1 + r_s/r + (e^2 G)/(r^2 k_1)}, r],
    A[r] -> {1 + r_s/r + (e^2 G)/(r^2 k_1)},
    A'[r] -> D[A[r] /. {A[r] -> 1 + r_s/r + (e^2 G)/(r^2 k_1)}, r]} /.
  Subscript^(1,0)[r, s] -> 0]
```

$$\left\{\Gamma^1{}_{11} == \frac{2\,e^2\,G + r\,k_1\,r_s}{2\,r\,(e^2\,G + r\,k_1\,(r + r_s))},\ r + \frac{e^2\,G}{r\,k_1} + r_s + \Gamma^1{}_{22} == 0,\right.$$

$$\Gamma^1{}_{33} == -r\,\text{Sin[theta]}^2\left(1 + \frac{e^2\,G}{r^2\,k_1} + \frac{r_s}{r}\right),$$

$$\Gamma^1{}_{44} == -\frac{(2\,e^2\,G + r\,k_1\,r_s)\,(e^2\,G + r\,k_1\,(r + r_s))}{2\,r^5\,k_1^2},$$

$$\Gamma^2{}_{12} == \frac{1}{r},\ \Gamma^2{}_{21} == \frac{1}{r},\ \Gamma^2{}_{33} == -\text{Cos[theta] Sin[theta]},$$

$$\Gamma^3{}_{13} == \frac{1}{r},\ \Gamma^3{}_{23} == \text{Cot[theta]},\ \Gamma^3{}_{31} == \frac{1}{r},$$

$$\Gamma^3{}_{32} == \text{Cot[theta]},\ \Gamma^4{}_{14} == \left\{-\frac{2\,e^2\,G + r\,k_1\,r_s}{2\,r\,(e^2\,G + r\,k_1\,(r + r_s))}\right\},$$

$$\left.\Gamma^4{}_{41} == \left\{-\frac{2\,e^2\,G + r\,k_1\,r_s}{2\,r\,(e^2\,G + r\,k_1\,(r + r_s))}\right\}\right\}$$

# Appendix B: Mathematica Calculations - Chapter 3

*Initialize MathTensor*

```
<<Mathtens.m
Dimension = 4;
SetDirectory["C:\Mathematica\Files\Dissertation\
   Tensor Calculations"];
```

*Define the first Gauge-Gravity Field Equation:*

```
H[lc_, lf_] = RiemannR[ua, ub, ud, lc]
   RiemannR[la, lb, ld, lf] - 1/4 Metricg[lc, lf]
   RiemannR[ua, ub, ug, ud] RiemannR[la, lb, lg, ld]
```

$-(R_{abdf})\,(R_c{}^{dab}) - \frac{1}{4}\,(g_{cf})\,(R_{abdg})\,(R^{abdg})$

*Define the second Gauge-Gravity Field Equation:*

```
Y1[la_, lb_, lc_] =
 2 Antisymmetrize[CD[RicciR[lb, lc], la], {la, lb}]
```

$-(R_{ac;b}) + R_{bc;a}$

*Expand out the covariant derivatives*
($G^p_{bc}$ *is used to denote the connection,* $\Gamma^p_{bc}$):

```
Y2[la, lb, lc] = CDtoOD[Y1[la, lb, lc]]
```

$-(R_{ac,b}) + R_{bc,a} + (G^p{}_{bc})\,(R_{pa}) - (G^p{}_{ac})\,(R_{pb})$

*Finally, this defines equation :*

```
Y[la_, lb_, lc_] = Y2[la, lb, lc]
```

$-(R_{ac,b}) + R_{bc,a} + (G^p{}_{bc})\,(R_{pa}) - (G^p{}_{ac})\,(R_{pb})$

*Define the Auxiliary Condition:*

```
ac[lc_, lf_] = RicciR[ua, lc] RicciR[la, lf] -
  RicciR[la, lb] RiemannR[lc, ua, lf, ub]
```

$(R_{af})\,(R_c{}^a) - (R_{ab})\,(R_c{}^a{}_f{}^b)$

*Evaluate Ordinary Derivatives:*

```
On[EvaluateODFlag]
```

## Tensor Calculations - Ricci, Riemann, Weyl Tensors, Field Equations

*Load the File:*

```
<<exp_sphere.m
MetricgFlag has been turned off.
```

*Display the Metric ("4" Labels the time coordinate):*

```
Table[Metricg[-i, -j], {i, 4}, {j, 4}]
```

$$\begin{pmatrix} -e^{2B(r)} & 0 & 0 & 0 \\ 0 & -r^2 & 0 & 0 \\ 0 & 0 & -r^2\sin^2(\text{theta}) & 0 \\ 0 & 0 & 0 & e^{2A(r)} \end{pmatrix}$$

*Here are the non-zero Christoffel Symbols :*

```
connection = {};
Do[
  If[
    AffineG[i, -j, -k] =!= 0,
     connection = {connection,
        SubscriptBox[
      SubscriptBox[SuperscriptBox[Γ, i], j], k] ==
     AffineG[i, -j, -k] // DisplayForm},
   Continue[]],

  {i, 4}, {j, 4}, {k, 4}]
Simplify[Flatten[connection]]
```

$\{\Gamma^1{}_{11} == B'[r],\ \Gamma^1{}_{22} == -E^{-2B[r]}\, r,$
$\Gamma^1{}_{33} == -E^{-2B[r]}\, r\, \text{Sin}[\text{theta}]^2,\ \Gamma^1{}_{44} == E^{2(A[r]-B[r])}\, A'[r],$
$\Gamma^2{}_{12} == \frac{1}{r},\ \Gamma^2{}_{21} == \frac{1}{r},\ \Gamma^2{}_{33} == -\text{Cos}[\text{theta}]\, \text{Sin}[\text{theta}],$
$\Gamma^3{}_{13} == \frac{1}{r},\ \Gamma^3{}_{23} == \text{Cot}[\text{theta}],\ \Gamma^3{}_{31} == \frac{1}{r},$
$\Gamma^3{}_{32} == \text{Cot}[\text{theta}],\ \Gamma^4{}_{14} == A'[r],\ \Gamma^4{}_{41} == A'[r]\}$

*Display the Ricci Tensor :*

```
Table[Simplify[RicciR[-i, -j]], {i, 4}, {j, 4}]
```

$\{\{-A'[r]^2 + \frac{2B'[r]}{r} + A'[r]\,B'[r] - A''[r],\ 0,\ 0,\ 0\},$
$\{0,\ E^{-2B[r]}\,(-1 + E^{2B[r]} - r\,A'[r] + r\,B'[r]),\ 0,\ 0\},\ \{0,\ 0,$
$E^{-2B[r]}\,\text{Sin}[\text{theta}]^2\,(-1 + E^{2B[r]} - r\,A'[r] + r\,B'[r]),\ 0\},\ \{$
$0,\ 0,\ \frac{E^{2(A[r]-B[r])}\,(r\,A'[r]^2 + A'[r]\,(2 - r\,B'[r]) + r\,A''[r])}{r}$

*Display the Ricci Scalar:*

```
Simplify[ScalarR]
```

$-\frac{1}{r^2}(2E^{-2B[r]}\,(-1 + E^{2B[r]} - r^2\,A'[r]^2 + 2\,r\,B'[r] +$
$r\,A'[r]\,(-2 + r\,B'[r]) - r^2\,A''[r]))$

*Non-zero Components of the Riemann Tensor :*

```
riem = {};
Do[
  If[
    RiemannR[i, -j, -k, -l] =!= 0,
     riem = {riem,
        SubscriptBox[SubscriptBox[
       SubscriptBox[SuperscriptBox[R, i], j], k], l]
     RiemannR[i, -j, -k, -l] // DisplayForm},
   Continue[]],

  {i, 4}, {j, 4}, {k, 4}, {l, 4}];
Simplify[Flatten[riem]]
```

$$
\begin{aligned}
&\{R^{1}{}_{212} == E^{-2B[r]}\, r\, B'[r],\\
&R^{1}{}_{221} == -E^{-2B[r]}\, r\, B'[r],\; R^{1}{}_{313} == E^{-2B[r]}\, r\, \mathrm{Sin}[theta]^{2}\, B'[\\
&R^{1}{}_{331} == -E^{-2B[r]}\, r\, \mathrm{Sin}[theta]^{2}\, B'[r],\\
&R^{1}{}_{414} == E^{2(A[r]-B[r])}\,(A'[r]^{2} - A'[r]\, B'[r] + A''[r]),\; R^{1}{}_{441} ==\\
&\quad -E^{2(A[r]-B[r])}\,(A'[r]^{2} - A'[r]\, B'[r] + A''[r]),\; R^{2}{}_{112} == -\frac{B'[}{r}\\
&R^{2}{}_{121} == \frac{B'[r]}{r},\; R^{2}{}_{323} == (1 - E^{-2B[r]})\,\mathrm{Sin}[theta]^{2},\; R^{2}{}_{332} =\\
&\quad (-1 + E^{-2B[r]})\,\mathrm{Sin}[theta]^{2},\; R^{2}{}_{424} == \frac{E^{2(A[r]-B[r])}\, A'[r]}{r},\\
&R^{2}{}_{442} == -\frac{E^{2(A[r]-B[r])}\, A'[r]}{r},\; R^{3}{}_{113} == -\frac{B'[r]}{r},\\
&R^{3}{}_{131} == \frac{B'[r]}{r},\; R^{3}{}_{223} == -1 + E^{-2B[r]},\; -1 + E^{-2B[r]} + R^{3}{}_{232} ==\\
&R^{3}{}_{434} == \frac{E^{2(A[r]-B[r])}\, A'[r]}{r},\; R^{3}{}_{443} == -\frac{E^{2(A[r]-B[r])}\, A'[r]}{r},\\
&R^{4}{}_{114} == A'[r]^{2} - A'[r]\, B'[r] + A''[r],\\
&R^{4}{}_{141} + A'[r]^{2} - A'[r]\, B'[r] + A''[r] == 0,\; R^{4}{}_{224} == E^{-2B[r]}\, r\, A'\\
&R^{4}{}_{242} == -E^{-2B[r]}\, r\, A'[r],\; R^{4}{}_{334} == E^{-2B[r]}\, r\, \mathrm{Sin}[theta]^{2}\, A'[\\
&R^{4}{}_{343} == -E^{-2B[r]}\, r\, \mathrm{Sin}[theta]^{2}\, A'[r]\}
\end{aligned}
$$

*Non-zero Components of the Weyl Tensor :*

```
weyl = {};
Do[
   If[
     WeylC[i, -j, -k, -l] =!= 0,
      weyl = {weyl,
         SubscriptBox[SubscriptBox[
        SubscriptBox[SuperscriptBox[C, i], j], k], l]
      WeylC[i, -j, -k, -l] // DisplayForm},
    Continue[]],


  {i, 4}, {j, 4}, {k, 4}, {l, 4}];
Simplify[Flatten[weyl]]
Collect[Simplify[% /.
   {A'[r] -> x1, A''[r] -> x2, B'[r] -> y1, B''[r] -> y2}
 r, Factor]
```

$$\{C^{1}{}_{212} == -\frac{1}{6}\,E^{-2B[r]}$$
$$(-1+E^{2B[r]}+r\,(x1-y1)-r^{2}\,(x1^{2}+x2-x1\,y1)),\ C^{1}{}_{221} ==$$
$$\frac{1}{6}\,E^{-2B[r]}\,(-1+E^{2B[r]}+r\,(x1-y1)-r^{2}\,(x1^{2}+x2-x1\,y1)),$$
$$C^{1}{}_{313} == -\frac{1}{6}\,E^{-2B[r]}$$
$$(-1+E^{2B[r]}+r\,(x1-y1)-r^{2}\,(x1^{2}+x2-x1\,y1))\ \mathrm{Sin[thet}$$
$$C^{1}{}_{331} == \frac{1}{6}\,E^{-2B[r]}$$
$$(-1+E^{2B[r]}+r\,(x1-y1)-r^{2}\,(x1^{2}+x2-x1\,y1))\ \mathrm{Sin[thet}$$
$$C^{1}{}_{414} ==$$
$$-\frac{E^{2(A[r]-B[r])}\,(-1+E^{2B[r]}+r\,(x1-y1)-r^{2}\,(x1^{2}+x2-x1\,y}{3\,r^{2}}$$
$$C^{1}{}_{441} ==$$
$$\frac{E^{2(A[r]-B[r])}\,(-1+E^{2B[r]}+r\,(x1-y1)-r^{2}\,(x1^{2}+x2-x1\,y1}{3\,r^{2}}$$
$$C^{2}{}_{112} == \frac{-1+E^{2B[r]}+r\,(x1-y1)-r^{2}\,(x1^{2}+x2-x1\,y1)}{6\,r^{2}},$$
$$C^{2}{}_{121} == \frac{1-E^{2B[r]}+r\,(-x1+y1)+r^{2}\,(x1^{2}+x2-x1\,y1)}{6\,r^{2}},$$
$$C^{2}{}_{323} == \frac{1}{3}\,E^{-2B[r]}$$
$$(-1+E^{2B[r]}+r\,(x1-y1)-r^{2}\,(x1^{2}+x2-x1\,y1))\ \mathrm{Sin[thet}$$
$$C^{2}{}_{332} == -\frac{1}{3}\,E^{-2B[r]}$$
$$(-1+E^{2B[r]}+r\,(x1-y1)-r^{2}\,(x1^{2}+x2-x1\,y1))\ \mathrm{Sin[thet}$$
$$C^{2}{}_{424} ==$$
$$\frac{E^{2(A[r]-B[r])}\,(-1+E^{2B[r]}+r\,(x1-y1)-r^{2}\,(x1^{2}+x2-x1\,y1}{6\,r^{2}}$$
$$C^{2}{}_{442} ==$$
$$-\frac{E^{2(A[r]-B[r])}\,(-1+E^{2B[r]}+r\,(x1-y1)-r^{2}\,(x1^{2}+x2-x1\,y}{6\,r^{2}}$$
$$C^{3}{}_{113} == \frac{-1+E^{2B[r]}+r\,(x1-y1)-r^{2}\,(x1^{2}+x2-x1\,y1)}{6\,r^{2}},$$
$$C^{3}{}_{131} == \frac{1-E^{2B[r]}+r\,(-x1+y1)+r^{2}\,(x1^{2}+x2-x1\,y1)}{6\,r^{2}},\ C^{3}{}_{2}$$
$$-\frac{1}{3}\,E^{-2B[r]}\,(-1+E^{2B[r]}+r\,(x1-y1)-r^{2}\,(x1^{2}+x2-x1\,y1)$$
$$C^{3}{}_{232} == \frac{1}{3}\,E^{-2B[r]}$$
$$(-1+E^{2B[r]}+r\,(x1-y1)-r^{2}\,(x1^{2}+x2-x1\,y1)),\ C^{3}{}_{434} ==$$
$$\frac{E^{2(A[r]-B[r])}\,(-1+E^{2B[r]}+r\,(x1-y1)-r^{2}\,(x1^{2}+x2-x1\,y1}{6\,r^{2}}$$
$$C^{3}{}_{443} ==$$
$$-\frac{E^{2(A[r]-B[r])}\,(-1+E^{2B[r]}+r\,(x1-y1)-r^{2}\,(x1^{2}+x2-x1\,y}{6\,r^{2}}$$
$$C^{4}{}_{114} == \frac{1-E^{2B[r]}+r\,(-x1+y1)+r^{2}\,(x1^{2}+x2-x1\,y1)}{3\,r^{2}},$$
$$C^{4}{}_{141} == \frac{-1+E^{2B[r]}+r\,(x1-y1)-r^{2}\,(x1^{2}+x2-x1\,y1)}{3\,r^{2}},$$
$$C^{4}{}_{224} == \frac{1}{6}\,E^{-2B[r]}$$
$$(-1+E^{2B[r]}+r\,(x1-y1)-r^{2}\,(x1^{2}+x2-x1\,y1)),\ C^{4}{}_{242} ==$$
$$-\frac{1}{6}\,E^{-2B[r]}\,(-1+E^{2B[r]}+r\,(x1-y1)-r^{2}\,(x1^{2}+x2-x1\,y1)$$
$$C^{4}{}_{334} == \frac{1}{6}\,E^{-2B[r]}$$
$$(-1+E^{2B[r]}+r\,(x1-y1)-r^{2}\,(x1^{2}+x2-x1\,y1))\ \mathrm{Sin[thet}$$
$$C^{4}{}_{343} == -\frac{1}{6}\,E^{-2B[r]}$$
$$(-1+E^{2B[r]}+r\,(x1-y1)-r^{2}\,(x1^{2}+x2-x1\,y1))\ \mathrm{Sin[thet}$$

# First gauge gravity field equation $H_{\mu\nu} = 0$

```
em = Table[MakeSum[H[-i, -j]], {i, 4}, {j, 4} ]
    Simplify[% /. {A'[r] -> x1, A''[r] -> x2, A'''[r] -> x3,
      B'[r] -> y1, B''[r] -> y2}]
    Collect[%, r, Factor]
```

$$\left\{\left\{\frac{E^{-2B[r]}\,(-1+E^{B[r]})^2\,(1+E^{B[r]})^2}{r^4} + \frac{2E^{-2B[r]}\,(x1-y1)\,(x1+y1)}{r^2} - E^{-2B[r]}\,(x1^2+x2-x1\,y1)^2,\ 0,\ 0,\ 0\right\},\right.$$

$$\left\{0,\ -\frac{E^{-4B[r]}\,(-1+E^{B[r]})^2\,(1+E^{B[r]})^2}{r^2} + E^{-4B[r]}\,r^2\,(x1^2+x2-x1\,y1)^2,\ 0,\ 0\right\},$$

$$\left\{0,\ 0,\ -\frac{E^{-4B[r]}\,(-1+E^{B[r]})^2\,(1+E^{B[r]})^2\,\mathrm{Sin}[theta]^2}{r^2} + E^{-4B[r]}\,r^2\,(x1^2+x2-x1\,y1)^2\,\mathrm{Sin}[theta]^2,\ 0\right\},$$

$$\left.\left\{0,\ 0,\ 0,\ -\frac{E^{2A[r]-4B[r]}\,(-1+E^{B[r]})^2\,(1+E^{B[r]})^2}{r^4} + \frac{2E^{2A[r]-4B[r]}\,(x1-y1)\,(x1+y1)}{r^2} + E^{2A[r]-4B[r]}\,(x1^2+x2-x1\,y1)^2\right\}\right\}$$

# Second gauge gravity field equation $Y_{\mu\nu\sigma} = 0$

```
yg = Table[Simplify[MakeSum[Y[-i, -j, -k]]], {i, 4},
   {j, 4}, {k, 4}]
Simplify[% /. {A'[r] -> x1, A''[r] -> x2, A'''[r] -> x3,
   B'[r] -> y1, B''[r] -> y2}]
Collect[%, r, Factor]
```

$$\{\{\{0, 0, 0, 0\}, \{0, -\frac{E^{-2B[r]}(-1+E^{B[r]})(1+E^{B[r]})}{r} + E^{-2B[r]}\, r\, (x1^2 + x1\, y1 - 2\, y1^2 + y2), 0, 0\},$$

$$\{0, 0, -\frac{E^{-2B[r]}(-1+E^{B[r]})(1+E^{B[r]})\, \mathrm{Sin}[\mathrm{theta}]^2}{r} + E^{-2B[r]}\, r\, (x1^2 + x1\, y1 - 2\, y1^2 + y2)\, \mathrm{Sin}[\mathrm{theta}]^2, 0\},$$

$$\{0, 0, 0, -\frac{2\, E^{2A[r]-2B[r]}\, x1}{r^2} + \frac{2\, E^{2A[r]-2B[r]}(x1^2 + x2 - x1\, y1)}{r} - E^{2A[r]-2B[r]}(-2\, x1\, x2 - x3 + 2\, x1^2\, y1 + 3\, x2\, y1 - 2\, x1\, y1^2 + x1\, y2)\}\},$$

$$\{\{0, \frac{E^{-2B[r]}(-1+E^{B[r]})(1+E^{B[r]})}{r} - E^{-2B[r]}\, r\, (x1^2 + x1\, y1 - 2\, y1^2 + y2), 0, 0\},$$

$$\{0, 0, 0, 0\}, \{0, 0, 0, 0\}, \{0, 0, 0, 0\}\},$$

$$\{\{0, 0, \frac{E^{-2B[r]}(-1+E^{B[r]})(1+E^{B[r]})\, \mathrm{Sin}[\mathrm{theta}]^2}{r} - E^{-2B[r]}\, r\, (x1^2 + x1\, y1 - 2\, y1^2 + y2)\, \mathrm{Sin}[\mathrm{theta}]^2, 0\},$$

$$\{0, 0, 0, 0\}, \{0, 0, 0, 0\}, \{0, 0, 0, 0\}\},$$

$$\{\{0, 0, 0, \frac{2\, E^{2A[r]-2B[r]}\, x1}{r^2} - \frac{2\, E^{2A[r]-2B[r]}(x1^2 + x2 - x1\, y1)}{r} + E^{2A[r]-2B[r]}(-2\, x1\, x2 - x3 + 2\, x1^2\, y1 + 3\, x2\, y1 - 2\, x1\, y1^2 + x1\, y2)\},$$

$$\{0, 0, 0, 0\}, \{0, 0, 0, 0\}, \{0, 0, 0, 0\}\}\}$$

# Different cases of the $H_{\mu\nu} = 0$ equations

*This is the (+)(+) case (Satisfies $Y_{\mu\nu\sigma} = 0$ equations too)*

```
s1 = 1; s2 = 1;
(((x1^2 + x2 - x1 y1) - s1 (E^(2 B[r]) - 1) / r^2) /.
  {x1 -> s2 y1, x2 -> s2 y2, x3 -> s 2 y3})
% /. {x1 -> A'[r], x2 -> A''[r], x3 -> A'''[r], y1 -> B'[r
  y2 -> B''[r], y3 -> B'''[r]}
```

$$-\frac{-1+E^{2B[r]}}{r^2} + B''[r]$$

```
Factor[% /. {B[r] -> -1/2 Log[K[r]],
   B'[r] -> D[-1/2 Log[K[r]], {r, 1}],
   B''[r] -> D[-1/2 Log[K[r]], {r, 2}]}]
```

$$-\frac{2\, K[r] - 2\, K[r]^2 - r^2\, K'[r]^2 + r^2\, K[r]\, K''[r]}{2\, r^2\, K[r]^2}$$

*Pavelle-Thompson works here (remember - solution goes as $1/f$):*

```
f = (1 + c[1] / r)^2;
Simplify[2 K[r] - 2 K[r]^2 - r^2 K'[r]^2 + r^2 K[r] K''[r] /.
  {K[r] -> f, K'[r] -> D[f, {r, 1}], K''[r] -> D[f, {r, 2
```

0

```
f = (1 + c1 r^2);
Simplify[2 K[r] - 2 K[r]^2 - r^2 K'[r]^2 + r^2 K[r] K''[r] /.
  {K[r] -> f, K'[r] -> D[f, {r, 1}], K''[r] -> D[f, {r, 2
```

```
-4 c1^2 r^4
```

*This is the (+)(-) case:*

```
s1 = 1; s2 = -1;
(((x1^2 + x2 - x1 y1) - s1 (E^(2 B[r]) - 1) / r^2) /.
  {x1 -> s2 y1, x2 -> s2 y2, x3 -> s 2 y3})
% /. {x1 -> A'[r], x2 -> A''[r], x3 -> A'''[r], y1 -> B'[r
  y2 -> B''[r], y3 -> B'''[r]}
```

$$-\frac{-1+E^{2B[r]}}{r^2}+2B'[r]^2-B''[r]$$

```
Factor[% /. {B[r] -> -1/2 Log[K[r]],
   B'[r] -> D[-1/2 Log[K[r]], {r, 1}],
   B''[r] -> D[-1/2 Log[K[r]], {r, 2}]}]
```

$$\frac{-2+2K[r]+r^2K''[r]}{2r^2K[r]}$$

```
DSolve[% == 0, K[r], r]
```

$$\left\{\left\{K[r]\to 1+\sqrt{r}\,C[2]\,\mathrm{Cos}\left[\frac{1}{2}\sqrt{7}\,\mathrm{Log}[r]\right]-\sqrt{r}\,C[1]\,\mathrm{Sin}\left[\frac{1}{2}\sqrt{7}\,\mathrm{Log}[r]\right]\right\}\right\}$$

*This is the (-)(+) case:*

```
s1 = -1; s2 = +1;
(((x1^2 + x2 - x1 y1) - s1 (E^(2 B[r]) - 1) / r^2) /.
  {x1 -> s2 y1, x2 -> s2 y2, x3 -> s 2 y3})
% /. {x1 -> A'[r], x2 -> A''[r], x3 -> A'''[r], y1 -> B'[r
  y2 -> B''[r], y3 -> B'''[r]}
```

$$\frac{-1+E^{2B[r]}}{r^2}+B''[r]$$

```
Factor[% /. {B[r] -> -1/2 Log[K[r]],
   B'[r] -> D[-1/2 Log[K[r]], {r, 1}],
   B''[r] -> D[-1/2 Log[K[r]], {r, 2}]}]
```

$$-\frac{-2K[r]+2K[r]^2-r^2K'[r]^2+r^2K[r]K''[r]}{2r^2K[r]^2}$$

*This is the (-)(-) case (Satisfies $Y_{\mu\nu\sigma}=0$ equations):*

```
s1 = -1; s2 = -1;
(((x1^2 + x2 - x1 y1) - s1 (E^(2 B[r]) - 1) / r^2) /.
  {x1 -> s2 y1, x2 -> s2 y2, x3 -> s 2 y3})
% /. {x1 -> A'[r], x2 -> A''[r], x3 -> A'''[r], y1 -> B'[r
  y2 -> B''[r], y3 -> B'''[r]}
```

$$\frac{-1+E^{2B[r]}}{r^2}+2B'[r]^2-B''[r]$$

```
Factor[% /. {B[r] -> -1/2 Log[K[r]],
   B'[r] -> D[-1/2 Log[K[r]], {r, 1}],
   B''[r] -> D[-1/2 Log[K[r]], {r, 2}]}]
```

$$\frac{2-2K[r]+r^2K''[r]}{2r^2K[r]}$$

```
DSolve[% == 0, K[r], r]
```

$$\left\{\left\{K[r]\to 1+\frac{C[1]}{r}+r^2\,C[2]\right\}\right\}$$

```
f = (1 + c1 r^2);
Simplify[2 - 2 K[r] + r^2 K''[r] /.
  {K[r] -> f, K'[r] -> D[f, {r, 1}], K''[r] -> D[f, {r, 2
0
```

# Different cases of the $Y_{\mu\nu\sigma} = 0$ equations

*This is the (+)(+) case (Satisfies $H_{\mu\nu}$ equations too):*

```
s1 = 1; s2 = 1;
Y1 =
 ((x1^2 + x2 - x1 y1) /. {x1^2 + x2 - x1 y1 -> s1 (E^(2 B[r]) - 1) / r^2})
  (A''[r] - B''[r]) + 2 B'[r] (A'[r] - B'[r]) - (E^(2 B[r]) - 1 / r^2
2 (A'[r] - B'[r]) B'[r] - A''[r] + B''[r]
Factor[
 Simplify[% /. {A'[r] -> s2 B'[r], A''[r] -> s2 B''[r]}]
0
Y2 = D[(x1^2 + x2 - x1 y1) /.
    {x1^2 + x2 - x1 y1 -> s1 (E^(2 B[r]) - 1) / r^2}, r] - 2 (B'[r] - 
  ((x1^2 + x2 - x1 y1) /. {x1^2 + x2 - x1 y1 -> s1 (E^(2 B[r]) - 1) / r^2
 2 A'[r] / r^2
```

$$-\frac{2\,(-1+E^{2 B[r]})}{r^3}-\frac{2\,A'[r]}{r^2}+\frac{2\,E^{2 B[r]}\,B'[r]}{r^2}-\frac{2\,(-1+E^{2 B[r]})\left(-\frac{1}{r}+B'[r]\right)}{r^2}$$

```
Simplify[% /. {A'[r] -> s2 B'[r], A''[r] -> s2 B''[r]}]
0
```

*This is the (+)(-) case (the only solution is B=constant):*

```
s1 = 1; s2 = -1;
Y1 =
 ((x1^2 + x2 - x1 y1) /. {x1^2 + x2 - x1 y1 -> s1 (E^(2 B[r]) - 1) / r^2})
  (A''[r] - B''[r]) + 2 B'[r] (A'[r] - B'[r]) - (E^(2 B[r]) - 1 / r^2
2 (A'[r] - B'[r]) B'[r] - A''[r] + B''[r]
Factor[
 Simplify[% /. {A'[r] -> s2 B'[r], A''[r] -> s2 B''[r]}]
-2 (2 B'[r]^2 - B''[r])
Y2 = D[(x1^2 + x2 - x1 y1) /.
    {x1^2 + x2 - x1 y1 -> s1 (E^(2 B[r]) - 1) / r^2}, r] - 2 (B'[r] - 
  ((x1^2 + x2 - x1 y1) /. {x1^2 + x2 - x1 y1 -> s1 (E^(2 B[r]) - 1) / r^2
 2 A'[r] / r^2
```

$$-\frac{2\,(-1+E^{2 B[r]})}{r^3}-\frac{2\,A'[r]}{r^2}+\frac{2\,E^{2 B[r]}\,B'[r]}{r^2}-\frac{2\,(-1+E^{2 B[r]})\left(-\frac{1}{r}+B'[r]\right)}{r^2}$$

```
Simplify[% /. {A'[r] -> s2 B'[r], A''[r] -> s2 B''[r]}]
```

```
4B'[r]
------
  r^2
DSolve[4 B'[r]/r^2 == 0, B[r], r]
{{B[r] → C[1]}}
```

*This is the (-)(+) case (the only solution is for (k=1) $B = \text{constant}$ )):*

```
s1 = -1; s2 = +1;
Y1 =
 ((x1^2 + x2 - x1 y1) /. {x1^2 + x2 - x1 y1 -> s1 (E^(2 B[r]) - 1)/r^2})
 (A''[r] - B''[r]) + 2 B'[r] (A'[r] - B'[r]) - (E^(2 B[r]) - 1)/r^2
Factor[
 Simplify[% /. {A'[r] -> s2 B'[r], A''[r] -> s2 B''[r]}]
-2 (-1 + E^B[r]) (1 + E^B[r]) / r^2
Factor[2 (-1 + E^B[r]) (1 + E^B[r]) / r^2 /. {B[r] -> -1/2 Log[K[r]
  B'[r] -> D[-1/2 Log[K[r]], {r, 1}],
  B''[r] -> D[-1/2 Log[K[r]], {r, 2}]}]
% /. {K[r] -> k}
-2 (-1 + Sqrt[k]) (1 + Sqrt[k]) / (k r^2)
Simplify[Solve[% == 0, k]]
{{k → 1}}
Factor[
 Simplify[% /. {A'[r] -> s2 B'[r], A''[r] -> s2 B''[r]}]
-2 (-1 + E^B[r]) (1 + E^B[r]) / r^2
Y2 = D[(x1^2 + x2 - x1 y1) /.
   {x1^2 + x2 - x1 y1 -> s1 (E^(2 B[r]) - 1)/r^2}, r] - 2 (B'[r] - 
   ((x1^2 + x2 - x1 y1) /. {x1^2 + x2 - x1 y1 -> s1 (E^(2 B[r]) - 1)/r^2
 2 A'[r]/r^2
2 (-1 + E^(2 B[r]))/r^3 - 2 A'[r]/r^2 - 2 E^(2 B[r]) B'[r]/r^2 +
 2 (-1 + E^(2 B[r])) (-1/r + B'[r]) / r^2
Simplify[% /. {A'[r] -> s2 B'[r], A''[r] -> s2 B''[r]}]
-4B'[r]/r^2
```

*This is the (-)(-) case (Satisfies $H_{\mu\nu}$ equations too):*

```
s1 = -1; s2 = -1;
Y1 =
 ((x1^2 + x2 - x1 y1) /. {x1^2 + x2 - x1 y1 -> s1 (E^(2 B[r]) - 1)/r^2})
 (A''[r] - B''[r]) + 2 B'[r] (A'[r] - B'[r]) - (E^(2 B[r]) - 1)/r^2
Factor[
 Simplify[% /. {A'[r] -> s2 B'[r], A''[r] -> s2 B''[r]}]
-2 (-1 + E^(2 B[r]) + 2 r^2 B'[r]^2 - r^2 B''[r]) / r^2
```

```
Factor[% /. {B[r] -> -1/2 Log[K[r]],
   B'[r] -> D[-1/2 Log[K[r]], {r, 1}],
   B''[r] -> D[-1/2 Log[K[r]], {r, 2}]}]
-(2 - 2 K[r] + r^2 K''[r])/(r^2 K[r])
DSolve[% == 0, K[r], r]
{{K[r] -> 1 + C[1]/r + r^2 C[2]}}
Y2 = D[(U) /. {U -> s1 (E^(2 B[r]) - 1)/r^2}, r] -
  2 (B'[r] - 1/r) ((U) /. {U -> s1 (E^(2 B[r]) - 1)/r^2}) - 2 A'[r]/r^2
2 (-1 + E^(2 B[r]))/r^3 - 2 A'[r]/r^2 - 2 E^(2 B[r]) B'[r]/r^2 +
  2 (-1 + E^(2 B[r])) (-1/r + B'[r])/r^2
Simplify[% /. {A'[r] -> s2 B'[r], A''[r] -> s2 B''[r]}]
0
```

*This is the case that A'' + $A'^2$ - A' B' = 0 (with A' = constant) - This also gives Ni's solution:*

```
Y2 = ((x1^2 + x2 - x1 y1) /. {x1^2 + x2 - x1 y1 -> 0}) -
  (A''[r] - B''[r]) + 2 B'[r] (A'[r] - B'[r]) - (E^(2 B[r]) - 1)/r^2
Factor[%]
-(-1 + E^(2 B[r]) - 2 r^2 A'[r] B'[r] + 2 r^2 B'[r]^2 + r^2 A''[r] - r^2 B''[
      r^2
Collect[
 -1 + E^(2 B[r]) - 2 r^2 A'[r] B'[r] + 2 r^2 B'[r]^2 + r^2 A''[r] - r^2 B''|
  r, Factor]
(-1 + E^B[r]) (1 + E^B[r]) +
 r^2 (-2 A'[r] B'[r] + 2 B'[r]^2 + A''[r] - B''[r])
Y1 = D[(U) /. {U -> 0}, r] - 2 (B'[r] - 1/r) ((U) /. {U ->
  2 A'[r]/r^2
-2 A'[r]/r^2
Simplify[%% /. {A'[r] -> 0, A''[r] -> 0}]
-(-1 + E^(2 B[r]))/r^2 - 2 B'[r]^2 + B''[r]
Factor[%]
-(-1 + E^(2 B[r]) + 2 r^2 B'[r]^2 - r^2 B''[r])/r^2
Collect[-1 + E^(2 B[r]) + 2 r^2 B'[r]^2 - r^2 B''[r], r, Factor]
(-1 + E^B[r]) (1 + E^B[r]) + r^2 (2 B'[r]^2 - B''[r])
Factor[% /. {B[r] -> -1/2 Log[K[r]],
   B'[r] -> D[-1/2 Log[K[r]], {r, 1}],
   B''[r] -> D[-1/2 Log[K[r]], {r, 2}]}]
-(2 - 2 K[r] + r^2 K''[r])/(2 r^2 K[r])
DSolve[% == 0, K[r], r]
{{K[r] -> 1 + C[1]/r + r^2 C[2]}}
Simplify[B'[r]^2 + 2 (E^(2 B[r]) - 1)^2/r^2]
```

$$\frac{2\,(-1+E^{2\,B[r]})^2}{r^2}+B'[r]^2$$

```
Factor[% /. {B[r] -> -1/2 Log[K[r]],
  B'[r] -> D[-1/2 Log[K[r]], {r, 1}],
  B''[r] -> D[-1/2 Log[K[r]], {r, 2}]}]
```

$$\frac{8-16\,K[r]+8\,K[r]^2+r^2\,K'[r]^2}{4\,r^2\,K[r]^2}$$

second gauge gravity field equation with A=0:

```
Table[Simplify[MakeSum[Y[-i, -j, -k]]], {i, 4},
 {j, 4}, {k, 4}]
Factor[
 % /. {A[r] -> 0, A'[r] -> 0, A''[r] -> 0, A'''[r] -> 0}]
```

$$\Big\{\Big\{\{0,0,0,0\},\Big\{0,-\frac{E^{-2\,B[r]}\,(-1+E^{2\,B[r]}+2\,r^2\,B'[r]^2-r^2\,B''[r])}{r},0,0\Big\},\Big\{0,-\frac{E^{-2\,B[r]}\,\text{Sin}[\text{theta}]^2\,(-1+E^{2\,B[r]}+2\,r^2\,B'[r]^2-r^2\,B''[r])}{r},0\Big\},\{0,0,0,0\}\Big\},$$

$$\Big\{\Big\{0,\frac{E^{-2\,B[r]}\,(-1+E^{2\,B[r]}+2\,r^2\,B'[r]^2-r^2\,B''[r])}{r},0,0\Big\},\{0,0,0,0\},\{0,0,0,0\},\{0,0,0,0\}\Big\},\Big\{\Big\{0,0,\frac{E^{-2\,B[r]}\,\text{Sin}[\text{theta}]^2\,(-1+E^{2\,B[r]}+2\,r^2\,B'[r]^2-r^2\,B''[r])}{r},0\Big\},$$

$$\{0,0,0,0\},\{0,0,0,0\},\{0,0,0,0\}\Big\},\{\{0,0,0,0\},\{0,0,0,0\},\{0,0,0,0\},\{0,0,0,0\}\}\Big\}$$

```
Factor[
 -1 + E^(2 B[r]) + 2 r^2 B'[r]^2 - r^2 B''[r] /. {B[r] -> -1/2 Log[K[r
  B'[r] -> D[-1/2 Log[K[r]], {r, 1}],
  B''[r] -> D[-1/2 Log[K[r]], {r, 2}]}]
```

$$-\frac{-2+2\,K[r]-r^2\,K''[r]}{2\,K[r]}$$

*This case is Ni's Solution:*

```
DSolve[% == 0, K[r], r]
```

$$\Big\{\Big\{K[r]\to 1+\frac{C[1]}{r}+r^2\,C[2]\Big\}\Big\}$$

*A Pavelle - Thompson Extraneous Solution - This is a special case of Ni's solution with* $c_2$ *= 0:*

```
f = (1 - c1/r);
Simplify[-(-2 + 2 K[r] - r^2 K''[r])/(2 K[r]) /.
  {K[r] -> f, K'[r] -> D[f, {r, 1}], K''[r] -> D[f, {r, 2
```

0

# Auxiliary Condition

```
Table[Factor[MakeSum[ac[-i, -j]]], {i, 4}, {j, 4}]
    Simplify[% /. {A'[r] -> x1, A''[r] -> x2, A'''[r] -> x3,
        B'[r] -> y1, B''[r] -> y2}]
    {{1/r^3 (2 E^(-2 B[r]) ((-1 + E^(2 B[r])) y1 -
          r y1 (x1 + y1) + r^2 (x1 + 2 y1) (x1^2 + x2 - x1 y1))),
      0, 0, 0},
     {0, 1/r (E^(-4 B[r])
        (r^2 x1^3 + y1 (1 - E^(2 B[r]) - r^2 x2 + r y1) + x1^2 (r - 2 r^2 y1) +
          x1 (-1 + E^(2 B[r]) + 2 r y1 + r^2 (x2 + y1^2)))),
      0, 0},
     {0, 0, 1/r (E^(-4 B[r]) (r^2 x1^3 + y1 (1 - E^(2 B[r]) - r^2 x2 + r y1) +
          x1^2 (r - 2 r^2 y1) + x1 (-1 + E^(2 B[r]) + 2 r y1 + r^2 (x2 + y1^2
        Sin[theta]^2), 0},
     {0, 0, 0, 1/r^3
      (2 E^(2 (A[r] - 2 B[r])) (2 r^2 x1^3 + r^2 x2 y1 + x1^2 (r - r^2 y1) +
          x1 (-1 + E^(2 B[r]) + r y1 + r^2 (2 x2 - y1^2))))}}
```

*00 Case:*

```
tt = Collect[(2 r^2 x1^3 + r^2 x2 y1 + x1^2 (r - r^2 y1) +
        x1 (-1 + E^(2 B[r]) + r y1 + r^2 (2 x2 - y1^2))), r, Factor] /.
    {x1 -> A'[r], x2 -> A''[r], x3 -> A'''[r], y1 -> B'[r],
     y2 -> B''[r], y3 -> B'''[r]}
(-1 + E^B[r]) (1 + E^B[r]) A'[r] + r A'[r] (A'[r] + B'[r]) +
 r^2 (2 A'[r] + B'[r]) (A'[r]^2 - A'[r] B'[r] + A''[r])
```

*11 Case:*

```
rr = Collect[(-1 + E^(2 B[r])) y1 - r y1 (x1 + y1) +
      r^2 (x1 + 2 y1) (x1^2 + x2 - x1 y1), r, Factor] /.
    {x1 -> A'[r], x2 -> A''[r], x3 -> A'''[r], y1 -> B'[r],
     y2 -> B''[r], y3 -> B'''[r]}
(-1 + E^B[r]) (1 + E^B[r]) B'[r] - r B'[r] (A'[r] + B'[r]) -
 r^2 (A'[r] + 2 B'[r]) (-A'[r]^2 + A'[r] B'[r] - A''[r])
```

*22 Case:*

```
φφ =
 Collect[r^2 x1^3 + y1 (1 - E^(2 B[r]) - r^2 x2 + r y1) + x1^2 (r - 2 r^2
      x1 (-1 + E^(2 B[r]) + 2 r y1 + r^2 (x2 + y1^2)), r, Factor] /.
    {x1 -> A'[r], x2 -> A''[r], x3 -> A'''[r], y1 -> B'[r],
     y2 -> B''[r], y3 -> B'''[r]}
(-1 + E^B[r]) (1 + E^B[r]) (A'[r] - B'[r]) + r (A'[r] + B'[r])^2 +
 r^2 (A'[r] - B'[r]) (A'[r]^2 - A'[r] B'[r] + A''[r])
```

*Linear Combinations:*

```
Simplify[Expand[tt / A'[r] - rr / B'[r]]]
-1/(A'[r] B'[r]) (r (A'[r] + B'[r]) (r A'[r]^3 - 2 r A'[r]^2 B'[r] -
      r B'[r] A''[r] + A'[r] (-2 B'[r] + r B'[r]^2 + r A''[r])))
```

```
Collect[(r A'[r]^3 - 2 r A'[r]^2 B'[r] -
    r B'[r] A''[r] + A'[r] (-2 B'[r] + r B'[r]^2 + r A''[r])),
  r, Factor]
-2 A'[r] B'[r] + r (A'[r] - B'[r]) (A'[r]^2 - A'[r] B'[r] + A''[r
```

*This is the (-) case for 00:*

```
s1 = 1; s2 = -1;
(-1 + E^B[r]) (1 + E^B[r]) x1 +
   r x1 (x1 + y1) + r^2 (2 x1 + y1) (x1^2 + x2 - x1 y1) /.
  {x1 -> s2 y1, x2 -> s2 y2, x3 -> s 2 y3}
% /. {x1 -> A'[r], x2 -> A''[r], x3 -> A'''[r], y1 -> B'[r
   y2 -> B''[r], y3 -> B'''[r]}
Factor[% /. {B[r] -> -1/2 Log[K[r]],
    B'[r] -> D[-1/2 Log[K[r]], {r, 1}],
    B''[r] -> D[-1/2 Log[K[r]], {r, 2}]}]
```

$$-\frac{K'[r]\ (-2+2K[r]-r^2 K''[r])}{4K[r]^2}$$

```
DSolve[K'[r] (-2 + 2 K[r] - r^2 K''[r]) / (4 K[r]^2) == 0, K[r], r]
```

$$\left\{\left\{K[r]\to 1+\frac{C[1]}{r}+r^2\,C[2]\right\}\right\}$$

*This is the (-) case for 11:*

```
s1 = 1; s2 = -1;
(-1 + E^(2 B[r])) y1 -
   r y1 (x1 + y1) + r^2 (x1 + 2 y1) (x1^2 + x2 - x1 y1) /.
  {x1 -> s2 y1, x2 -> s2 y2, x3 -> s 2 y3}
% /. {x1 -> A'[r], x2 -> A''[r], x3 -> A'''[r], y1 -> B'[r
   y2 -> B''[r], y3 -> B'''[r]}
Factor[% /. {B[r] -> -1/2 Log[K[r]],
    B'[r] -> D[-1/2 Log[K[r]], {r, 1}],
    B''[r] -> D[-1/2 Log[K[r]], {r, 2}]}]
```

$$\frac{K'[r]\ (-2+2K[r]-r^2 K''[r])}{4K[r]^2}$$

*This is the (-) case for 22:*

```
s1 = 1; s2 = -1;
(-1 + E^B[r]) (1 + E^B[r]) (x1 - y1) +
   r (x1 + y1)^2 + r^2 (x1 - y1) (x1^2 + x2 - x1 y1) /.
  {x1 -> s2 y1, x2 -> s2 y2, x3 -> s 2 y3}
% /. {x1 -> A'[r], x2 -> A''[r], x3 -> A'''[r], y1 -> B'[r
   y2 -> B''[r], y3 -> B'''[r]}
Factor[% /. {B[r] -> -1/2 Log[K[r]],
    B'[r] -> D[-1/2 Log[K[r]], {r, 1}],
    B''[r] -> D[-1/2 Log[K[r]], {r, 2}]}]
```

$$-\frac{K'[r]\ (-2+2K[r]-r^2 K''[r])}{2K[r]^2}$$

-----------------------------------------------------

*This is the (+) case for 00:*

```
s1 = 1; s2 = 1;
(-1 + E^B[r]) (1 + E^B[r]) x1 +
   r x1 (x1 + y1) + r^2 (2 x1 + y1) (x1^2 + x2 - x1 y1) /.
  {x1 -> s2 y1, x2 -> s2 y2, x3 -> s 2 y3}
% /. {x1 -> A'[r], x2 -> A''[r], x3 -> A'''[r], y1 -> B'[r
   y2 -> B''[r], y3 -> B'''[r]}
Factor[% /. {B[r] -> (-1/2) Log[K[r]],
    B'[r] -> D[(-1/2) Log[K[r]], {r, 1}],
    B''[r] -> D[(-1/2) Log[K[r]], {r, 2}]}]
```

```
1/(4 K[r]^3) (K'[r] (-2 K[r] + 2 K[r]^2 + 2 r K[r] K'[r] -
     3 r^2 K'[r]^2 + 3 r^2 K[r] K''[r]))
```

*This is the (+) case for 11:*

```
s1 = 1; s2 = 1;
(-1 + E^(2 B[r])) y1 -
   r y1 (x1 + y1) + r^2 (x1 + 2 y1) (x1^2 + x2 - x1 y1) /.
  {x1 -> s2 y1, x2 -> s2 y2, x3 -> s 2 y3}
% /. {x1 -> A'[r], x2 -> A''[r], x3 -> A'''[r], y1 -> B'[r
   y2 -> B''[r], y3 -> B'''[r]}
Factor[% /. {B[r] -> (-1/2) Log[K[r]],
    B'[r] -> D[(-1/2) Log[K[r]], {r, 1}],
    B''[r] -> D[(-1/2) Log[K[r]], {r, 2}]}]
```

```
1/(4 K[r]^3) (K'[r] (-2 K[r] + 2 K[r]^2 - 2 r K[r] K'[r] -
     3 r^2 K'[r]^2 + 3 r^2 K[r] K''[r]))
```

*This is the (+) case for 22:*

```
s1 = 1; s2 = 1;
(-1 + E^B[r]) (1 + E^B[r]) (x1 - y1) +
   r (x1 + y1)^2 + r^2 (x1 - y1) (x1^2 + x2 - x1 y1) /.
  {x1 -> s2 y1, x2 -> s2 y2, x3 -> s 2 y3}
% /. {x1 -> A'[r], x2 -> A''[r], x3 -> A'''[r], y1 -> B'[r
   y2 -> B''[r], y3 -> B'''[r]}
Factor[% /. {B[r] -> (-1/2) Log[K[r]],
    B'[r] -> D[(-1/2) Log[K[r]], {r, 1}],
    B''[r] -> D[(-1/2) Log[K[r]], {r, 2}]}]
```

```
r K'[r]^2 / K[r]^2
```

```
DSolve[r K'[r]^2 / K[r]^2 == 0, K[r], r]
```

```
{{K[r] → C[1]}}
```

```
(-2 K[r] + 2 K[r]^2 -
    2 r K[r] K'[r] - 3 r^2 K'[r]^2 + 3 r^2 K[r] K''[r]) /.
  {K[r] -> c1, K'[r] -> D[c1, {r, 1}],
   K''[r] -> D[c1, {r, 2}]}
```

```
-2 c1 + 2 c1^2
```

```
Solve[% == 0, c1]
```

```
{{c1 → 0}, {c1 → 1}}
-----------------------------------------------------
```

*This is the A - 0 case for 00:*

```
s1 = 1; s2 = 0;
(-1 + E^B[r]) (1 + E^B[r]) x1 +
  r x1 (x1 + y1) + r^2 (2 x1 + y1) (x1^2 + x2 - x1 y1) /.
 {x1 -> s2 y1, x2 -> s2 y2, x3 -> s 2 y3}
0
```

*This is the A - 0 case for 11:*

```
s1 = 1; s2 = 0;
(-1 + E^(2 B[r])) y1 -
  r y1 (x1 + y1) + r^2 (x1 + 2 y1) (x1^2 + x2 - x1 y1) /.
 {x1 -> s2 y1, x2 -> s2 y2, x3 -> s 2 y3}
% /. {x1 -> A'[r], x2 -> A''[r], x3 -> A'''[r], y1 -> B'[r
  y2 -> B''[r], y3 -> B'''[r]}
Factor[% /. {B[r] -> -1/2 Log[K[r]],
   B'[r] -> D[-1/2 Log[K[r]], {r, 1}],
   B''[r] -> D[-1/2 Log[K[r]], {r, 2}]}]
```

$$\frac{K'[r]\ (-2 + 2\,K[r] - r\,K'[r])}{4\,K[r]^2}$$

*This is the A - 0 case for 22:*

```
s1 = 1; s2 = 0;
(-1 + E^B[r]) (1 + E^B[r]) (x1 - y1) +
  r (x1 + y1)^2 + r^2 (x1 - y1) (x1^2 + x2 - x1 y1) /.
 {x1 -> s2 y1, x2 -> s2 y2, x3 -> s 2 y3}
% /. {x1 -> A'[r], x2 -> A''[r], x3 -> A'''[r], y1 -> B'[r
  y2 -> B''[r], y3 -> B'''[r]}
Factor[% /. {B[r] -> -1/2 Log[K[r]],
   B'[r] -> D[-1/2 Log[K[r]], {r, 1}],
   B''[r] -> D[-1/2 Log[K[r]], {r, 2}]}]
```

$$\frac{K'[r]\ (-2 + 2\,K[r] - r\,K'[r])}{4\,K[r]^2}$$

```
Factor[% /. {B[r] -> -1/2 Log[K[r]],
   B'[r] -> D[-1/2 Log[K[r]], {r, 1}],
   B''[r] -> D[-1/2 Log[K[r]], {r, 2}]}]
```

*Each term has this differential equation as a factor - which has the Einstein Universe as a solution:*

```
DSolve[K'[r] (-2 + 2 K[r] - r K'[r]) / (4 K[r]^2) == 0, K[r], r]
{{K[r] → C[1]}, {K[r] → 1 + r^2 C[1]}}
-----------------------------------------------------
```

*This is the B - 0 case for 00:*

```
s1 = 1; s2 = 0;
(-1 + E^B[r]) (1 + E^B[r]) x1 +
  r x1 (x1 + y1) + r^2 (2 x1 + y1) (x1^2 + x2 - x1 y1) /.
 {y1 -> s2, y2 -> s2}
% /. {x1 -> A'[r], x2 -> A''[r], x3 -> A'''[r], y1 -> B'[r
  y2 -> B''[r], y3 -> B'''[r]}
Factor[% /. {A[r] -> (-1/2) Log[K[r]],
   A'[r] -> D[(-1/2) Log[K[r]], {r, 1}],
   A''[r] -> D[(-1/2) Log[K[r]], {r, 2}]}]
```

$$-\frac{1}{4K[r]^3}(K'[r]\ (-2K[r]^2+2E^{2B[r]}K[r]^2-rK[r]K'[r]+3r^2K'[r]^2-2r^2K[r]K''[r]))$$

*This is the B - 0 case for 11:*

```
s1 = 1; s2 = 0;
(-1 + E^(2 B[r])) y1 -
  r y1 (x1 + y1) + r^2 (x1 + 2 y1) (x1^2 + x2 - x1 y1) /.
 {y1 -> s2, y2 -> s2}
% /. {x1 -> A'[r], x2 -> A''[r], x3 -> A'''[r], y1 -> B'[r
  y2 -> B''[r], y3 -> B'''[r]}
Factor[%11 /. {A[r] -> (-1/2) Log[K[r]],
   A'[r] -> D[(-1/2) Log[K[r]], {r, 1}],
   A''[r] -> D[(-1/2) Log[K[r]], {r, 2}]}]
```

$$\frac{r^2K'[r]\ (-3K'[r]^2+2K[r]K''[r])}{8K[r]^3}$$

*This is the B - 0 case for 22:*

```
s1 = 1; s2 = 0;
(-1 + E^B[r]) (1 + E^B[r]) (x1 - y1) +
  r (x1 + y1)^2 + r^2 (x1 - y1) (x1^2 + x2 - x1 y1) /.
 {y1 -> s2, y2 -> s2}
% /. {x1 -> A'[r], x2 -> A''[r], x3 -> A'''[r], y1 -> B'[r
  y2 -> B''[r], y3 -> B'''[r]}
Factor[% /. {A[r] -> (-1/2) Log[K[r]],
   A'[r] -> D[(-1/2) Log[K[r]], {r, 1}],
   A''[r] -> D[(-1/2) Log[K[r]], {r, 2}]}]
```

$$-\frac{1}{8K[r]^3}(K'[r]\ (-4K[r]^2+4E^{2B[r]}K[r]^2-2rK[r]K'[r]+3r^2K'[r]^2-2r^2K[r]K''[r]))$$

*Solution for the 22 case:*

```
DSolve[(r^2 K'[r] (-3 K'[r]^2 + 2 K[r] K''[r]))/(8 K[r]^3) == 0, K[r], r]
```

$$\left\{\left\{K[r]\to\frac{4}{C[1]^2\ (r^2-2rC[2]+C[2]^2)}\right\}\right\}$$

*But does not satisfy the other two equations:*

```
f = 4/(C[1]^2 (r^2 - 2 r C[2] + C[2]^2));
```

```
(-4 K[r]^2 + 4 E^(2 B[r]) K[r]^2 -
   2 r K[r] K'[r] + 3 r^2 K'[r]^2 - 2 r^2 K[r] K''[r]) /.
 {K[r] -> f, K'[r] -> D[f, r], K''[r] -> D[f, {r, 2}]}
Simplify[%]
```

$$\frac{64\,(E^{2B[r]}\,r + C[2] - E^{2B[r]}\,C[2])}{C[1]^4\,(r - C[2])^5}$$

Additional Vanishing Ricci Scalar Analysis :

*Vanishing Ricci Scalar Differential equations (A = B case (+)(+)):*

```
((-1 + E^(2 B[r]) - 2 r (x1 - y1) - r^2 (x1^2 + x2 - x1 y1)) /.
  {x1 -> A'[r], x2 -> A''[r], x3 -> A'''[r], y1 -> B'[r],
   y2 -> B''[r], y3 -> B'''[r]})
A[r] = B[r];
Factor[
 % /. {A'[r] -> D[A[r], {r, 1}], A''[r] -> D[A[r], {r, 2}
Factor[% /. {B[r] -> -1/2 Log[K[r]],
   B'[r] -> D[-1/2 Log[K[r]], {r, 1}],
   B''[r] -> D[-1/2 Log[K[r]], {r, 2}]}]
```

$$-\frac{-2K[r] + 2K[r]^2 + r^2 K'[r]^2 - r^2 K[r]\,K''[r]}{2K[r]^2}$$

*Pavelle-Thompson works here (Schwarzschild does not):*

```
f = (1 + k / r )^2;
Simplify[-2 K[r] + 2 K[r]^2 + r^2 K'[r]^2 - r^2 K[r] K''[r] /.
  {K[r] -> f, K'[r] -> D[f, {r, 1}], K''[r] -> D[f, {r, 2
```

$0$

*Vanishing Ricci Scalar Differential equations (A=const. case):*

```
((-1 + E^(2 B[r]) - 2 r (x1 - y1) - r^2 (x1^2 + x2 - x1 y1)) /.
  {x1 -> A'[r], x2 -> A''[r], x3 -> A'''[r], y1 -> B'[r],
   y2 -> B''[r], y3 -> B'''[r]})
g = c[1];
Factor[%% /. {A'[r] -> D[g, {r, 1}], A''[r] -> D[g, {r, 2
Factor[% /. {B[r] -> -1/2 Log[K[r]],
   B'[r] -> D[-1/2 Log[K[r]], {r, 1}],
   B''[r] -> D[-1/2 Log[K[r]], {r, 2}]}]
```

$$-\frac{-1 + K[r] + r\,K'[r]}{K[r]}$$

*Thompson Solution works:*

```
Simplify[DSolve[-(-1 + K[r] + r K'[r])/K[r] == 0, K[r], r]]
```

$$\left\{\left\{K[r] \to \frac{r + C[1]}{r}\right\}\right\}$$

*Vanishing Ricci Scalar Differential equations (A' = B' $+\frac{(-1 + E^{2B[r]})}{2r}$ case):*

```
((-1 + E^(2 B[r]) - 2 r (x1 - y1) - r^2 (x1^2 + x2 - x1 y1)) /.
  {x1 -> A'[r], x2 -> A''[r], x3 -> A'''[r], y1 -> B'[r],
   y2 -> B''[r], y3 -> B'''[r]})
g = B'[r] + (-1 + E^(2 B[r]))/(2 r);
```

```
Factor[%% /. {A'[r] -> g, A''[r] -> D[g, {r, 1}]}]
Collect[%270, r, Factor]
Factor[% /. {B[r] -> -1/2 Log[K[r]],
   B'[r] -> D[-1/2 Log[K[r]], {r, 1}],
   B''[r] -> D[-1/2 Log[K[r]], {r, 2}]}]
Simplify[1 - 4 K[r] + 3 K[r]^2 - 3 r K'[r] + r K[r] K'[r] +
  2 r^2 K'[r]^2 - 2 r^2 K[r] K''[r]]
1 + 3 K[r]^2 - 3 r K'[r] + 2 r^2 K'[r]^2 +
 K[r] (-4 + r K'[r] - 2 r^2 K''[r])
```

*Vanishing Ricci Scalar Differential equations (A = - B case):*

```
((-1 + E^(2 B[r]) - 2 r (x1 - y1) - r^2 (x1^2 + x2 - x1 y1)) /.
  {x1 -> A'[r], x2 -> A''[r], x3 -> A'''[r], y1 -> B'[r],
   y2 -> B''[r], y3 -> B'''[r]})
g = -B[r];
Factor[% /. {A'[r] -> D[g, {r, 1}], A''[r] -> D[g, {r, 2]
Collect[%, r]
Factor[% /. {B[r] -> -1/2 Log[K[r]],
   B'[r] -> D[-1/2 Log[K[r]], {r, 1}],
   B''[r] -> D[-1/2 Log[K[r]], {r, 2}]}]
-(-2 + 2 K[r] + 4 r K'[r] + r^2 K''[r])/(2 K[r])
DSolve[-2 + 2 K[r] + 4 r K'[r] + r^2 K''[r] == 0, K[r], r]
f = 1 + k / r;
Simplify[-2 + 2 K[r] + 4 r K'[r] + r^2 K''[r] /.
  {K[r] -> f, K'[r] -> D[f, {r, 1}], K''[r] -> D[f, {r, 2
0
f = 1 + k / r^2;
Simplify[-2 + 2 K[r] + 4 r K'[r] + r^2 K''[r] /.
  {K[r] -> f, K'[r] -> D[f, {r, 1}], K''[r] -> D[f, {r, 2
0
```

Additional Vanishing Weyl Tensor Calculations :

*Vanishing Weyl Tensor Differential equations (A' = B' $-\frac{(-1+E^{2B[r]})}{r}$ case):*

```
(((-1 + E^(2 B[r]) + r (x1 - y1) - r^2 (x1^2 + x2 - x1 y1))) /.
  {x1 -> A'[r], x2 -> A''[r], x3 -> A'''[r], y1 -> B'[r],
   y2 -> B''[r], y3 -> B'''[r]})
g = B'[r] - (-1 + E^(2 B[r]))/r;
Factor[%% /. {A'[r] -> g, A''[r] -> D[g, {r, 1}]}]
Collect[%, r, Factor]
-E^(2 B[r]) (-1 + E^B[r]) (1 + E^B[r]) + (-1 + 3 E^(2 B[r])) r B'[r] - r^2 B''
Factor[% /. {B[r] -> -1/2 Log[K[r]],
   B'[r] -> D[-1/2 Log[K[r]], {r, 1}],
   B''[r] -> D[-1/2 Log[K[r]], {r, 2}]}]
(-2 + 2 K[r] - 3 r K'[r] + r K[r] K'[r] - r^2 K'[r]^2 + r^2 K[r] K''[
 )/(2 K[r]^2)
```

```
f = (1 + k3 r);
Simplify[-2 + 2 K[r] -
   3 r K'[r] + r K[r] K'[r] - r^2 K'[r]^2 + r^2 K[r] K''[r] /.
  {K[r] -> f, K'[r] -> D[f, {r, 1}], K''[r] -> D[f, {r, 2
0
```

*Vanishing Weyl Tensor Differential equations (A=const. case):*

```
g = B'[r] - (-1 + E^(2 B[r]))/r ;
Factor[% /. {A'[r] -> g, A''[r] -> D[g, {r, 1}]}]
Factor[% /. {B[r] -> -1/2 Log[K[r]],
   B'[r] -> D[-1/2 Log[K[r]], {r, 1}],
   B''[r] -> D[-1/2 Log[K[r]], {r, 2}]}]
-(2 - 2 K[r] + r K'[r])/(2 r K[r])
DSolve[-2 + 2 K[r] - r K'[r] == 0, K[r], r]
{{K[r] → 1 + r^2 C[1]}}
```

*Vanishing Weyl Tensor Differential equations (A = - B case):*

```
((-1 + E^(2 B[r]) + r (x1 - y1) - r^2 (x1^2 + x2 - x1 y1)) /.
  {x1 -> A'[r], x2 -> A''[r], x3 -> A'''[r], y1 -> B'[r],
   y2 -> B''[r], y3 -> B'''[r]})
g = -B[r];
Factor[%% /. {A'[r] -> D[g, {r, 1}], A''[r] -> D[g, {r, 2
Factor[% /. {B[r] -> -1/2 Log[K[r]],
   B'[r] -> D[-1/2 Log[K[r]], {r, 1}],
   B''[r] -> D[-1/2 Log[K[r]], {r, 2}]}]
-(-2 + 2 K[r] - 2 r K'[r] + r^2 K''[r])/(2 K[r])
f = 1 + k3 r + k4 r^2;
Simplify[-2 + 2 K[r] - 2 r K'[r] + r^2 K''[r] /.
  {K[r] -> f, K'[r] -> D[f, {r, 1}], K''[r] -> D[f, {r, 2
0
```

*Vanishing Weyl Tensor Differential equations (A = B case) (this is also the (+)(+) case - satisfies $Y_{\sigma\mu\nu}$ = 0 equations, and Vanishing Ricci Scalar and Weyl Tensor too)):*

```
((-1 + E^(2 B[r]) + r (x1 - y1) - r^2 (x1^2 + x2 - x1 y1)) /.
  {x1 -> A'[r], x2 -> A''[r], x3 -> A'''[r], y1 -> B'[r],
   y2 -> B''[r], y3 -> B'''[r]})
g = B[r];
Factor[%% /. {A'[r] -> D[g, {r, 1}], A''[r] -> D[g, {r, 2
Factor[% /. {B[r] -> -1/2 Log[K[r]],
   B'[r] -> D[-1/2 Log[K[r]], {r, 1}],
   B''[r] -> D[-1/2 Log[K[r]], {r, 2}]}]
-(-2 K[r] + 2 K[r]^2 + r^2 K'[r]^2 - r^2 K[r] K''[r])/(2 K[r]^2)
```

*Pavelle-Thompson works here (reminder - solution goes as 1 / f ):*

```
f = (1 + k1 / r)^2;
Simplify[-2 K[r] + 2 K[r]^2 + r^2 K'[r]^2 - r^2 K[r] K''[r] /.
  {K[r] -> f, K'[r] -> D[f, {r, 1}], K''[r] -> D[f, {r, 2
0
```

# Gauss-Bonnet Variation Results

*A separate MathTensor Session:*

```
<< Mathtens.m
Dimension = 4;
SetDirectory["C:\Mathematica\Files\Dissertation\
   Tensor Calculations"];
```

# Euler-Gauss-Bonnet Linear combination

```
gb = -1/4 Epsilon[ua, ub, uc, ud] Epsilon[li, lj, lk, ll]
  RiemannR[ui, uj, la, lb] RiemannR[uk, ul, lc, ld]
```

$$-\frac{1}{4}\,(\mathrm{Epsilon}_{ijkl})\,(\mathrm{Epsilon}^{abcd})\,(R_{ab}{}^{ij})\,(R_{cd}{}^{kl})$$

```
ExpandAll[% /. EpsilonProductTensorRule]
```

$$R^2 - (R_i{}^k)\,(R_k{}^i) - (R_j{}^k)\,(R_k{}^j) - (R_i{}^l)\,(R_l{}^i) - (R_j{}^l)\,(R_l{}^j) + (R_{ij}{}^{kl})\,(R_{kl}{}^{ij})$$

```
Tsimplify[%]
```

$$R^2 - 4\,(R_i{}^k)\,(R_k{}^i) + (R_{ij}{}^{kl})\,(R_{kl}{}^{ij})$$

This is $\delta S_1$:

*Perform the variation:*

```
Variation[Sqrt[Detg] ScalarR^2, Metricg]
```

$$2\sqrt{g}\,R\,(h_{pq;}{}^{pq}) - 2\sqrt{g}\,R\,(h_p{}^p{}_{;q}{}^q) + \frac{1}{2}\sqrt{g}\,R^2\,(g^{pq})\,(h_{pq}) - 2\sqrt{g}\,R\,(R_{pq})\,(h^{pq})$$

*Remove derivatives from $\delta g$ and factor det g:*

```
Expand[PIntegrate[%, Metricg] / Sqrt[Detg]]
```

$$-2\,(R_{;}{}^q)\,(h_{pq;}{}^p) + 2\,(R_{;}{}^q)\,(h_p{}^p{}_{;q}) + \frac{1}{2}R^2\,(g^{pq})\,(h_{pq}) - 2R\,(R_{pq})\,(h^{pq})$$

*Remove derivatives from $\delta g$ :*

```
PIntegrate[%, Metricg]
```

$$2\,(R_{;}{}^{pq})\,(h_{pq}) + \frac{1}{2}R^2\,(g^{pq})\,(h_{pq}) - 2\,(R_{;q}{}^q)\,(h_p{}^p) - 2R\,(R_{pq})$$

*Factor out $\delta g$ :*

```
VariationalDerivative[%, Metricg, ua, ub]
```

$$2\,(R_{;ab}) + \frac{1}{2}R^2\,(g_{ab}) - 2\,(R_{;q}{}^q)\,(g_{ab}) - 2R\,(R_{ab})$$

*This is the result for $\delta S_1$:*

```
δS1 = Canonicalize[%]
```

$$2\,(R_{;ab}) + \frac{1}{2}R^2\,(g_{ab}) - 2\,(R_{;p}{}^p)\,(g_{ab}) - 2R\,(R_{ab})$$

This is $\delta S_2$:

```
RicciR[la, lb] RicciR[ua, ub]
```

$(R_{ab})\ (R^{ab})$

*Perform the variation:*

```
Variation[Sqrt[Detg] %, Metricg]
```

$$-\frac{1}{2}\sqrt{g}\ (h^{pq}{}_{;r}{}^{r})\ (R_{pq}) + \sqrt{g}\ (h^{p}{}_{q;}{}^{rq})\ (R_{pr}) - \frac{1}{2}\sqrt{g}\ (h_{p}{}^{p}{}_{;}{}^{qr})\ (R_{qr}) - \frac{1}{2}\sqrt{g}\ (h_{pq;r}{}^{r})\ (R^{pq}) + \sqrt{g}\ (h_{pq;r}{}^{q})\ (R^{pr}) - \frac{1}{2}\sqrt{g}\ (h_{p}{}^{p}{}_{;qr})\ (R^{qr}) + \frac{1}{2}\sqrt{g}\ (g^{pq})\ (R_{rs})\ (R^{rs})\ (h_{pq}) - \sqrt{g}\ (R_{pq})\ (R^{qr})\ (h^{p}{}_{r}) - \sqrt{g}\ (R_{pq})\ (R^{pr})\ (h^{q}{}_{r})$$

*Remove derivatives from $\delta g$ and factor det g:*

```
Expand[PIntegrate[%, Metricg] / Sqrt[Detg]]
```

$$\frac{1}{2}(R^{pq}{}_{;}{}^{r})\ (h_{pq;r}) - (R^{pr}{}_{;}{}^{q})\ (h_{pq;r}) + \frac{1}{2}(R^{qr}{}_{;r})\ (h_{p}{}^{p}{}_{;q}) + \frac{1}{2}(R_{qr;}{}^{r})\ (h_{p}{}^{p}{}_{;}{}^{q}) - (R_{pr;}{}^{q})\ (h^{p}{}_{q;}{}^{r}) + \frac{1}{2}(R_{pq;}{}^{r})\ (h^{pq}{}_{;r}) + \frac{1}{2}(g^{pq})\ (R_{rs})\ (R^{rs})\ (h_{pq}) - (R_{pq})\ (R^{qr})\ (h^{p}{}_{r}) - (R_{pq})\ (R^{pr})$$

*Remove derivatives from $\delta g$ :*

```
PIntegrate[%, Metricg]
```

$$-\frac{1}{2}(R^{pq}{}_{;}{}^{r}{}_{r})\ (h_{pq}) + (R^{pr}{}_{;}{}^{q}{}_{r})\ (h_{pq}) + \frac{1}{2}(g^{pq})\ (R_{rs})\ (R^{rs})\ (h_{p\ldots} \ \frac{1}{2}(R_{qr;}{}^{rq})\ (h_{p}{}^{p}) - \frac{1}{2}(R^{qr}{}_{;rq})\ (h_{p}{}^{p}) + (R_{pr;}{}^{qr})\ (h^{p}{}_{q}) - (R_{pq})\ (R^{qr})\ (h^{p}{}_{r}) - \frac{1}{2}(R_{pq;}{}^{r}{}_{r})\ (h^{pq}) - (R_{pq})\ (R^{pr})\ (h^{q}{}_{r})$$

*Factor out $\delta g$ :*

```
VariationalDerivative[%, Metricg, ua, ub]
```

$$R_{ra;b}{}^{r} - R_{ab;}{}^{r}{}_{r} + R_{a}{}^{r}{}_{;br} - \frac{1}{2}(R_{qr;}{}^{rq})\ (g_{ab}) - \frac{1}{2}(R^{qr}{}_{;rq})\ (g_{ab}) - (R_{pa})\ (R_{b}{}^{p}) - (R_{qa})\ (R_{b}{}^{q}) + \frac{1}{2}(g_{ab})\ (R_{rs})\ (R^{rs})$$

*This is the result for $\delta S_2$:*

```
Tsimplify[Canonicalize[%]]
```

$$2\ (R_{pa;b}{}^{p}) - R_{ab;p}{}^{p} - (R_{pq;}{}^{qp})\ (g_{ab}) - 2\ (R_{pa})\ (R_{b}{}^{p}) + \frac{1}{2}(g_{ab})\ (R_{pq})\ (R^{pq})$$

*Express in a different form by removing derivatives from the Ricci tensor:*

```
δS2 = ApplyRulesRepeated[%, RiemannRules]
```

$$R_{;ab} - R_{ab;p}{}^{p} - \frac{1}{2}(R_{;p}{}^{p})\ (g_{ab}) + \frac{1}{2}(g_{ab})\ (R_{pq})\ (R^{pq}) - 2\ (R_{pq})\ (R_{a}{}^{q}{}_{b}{}^{p})$$

This is $\delta S_3$:

```
RiemannR[ua, ub, ug, ud] RiemannR[la, lb, lg, ld]
```

$(R_{abdg})\ (R^{abdg})$

*Perform the variation:*

**Variation[Sqrt[Detg] %, Metricg]**

$$-2\sqrt{g}\,(h^{pq;rs})\,(R_{prqs}) - 2\sqrt{g}\,(h_{pq;rs})\,(R^{prqs}) + \frac{1}{2}\sqrt{g}\,(g^{pq})\,(R_{rstu})\,(R^{rstu})\,(h_{pq}) - \frac{1}{2}\sqrt{g}\,(R_{pqrs})\,(R_t{}^{qrs})\,(h^{pt}) + \sqrt{g}\,(R_{pqrs})\,(R^{ptrs})\,(h^q{}_t) + \frac{1}{2}\sqrt{g}\,(R_{pqrs})\,(R_t{}^{prs})\,(h^{qt}) - \sqrt{g}\,(R_{pqrs})\,(R_t{}^{spq})\,(h^{rt}) + \sqrt{g}\,(R_{pqrs})\,(R_t{}^{rpq})\,(h^{st})$$

*Remove derivatives from $\delta g$ and factor det g:*

**Expand[PIntegrate[%, Metricg] / Sqrt[Detg]]**

$$2\,(R^{prqs}{}_{;s})\,(h_{pq;r}) + 2\,(R_{prqs;}{}^{s})\,(h^{pq;r}) + \frac{1}{2}(g^{pq})\,(R_{rstu})\,(R^{rstu})\,(h_{pq}) - \frac{1}{2}(R_{pqrs})\,(R_t{}^{qrs})\,(h^{pt}) + (R_{pqrs})\,(R^{ptrs})\,(h^q{}_t) + \frac{1}{2}(R_{pqrs})\,(R_t{}^{prs})\,(h^{qt}) - (R_{pqrs})\,(R_t{}^{spq})\,(h^{rt}) + (R_{pqrs})\,(R_t{}^{rpq})\,(h^{st})$$

*Remove derivatives from $\delta g$ :*

**PIntegrate[%, Metricg]**

$$-2\,(R^{prqs}{}_{;sr})\,(h_{pq}) + \frac{1}{2}(g^{pq})\,(R_{rstu})\,(R^{rstu})\,(h_{pq}) - 2\,(R_{prqs;}{}^{sr})\,(h^{pq}) - \frac{1}{2}(R_{pqrs})\,(R_t{}^{qrs})\,(h^{pt}) + (R_{pqrs})\,(R^{ptrs})\,(h^q{}_t) + \frac{1}{2}(R_{pqrs})\,(R_t{}^{prs})\,(h^{qt}) - (R_{pqrs})\,(R_t{}^{spq})\,(h^{rt}) + (R_{pqrs})\,(R_t{}^{rpq})\,(h^{st})$$

*Factor out $\delta g$ :*

**VariationalDerivative[%, Metricg, ua, ub]**

$$-2\,(R_{rasb;}{}^{sr}) - 2\,(R_a{}^r{}_b{}^s{}_{;sr}) - \frac{1}{2}(R_{pars})\,(R_b{}^{prs}) + \frac{1}{2}(R_{qars})\,(R_b{}^{qrs}) + (R_{pqra})\,(R_b{}^{rpq}) + (R_{pqsa})\,(R_b{}^{spq}) + \frac{1}{2}(g_{ab})\,(R_{rstu})\,(R^{rstu})$$

*This is the result for $\delta S_3$:*

**Tsimplify[Canonicalize[%]]**

$$-4\,(R_{paqb;}{}^{qp}) + 2\,(R_{pqra})\,(R_b{}^{rpq}) + \frac{1}{2}(g_{ab})\,(R_{pqrs})\,(R^{pqrs})$$

*Express in a different form by removing derivatives from the Riemann tensor:*

**$\delta S_3$ = ApplyRulesRepeated[%, RiemannRules]**

$$2\,(R_{;ab}) - 4\,(R_{ab;p}{}^{p}) + 4\,(R_{pb})\,(R_a{}^p) - 4\,(R_{pq})\,(R_a{}^p{}_b{}^q) + 2\,(R_{pqra})\,(R_b{}^{rpq}) + \frac{1}{2}(g_{ab})\,(R_{pqrs})\,(R^{pqrs})$$

*Lanczos Linear Combinations:*

**$\delta S_1 - 4\,\delta S_2 + \delta S_3$**

$$4\,(R_{;ab}) - 4\,(R_{ab;p}{}^{p}) + \frac{1}{2}R^2\,(g_{ab}) - 2\,(R_{;p}{}^{p})\,(g_{ab}) - 2R\,(R_{ab}) + 4\,(R_{pb})\,(R_a{}^p) - 4\,(R_{pq})\,(R_a{}^p{}_b{}^q) - 4\left(R_{;ab} - R_{ab;p}{}^{p} - \frac{1}{2}(R_{;p}{}^{p})\,(g_{ab}) + \frac{1}{2}(g_{ab})\,(R_{pq})\,(R^{pq}) - 2\,(R_{pq})\,(R_a{}^q{}_b{}^p)\right) + 2\,(R_{pqra})\,(R_b{}^{rpq}) + \frac{1}{2}(g_{ab})\,(R_{pqrs})\,(R^{pqrs})$$

**Expand[%]**

$\frac{1}{2} R^2 (g_{ab}) - 2 R (R_{ab}) + 4 (R_{pb}) (R_a{}^p) - 2 (g_{ab}) (R_{pq}) (R^{pq}) -$
$4 (R_{pq}) (R_a{}^p{}_b{}^q) + 8 (R_{pq}) (R_a{}^q{}_b{}^p) + 2 (R_{pqra}) (R_b{}^{rpq}) +$
$\frac{1}{2} (g_{ab}) (R_{pqrs}) (R^{pqrs})$

*This gives the Gauss-Bonnet Variation result:*

```
Tsimplify[CanAll[Expand[-%58 / 2]]]
```

$-\frac{1}{4} R^2 (g_{ab}) + R (R_{ab}) - 2 (R_{pa}) (R_b{}^p) + (g_{ab}) (R_{pq}) (R^{pq}) -$
$2 (R_{pq}) (R_a{}^p{}_b{}^q) - (R_{pqra}) (R_b{}^{rpq}) - \frac{1}{4} (g_{ab}) (R_{pqrs}) (R^{pqrs})$

```
% /. RiemannToWeylRule
```

$-\frac{1}{4} R^2 (g_{ab}) + R (R_{ab}) - 2 (R_{pa}) (R_b{}^p) + (g_{ab}) (R_{pq}) (R^{pq}) -$
$2 (R_{pq}) \Big(-\frac{1}{6} R (-(g_a{}^q) (g_b{}^p) + (g_{ab}) (g^{pq})) + \frac{1}{2} ((g^{pq}) (R_{ab}$
$(g_b{}^p) (R_a{}^q) - (g_a{}^q) (R_b{}^p) + (g_{ab}) (R^{pq})) + C_a{}^p{}_b{}^q\Big) -$
$\Big(-\frac{1}{6} R (-(g_{pa}) (g_{qr}) + (g_{pr}) (g_{qa})) + \frac{1}{2} ((g_{qa}) (R_{pr}) -$
$(g_{qr}) (R_{pa}) - (g_{pa}) (R_{qr}) + (g_{pr}) (R_{qa})) + C_{pqra}\Big)$
$\Big(-\frac{1}{6} R (-(g_b{}^q) (g^{pr}) + (g_b{}^p) (g^{qr})) + \frac{1}{2} ((g^{qr}) (R_b{}^p) -$
$(g^{pr}) (R_b{}^q) - (g_b{}^q) (R^{pr}) + (g_b{}^p) (R^{qr})) + C_b{}^{rpq}\Big) -$
$\frac{1}{4} (g_{ab}) \Big(-\frac{1}{6} R (-(g_{ps}) (g_{qr}) + (g_{pr}) (g_{qs})) + \frac{1}{2} ((g_{qs}) (R_p$
$(g_{qr}) (R_{ps}) - (g_{ps}) (R_{qr}) + (g_{pr}) (R_{qs})) + C_{pqrs}\Big)$
$\Big(-\frac{1}{6} R (-(g^{ps}) (g^{qr}) + (g^{pr}) (g^{qs})) +$
$\frac{1}{2} ((g^{qs}) (R^{pr}) - (g^{qr}) (R^{ps}) - (g^{ps}) (R^{qr}) + (g^{pr}) (R^{qs}))$
$C^{pqrs}\Big)$

```
On[MetricgFlag]
CanAll[%%]
```

$\frac{1}{6} R (g_{ab}) (C_{pq}{}^{pq}) + \frac{2}{3} R (C_{pab}{}^p) - (R_{pa}) (C_{qb}{}^{pq}) - (R_{pb}) (C_q{}^p{}_a{}^q$
$\frac{1}{2} (g_{ab}) (R_{pq}) (C_r{}^{pqr}) + \frac{1}{2} (g_{ab}) (R_{pq}) (C_r{}^{qpr}) - (R_{pq}) (C_a{}^p{}_b{}^q$
$(R_{pq}) (C_a{}^q{}_b{}^p) - (C_{pqra}) (C_b{}^{rpq}) - \frac{1}{4} (g_{ab}) (C_{pqrs}) (C^{pqrs})$

```
Tsimplify[%]
```

$\frac{1}{6} R (g_{ab}) (C_{pq}{}^{pq}) + \frac{2}{3} R (C_{pab}{}^p) - (R_{pa}) (C_{qb}{}^{pq}) - (R_{pb}) (C_q{}^p{}_a{}^q$
$(g_{ab}) (R_{pq}) (C_r{}^{pqr}) - (C_{pqra}) (C_b{}^{rpq}) - \frac{1}{4} (g_{ab}) (C_{pqrs}) (C^{pqr}$

*Define a Rule Making the Weyl Tensor Trace-Free on any Index Pair:*

```
RuleUnique[WeylTraceRule, WeylC[la_, lb_, lc_, ld_],
 0, PairQ[la, lb] || PairQ[la, lc] || PairQ[la, ld] ||
  PairQ[lb, lc] || PairQ[lb, ld] || PairQ[lc, ld]]
```

*Apply the rule:*

```
ApplyRules[%%, WeylTraceRule]
```

$-(C_{pqra}) (C_b{}^{rpq}) - \frac{1}{4} (g_{ab}) (C_{pqrs}) (C^{pqrs})$

# Auxiliary Condition for Miscellaneous Metrics

*Initialize a Separate MathTensor Session:*

```
<< Mathtens.m
Dimension = 4;
SetDirectory["F:\Mathematica\Files\Dissertation\
   Tensor Calculations"];
```

*Define the Auxiliary Condition:*

```
X[lc_, lf_] = RicciR[ua, lc] RicciR[la, lf] -
  RicciR[la, lb] RiemannR[lc, ua, lf, ub]
```

$(R_{af})\ (R_c{}^a) - (R_{ab})\ (R_c{}^a{}_f{}^b)$

*Evaluate Ordinary Derivatives:*

```
On[EvaluateODFlag]
```

*For the Pavelle-Thompson Metric:*

```
<< Pavelle_Thompson.m
MetricgFlag has been turned off.
Table[Metricg[-i, -j], {i, 4}, {j, 4}]
```

$$\begin{pmatrix} -\frac{1}{\left(\frac{cl}{r}+1\right)^2} & 0 & 0 & 0 \\ 0 & -r^2 & 0 & 0 \\ 0 & 0 & -r^2\sin^2(\text{theta}) & 0 \\ 0 & 0 & 0 & \frac{1}{\left(\frac{cl}{r}+1\right)^2} \end{pmatrix}$$

```
Table[MakeSum[X[-i, -j]], {i, 4}, {j, 4}]
Simplify[%]
```

$$\left\{\left\{-\frac{4\,cl^2\,(3\,cl+5\,r)}{r^6\,(cl+r)},\ 0,\ 0,\ 0\right\},\ \left\{0,\ \frac{4\,cl^2\,(cl+r)^2}{r^6},\ 0,\ 0\right\},\ \left\{0,\ 0,\ \frac{4\,cl^2\,(cl+r)^2\,\text{Sin[theta]}^2}{r^6},\ 0\right\},\ \left\{0,\ 0,\ 0,\ -\frac{4\,cl^2\,(cl+3\,r)}{r^6\,(cl+r)}\right\}\right\}$$

```
Simplify[ScalarR]
0
```

*For the Thompson Metric:*

```
<< Thompson.m
MetricgFlag has been turned off.
Table[Metricg[-i, -j], {i, 4}, {j, 4}]
```

$$\begin{pmatrix} -\frac{1}{\frac{cl}{r}+1} & 0 & 0 & 0 \\ 0 & -r^2 & 0 & 0 \\ 0 & 0 & -r^2\sin^2(\text{theta}) & 0 \\ 0 & 0 & 0 & 1 \end{pmatrix}$$

```
Table[MakeSum[X[-i, -j]], {i, 4}, {j, 4}]
```

$$\left\{\left\{-\frac{3\,cl^2}{2\,r^5\,(cl+r)},\ 0,\ 0,\ 0\right\},\ \left\{0,\ \frac{3\,cl^2}{4\,r^4},\ 0,\ 0\right\},\ \left\{0,\ 0,\ \frac{3\,cl^2\,\text{Sin[theta]}^2}{4\,r^4},\ 0\right\},\ \{0,\ 0,\ 0,\ 0\}\right\}$$

```
Simplify[ScalarR]
0
```

*For Kottler's Metric:*

```
<<Kottler.m
Table[Metricg[-i, -j], {i, 4}, {j, 4}]
```

$$\begin{pmatrix} -\frac{1}{c2\, r^2+1+\frac{c1}{r}} & 0 & 0 & 0 \\ 0 & -r^2 & 0 & 0 \\ 0 & 0 & -r^2 \sin^2(\text{theta}) & 0 \\ 0 & 0 & 0 & c2\, r^2+1+\frac{c1}{r} \end{pmatrix}$$

```
Table[MakeSum[X[-i, -j]], {i, 4}, {j, 4}]
Simplify[%]
{{0, 0, 0, 0}, {0, 0, 0, 0}, {0, 0, 0, 0}, {0, 0, 0, 0}}
Simplify[ScalarR]
12 c2
```

*For the Einstein Universe Metric:*

```
<<Einstein.m
MetricgFlag has been turned off.
Table[Metricg[-i, -j], {i, 4}, {j, 4}]
```

$$\begin{pmatrix} -\frac{1}{c1\, r^2+1} & 0 & 0 & 0 \\ 0 & -r^2 & 0 & 0 \\ 0 & 0 & -r^2 \sin^2(\text{theta}) & 0 \\ 0 & 0 & 0 & 1 \end{pmatrix}$$

```
Table[MakeSum[X[-i, -j]], {i, 4}, {j, 4}]
{{0, 0, 0, 0}, {0, 0, 0, 0}, {0, 0, 0, 0}, {0, 0, 0, 0}}
Simplify[ScalarR]
6 c1
Table[RicciR[-i, -j], {i, 4}, {j, 4}]
```

$$\begin{pmatrix} -\frac{2\, c1}{c1\, r^2+1} & 0 & 0 & 0 \\ 0 & -2\, c1\, r^2 & 0 & 0 \\ 0 & 0 & -2\, c1\, r^2 \sin^2(\text{theta}) & 0 \\ 0 & 0 & 0 & 0 \end{pmatrix}$$

```
Table[RicciR[-i, -j] - 1/4 ScalarR Metricg[-i, -j],
  {i, 4}, {j, 4}]
```

$$\begin{pmatrix} -\frac{c1}{2\,(c1\, r^2+1)} & 0 & 0 & 0 \\ 0 & -\frac{1}{2}(c1\, r^2) & 0 & 0 \\ 0 & 0 & -\frac{1}{2} c1\, r^2 \sin^2(\text{theta}) & 0 \\ 0 & 0 & 0 & -\frac{1}{2}(3\, c1) \end{pmatrix}$$

```
Table[WeylC[-i, -j, -k, -l], {i, 4}, {j, 4}, {k, 4}, {l,
```

*For Ni's Metric:*

```
<<Ni.m
Table[Metricg[-i, -j], {i, 4}, {j, 4}]
```

$$\begin{pmatrix} -\frac{1}{c2\, r^2+1+\frac{c1}{r}} & 0 & 0 & 0 \\ 0 & -r^2 & 0 & 0 \\ 0 & 0 & -r^2 \sin^2(\text{theta}) & 0 \\ 0 & 0 & 0 & 1 \end{pmatrix}$$

```
Table[MakeSum[X[-i, -j]], {i, 4}, {j, 4}]
Simplify[%]
```

$$\left\{\left\{-\frac{3\,c1\,(c1-2\,c2\,r^3)}{2\,r^5\,(c1+r+c2\,r^3)},\,0,\,0,\,0\right\},\right.$$
$$\left\{0,\,\frac{3\,c1\,(c1-2\,c2\,r^3)}{4\,r^4},\,0,\,0\right\},$$
$$\left.\left\{0,\,0,\,\frac{3\,c1\,(c1-2\,c2\,r^3)\,\mathrm{Sin[theta]}^2}{4\,r^4},\,0\right\},\,\{0,\,0,\,0,\,0\}\right\}$$

**Simplify[ScalarR]**

6 c2

*For the Vanishing Ricci Scalar Metric:*

**<< VanishingRicci.m**

MetricgFlag has been turned off.

**Table[Metricg[-i, -j], {i, 4}, {j, 4}]**

$$\begin{pmatrix} -\frac{1}{\frac{c1}{r^2}+1} & 0 & 0 & 0 \\ 0 & -r^2 & 0 & 0 \\ 0 & 0 & -r^2\sin^2(\mathrm{theta}) & 0 \\ 0 & 0 & 0 & \frac{c1}{r^2}+1 \end{pmatrix}$$

**Table[RicciR[-i, -j], {i, 4}, {j, 4}]**

$$\left\{\left\{-\frac{c1}{r^2\,(c1+r^2)},\,0,\,0,\,0\right\},\,\left\{0,\,\frac{c1}{r^2},\,0,\,0\right\},\right.$$
$$\left.\left\{0,\,0,\,\frac{c1\,\mathrm{Sin[theta]}^2}{r^2},\,0\right\},\,\left\{0,\,0,\,0,\,\frac{c1^2+c1\,r^2}{r^6}\right\}\right\}$$

**Simplify[Table[MakeSum[X[-i, -j]], {i, 4}, {j, 4}]]**

$$\left\{\left\{\frac{4\,c1^2}{r^6\,(c1+r^2)},\,0,\,0,\,0\right\},\,\left\{0,\,-\frac{4\,c1^2}{r^6},\,0,\,0\right\},\right.$$
$$\left\{0,\,0,\,-\frac{4\,c1^2\,\mathrm{Sin[theta]}^2}{r^6},\,0\right\},\,\left\{0,\,0,\,0,\,-\frac{4\,c1^2\,(c1+r^2}{r^{10}}\right.$$

**Simplify[ScalarR]**

6 c2

# Tensor Calculations for the Auxiliary Condition:

Define the Auxiliary Condition:

**ac[lc_, lf_] = RicciR[ua, lc] RicciR[la, lf] -**
**RicciR[la, lb] RiemannR[lc, ua, lf, ub]**

$$(R_{af})\,(R_c{}^a) - (R_{ab})\,(R_c{}^a{}_f{}^b)$$

**% /. RiemannToWeylRule**

$$(R_{af})\,(R_c{}^a) - (R_{ab})\left(-\frac{1}{6}\,R\,(-(g_c{}^b)\,(g_f{}^a) + (g_{cf})\,(g^{ab})) +\right.$$
$$\frac{1}{2}\,((g^{ab})\,(R_{cf}) - (g_f{}^a)\,(R_c{}^b) - (g_c{}^b)\,(R_f{}^a) + (g_{cf})\,(R^{ab}))$$
$$\left.C_c{}^a{}_f{}^b\right)$$

**% /. RicciToTraceFreeRicciRule**

$$\left(\frac{1}{4}\,R\,(g_{af}) + TFR_{af}\right)\left(\frac{1}{4}\,R\,(g_c{}^a) + TFR_c{}^a\right) -$$
$$\left(\frac{1}{4}\,R\,(g_{ab}) + TFR_{ab}\right)\left(-\frac{1}{6}\,R\,(-(g_c{}^b)\,(g_f{}^a) + (g_{cf})\,(g^{ab})) +\right.$$
$$\frac{1}{2}\left((g^{ab})\left(\frac{1}{4}\,R\,(g_{cf}) + TFR_{cf}\right) - (g_f{}^a)\left(\frac{1}{4}\,R\,(g_c{}^b) + TFR_c{}^b\right.\right.$$
$$(g_c{}^b)\left(\frac{1}{4}\,R\,(g_f{}^a) + TFR_f{}^a\right) + (g_{cf})\left(\frac{1}{4}\,R\,(g^{ab}) + TFR^{ab}\right.$$
$$\left.C_c{}^a{}_f{}^b\right)$$

**CanAll[Expand[%]]**

$$-\frac{1}{12} R\,(g_{cf})\,(TFR_a{}^a) + \frac{1}{8} R\,(TFR_{cf}) -$$
$$\frac{1}{2}\,(TFR_a{}^a)\,(TFR_{cf}) + (TFR_{af})\,(TFR_c{}^a) + \frac{1}{2}\,(TFR_c{}^b)\,(TFR_{fb}) +$$
$$\frac{5}{24} R\,(TFR_{fc}) + \frac{1}{2}\,(TFR_{ac})\,(TFR_f{}^a) - \frac{1}{8} R\,(g_{cf})\,(TFR^a{}_a) -$$
$$\frac{1}{2}\,(g_{cf})\,(TFR_{ab})\,(TFR^{ab}) + \frac{1}{4} R\,(C_{afc}{}^a) - (TFR_{ab})\,(C_c{}^a{}_f{}^b)$$

```
Tsimplify[%]
```

$$-\frac{5}{24} R\,(g_{cf})\,(TFR_p{}^p) + \frac{1}{3} R\,(TFR_{cf}) - \frac{1}{2}\,(TFR_p{}^p)\,(TFR_{cf}) +$$
$$2\,(TFR_{pf})\,(TFR_c{}^p) - \frac{1}{2}\,(g_{cf})\,(TFR_{pq})\,(TFR^{pq}) + \frac{1}{4} R\,(C_{pcf}{}^p) -$$
$$(TFR_{pq})\,(C_c{}^p{}_f{}^q)$$

*Define a Rule Making the Weyl Tensor Trace-Free on any Index Pair:*

```
RuleUnique[WeylTraceRule, WeylC[la_, lb_, lc_, ld_],
 0, PairQ[la, lb] || PairQ[la, lc] || PairQ[la, ld] ||
  PairQ[lb, lc] || PairQ[lb, ld] || PairQ[lc, ld]]
```

*Apply the rule:*

```
ApplyRules[%%, WeylTraceRule]
```

$$-\frac{5}{24} R\,(g_{cf})\,(TFR_p{}^p) + \frac{1}{3} R\,(TFR_{cf}) - \frac{1}{2}\,(TFR_p{}^p)\,(TFR_{cf}) +$$
$$2\,(TFR_{pf})\,(TFR_c{}^p) - \frac{1}{2}\,(g_{cf})\,(TFR_{pq})\,(TFR^{pq}) - (TFR_{pq})\,(C_c{}^p$$

*Define a Rule to Make the Trace Free Ricci Part Vanish on Contraction:*

```
RuleUnique[RicciTraceRule, TraceFreeRicciR[la_, lb_],
 0, PairQ[la, lb]]
```

*Apply the rule:*

```
ApplyRules[%%, RicciTraceRule]
```

$$\frac{1}{3} R\,(TFR_{cf}) + 2\,(TFR_{pf})\,(TFR_c{}^p) - \frac{1}{2}\,(g_{cf})\,(TFR_{pq})\,(TFR^{pq}) -$$
$$(TFR_{pq})\,(C_c{}^p{}_f{}^q)$$

*This is the result:*

```
Tsimplify[%]
```

$$\frac{1}{3} R\,(TFR_{cf}) + 2\,(TFR_{pf})\,(TFR_c{}^p) - \frac{1}{2}\,(g_{cf})\,(TFR_{pq})\,(TFR^{pq}) -$$
$$(TFR_{pq})\,(C_c{}^p{}_f{}^q)$$

# Appendix C: Gauge Kinematics in Classical Mechanics

In the first section of Chapter 4, the prototype $SU(2)$ gauge theory formalism developed by Yang and Mills has been considered. The general pattern in this approach may be identified:

> *starting from a global symmetry - local dynamical invariance is established by introducing a "connection" - expressed as a linear combination of the algebra basis for the symmetry group.*

Therefore, as an exercise following this pattern - consider the equations resulting from this general prescription outlined above - applied to the invariance of Lagrange's equations under local† rotations.

To this end, let us consider the free motion of a point mass in rectilinear coordinates so that

$$L = \tfrac{1}{2} m \delta_{ij} \dot{x}^i \dot{x}^j \ ; \ i, j = 1,2,3 . \tag{C.1}$$

Obviously, (C.1) is (globally) invariant under a constant rotation:

$$\begin{gathered} \dot{x}^i \to \dot{x}^{i'} = d_t(x^{i'}) = d_t(R^{i'}_{\cdot j} x^j) = R^{i'}_{\cdot j} \dot{x}^j \\ \Rightarrow \ L = L' , \end{gathered} \tag{C.2}$$

since $R^i_{\cdot j}$ is orthogonal where

$$R^i_{\cdot j} = [e^{\theta^k \lambda_k}]^i_{\cdot j} \in SO(3,\mathbb{R}) , \tag{C.3}$$

using the adjoint representation ($\Rightarrow$ $i$, $j$ and $k$ have the same index range) with $\theta^k \equiv$ group parameters and $\lambda_k$ generating the $so(3)$ Lie algebra:

$$[\lambda_j , \lambda_k] = \varepsilon^i_{\cdot jk} \lambda_i . \tag{C.4}$$

---

† "local" in this sense means that the group parameters are time dependent.

However, under local rotations (i.e. $\theta^k = \theta^k(t)$), $L$ is no longer invariant as may be checked substituting (C.2) into (C.1) for the case that $R^{i'}_{\cdot j} = R^{i'}_{\cdot j}(t)$. Therefore, let us construct a gauge covariant derivative from $so(3,\mathbb{R})$:

$$\nabla_t x^i = d_t x^i + \Gamma^i_{\cdot jt} x^j , \tag{C.5}$$

to obtain local invariance under (C.3). Following the prescription outlined above, the $\Gamma$'s are

$$\Gamma^i_{\cdot jt} \equiv [\Gamma^k_{\cdot t} \lambda_k]^i_{\cdot j} = \Gamma^k_{\cdot t} \lambda^i_{\cdot kj} \equiv \Gamma^k_{\cdot t} \varepsilon^i_{\cdot kj} , \tag{C.6}$$

and with this identification, (C.5) becomes

$$\nabla_t x^i = d_t x^i + (\Gamma^k_{\cdot t} \wedge x^j)^i . \tag{C.7}$$

But this already looks familiar recalling the usual replacement in a (non-inertial) rotating reference frame:

$$\dot{x}'^i = \dot{x}^i + (\omega^k \wedge x^j)^i , \tag{C.8}$$

and then comparing with (C.7) we identify

$$\Gamma^k_{\cdot t} \equiv \omega^k = d_t \theta^k \;\Rightarrow\; \Gamma^i_{\cdot jt} \equiv \varepsilon^i_{\cdot kj} \omega^k = \begin{pmatrix} 0 & \omega^3 & -\omega^2 \\ -\omega^3 & 0 & \omega^1 \\ \omega^2 & -\omega^1 & 0 \end{pmatrix} . \tag{C.9}$$

Carrying through with the analogy, we derive the transformation law of the $\Gamma$'s:

$$\Gamma^{i'}_{\cdot j't} = R^{i'}_{\cdot k} \Gamma^k_{\cdot lt} R^{\cdot l}_{j'} - (d_t R^{i'}_{\cdot l}) R^{\cdot l}_{j'} ,$$

requiring as before that

$$\nabla_t x^i \rightarrow \nabla_t x^{i'} = R^{i'}_{\cdot j} \nabla_t x^j . \tag{C.10}$$

Given the transformation in (C.10), $\nabla_t x^{i'}$ and $L'$ are now invariant under time-dependent $SO(3,\mathbb{R})$ transformations when $L$ is expressed in terms of $\nabla_t$:

$$L = \tfrac{1}{2} m \delta_{ij} (\nabla_t x^i)(\nabla_t x^j) ,$$

as are the dynamical equations:

$$\nabla_t \left( \partial_{\nabla_t x^i} L \right) - \partial_{x^i} L = 0 . \tag{C.11}$$

Substituting for $\nabla_t$ in the above gives the form of $L$:

$$L = \tfrac{1}{2} m \left[ (d_t \vec{x})^2 + 2 d_t \vec{x} \cdot (\vec{\omega} \wedge \vec{x}) + (\vec{\omega} \wedge \vec{x})^2 \right] , \tag{C.12}$$

and then (C.11) becomes

$$m(\nabla_t^2 \vec{x})^i \;=\; -\,m\Big[(d_t^2 \vec{x})^i \;+\; 2(\vec{\omega} \wedge d_t \vec{x})^i \;+\; (d_t \vec{\omega} \wedge \vec{x})^i \;+\; (\vec{\omega} \wedge \vec{\omega} \wedge \vec{x})^i\Big], \quad \text{(C.13)}$$

showing that local gauge invariance of Lagrange's equations under $SO(3,\mathbb{R})$ leads to the usual dynamics in a (non-inertial) rotating frame of reference. However, this is as far as we go with the analogy, since by constructing

$$\varphi^{i}_{\cdot jtt} \;=\; \partial_t \Gamma^{i}_{\cdot jt} \;-\; \partial_t \Gamma^{i}_{\cdot jt} \;+\; \Gamma^{i}_{\cdot kt} \Gamma^{k}_{\cdot jt} \;-\; \Gamma^{i}_{\cdot kt} \Gamma^{k}_{\cdot jt}\,, \qquad \text{(C.14)}$$

we see immediately that the gauge field is identically zero. As a result, one could not expect to obtain (C.11) from an action constructed like (3.13). Therefore, (C.12) and (C.13) are essentially gauge kinematic equations based on $SO(3,\mathbb{R})$, not the gauge dynamical equations usually considered in standard gauge theory.

One final comment: the connection in (C.9) is given in the adjoint representation. But we might also consider the spinor representation and therefore obtain a "spinor formulation" of mechanics as discussed by Hestenes [154]. However, as seen above, this spinor form would be merely an arbitrary representation of a gauge-type (kinematic) theory based on $SO(3,\mathbb{R})$ (giving (C.12) and (C.13) in the adjoint rep). Therefore, it would seem unnecessary to view these equations as deriving solely from a Clifford algebra reformulation of classical mechanics - as motivated by Hestenes.

# Appendix D: Orthogonal Groups

A central theme of the analysis considered in the early sections of Chapter 3 is invariance under a given transformation group. The general gauge theory formalism reviewed in Chapter 2 provides a systematic method for deriving dynamical equations by postulating invariance under a given symmetry group. This technique is applied in Appendix C to a familiar example from classical mechanics. The purpose of the present Appendix is to provide a brief introduction to the orthogonal group using low dimensional examples and to define the notation used in earlier Chapters.

## Axioms

A *group* is the pair: $(\mathtt{G},\circ)$, where $\mathtt{G}$ is a set and $\circ$ is the binary operation of group multiplication satisfying (using the usual quantifiers: $\forall \equiv$ for all, for every; $\exists \equiv$ there exists; and the abbreviations: *s.t.* $\equiv$ such that; ! $\equiv$ unique).

(1) $g_i \circ g_j \in \mathtt{G},\ \forall\, g_i, g_j \in \mathtt{G}$ (*closure*)

(2) $(g_i \circ g_j) \circ g_k = g_i \circ (g_j \circ g_k),\ \forall\, g_i, g_j, g_k \in \mathtt{G}$ (*associatively*)

(3) $\exists$ ! element $1 \in \mathtt{G}$, *s.t.* $g_i \circ 1 = 1 \circ g_i = g_i,\ \forall\, g_i \in \mathtt{G}$ (*existence of the identity*)

(4) $\forall g_i \in \mathtt{G}$, $\exists$ ! element $g_i^{-1}$, *s.t.* $g_i^{-1} \circ g_i = g_i \circ g_i^{-1} = 1,\ \forall\, g_i \in \mathtt{G}$ (*existence of the inverse*)

An *algebra* is the triple: $(\mathtt{A},\mathtt{V},\circ)$, where $\mathtt{A}$ is a set, $\mathtt{V}$ is a linear vector space, and $\circ$ is a map, $\circ : \mathtt{V} \times \mathtt{V} \to \mathtt{V}$, defined between elements of $\mathtt{A}$ and satisfying:

(1) $a_i \circ a_j \in \mathtt{A},\ \forall\, a_i, a_j \in \mathtt{A}$ (*closure*)

(2) $a_i \circ (a_j + a_k) = a_i \circ a_j + a_i \circ a_k\ ,\ \forall\, a_i, a_j, a_k \in \mathtt{A}$ (*bilinearity*)

$(a_j + a_k) \circ a_i = a_j \circ a_i + a_k \circ a_i\ ,\ \forall\, a_i, a_j, a_k \in \mathtt{A}$

Other varieties of algebra's may be obtained depending on which additional postulates are satisfied, for example:

(3) $(a_i \circ a_j) \circ a_k = a_i \circ (a_j \circ a_k)$ , $\forall\ a_i, a_j, a_k \in \mathtt{A}$ (*associativity*)

(4) $\exists\,!$ element $1 \in$ !, $s.t.$ $a_i \circ 1 = a_i$ , $\forall\ a_i \in \mathtt{A}$ (*existence of the identity*)

(5) $a_i \circ a_j = a_i \circ a_j$ , $\forall\ a_i, a_j \in \mathtt{A}$ (*commutative*)

(6) $a_i \circ a_j = -\ a_j \circ a_i$ , $\forall\ a_i, a_j \in \mathtt{A}$ (*anti-commutative*)

(7) $(a_i \circ a_j) \circ a_k = a_i \circ (a_j \circ a_k)\ +\ a_j \circ (a_i \circ a_k)$ (*a derivation*)

Along with (3), (4), (5), (6), (7); the properties of $\circ$ define the algebra and in the case of a Lie algebra (for instance, Gilmore [155]) the product is given by the Lie bracket: $[a_i, a_j]\ =\ -\,[a_j, a_i]$. With this anti-commutative product, property (7) may be written:

$$[[a_i\ , a_j]\ , a_k]\ +\ [[a_j\ , a_k]\ , a_i]\ +\ [[a_k\ , a_i]\ , a_j]\ =\ 0\,. \qquad \text{(D.1)}$$

which is more commonly referred to as the Jacobi identity. There are other properties of Lie algebra's that classify them as topological spaces, but these topics are beyond the scope of this Appendix. We will, however, make use of the exponential mapping to construct elements of the Lie group, which is dependent upon the property (simply assumed here) that each element of the group may be continuously connected to the identity element in group space.

## Orthogonal Groups

Consider the group of transformations preserving the line element of an $R_n$ (i.e. an $E_n$ with a symmetric fundamental tensor $g$; *ordinary* if $g$ is positive definite, *indefinite* otherwise) defined in this case using a rectilinear coordinate system, $(i)$ :

$$ds^2\ =\ g_{ij}\,dx^i dx^j\ \ ;\ i, j, k\ldots\ i', j', k'\ldots\ etc.\ =1, \ldots, n\ . \qquad \text{(D.2)}$$

The kernel symbol, $g$, denotes a general metric quantity but in the rectilinear system, a representation of $g$ is $g_{ij}$, identical to the Kronecker delta ($\overset{*}{=}$ emphasizes that the equation is only valid in the specified coordinate system):

$$g_{ij} \stackrel{*}{=} \delta_{ij} \ . \tag{D.3}$$

Using kernel-index notation the transformation is expressed:

$$ds^2 \to ds'^2 = ds^2 = g_{i'j'}\, dx^{i'} dx^{j'} \ , \tag{D.4}$$

and equality with (D.2) means that $ds^2$ is a scalar quantity. Substituting the contravariant transformation rule:

$$dx^{i'} = \frac{\partial x^{i'}}{\partial x^{i}} dx^{i} \equiv R^{i'}_{\ i}\, dx^{i} \ , \tag{D.5}$$

into (D.2) and then combining with (D.4) gives a condition on the invariance of $g_{ij}$:

$$R^{i'}_{\ i} R^{j'}_{\ j}\, g_{i'j'} = g_{ij} \ , \tag{D.6}$$

or equivalently a condition on the $R^{i'}_{\ i}$:

$$R^{i'}_{\ i} \in O(n) \, . \tag{D.7}$$

I.e. $R^{i'}_{\ i}$ is an element of the orthogonal group $O(n)$, a subgroup of the general linear group, $Gl(n)$ (the group of all $n \times n$ nonsingular transformations). But note that no number field ($\mathrm{F}$) has been specified to this point for the entries of $R^{i'}_{\ i}$, e.g. the real ($\mathbb{R}$), or complex ($\mathbb{C}$) number fields might be used to define entries of $R^{i'}_{\ i}$. But rather than specifying a pre-determined number field for the undetermined group elements, for generality let us solve (D.6) for several low dimensional cases and then let the possible ranges of these entries determine $\mathrm{F}$.

Returning to (D.6), multiplication by $g^{kj}$ gives

$$R^{i'}_{\ i} R_{i'}^{\ \ k} = \delta_i^{\ k} \ , \tag{D.8}$$

and the quantity

$$R_{i'}^{\ \ k} = g_{i'j'} R^{j'}_{\ j}\, g^{kj} \ , \tag{D.9}$$

is defined as the *adjoint* of $R^{j'}_{\ j}$ through the raising and lowering of appropriate indices as shown above. But note that for proper matrix multiplication the ordering of indices should be switched on $R_{i'}^{\ \ k}$ to $R^{k}_{\ i'}$, so that in real applications (D.8) and (D.9) should read:

$$R^{i'}_{\ i} R^{k}_{\ i'} = \delta^{k}_{i}, \tag{D.10}$$

$$R^{k}_{\ i'} = g_{i'j'} R^{j'}_{j} g^{kj}, \tag{D.11}$$

respectively. In rectilinear coordinates the adjoint is identical to the transpose, however, in cases involving an indefinite metric, the transpose and adjoint are not necessarily identical (e.g. see $O(1,1)$ below).

$O(2)$

Consider a special of case of the equations when the dimension of $R^{i}_{\ j}$ is $2 \times 2$ and defined with entries of an arbitrary field $\mathrm{F}$:

$$R^{i}_{\ j} = \begin{pmatrix} a & b \\ c & d \end{pmatrix}. \tag{D.12}$$

Equation (D.10) is then:

$$\begin{pmatrix} a & b \\ c & d \end{pmatrix}\begin{pmatrix} a & c \\ b & d \end{pmatrix} = \begin{pmatrix} a^2+b^2 & ac+bd \\ ac+bd & c^2+d^2 \end{pmatrix} = \begin{pmatrix} 1 & 0 \\ 0 & 1 \end{pmatrix}, \tag{D.13}$$

giving a system of 3 equations in 4 unknowns, and therefore the solution will have only 1 independent parameter. The general solution satisfying (D.13) gives 4 possible elements for the $R^{i}_{\ j}$ (expressed arbitrarily in terms of the parameter $c$):

$$\begin{aligned} R^{i}_{\ j} = \Bigg\{ &\begin{pmatrix} \sqrt{1-c^2} & -c \\ c & \sqrt{1-c^2} \end{pmatrix}, \begin{pmatrix} -\sqrt{1-c^2} & -c \\ c & -\sqrt{1-c^2} \end{pmatrix}, \\ &\begin{pmatrix} -\sqrt{1-c^2} & c \\ c & +\sqrt{1-c^2} \end{pmatrix}, \begin{pmatrix} +\sqrt{1-c^2} & c \\ c & -\sqrt{1-c^2} \end{pmatrix} \Bigg\}. \end{aligned} \tag{D.14}$$

The first two correspond to transformations with $\det R^{i}_{\ j} = +1$, i.e. "special" transformations or $SO(2)$, while the second two correspond to reflections about the $x$ and $y$ axes, respectively; with $\det R^{i}_{\ j} = -1$. In addition note from (D.14) that every solution to (D.13) satisfies $\det R^{i}_{\ j} = \pm 1$, independent of $c$ which is a general property that can be derived directly from (D.8)).

$SO(2,\mathbb{R})$

To construct the Lie subgroup $SO(2,\mathbb{R})$ of $O(2,\mathbb{R})$, consider an exponential expansion (since by definition of the Lie group these elements are continuously connected to the identity), parametrized in terms of $c$ ($c$ is assumed real) and the undetermined generator $L^i{}_j$:

$$R^i{}_j = e^{c L^i{}_j} . \tag{D.15}$$

Series expanding both sides in the parameter $c$ about $c = 0$ to first order gives

$$R^i{}_j\Big|_{c=0} + \partial_c R^i{}_j\Big|_{c=0} c = \delta^i{}_j + c L^i{}_j , \tag{D.16}$$

and then substituting explicitly into (D.13) for the first solution of (D.14) results in the system:

$$\begin{pmatrix} 0 & -1 \\ 1 & 0 \end{pmatrix} c = \begin{pmatrix} f & h \\ j & k \end{pmatrix} c , \tag{D.17}$$

assuming arbitrary elements, $\{f, h, j, k\}$, for the $L^i{}_j$. Therefore to first order in $c$ the solution is:

$$L^i{}_j = \begin{pmatrix} 0 & -1 \\ 1 & 0 \end{pmatrix} , \tag{D.18}$$

and then substituting into (D.15) (series expanding and changing $c \to \theta$) gives the result:

$$R^i{}_j = \begin{pmatrix} 1 - \frac{\theta^2}{2!} + \frac{\theta^4}{4!} + \ldots & -\theta + \frac{\theta^3}{3!} - \frac{\theta^5}{5!} + \ldots \\ \theta - \frac{\theta^3}{3!} + \frac{\theta^5}{5!} - \ldots & 1 - \frac{\theta^2}{2!} + \frac{\theta^4}{4!} + \ldots \end{pmatrix} = \begin{pmatrix} \cos\theta & -\sin\theta \\ \sin\theta & \cos\theta \end{pmatrix} , \tag{D.19}$$

or by comparison with (D.17), $c \to \sin\theta$ .

For the third element of (D.14), (D.16) becomes:

$$\begin{pmatrix} -1 & 0 \\ 0 & +1 \end{pmatrix} + \begin{pmatrix} 0 & 1 \\ 1 & 0 \end{pmatrix} c = \begin{pmatrix} 1 & 0 \\ 0 & 1 \end{pmatrix} + \begin{pmatrix} f & h \\ j & k \end{pmatrix} c , \tag{D.20}$$

showing that no Lie algebra exists in this case because the solution is "discontinuously" connected to the identity by inspection. Similar results are obtained for the fourth element. As discussed below the second group element differs from the first by a rotation of $\pi$. As a result, the entire Lie algebra is given by (D.18), corresponding to the Lie subgroup (D.19) for $O(2,\mathbb{R})$.

Substituting the solution $c = \sin\theta$ into (D.14) gives an alternative expression for the group elements:

$$R^i_{\ j} = \left\{ \begin{pmatrix} \cos\theta & -\sin\theta \\ \sin\theta & \cos\theta \end{pmatrix}, \begin{pmatrix} -\cos\theta & -\sin\theta \\ \sin\theta & -\cos\theta \end{pmatrix}, \begin{pmatrix} -\cos\theta & \sin\theta \\ \sin\theta & \cos\theta \end{pmatrix}, \begin{pmatrix} \cos\theta & \sin\theta \\ \sin\theta & -\cos\theta \end{pmatrix} \right\}. \quad \text{(D.21)}$$

The first group element is (D.19) (a counter-clockwise rotation - denoted by $R_{\uparrow}$ (clockwise will be $R_{\downarrow}$)); the second is a clockwise rotation followed by a rotation $\pi$ ($R_{\pi} \times R_{\downarrow}$, and is therefore a member of the Lie group); the third is a counter-clockwise rotation followed by a reflection about the *y*-axis, then a rotation of $\pi$ $(R_{\pi} \times R_{fy} \times R_{\downarrow})$; the fourth is a clockwise rotation followed by a reflection about the *y*-axis $(R_{fy} \times R_{\downarrow})$. Therefore, any element of (D.21) may be obtained from a minimal set consisting of an arbitrary rotation and a single reflection (in the above example the third element results from: $R_{\pi} \times R_{fy} = R_{fx}$). Hence, one choice would be

$$R^i_{\ j} = \left\{ \begin{pmatrix} \cos\theta & -\sin\theta \\ \sin\theta & \cos\theta \end{pmatrix}, \begin{pmatrix} 1 & 0 \\ 0 & -1 \end{pmatrix} \right\} \equiv \left\{ R_{\theta}, R_{fy} \right\}. \quad \text{(D.22)}$$

The base structure of $O(2)$ (defined over the real number field) is then apparent from (D.21):

$$\begin{aligned} O(2,\mathbb{R}) &= SO(2,\mathbb{R}) \oplus R_f(2) \times SO(2,\mathbb{R}) \\ &\equiv \left\{ I \oplus R_f(2) \right\} \times SO(2,\mathbb{R}) \,. \end{aligned} \quad \text{(D.23)}$$

The first two elements of (D.21) are $\in SO(2,\mathbb{R})$ while the second two are obtained as the product of a single reflection and a rotation, i.e. these elements are $\in R_f(2) \times SO(2,\mathbb{R})$, where $R_f(2)$ denotes a reflection in 2 dimensions; the Lie sub-algebra, $so(2,\mathbb{R})$, is given by (D.18).

### $O(2,\mathbb{C})$

Consider now $O(2,\mathbb{C})$, i.e. (D.14) with $c \in \mathbb{C}$. In this case, $\theta$ will have both real and imaginary components. Substituting $\theta \rightarrow \theta + \mathrm{i}\varphi$ into (D.19) gives for the Lie group $SO(2,\mathbb{C})$:

$$\begin{aligned}\begin{pmatrix}\cos(\theta+\mathrm{i}\varphi) & -\sin(\theta+\mathrm{i}\varphi)\\ \sin(\theta+\mathrm{i}\varphi) & \cos(\theta+\mathrm{i}\varphi)\end{pmatrix} &= \begin{pmatrix}\cos\theta\cosh\varphi-\mathrm{i}\sin\theta\sinh\varphi & -\sin\theta\cosh\varphi-\mathrm{i}\cos\theta\sinh\varphi\\ \sin\theta\cosh\varphi+\mathrm{i}\cos\theta\sinh\varphi & \cos\theta\cosh\varphi-\mathrm{i}\sin\theta\sinh\varphi\end{pmatrix}\\ &= \begin{pmatrix}\cos\theta & -\sin\theta\\ \sin\theta. & \cos\theta\end{pmatrix}\begin{pmatrix}\cosh\varphi & -\mathrm{i}\sinh\varphi\\ \mathrm{i}\sinh\varphi & \cosh\varphi\end{pmatrix},\end{aligned} \tag{D.24}$$

having used the identities:

$$\begin{aligned}\cos(\mathrm{i}\varphi) &= \cosh\varphi\\ \sin(\mathrm{i}\varphi) &= \mathrm{i}\sinh\varphi\\ \cos(\theta+\mathrm{i}\varphi) &= \cos\theta\cosh\varphi-\mathrm{i}\sin\theta\sinh\varphi\\ \sin(\theta+\mathrm{i}\varphi) &= \sin\theta\cosh\varphi+\mathrm{i}\cos\theta\sinh\varphi\ .\end{aligned} \tag{D.25}$$

Therefore the Lie subgroup of $O(2,\mathrm{C})$ is a 2-parameter group that may be factored as the product of 2 single parameter groups $SO(2,\mathbb{R})$ and $SO(2,\mathrm{i})$, noting that the secondary element of (D.24):

$$\begin{pmatrix}\cosh\varphi & -\mathrm{i}\sinh\varphi\\ \mathrm{i}\sinh\varphi & \cosh\varphi\end{pmatrix}, \tag{D.26}$$

is an element of $SO(2,\mathrm{i})$. Similarly, this second element of (D.21) factors as:

$$\begin{pmatrix}-\cos(\theta+\mathrm{i}\varphi) & -\sin(\theta+\mathrm{i}\varphi)\\ \sin(\theta+\mathrm{i}\varphi) & -\cos(\theta+\mathrm{i}\varphi)\end{pmatrix} = \begin{pmatrix}-\cos\theta & -\sin\theta\\ \sin\theta & -\cos\theta\end{pmatrix}\begin{pmatrix}\cosh\varphi & \mathrm{i}\sinh\varphi\\ -\mathrm{i}\sinh\varphi & \cosh\varphi\end{pmatrix}, \tag{D.27}$$

and for the third and fourth elements, respectively:

$$\begin{pmatrix}-\cos(\theta+\mathrm{i}\varphi) & \sin(\theta+\mathrm{i}\varphi)\\ \sin(\theta+\mathrm{i}\varphi) & \cos(\theta+\mathrm{i}\varphi)\end{pmatrix} = \begin{pmatrix}-\cos\theta & \sin\theta\\ \sin\theta & \cos\theta\end{pmatrix}\begin{pmatrix}\cosh\varphi & -\mathrm{i}\sinh\varphi\\ \mathrm{i}\sinh\varphi & \cosh\varphi\end{pmatrix}, \tag{D.28}$$

$$\begin{pmatrix}\cos(\theta+\mathrm{i}\varphi) & \sin(\theta+\mathrm{i}\varphi)\\ \sin(\theta+\mathrm{i}\varphi) & -\cos(\theta+\mathrm{i}\varphi)\end{pmatrix} = \begin{pmatrix}\cos\theta & \sin\theta\\ \sin\theta & -\cos\theta\end{pmatrix}\begin{pmatrix}\cosh\varphi & \mathrm{i}\sinh\varphi\\ -\mathrm{i}\sinh\varphi & \cosh\varphi\end{pmatrix}. \tag{D.29}$$

Factoring the elements of $O(2,\mathbb{R})$ in this manner, it is apparent that the group relation (D.23) generalizes in this case to

$$O(2,\mathrm{C}) = SO(2,\mathbb{R})\times SO(2,\mathrm{i}) \oplus R_f(2)\times SO(2,\mathbb{R})\times SO(2,\mathrm{i})\,, \tag{D.30}$$

or simply

$$O(2,\mathrm{C}) = O(2,\mathbb{R})\times SO(2,\mathrm{i})\,. \tag{D.31}$$

$O(1,1)$

For the group $O(1,1)$ consider

$$R^i{}_j = \begin{pmatrix}a & b\\ c & d\end{pmatrix}, \tag{D.32}$$

but in this case the metric is defined

$$g_{ij} = \begin{pmatrix} 1 & 0 \\ 0 & -1 \end{pmatrix}, \tag{D.33}$$

and the adjoint is not simply the transpose as it was for $O(2)$, but rather:

$$R^{k}_{\ i'} = g_{i'j'} R_{j}^{\ j'} g^{kj} = \begin{pmatrix} a & -c \\ -b & d \end{pmatrix}. \tag{D.34}$$

The system (D.13) is then

$$\begin{pmatrix} a & b \\ c & d \end{pmatrix}\begin{pmatrix} a & -c \\ -b & d \end{pmatrix} = \begin{pmatrix} a^2 - b^2 & bd - ac \\ ac - bd & d^2 - c^2 \end{pmatrix} = \begin{pmatrix} 1 & 0 \\ 0 & 1 \end{pmatrix}, \tag{D.35}$$

a system of 3 equations in 4 unknowns, and therefore the solution will again have only 1 independent parameter. The general solution satisfying (D.35) gives 4 possible elements for the $R^{i}_{\ j}$ (expressed arbitrarily in terms of the parameter *c*):

$$R^{i}_{\ j} = \left\{ \begin{pmatrix} \sqrt{1-c^2} & c \\ c & \sqrt{1-c^2} \end{pmatrix}, \begin{pmatrix} -\sqrt{1-c^2} & c \\ c & -\sqrt{1-c^2} \end{pmatrix}, \begin{pmatrix} -\sqrt{1-c^2} & -c \\ c & +\sqrt{1-c^2} \end{pmatrix}, \begin{pmatrix} +\sqrt{1-c^2} & -c \\ c & -\sqrt{1-c^2} \end{pmatrix} \right\}. \tag{D.36}$$

The first two correspond to transformations with $\det R^{i}_{\ j} = +1$, i.e. "special" transformations or $SO(1,1)$, while the second two correspond to $\det R^{i}_{\ j} = -1$.

$O(1,1,\mathbb{R})$

To construct the Lie subgroup $SO(1,1,\mathbb{R})$ of $O(1,1,\mathbb{R})$, consider an exponential expansion parametrized in terms of *c* and the undetermined generator $L^{i}_{\ j}$:

$$R^{i}_{\ j} = e^{c L^{i}_{\ j}}. \tag{D.37}$$

Series expanding both sides in the parameter *c* about $c = 0$ to first order gives

$$R^{i}_{\ j}\big|_{c=0} + \partial_c R^{i}_{\ j}\big|_{c=0} c = \delta^{i}_{\ j} + c L^{i}_{\ j}, \tag{D.38}$$

and then substituting explicitly into (D.38) for the first solution of (D.36) results in the system:

$$\begin{pmatrix} 0 & 1 \\ 1 & 0 \end{pmatrix} c = \begin{pmatrix} f & h \\ j & k \end{pmatrix} c, \tag{D.39}$$

assuming undetermined elements, $\{f, h, j, k\}$, for the $L^i{}_j$. The solution is thus:

$$L^i{}_j = \begin{pmatrix} 0 & 1 \\ 1 & 0 \end{pmatrix}. \tag{D.40}$$

Substituting into (D.37) (series expanding and changing $c \to \theta$) gives the result:

$$R^i{}_j = \begin{pmatrix} 1 + \frac{\theta^2}{2!} + \frac{\theta^4}{4!} + \ldots & \theta + \frac{\theta^3}{3!} + \frac{\theta^5}{5!} + \ldots \\ \theta + \frac{\theta^3}{3!} + \frac{\theta^5}{5!} + \ldots & 1 + \frac{\theta^2}{2!} + \frac{\theta^4}{4!} + \ldots \end{pmatrix} = \begin{pmatrix} \cosh\theta & \sinh\theta \\ \sinh\theta & \cosh\theta \end{pmatrix}. \tag{D.41}$$

For the second element of (D.36), (D.38) becomes:

$$\begin{pmatrix} -1 & 0 \\ 0 & -1 \end{pmatrix} + \begin{pmatrix} 0 & 1 \\ 1 & 0 \end{pmatrix} c = \begin{pmatrix} 1 & 0 \\ 0 & 1 \end{pmatrix} + \begin{pmatrix} f & h \\ j & k \end{pmatrix} c, \tag{D.42}$$

again showing that no Lie algebra exists here because the solution is "discontinuously" connected to the identity. Similar results are obtained for the third and fourth elements of (D.36), therefore the entire Lie algebra is given by (D.40), corresponding to the Lie subgroup (D.41) for $O(1,1,\mathbb{R})$.

Substituting the solution $c = \sinh\theta$ into (D.36) gives an alternative expression for the group elements:

$$R^i{}_j = \left\{ \begin{pmatrix} \cosh\theta & \sinh\theta \\ \sinh\theta & \cosh\theta \end{pmatrix}, \begin{pmatrix} -\cosh\theta & \sinh\theta \\ \sinh\theta & -\cosh\theta \end{pmatrix}, \right.$$
$$\left. \begin{pmatrix} -\cosh\theta & -\sinh\theta \\ \sinh\theta & \cosh\theta \end{pmatrix}, \begin{pmatrix} \cosh\theta & -\sinh\theta \\ \sinh\theta & -\cosh\theta \end{pmatrix} \right\}. \tag{D.43}$$

For purposes of illustration, the significance of each element in (D.43) may be understood within the context of a 2-d Minkowski spacetime (Schouten [83], p. 43). Equation (D.44) provides the (indefinite) metric for this space with coordinates $(x^0, x^1) \equiv (ct, x)$, and scalar product:

$$x_i x^i \equiv g_{ij} x^i x^j = (x^0)^2 - (x^1)^2. \tag{D.44}$$

The first element is a "boost" or transformation to an inertial frame moving with velocity $v$ in the -$x$ direction ($\Lambda_{-x}$) (the boost parameter $\theta$ is related to $v$ through: $\cosh\varphi = \gamma = 1/\sqrt{1-\beta}$; $\sinh\varphi = \gamma\beta$; $\beta = \frac{v}{c}$); the second is a boost in the +$x$ direction followed by both space ($P$) and time ($T$) inversions ($P \times T \times \Lambda_{+x}$); the third is a boost in the -$x$ direction followed by a spatial reflection $P \times \Lambda_{-x}$; while the fourth is a boost in the +$x$ direction followed by a time inversion $T \times \Lambda_{+x}$.

The 4 elements of (D.43) are discontinuous in the sense that any 2 group elements may not be connected by continuously varying the group parameter $\theta$ - but rather must be obtained from the following separate discrete transformations: $P \times T \times$, $P \times$, $T \times$, times an arbitrary boost. Hence, the base structure of $O(1,1,\mathbb{R})$ is apparent from the combinations that must be used to obtain (D.43):

$$\begin{aligned} O(1,1,\mathbb{R}) &= SO(1,1,\mathbb{R}) \oplus P \times T \times SO(1,1,\mathbb{R}) \\ &\quad \oplus P \times SO(1,1,\mathbb{R}) \oplus T \times SO(1,1,\mathbb{R}) \\ &\equiv \{I \oplus P \oplus T \oplus P \times T\} \times SO(1,1,\mathbb{R}) \, . \end{aligned} \tag{D.45}$$

An example of a minimal set would therefore be given by:

$$\begin{aligned} R^i{}_j &= \left\{ \begin{pmatrix} \cosh\theta & \sinh\theta \\ \sinh\theta & \cosh\theta \end{pmatrix}, \begin{pmatrix} -1 & 0 \\ 0 & -1 \end{pmatrix}, \begin{pmatrix} -1 & 0 \\ 0 & +1 \end{pmatrix}, \begin{pmatrix} +1 & 0 \\ 0 & -1 \end{pmatrix} \right\} \\ &\equiv \{\Lambda_x, P \times T, T, P\} \, , \end{aligned} \tag{D.46}$$

assuming $\theta$ arbitrary.

## $O(1,1,\mathrm{C})$

Consider $O(1,1,\mathrm{C})$, i.e. (D.36) with $c \in \mathrm{C}$. Substituting $\theta \to \theta + \mathrm{i}\varphi$ into (D.43) gives the corresponding Lie group element:

$$\begin{pmatrix} \cosh(\theta + \mathrm{i}\varphi) & \sinh(\theta + \mathrm{i}\varphi) \\ \sinh(\theta + \mathrm{i}\varphi) & \cosh(\theta + \mathrm{i}\varphi) \end{pmatrix} = \begin{pmatrix} \cosh\theta & \sinh\theta \\ \sinh\theta & \cosh\theta \end{pmatrix} \begin{pmatrix} \cos\varphi & \mathrm{i}\sin\varphi \\ \mathrm{i}\sin\varphi & \cos\varphi \end{pmatrix}, \tag{D.47}$$

having used the identities:

$$\begin{aligned} \cosh(\mathrm{i}\varphi) &= \cos\varphi \\ \sinh(\mathrm{i}\varphi) &= \mathrm{i}\sin\varphi \\ \cosh(\theta + \mathrm{i}\varphi) &= \cosh\theta\cos\varphi + \mathrm{i}\sinh\theta\sin\varphi \\ \sinh(\theta + \mathrm{i}\varphi) &= \sinh\theta\cos\varphi + \mathrm{i}\cosh\theta\sin\varphi \, . \end{aligned} \tag{D.48}$$

Therefore the Lie subgroup of $O(1,1,\mathrm{C})$ is a 2-parameter group that may be factored as the product of 2 single parameter groups $SO(1,1,\mathbb{R})$ and $SO(1,1,\mathrm{i})$, noting that the secondary element of (D.47):

$$\begin{pmatrix} \cos\varphi & \mathrm{i}\sin\varphi \\ \mathrm{i}\sin\varphi & \cos\varphi \end{pmatrix}, \tag{D.49}$$

gives an element of $SO(1,1,\mathrm{i})$. Similarly, the second element of (D.43) factors as:

$$\begin{pmatrix} -\cosh(\theta+\mathrm{i}\varphi) & \sinh(\theta+\mathrm{i}\varphi) \\ \sinh(\theta+\mathrm{i}\varphi) & -\cosh(\theta+\mathrm{i}\varphi) \end{pmatrix} = \begin{pmatrix} -\cosh\theta & \sinh\theta \\ \sinh\theta & -\cosh\theta \end{pmatrix}\begin{pmatrix} \cos\varphi & \mathrm{i}\sin\varphi \\ \mathrm{i}\sin\varphi & \cos\varphi \end{pmatrix}, \tag{D.50}$$

and for the third and fourth elements, respectively:

$$\begin{pmatrix} -\cosh(\theta+\mathrm{i}\varphi) & -\sinh(\theta+\mathrm{i}\varphi) \\ \sinh(\theta+\mathrm{i}\varphi) & \cosh(\theta+\mathrm{i}\varphi) \end{pmatrix} = \begin{pmatrix} -\cosh\theta & -\sinh\theta \\ \sinh\theta & \cosh\theta \end{pmatrix}\begin{pmatrix} \cos\varphi & \mathrm{i}\sin\varphi \\ \mathrm{i}\sin\varphi & \cos\varphi \end{pmatrix}, \tag{D.51}$$

$$\begin{pmatrix} \cosh(\theta+\mathrm{i}\varphi) & -\sinh(\theta+\mathrm{i}\varphi) \\ \sinh(\theta+\mathrm{i}\varphi) & -\cosh(\theta+\mathrm{i}\varphi) \end{pmatrix} = \begin{pmatrix} \cosh\theta & -\sinh\theta \\ \sinh\theta & -\cosh\theta \end{pmatrix}\begin{pmatrix} \cos\varphi & \mathrm{i}\sin\varphi \\ \mathrm{i}\sin\varphi & \cos\varphi \end{pmatrix}, \tag{D.52}$$

Hence, it is apparent that the group relation (D.45) generalizes in this case to

$$O(1,1,\mathrm{C}) = O(1,1,\mathbb{R}) \times SO(1,1,\mathrm{i}) . \tag{D.53}$$

# Appendix E: Mathematica Calculations - Chapter 4

## Conformal Transformations

*Initialize MathTensor*

```
<< Mathtens.m
```

*Conformal Transformation of the Metric:*

```
RuleUnique[MetricConformalRule1, Metricg[a_, b_],
 λ^2 Metricg[a, b], LowerIndexQ[a] && LowerIndexQ[b]]
ApplyRules[Metricg[la, lb], MetricConformalRule1]
```

$\lambda^2 (g_{ab})$

```
RuleUnique[MetricConformalRule2, Metricg[a_, b_],
 λ^-2 Metricg[a, b], UpperIndexQ[a] && UpperIndexQ[b]]
ApplyRules[Metricg[ua, ub], MetricConformalRule2]
```

$\frac{g^{ab}}{\lambda^2}$

*Conformal Transformation of Connection:*

```
Off[MetricgFlag]
rhs = AffineG[ua, lb, lc] + λ^-1
   (Metricg[ua, lb] OD[λ, lc] + Metricg[ua, lc] OD[λ, l
    Metricg[lb, lc] Metricg[ua, ud] OD[λ, ld])
```

$G^a{}_{bc} + \frac{(g_c{}^a)(\lambda_{,b}) + (g_b{}^a)(\lambda_{,c}) - (g_{bc})(g^{ad})(\lambda_{,d})}{\lambda}$

```
RuleUnique[
 AffineConformalRule, AffineG[ua_, lb_, lc_], rhs,
 UpperIndexQ[ua] && LowerIndexQ[lb] && LowerIndexQ[lc]
ApplyRules[AffineG[ua, lb, lc], AffineConformalRule]
```

$G^a{}_{bc} - \frac{(g_{bc})(g^{pa})(\lambda_{,p})}{\lambda} + \frac{(g_c{}^a)(\lambda_{,b})}{\lambda} + \frac{(g_b{}^a)(\lambda_{,c})}{\lambda}$

*Make Rules to Express in Normal Coordinates:*

```
AffineZeroRule :=
 {OD[AffineG[a_, b_, c_], d_] -> OD[AffineG[a, b, c], d
  AffineG[a_, b_, c_] -> 0}
ApplyRules[OD[AffineG[ua, lb, lc], ld], AffineZeroRule
```

$G^a{}_{bc,d}$

```
MetricDerivZeroRule := {OD[Metricg[a_, b_], c_] -> 0}
NormCoordinateRule :=
 Flatten[{AffineZeroRule, MetricDerivZeroRule}]
```

*Apply this to the Riemann Tensor:*

```
ApplyRules[AffineG[ua, lb, lc], AffineConformalRule]
```

$G^a{}_{bc} - \frac{(g_{bc})(g^{pa})(\lambda_{,p})}{\lambda} + \frac{(g_c{}^a)(\lambda_{,b})}{\lambda} + \frac{(g_b{}^a)(\lambda_{,c})}{\lambda}$

```
ApplyRules[RiemannR[ua, lb, lc, ld], RiemannToAffineR
```

$(G^p{}_{bd})(G^a{}_{pc}) - (G^p{}_{bc})(G^a{}_{pd}) - G^a{}_{bc,d} + G^a{}_{bd,c}$

```
Expand[ApplyRules[%, AffineConformalRule]]
ApplyRules[%, NormCoordinateRule]
```

$$-\frac{(g_{bd})\ (g_c{}^a)\ (g^{pq})\ (\lambda_{,p})\ (\lambda_{,q})}{\lambda^2}+\frac{(g_{bc})\ (g_d{}^a)\ (g^{pq})\ (\lambda_{,p})\ (\lambda_{,}}{\lambda^2}$$

$$\frac{(g_{pd})\ (g_{bc})\ (g^{pq})\ (g^{ra})\ (\lambda_{,q})\ (\lambda_{,r})}{\lambda^2}+$$

$$\frac{(g_{pc})\ (g_{bd})\ (g^{pq})\ (g^{ra})\ (\lambda_{,q})\ (\lambda_{,r})}{\lambda^2}+\frac{(g_{bd})\ (g^{pa})\ (\lambda_{,p})\ (\lambda_{,}}{\lambda^2}$$

$$\frac{2\ (g_d{}^a)\ (\lambda_{,b})\ (\lambda_{,c})}{\lambda^2}-\frac{(g_{bc})\ (g^{pa})\ (\lambda_{,p})\ (\lambda_{,d})}{\lambda^2}+$$

$$\frac{2\ (g_c{}^a)\ (\lambda_{,b})\ (\lambda_{,d})}{\lambda^2}-G^a{}_{bc,d}+G^a{}_{bd,c}-\frac{(g_{bd})\ (g^{pa})\ (\lambda_{,pc})}{\lambda}+$$

$$\frac{(g_{bc})\ (g^{pa})\ (\lambda_{,pd})}{\lambda}+\frac{(g_d{}^a)\ (\lambda_{,bc})}{\lambda}-\frac{(g_c{}^a)\ (\lambda_{,bd})}{\lambda}$$

```
On[MetricgFlag]
Dum[%%]
```

$$-\frac{2\ (g_d{}^a)\ (\lambda_{,b})\ (\lambda_{,c})}{\lambda^2}+\frac{2\ (g_c{}^a)\ (\lambda_{,b})\ (\lambda_{,d})}{\lambda^2}-$$

$$\frac{(g_{bd})\ (g_c{}^a)\ (\lambda_{,p})\ (\lambda_{,}{}^{p})}{\lambda^2}+\frac{(g_{bc})\ (g_d{}^a)\ (\lambda_{,p})\ (\lambda_{,}{}^{p})}{\lambda^2}+$$

$$\frac{2\ (g_{bd})\ (\lambda_{,c})\ (\lambda_{,}{}^{a})}{\lambda^2}-\frac{2\ (g_{bc})\ (\lambda_{,d})\ (\lambda_{,}{}^{a})}{\lambda^2}-G^a{}_{bc,d}+G^a{}_{bd,c}+$$

$$\frac{(g_d{}^a)\ (\lambda_{,bc})}{\lambda}-\frac{(g_c{}^a)\ (\lambda_{,bd})}{\lambda}-\frac{(g_{bd})\ (\lambda_{,c}{}^{a})}{\lambda}+\frac{(g_{bc})\ (\lambda_{,d}{}^{a})}{\lambda}$$

```
% - %[[7]] - %[[8]] + RiemannR[ua, lb, lc, ld]
```

$$-\frac{2\ (g_d{}^a)\ (\lambda_{,b})\ (\lambda_{,c})}{\lambda^2}+\frac{2\ (g_c{}^a)\ (\lambda_{,b})\ (\lambda_{,d})}{\lambda^2}-$$

$$\frac{(g_{bd})\ (g_c{}^a)\ (\lambda_{,p})\ (\lambda_{,}{}^{p})}{\lambda^2}+\frac{(g_{bc})\ (g_d{}^a)\ (\lambda_{,p})\ (\lambda_{,}{}^{p})}{\lambda^2}+$$

$$\frac{2\ (g_{bd})\ (\lambda_{,c})\ (\lambda_{,}{}^{a})}{\lambda^2}-\frac{2\ (g_{bc})\ (\lambda_{,d})\ (\lambda_{,}{}^{a})}{\lambda^2}+\frac{(g_d{}^a)\ (\lambda_{,bc})}{\lambda}-$$

$$\frac{(g_c{}^a)\ (\lambda_{,bd})}{\lambda}-\frac{(g_{bd})\ (\lambda_{,c}{}^{a})}{\lambda}+\frac{(g_{bc})\ (\lambda_{,d}{}^{a})}{\lambda}-R_b{}^a{}_{cd}$$

```
% /. OD -> CD
```

$$-\frac{2\ (\lambda_{;d})\ (\lambda_{;}{}^{a})\ (g_{bc})}{\lambda^2}+\frac{(\lambda_{;d}{}^{a})\ (g_{bc})}{\lambda}+\frac{2\ (\lambda_{;c})\ (\lambda_{;}{}^{a})\ (g_{bd})}{\lambda^2}-$$

$$\frac{(\lambda_{;c}{}^{a})\ (g_{bd})}{\lambda}+\frac{2\ (\lambda_{;b})\ (\lambda_{;d})\ (g_c{}^a)}{\lambda^2}-\frac{(\lambda_{;bd})\ (g_c{}^a)}{\lambda}-$$

$$\frac{(\lambda_{;p})\ (\lambda_{;}{}^{p})\ (g_{bd})\ (g_c{}^a)}{\lambda^2}-\frac{2\ (\lambda_{;b})\ (\lambda_{;c})\ (g_d{}^a)}{\lambda^2}+$$

$$\frac{(\lambda_{;bc})\ (g_d{}^a)}{\lambda}+\frac{(\lambda_{;p})\ (\lambda_{;}{}^{p})\ (g_{bc})\ (g_d{}^a)}{\lambda^2}-R_b{}^a{}_{cd}$$

```
Collect[%, λ]
```

$$\frac{(\lambda_{;d}{}^{a})\ (g_{bc})-(\lambda_{;c}{}^{a})\ (g_{bd})-(\lambda_{;bd})\ (g_c{}^a)+(\lambda_{;bc})\ (g_d{}^a)}{\lambda}+$$

$$\frac{1}{\lambda^2}\ (-2\ (\lambda_{;d})\ (\lambda_{;}{}^{a})\ (g_{bc})+2\ (\lambda_{;c})\ (\lambda_{;}{}^{a})\ (g_{bd})+$$

$$2\ (\lambda_{;b})\ (\lambda_{;d})\ (g_c{}^a)-(\lambda_{;p})\ (\lambda_{;}{}^{p})\ (g_{bd})\ (g_c{}^a)-$$

$$2\ (\lambda_{;b})\ (\lambda_{;c})\ (g_d{}^a)+(\lambda_{;p})\ (\lambda_{;}{}^{p})\ (g_{bc})\ (g_d{}^a))-$$

$$R_b{}^a{}_{cd}$$

```
RuleUnique[RiemannConformalRule,
 RiemannR[ua_, lb_, lc_, ld_], %]
ApplyRules[RiemannR[ua, lb, lc, ld],
 RiemannConformalRule]
```

$$-\frac{2\ (\lambda_{;d})\ (\lambda_{;}{}^{a})\ (g_{bc})}{\lambda^2}+\frac{(\lambda_{;d}{}^{a})\ (g_{bc})}{\lambda}+\frac{2\ (\lambda_{;c})\ (\lambda_{;}{}^{a})\ (g_{bd})}{\lambda^2}-$$

$$\frac{(\lambda_{;c}{}^{a})\ (g_{bd})}{\lambda}+\frac{2\ (\lambda_{;b})\ (\lambda_{;d})\ (g_c{}^a)}{\lambda^2}-\frac{(\lambda_{;bd})\ (g_c{}^a)}{\lambda}-$$

$$\frac{(\lambda_{;p})\ (\lambda_{;}{}^{p})\ (g_{bd})\ (g_c{}^a)}{\lambda^2}-\frac{2\ (\lambda_{;b})\ (\lambda_{;c})\ (g_d{}^a)}{\lambda^2}+$$

$$\frac{(\lambda_{;bc})\ (g_d{}^a)}{\lambda}+\frac{(\lambda_{;p})\ (\lambda_{;}{}^{p})\ (g_{bc})\ (g_d{}^a)}{\lambda^2}-R_b{}^a{}_{cd}$$

```
Dimension = 4;
ApplyRules[Metricg[uc, ui] Metricg[ud, uj],
  MetricConformalRule2] ApplyRules[
  RiemannR[ua, lb, lc, ld], RiemannConformalRule]
 ApplyRules[RiemannR[ub, la, li, lj],
  RiemannConformalRule]
CanAll[Expand[%]]
```

$$\frac{4R(\lambda_{;p})(\lambda_{;}{}^{p})}{\lambda^6} - \frac{24(\lambda_{;p})^2(\lambda_{;}{}^{p})^2}{\lambda^8} - \frac{8(\lambda_{;p})(\lambda_{;}{}^{p})(\lambda_{;q}{}^{q})}{\lambda^7} - \frac{4(\lambda_{;p}{}^{p})(\lambda_{;q}{}^{q})}{\lambda^6} + \frac{32(\lambda_{;p})(\lambda_{;q})(\lambda_{;}{}^{pq})}{\lambda^7} - \frac{8(\lambda_{;pq})(\lambda_{;}{}^{pq})}{\lambda^6} - \frac{16(\lambda_{;p})(\lambda_{;q})(R^{pq})}{\lambda^6} + \frac{8(\lambda_{;pq})(R^{pq})}{\lambda^5} - \frac{(R_{pqrs})(R^{pqrs})}{\lambda^4}$$

```
Collect[-%%%, λ]
```

$$\frac{24(\lambda_{;p})^2(\lambda_{;}{}^{p})^2}{\lambda^8} + \frac{8(\lambda_{;p})(\lambda_{;}{}^{p})(\lambda_{;q}{}^{q}) - 32(\lambda_{;p})(\lambda_{;q})(\lambda_{;}{}^{p}}{\lambda^7}$$

$$\frac{8(\lambda_{;pq})(R^{pq})}{\lambda^5} + \frac{1}{\lambda^6}\big(-4R(\lambda_{;p})(\lambda_{;}{}^{p}) + 4(\lambda_{;p}{}^{p})(\lambda_{;q}{}^{q}) + 8(\lambda_{;pq})(\lambda_{;}{}^{pq}) + 16(\lambda_{;p})(\lambda_{;q})(R^{pq})\big) + \frac{(R_{pqrs})(R^{pqrs})}{\lambda^4}$$

*Construct the Ricci Tensor:*

```
Expand[ApplyRules[
   RiemannR[ua, lb, lc, ld], RiemannConformalRule]
  Metricg[uc, la]]
```

$$\frac{4(\lambda_{;b})(\lambda_{;d})}{\lambda^2} - \frac{2(\lambda_{;bd})}{\lambda} - \frac{3(\lambda_{;p})(\lambda_{;}{}^{p})(g_{bd})}{\lambda^2} + \frac{2(\lambda_{;a})(\lambda_{;}{}^{a})(g_{bd})}{\lambda^2} - \frac{(\lambda_{;a}{}^{a})(g_{bd})}{\lambda} + R_{bd}$$

```
Collect[Tsimplify[%], λ]
```

$$\frac{4(\lambda_{;b})(\lambda_{;d}) - (\lambda_{;p})(\lambda_{;}{}^{p})(g_{bd})}{\lambda^2} + \frac{-2(\lambda_{;bd}) - (\lambda_{;a}{}^{a})(g_{bd})}{\lambda}$$

*Apply this to Ricci Tensor:*

```
ApplyRules[RicciR[la, lb], RicciToAffineRule]
```

$$(G^{p}{}_{ab})(G^{q}{}_{pq}) - (G^{p}{}_{qa})(G^{q}{}_{pb}) - G^{p}{}_{pa,b} + G^{p}{}_{ab,p}$$

```
Expand[ApplyRules[%, AffineConformalRule]]
ApplyRules[%, NormCoordinateRule]
```

$$\frac{2(g_{ab})(g^{pq})(\lambda_{,p})(\lambda_{,q})}{\lambda^2} - \frac{(g_{p}{}^{p})(g_{ab})(g^{qr})(\lambda_{,q})(\lambda_{,r})}{\lambda^2} - \frac{(g_{pb})(g_{qa})(g^{pr})(g^{qs})(\lambda_{,r})(\lambda_{,s})}{\lambda^2} + \frac{(g_{pq})(g_{ab})(g^{pr})(g^{qs})(\lambda_{,r})(\lambda_{,s})}{\lambda^2} - \frac{(g_{pa})(g^{pq})(\lambda_{,q})(\lambda_{,}}{\lambda^2}$$

$$\frac{2(\lambda_{,a})(\lambda_{,b})}{\lambda^2} + \frac{2(g_{p}{}^{p})(\lambda_{,a})(\lambda_{,b})}{\lambda^2} - G^{p}{}_{pa,b} + G^{p}{}_{ab,p} - \frac{(g_{ab})(g^{pq})(\lambda_{,pq})}{\lambda} + \frac{(g_{pa})(g^{pq})(\lambda_{,qb})}{\lambda} + \frac{\lambda_{,ab}}{\lambda} - \frac{(g_{p}{}^{p})(\lambda_{,}}{\lambda}$$

```
On[MetricgFlag]
Dum[%%]
```

$$\frac{4(\lambda_{,a})(\lambda_{,b})}{\lambda^2} - \frac{(g_{ab})(\lambda_{,p})(\lambda_{,}{}^{p})}{\lambda^2} - G^{p}{}_{pa,b} + G^{p}{}_{ab,p} - \frac{(g_{ab})(\lambda_{,p}{}^{p})}{\lambda} - \frac{2(\lambda_{,ab})}{\lambda}$$

```
% - %[[3]] - %[[4]] + RicciR[la, lb]
```

$$\frac{4\,(\lambda_{,a})\,(\lambda_{,b})}{\lambda^2} - \frac{(g_{ab})\,(\lambda_{,p})\,(\lambda^{,p})}{\lambda^2} - \frac{(g_{ab})\,(\lambda_{,p}{}^{p})}{\lambda} - \frac{2\,(\lambda_{,ab})}{\lambda} + R_{ab}$$

```
% /. OD -> CD
```

$$\frac{4\,(\lambda_{;a})\,(\lambda_{;b})}{\lambda^2} - \frac{2\,(\lambda_{;ab})}{\lambda} - \frac{(\lambda_{;p})\,(\lambda^{;p})\,(g_{ab})}{\lambda^2} - \frac{(\lambda_{;p}{}^{p})\,(g_{ab})}{\lambda} R_{ab}$$

```
Collect[Tsimplify[%], λ]
```

$$\frac{4\,(\lambda_{;a})\,(\lambda_{;b}) - (\lambda_{;p})\,(\lambda^{;p})\,(g_{ab})}{\lambda^2} + \frac{-2\,(\lambda_{;ab}) - (\lambda_{;p}{}^{p})\,(g_{ab})}{\lambda}$$

```
RuleUnique[RicciRConformalRule, RicciR[la_, lb_],
 %, LowerIndexQ[la] && LowerIndexQ[lb]]
ApplyRules[RicciR[la, lb], RicciRConformalRule]
```

$$\frac{4\,(\lambda_{;a})\,(\lambda_{;b})}{\lambda^2} - \frac{2\,(\lambda_{;ab})}{\lambda} - \frac{(\lambda_{;p})\,(\lambda^{;p})\,(g_{ab})}{\lambda^2} - \frac{(\lambda_{;p}{}^{p})\,(g_{ab})}{\lambda} R_{ab}$$

*Ricci Conformal Rule2:*

```
ApplyRules[
  Metricg[uc, ua] Metricg[ud, ub], MetricConformalRul(
 ApplyRules[RicciR[la, lb] , RicciRConformalRule]
 ApplyRules[ RicciR[lc, ld], RicciRConformalRule]
```

$$\frac{1}{\lambda^4}\left((g^{ac})\,(g^{bd})\left(\frac{4\,(\lambda_{;a})\,(\lambda_{;b})}{\lambda^2} - \frac{2\,(\lambda_{;ab})}{\lambda} - \frac{(\lambda_{;p})\,(\lambda^{;p})\,(g_{ab})}{\lambda^2} - \frac{(\lambda_{;p}{}^{p})\,(g_{ab})}{\lambda} + R_{ab}\right)\left(\frac{4\,(\lambda_{;c})\,(\lambda_{;d})}{\lambda^2} - \frac{2\,(\lambda_{;cd})}{\lambda} - \frac{(\lambda_{;p})\,(\lambda^{;p})\,(g_{cd})}{\lambda^2} - \frac{(\lambda_{;p}{}^{p})\,(g_{cd})}{\lambda} + R_{cd}\right)\right)$$

```
CanAll[Expand[%]]
Collect[%, λ]
```

$$\frac{4\,(\lambda_{;p})^2\,(\lambda^{;p})^2 + 8\,(\lambda_{;p})\,(\lambda_{;q})\,(\lambda^{;p})\,(\lambda^{;q})}{\lambda^8} + \frac{1}{\lambda^7}\left(8\,(\lambda_{;p})\,(\lambda_{;q})\,(\lambda_{;pq}) - 4\,(\lambda_{;p})\,(\lambda^{;p})\,(\lambda_{;q}{}^{q}) - 16\,(\lambda_{;p})\,(\lambda_{;q})\,(\lambda^{;pq})\right) + \frac{(R_{pq})\,(R^{pq})}{\lambda^4} + \frac{1}{\lambda^6}\left(-2\,R\,(\lambda_{;p})\,(\lambda^{;p}) + 4\,(\lambda_{;p}{}^{p})^2 + 4\,(\lambda_{;p}{}^{p})\,(\lambda_{;q}{}^{q}) + 4\,(\lambda_{;pq})\,(\lambda^{;pq}) + 8\,(\lambda_{;p})\,(\lambda_{;q})\,(R^{pq})\right) + \frac{-2\,R\,(\lambda_{;p}{}^{p}) - 4\,(\lambda_{;pq})\,(R^{pq})}{\lambda^5}$$

*Construct the Ricci Scalar:*

```
ApplyRules[Metricg[ua, ub], MetricConformalRule2]
 ApplyRules[RicciR[la, lb] , RicciRConformalRule]
```

$$\frac{1}{\lambda^2}\left((g^{ab})\left(\frac{4\,(\lambda_{;a})\,(\lambda_{;b})}{\lambda^2} - \frac{2\,(\lambda_{;ab})}{\lambda} - \frac{(\lambda_{;p})\,(\lambda^{;p})\,(g_{ab})}{\lambda^2} - \frac{(\lambda_{;p}{}^{p})\,(g_{ab})}{\lambda} + R_{ab}\right)\right)$$

```
Expand[%]
Tsimplify[%]
```

$$\frac{R}{\lambda^2} - \frac{6\,(\lambda_{;p}{}^{p})}{\lambda^3}$$

```
RuleUnique[ScalarRConformalRule, ScalarR, %]
ApplyRules[ScalarR, ScalarRConformalRule]
```

$$\frac{R}{\lambda^2} - \frac{6\,(\lambda_{;p}{}^{p})}{\lambda^3}$$

**Expand[%^2]**

$$\frac{R^2}{\lambda^4} - \frac{12 R (\lambda_{;p}{}^{p})}{\lambda^5} + \frac{36 (\lambda_{;p}{}^{p})^2}{\lambda^6}$$

# Invariance of the Weyl Tensor

**WeylC[ua, lb, lc, ld] /. WeylToRiemannRule**

$$-\frac{1}{6} R (-(g_{bd}) (g_c{}^a) + (g_{bc}) (g_d{}^a)) +$$
$$\frac{1}{2} ((g_d{}^a) (R_{bc}) - (g_c{}^a) (R_{bd}) - (g_{bd}) (R_c{}^a) + (g_{bc}) (R_d{}^a)) - F$$

**ApplyRules[%87, {MetricConformalRule1,**
**ScalarRConformalRule, RiemannConformalRule,**
**RicciRConformalRule, RicciRConformalRule2}]**

$$\frac{1}{6} R (g_{bd}) (g_c{}^a) - \frac{1}{6} R (g_{bc}) (g_d{}^a) + \frac{1}{2} (g_d{}^a) (R_{bc}) -$$
$$\frac{1}{2} (g_c{}^a) (R_{bd}) - \frac{1}{2} (g_{bd}) (R_c{}^a) + \frac{1}{2} (g_{bc}) (R_d{}^a) - R_b{}^a{}_{cd}$$

*These 2 Results are the Same:*

**Expand[% - %%]**

0

# Derivation of the Conformal Field Equations

## *Conformal Quadratic Term:*

**WeylC[ua, lb, lc, ld]**

$$-(C_b{}^a{}_{cd})$$

**% /. WeylToRiemannRule**

$$-\frac{1}{6} R (-(g_{bd}) (g_c{}^a) + (g_{bc}) (g_d{}^a)) +$$
$$\frac{1}{2} ((g_d{}^a) (R_{bc}) - (g_c{}^a) (R_{bd}) - (g_{bd}) (R_c{}^a) + (g_{bc}) (R_d{}^a)) - F$$

**WeylC[ua, lb, lg, ld] WeylC[ub, la, ug, ud]**

$$(C_a{}^{bdg}) (C_b{}^a{}_{dg})$$

**CanAll[Expand[% /. WeylToRiemannRule]]**

$$-\frac{R^2}{3} + 2 (R_{pq}) (R^{pq}) - (R_{pqrs}) (R^{pqrs})$$

**Factor[%]**

$$\frac{1}{3} (R^2 - 6 (R_{pq}) (R^{pq}) + 3 (R_{pqrs}) (R^{pqrs}))$$

**Tsimplify[%]**

$$\frac{R^2}{3} - 2 (R_{pq}) (R^{pq}) + (R_{pqrs}) (R^{pqrs})$$

-------

**WeylC[ua, lb, lg, ld] WeylC[la, ub, ug, ud]**

$$-(C_a{}^{bdg}) (C_b{}^a{}_{dg})$$

**CanAll[Expand[% /. WeylToRiemannRule]]**

$$\frac{R^2}{3} - 2 (R_{pq}) (R^{pq}) + (R_{pqrs}) (R^{pqrs})$$

**Factor[%]**

$$\frac{1}{3} (R^2 - 6 (R_{pq}) (R^{pq}) + 3 (R_{pqrs}) (R^{pqrs}))$$

**Tsimplify[%]**

$$\frac{R^2}{3} - 2 (R_{pq}) (R^{pq}) + (R_{pqrs}) (R^{pqrs})$$

*A Topological-Type term based on the Weyl Tensor:*

```
1/4 Epsilon[ua, ub, uc, ud] Epsilon[li, lj, lk, ll]
 WeylC[ui, uj, la, lb] WeylC[uk, ul, lc, ld]
```

$$(\mathrm{Epsilon}_{ijkl})\ (\mathrm{Epsilon}^{abcd})\ (C_{ab}{}^{ij})\ (C_{cd}{}^{kl})$$

*Expand Levi-Civita product:*

```
Tsimplify[CanAll[Expand[% /. WeylToRiemannRule] /.
   EpsilonProductTensorRule]]
```

$$-\frac{4R^2}{3} + 8\ (R_{pq})\ (R^{pq}) - 4\ (R_{pqrs})\ (R^{pqrs})$$

```
1/4 Epsilon[ua, ub, uc, ud] Epsilon[li, lj, lk, ll]
 RiemannR[ui, uj, la, lb] RiemannR[uk, ul, lc, ld]
```

$$\frac{1}{4}\ (\mathrm{Epsilon}_{ijkl})\ (\mathrm{Epsilon}^{abcd})\ (R_{ab}{}^{ij})\ (R_{cd}{}^{kl})$$

*Expand Levi-Civita product:*

```
Tsimplify[CanAll[Expand[% /. WeylToRiemannRule] /.
   EpsilonProductTensorRule]]
```

$$-R^2 + 4\ (R_{pq})\ (R^{pq}) - (R_{pqrs})\ (R^{pqrs})$$

Perform the variation:

```
Variation[Sqrt[Detg] %, Metricg]
```

*Remove derivatives from $\delta g$ and factor det g:*

```
Expand[PIntegrate[%, Metricg] / Sqrt[Detg]]
```

*Remove derivatives from $\delta g$ :*

```
PIntegrate[%, Metricg]
```

*Factor out $\delta g$ :*

```
VariationalDerivative[%, Metricg, ua, ub]
```

$$\frac{2}{3}\ (R_{;ab}) - 2\ (R_{ra;b}{}^{r}) + 2\ (R_{ab;}{}^{r}{}_{r}) - 2\ (R_{a}{}^{r}{}_{;br}) - 2\ (R_{rasb;}{}^{sr}) -$$
$$2\ (R_{a}{}^{r}{}_{b}{}^{s}{}_{;sr}) + \frac{1}{6} R^2\ (g_{ab}) - \frac{2}{3}\ (R_{;q}{}^{q})\ (g_{ab}) + (R_{qr;}{}^{rq})\ (g_{ab}) +$$
$$(R^{qr}{}_{;rq})\ (g_{ab}) - \frac{2}{3} R\ (R_{ab}) + 2\ (R_{pa})\ (R_{b}{}^{p}) + 2\ (R_{qa})\ (R_{b}{}^{q}) -$$
$$(g_{ab})\ (R_{rs})\ (R^{rs}) - \frac{1}{2}\ (R_{pars})\ (R_{b}{}^{prs}) + \frac{1}{2}\ (R_{qars})\ (R_{b}{}^{qrs}) +$$
$$(R_{pqra})\ (R_{b}{}^{rpq}) + (R_{pqsa})\ (R_{b}{}^{spq}) + \frac{1}{2}\ (g_{ab})\ (R_{rstu})\ (R^{rstu})$$

*This is the result for $\delta C^2$:*

```
Tsimplify[Canonicalize[%]]
```

$$\frac{2}{3}\ (R_{;ab}) - 4\ (R_{pa;b}{}^{p}) + 2\ (R_{ab;p}{}^{p}) - 4\ (R_{paqb;}{}^{qp}) + \frac{1}{6} R^2\ (g_{ab})\ \cdot$$
$$\frac{2}{3}\ (R_{;p}{}^{p})\ (g_{ab}) + 2\ (R_{pq;}{}^{qp})\ (g_{ab}) - \frac{2}{3} R\ (R_{ab}) + 4\ (R_{pa})\ (R_{b}{}^{p})$$
$$(g_{ab})\ (R_{pq})\ (R^{pq}) + 2\ (R_{pqra})\ (R_{b}{}^{rpq}) + \frac{1}{2}\ (g_{ab})\ (R_{pqrs})\ (R^{pqr}$$

*Substitute in terms of the Weyl tensor:*

```
% /. RiemannToWeylRule
CanAll[Absorbg[%]]
```

*Define the trace - free property of Conformal tensor:*

```
RuleUnique[WeylTraceRule, WeylC[la_, lb_, lc_, ld_],
 0, PairQ[la, lb] || PairQ[la, lc] || PairQ[la, ld] ||
  PairQ[lb, lc] || PairQ[lb, ld] || PairQ[lc, ld]]
ApplyRules[%%, WeylTraceRule]
```

$-4\,(R_{pa;b}{}^{p}) + 2\,(R_{pa;}{}^{p}{}_{b}) + 2\,(R_{pb;a}{}^{p}) - 4\,(C_{paqb;}{}^{qp}) + \frac{1}{3}R^{2}\,(g_{ab}$

$\frac{4}{3}R\,(R_{ab}) + 4\,(R_{pa})\,(R_{b}{}^{p}) - (g_{ab})\,(R_{pq})\,(R^{pq}) - 2\,(R_{pq})\,(C_{a}{}^{p}{}_{b}{}^{q}$

$2\,(R_{pq})\,(C_{a}{}^{q}{}_{b}{}^{p}) + 2\,(C_{pqra})\,(C_{b}{}^{rpq}) + \frac{1}{2}\,(g_{ab})\,(C_{pqrs})\,(C^{pqrs})$

*Define Pirani's Conformal Identity:*

```
Conformal[lc_, lf_] =
 -CanAll[2 WeylC[ua, ub, ud, lc] WeylC[la, lb, ld, lf] -
   1/2 Metricg[lc, lf] WeylC[ua, ub, ug, ud]
    WeylC[la, lb, lg, ld]]
```

$2\,(C_{pqrc})\,(C_{f}{}^{rpq}) + \frac{1}{2}\,(g_{cf})\,(C_{pqrs})\,(C^{pqrs})$

*Use Pirani's Identity:*

```
%% /. (Conformal[la, lb]) → 0
```

$-4\,(R_{pa;b}{}^{p}) + 2\,(R_{pa;}{}^{p}{}_{b}) + 2\,(R_{pb;a}{}^{p}) - 4\,(C_{paqb;}{}^{qp}) +$

$\frac{1}{3}R^{2}\,(g_{ab}) - \frac{4}{3}R\,(R_{ab}) + 4\,(R_{pa})\,(R_{b}{}^{p}) - (g_{ab})\,(R_{pq})\,(R^{pq}) -$

$2\,(R_{pq})\,(C_{a}{}^{p}{}_{b}{}^{q}) - 2\,(R_{pq})\,(C_{a}{}^{q}{}_{b}{}^{p})$

```
%% /. WeylToRiemannRule
Tsimplify[CanAll[Absorbg[%]]]
```

$\frac{2}{3}\,(R_{;ab}) - 4\,(R_{pa;b}{}^{p}) + 2\,(R_{ab;p}{}^{p}) - 4\,(R_{paqb;}{}^{qp}) -$

$\frac{1}{3}R^{2}\,(g_{ab}) - \frac{2}{3}\,(R_{;p}{}^{p})\,(g_{ab}) + 2\,(R_{pq;}{}^{qp})\,(g_{ab}) + \frac{4}{3}R\,(R_{ab}) +$

$(g_{ab})\,(R_{pq})\,(R^{pq}) - 4\,(R_{pq})\,(R_{a}{}^{p}{}_{b}{}^{q})$

```
ApplyRulesRepeated[%, RiemannRules]
```

$\frac{2}{3}\,(R_{;ab}) - 2\,(R_{ab;p}{}^{p}) - \frac{1}{3}R^{2}\,(g_{ab}) + \frac{1}{3}\,(R_{;p}{}^{p})\,(g_{ab}) + \frac{4}{3}R\,(R$

$4\,(R_{pb})\,(R_{a}{}^{p}) - 4\,(R_{pa})\,(R_{b}{}^{p}) + (g_{ab})\,(R_{pq})\,(R^{pq}) - 4\,(R_{pq})\,(R$

*This is the final result:*

```
WeylField[la_, lb_] = Tsimplify[CanAll[Absorbg[%]]]
```

$\frac{2}{3}\,(R_{;ab}) - 2\,(R_{ab;p}{}^{p}) - \frac{1}{3}R^{2}\,(g_{ab}) + \frac{1}{3}\,(R_{;p}{}^{p})\,(g_{ab}) +$

$\frac{4}{3}R\,(R_{ab}) + (g_{ab})\,(R_{pq})\,(R^{pq}) - 4\,(R_{pq})\,(R_{a}{}^{p}{}_{b}{}^{q})$

*Check That it is Traceless:*

```
Metricg[ua, ub] WeylField[la, lb]
```

$(g^{ab})\left(\frac{2}{3}\,(R_{;ab}) - 2\,(R_{ab;p}{}^{p}) - \frac{1}{3}R^{2}\,(g_{ab}) + \frac{1}{3}\,(R_{;p}{}^{p})\,(g_{ab}) + \frac{4}{3}R\,(R_{ab}) + (g_{ab})\,(R_{pq})\,(R^{pq}) - 4\,(R_{pq})\,(R_{a}{}^{q}{}_{b}{}^{p})\right)$

```
Expand[%]
```

$\frac{2}{3}\,(R_{;ab})\,(g^{ab}) - 2\,(R_{ab;p}{}^{p})\,(g^{ab}) -$

$\frac{1}{3}R^{2}\,(g_{ab})\,(g^{ab}) + \frac{1}{3}\,(R_{;p}{}^{p})\,(g_{ab})\,(g^{ab}) + \frac{4}{3}R\,(g^{ab})\,(R_{ab}) +$

$(g_{ab})\,(g^{ab})\,(R_{pq})\,(R^{pq}) - 4\,(g^{ab})\,(R_{pq})\,(R_{a}{}^{q}{}_{b}{}^{p})$

```
Absorbg[%]
```

$-\frac{2}{3}\,(R_{;p}{}^{p}) + \frac{2}{3}\,(R_{;a}{}^{a})$

```
Tsimplify[%]
```

0

*Input the Mannheim - Kazanas Result:*

```
W1[la_, lb_] =
 Absorbg[2 Metricg[la, lb] CD[ScalarR, li, ui] -
   2 CD[ScalarR, la, lb] - 2 ScalarR RicciR[la, lb] +
   1/2 Metricg[la, lb] ScalarR^2]
```

$-2\,(R_{;ab}) + \frac{1}{2}\,R^2\,(g_{ab}) + 2\,(R_{;i}{}^{i})\,(g_{ab}) - 2\,R\,(R_{ab})$

```
W2[la_, lb_] =
 Absorbg[1/2 Metricg[la, lb] CD[ScalarR, li, ui] +
   CD[RicciR[la, lb], li, ui] -
   CD[RicciR[ui, la], lb, li] - CD[RicciR[ui, lb], la, l
   2 RicciR[ui, la] RicciR[lb, li] +
   1/2 Metricg[la, lb] RicciR[li, lj] RicciR[ui, uj]]
```

$R_{ab;i}{}^{i} - R_{a}{}^{i}{}_{;bi} - R_{b}{}^{i}{}_{;ai} + \frac{1}{2}\,(R_{;i}{}^{i})\,(g_{ab}) - 2\,(R_{a}{}^{i})\,(R_{bi}) +$

$\frac{1}{2}\,(g_{ab})\,(R_{ij})\,(R^{ij})$

```
W2[la, lb] - 1/3 W1[la, lb]
```

$R_{ab;i}{}^{i} - R_{a}{}^{i}{}_{;bi} - R_{b}{}^{i}{}_{;ai} + \frac{1}{2}\,(R_{;i}{}^{i})\,(g_{ab}) +$

$\frac{1}{3}\left(2\,(R_{;ab}) - \frac{1}{2}\,R^2\,(g_{ab}) - 2\,(R_{;i}{}^{i})\,(g_{ab}) + 2\,R\,(R_{ab})\right) -$

$2\,(R_{a}{}^{i})\,(R_{bi}) + \frac{1}{2}\,(g_{ab})\,(R_{ij})\,(R^{ij})$

*This is Mannheim (Open Questions in ...) - Eq. (15):*

```
Tsimplify[CanAll[%]]
```

$\frac{2}{3}\,(R_{;ab}) - R_{pa;b}{}^{p} - R_{pb;a}{}^{p} + R_{ab;p}{}^{p} - \frac{1}{6}\,R^2\,(g_{ab}) - \frac{1}{6}\,(R_{;p}{}^{p})\,(g_{a}$

$\frac{2}{3}\,R\,(R_{ab}) - 2\,(R_{pa})\,(R_{b}{}^{p}) + \frac{1}{2}\,(g_{ab})\,(R_{pq})\,(R^{pq})$

```
ApplyRulesRepeated[%48, RiemannRules]
```

$-\frac{1}{3}\,(R_{;ab}) + R_{ab;p}{}^{p} - \frac{1}{6}\,R^2\,(g_{ab}) - \frac{1}{6}\,(R_{;p}{}^{p})\,(g_{ab}) + \frac{2}{3}\,R\,(R_{ab})$

$(R_{pb})\,(R_{a}{}^{p}) - 3\,(R_{pa})\,(R_{b}{}^{p}) + \frac{1}{2}\,(g_{ab})\,(R_{pq})\,(R^{pq}) +$

$(R_{pq})\,(R_{a}{}^{p}{}_{b}{}^{q}) + (R_{pq})\,(R_{a}{}^{q}{}_{b}{}^{p})$

```
Tsimplify[CanAll[%]]
```

$-\frac{1}{3}\,(R_{;ab}) + R_{ab;p}{}^{p} - \frac{1}{6}\,R^2\,(g_{ab}) - \frac{1}{6}\,(R_{;p}{}^{p})\,(g_{ab}) +$

$\frac{2}{3}\,R\,(R_{ab}) - 4\,(R_{pa})\,(R_{b}{}^{p}) + \frac{1}{2}\,(g_{ab})\,(R_{pq})\,(R^{pq}) + 2\,(R_{pq})\,(R_{\cdot}$

*Check That (MK) is Traceless:*

```
Metricg[ua, ub] %
```

$(g^{ab})$

$\left(-\frac{1}{3}\,(R_{;ab}) + R_{ab;p}{}^{p} - \frac{1}{6}\,R^2\,(g_{ab}) - \frac{1}{6}\,(R_{;p}{}^{p})\,(g_{ab}) + \frac{2}{3}\,R\,(R_{a}\right.$

$\left.4\,(R_{pa})\,(R_{b}{}^{p}) + \frac{1}{2}\,(g_{ab})\,(R_{pq})\,(R^{pq}) + 2\,(R_{pq})\,(R_{a}{}^{p}{}_{b}{}^{q})\right)$

```
Expand[%]
```

$$-\frac{1}{3}(R_{;ab})(g^{ab}) + (R_{ab;p}{}^{p})(g^{ab}) - \frac{1}{6}R^{2}(g_{ab})(g^{ab}) -$$
$$\frac{1}{6}(R_{;p}{}^{p})(g_{ab})(g^{ab}) + \frac{2}{3}R(g^{ab})(R_{ab}) - 4(g^{ab})(R_{pa})(R_{b}{}^{p})$$
$$\frac{1}{2}(g_{ab})(g^{ab})(R_{pq})(R^{pq}) + 2(g^{ab})(R_{pq})(R_{a}{}^{p}{}_{b}{}^{q})$$

```
Absorbg[%]
```

$$\frac{1}{3}(R_{;p}{}^{p}) - \frac{1}{3}(R_{;a}{}^{a}) + 4(R_{pq})(R^{pq}) - 4(R_{pa})(R^{pa})$$

```
Tsimplify[%]
0
```

# Tensor Calculations for the Conformal Spherical Metric

*Evaluate Ordinary Derivatives:*

```
On[EvaluateODFlag]
```

*Load the File:*

```
<< MK.m
MetricgFlag has been turned off.
```

*Display the Metric ("4" Labels the time coordinate):*

```
Table[Metricg[-i, -j], {i, 4}, {j, 4}]
```

$$\begin{pmatrix} -\frac{1}{k3\,r - k2 - \frac{k1}{r} - k4\,r^{2} + 1} & 0 & 0 & 0 \\ 0 & -r^{2} & 0 & 0 \\ 0 & 0 & -r^{2}\sin^{2}(\text{theta}) & 0 \\ 0 & 0 & 0 & k3\,r - k2 - \frac{k1}{r} - k4\,r^{2} + \end{pmatrix}$$

*Here are the non-zero Christoffel Symbols $\Gamma^{i}{}_{jk}$ (Schwarzschild):*

```
connection = {};
Do[
   If[
     AffineG[i, -j, -k] =!= 0,
      connection = {connection,
         SubscriptBox[
       SubscriptBox[SuperscriptBox[Γ, i], j], k] ==
      AffineG[i, -j, -k] // DisplayForm},
    Continue[]],

  {i, 4}, {j, 4}, {k, 4}];
Simplify[Flatten[connection]]
```

$$\left\{\Gamma^1{}_{11} == \frac{k1 + r^2\,(k3 - 2\,k4\,r)}{2\,r\,(k1 + r\,(-1 + k2 - k3\,r + k4\,r^2))},\right.$$

$$\Gamma^1{}_{22} == k1 + r\,(-1 + k2 - k3\,r + k4\,r^2),$$

$$\Gamma^1{}_{33} == (k1 + r\,(-1 + k2 - k3\,r + k4\,r^2))\,\mathrm{Sin}[\mathrm{theta}]^2,$$

$$\Gamma^1{}_{44} == -\frac{(k1 + r^2\,(k3 - 2\,k4\,r))\,(k1 + r\,(-1 + k2 - k3\,r + k4\,r}{2\,r^3}$$

$$\Gamma^2{}_{12} == \frac{1}{r},\ \Gamma^2{}_{21} == \frac{1}{r},$$

$$\Gamma^2{}_{33} == -\mathrm{Cos}[\mathrm{theta}]\,\mathrm{Sin}[\mathrm{theta}],\ \Gamma^3{}_{13} == \frac{1}{r},$$

$$\Gamma^3{}_{23} == \mathrm{Cot}[\mathrm{theta}],\ \Gamma^3{}_{31} == \frac{1}{r},\ \Gamma^3{}_{32} == \mathrm{Cot}[\mathrm{theta}],$$

$$\Gamma^4{}_{14} == -\frac{k1 + r^2\,(k3 - 2\,k4\,r)}{2\,r\,(k1 + r\,(-1 + k2 - k3\,r + k4\,r^2))},$$

$$\left.\Gamma^4{}_{41} == -\frac{k1 + r^2\,(k3 - 2\,k4\,r)}{2\,r\,(k1 + r\,(-1 + k2 - k3\,r + k4\,r^2))}\right\}$$

*Display the Ricci Tensor :*

```
Table[Simplify[RicciR[-i, -j]], {i, 4}, {j, 4}]
```

$$\left\{\left\{\frac{k3 - 3\,k4\,r}{k1 + r\,(-1 + k2 - k3\,r + k4\,r^2)},\ 0,\ 0,\ 0\right\},\right.$$

$$\{0,\ k2 + r\,(-2\,k3 + 3\,k4\,r),\ 0,\ 0\},$$

$$\{0,\ 0,\ (k2 + r\,(-2\,k3 + 3\,k4\,r))\,\mathrm{Sin}[\mathrm{theta}]^2,\ 0\},$$

$$\left.\left\{0,\ 0,\ 0,\ \frac{(-k3 + 3\,k4\,r)\,(k1 + r\,(-1 + k2 - k3\,r + k4\,r^2))}{r^2}\right\}\right\}$$

*Display the Ricci Scalar:*

```
ScalarR
```

$$-\frac{2\,(k2 - 3\,k3\,r + 6\,k4\,r^2)}{r^2}$$

*Non-zero Components of the Weyl Tensor* $C^i{}_{jkl}$*:*

```
weyl = {};
Do[
   If[
     WeylC[i, -j, -k, -l] =!= 0,
      weyl = {weyl,
         SubscriptBox[SubscriptBox[
        SubscriptBox[SuperscriptBox[C, i], j], k], l]
      WeylC[i, -j, -k, -l] // DisplayForm},
    Continue[]],

  {i, 4}, {j, 4}, {k, 4}, {l, 4}];
Simplify[Flatten[weyl]]
```

$$\Big\{C^1{}_{212} == -\frac{3\,k1 + k2\,r}{6\,r},$$

$$C^1{}_{221} == \frac{k2}{6} + \frac{k1}{2\,r},\; C^1{}_{313} == -\frac{(3\,k1 + k2\,r)\,\mathrm{Sin}[\mathrm{theta}]^2}{6\,r},$$

$$C^1{}_{331} == \frac{(3\,k1 + k2\,r)\,\mathrm{Sin}[\mathrm{theta}]^2}{6\,r},$$

$$C^1{}_{414} == \frac{(3\,k1 + k2\,r)\,(k1 + r\,(-1 + k2 - k3\,r + k4\,r^2))}{3\,r^4},$$

$$C^1{}_{441} == -\frac{(3\,k1 + k2\,r)\,(k1 + r\,(-1 + k2 - k3\,r + k4\,r^2))}{3\,r^4},$$

$$C^2{}_{112} == -\frac{3\,k1 + k2\,r}{6\,r^2\,(k1 + r\,(-1 + k2 - k3\,r + k4\,r^2))},$$

$$C^2{}_{121} == \frac{3\,k1 + k2\,r}{6\,r^2\,(k1 + r\,(-1 + k2 - k3\,r + k4\,r^2))},$$

$$C^2{}_{323} == \frac{(3\,k1 + k2\,r)\,\mathrm{Sin}[\mathrm{theta}]^2}{3\,r},$$

$$C^2{}_{332} == -\frac{(3\,k1 + k2\,r)\,\mathrm{Sin}[\mathrm{theta}]^2}{3\,r},$$

$$C^2{}_{424} == -\frac{(3\,k1 + k2\,r)\,(k1 + r\,(-1 + k2 - k3\,r + k4\,r^2))}{6\,r^4},$$

$$C^2{}_{442} == \frac{(3\,k1 + k2\,r)\,(k1 + r\,(-1 + k2 - k3\,r + k4\,r^2))}{6\,r^4},$$

$$C^3{}_{113} == -\frac{3\,k1 + k2\,r}{6\,r^2\,(k1 + r\,(-1 + k2 - k3\,r + k4\,r^2))},$$

$$C^3{}_{131} == \frac{3\,k1 + k2\,r}{6\,r^2\,(k1 + r\,(-1 + k2 - k3\,r + k4\,r^2))},$$

$$\frac{k2}{3} + \frac{k1}{r} + C^3{}_{223} == 0,\; C^3{}_{232} == \frac{k2}{3} + \frac{k1}{r},$$

$$C^3{}_{434} == -\frac{(3\,k1 + k2\,r)\,(k1 + r\,(-1 + k2 - k3\,r + k4\,r^2))}{6\,r^4},$$

$$C^3{}_{443} == \frac{(3\,k1 + k2\,r)\,(k1 + r\,(-1 + k2 - k3\,r + k4\,r^2))}{6\,r^4},$$

$$C^4{}_{114} == \frac{3\,k1 + k2\,r}{3\,r^2\,(k1 + r\,(-1 + k2 - k3\,r + k4\,r^2))},\; C^4{}_{141} ==$$

$$-\frac{3\,k1 + k2\,r}{3\,r^2\,(k1 + r\,(-1 + k2 - k3\,r + k4\,r^2))},\; C^4{}_{224} == \frac{k2}{6} + \frac{k1}{2\,r},$$

$$C^4{}_{242} == -\frac{3\,k1 + k2\,r}{6\,r},\; C^4{}_{334} == \frac{(3\,k1 + k2\,r)\,\mathrm{Sin}[\mathrm{theta}]^2}{6\,r},$$

$$C^4{}_{343} == -\frac{(3\,k1 + k2\,r)\,\mathrm{Sin}[\mathrm{theta}]^2}{6\,r}\Big\}$$

Substitute into the Conformal Field Equation:

```
Table[Simplify[MakeSum[WeylField[-i, -j]]], {i, 4},
 {j, 4}]
```

$$\Big\{\Big\{-\frac{2\,(-2\,k2 + k2^2 + 3\,k1\,k3)}{3\,r^3\,(k1 + r\,(-1 + k2 - k3\,r + k4\,r^2))},\,0,\,0,\,0\Big\},$$

$$\Big\{0,\,-\frac{2\,(-2\,k2 + k2^2 + 3\,k1\,k3)}{3\,r^2},\,0,\,0\Big\},$$

$$\Big\{0,\,0,\,-\frac{2\,(-2\,k2 + k2^2 + 3\,k1\,k3)\,\mathrm{Sin}[\mathrm{theta}]^2}{3\,r^2},\,0\Big\},\,\Big\{0,\,0,$$

$$0,\,\frac{2\,(-2\,k2 + k2^2 + 3\,k1\,k3)\,(k1 + r\,(-1 + k2 - k3\,r + k4\,r^2))}{3\,r^5}$$

# Appendix F: Newtonian Phase-Plane Analysis

As discussed earlier in Chapter 5, the standard analysis of the Newtonian orbital dynamics is based on the change of independent variable, $t \to \varphi$, for the purpose of finding a closed form solution describing the orbital geometry. But a phase-plane analysis of the differential equations using time as the independent variable is no more complicated in principle than using $\varphi$. Furthermore, there are results shared by the relativistic case (discussed in Section VI) that are clarified in this analysis.

To begin, consider the Newtonian limit of the equations derived in Section II. The effective potential/(unit rest energy) is listed in (5.14), and is defined as $\hat{V}_{eff}^2$ which gives the proper Newtonian limit for $\hat{V}_{eff}$ (to within an additive constant) in the limit of large *r*. As a result, the Newtonian limit of (5.14) is given by

$$\hat{V}_{eff} = \left[1 - x + (x^2 - x^3)/2\sigma\right]^{1/2} \approx 1 - x/2 + x^2/4\sigma, \tag{F.1}$$

which differs from the standard Newtonian form by the addition of an additive constant (corresponding to the rest mass energy of $m_0$). The standard Newtonian effective potential energy is chosen to be zero at infinity which gives the usual expression:

$$\hat{V}_{eff} = x^2/4\sigma - x/2, \tag{F.2}$$

compared to the relativistic limit where the energy at infinity corresponds to the rest mass energy. But this additive constant is of little consequence insofar as the dynamics are concerned, and so we adopt (F.2) for the remaining discussion.

The corresponding Newtonian expression for (5.44) is derived using the standard Lagrangian and Hamiltonian results:

$$(r_s/c)^2 \dot{x}^2 = 2x^4\left[\hat{E} - \hat{V}_{eff}\right], \tag{F.3}$$

where $\hat{V}_{eff}$ is given by (F.2) and $x = r_s/r$. Although this choice of units seems odd at first, the most straightforward comparison with the relativistic case is obtained in this form. As a check, (F.3) reduces to the equation (after substituting (5.2), (F.2), and then (5.13)):

$$(du/dt)^2 = -u^4(J/m)^2[u^2 - 2u_0 u - b_u^2], \tag{F.4}$$

where $u = 1/r$ and $u_0 = GMm^2 / J^2$ gives the standard radius of a circular orbit. The constant: $b_u^2 \equiv 1/b^2 = 2mE / J^2$, expresses the impact parameter (for a particle approaching from infinity) in terms of $E$ and $J$. The zeroes of (F.4) give the standard turning points of the effective potential (aside from $u = 0$). Furthermore, substituting $u = u(\varphi)$ and then (5.6) into (F.4) (the Newtonian expression for $J$ is identical in form to the relativistic case) leads to the standard second order differential equation that is commonly evaluated for the analysis of these orbits.

Continuing with the analysis, differentiating (F.3) gives the dimensionless phase-plane equations expressed using $t$ as the independent variable:

$$\begin{aligned} \dot{x} &= y = \pm x^2 c[4\sigma\hat{E} + 2\sigma x - x^2)]^{1/2} / r_s\sqrt{2\sigma} \\ \dot{y} &= -x^3 c^2 (3x^2 - 5\sigma x - 8\sigma\hat{E}) / 2\sigma r_s^{\,2}. \end{aligned} \tag{F.5}$$

Solving simultaneously, $\dot{x} = \dot{y} = 0$ for $\hat{E}^2$ and $x$ then gives the two fixed points:

$$\{x_1 = \sigma\,;\ \hat{E} = -\sigma/4\} \text{ and } x_2 = 0. \tag{F.6}$$

The first gives the standard results: a center node corresponding to a Newtonian circular orbit with radius $r_1$ and energy given by

$$\begin{aligned} x_1 = \sigma \ &\Rightarrow\ r_1 = r_s/\sigma = J^2 / GMm^2 \\ \hat{E} = -\sigma/4 \ &\Rightarrow\ E = -m(GMm/J)^2/2. \end{aligned} \tag{F.7}$$

The second fixed point at infinity simply expresses the fact that it takes an infinite amount of time for the orbiting particle, $m$, to reach the turning point at infinity (in the case of parabolic and hyperbolic orbits) - a fixed point that is shared in the relativistic orbital dynamics. For comparison with the relativistic phase-plane results the Newtonian phase diagram for (F.5) is shown in Figure 35.

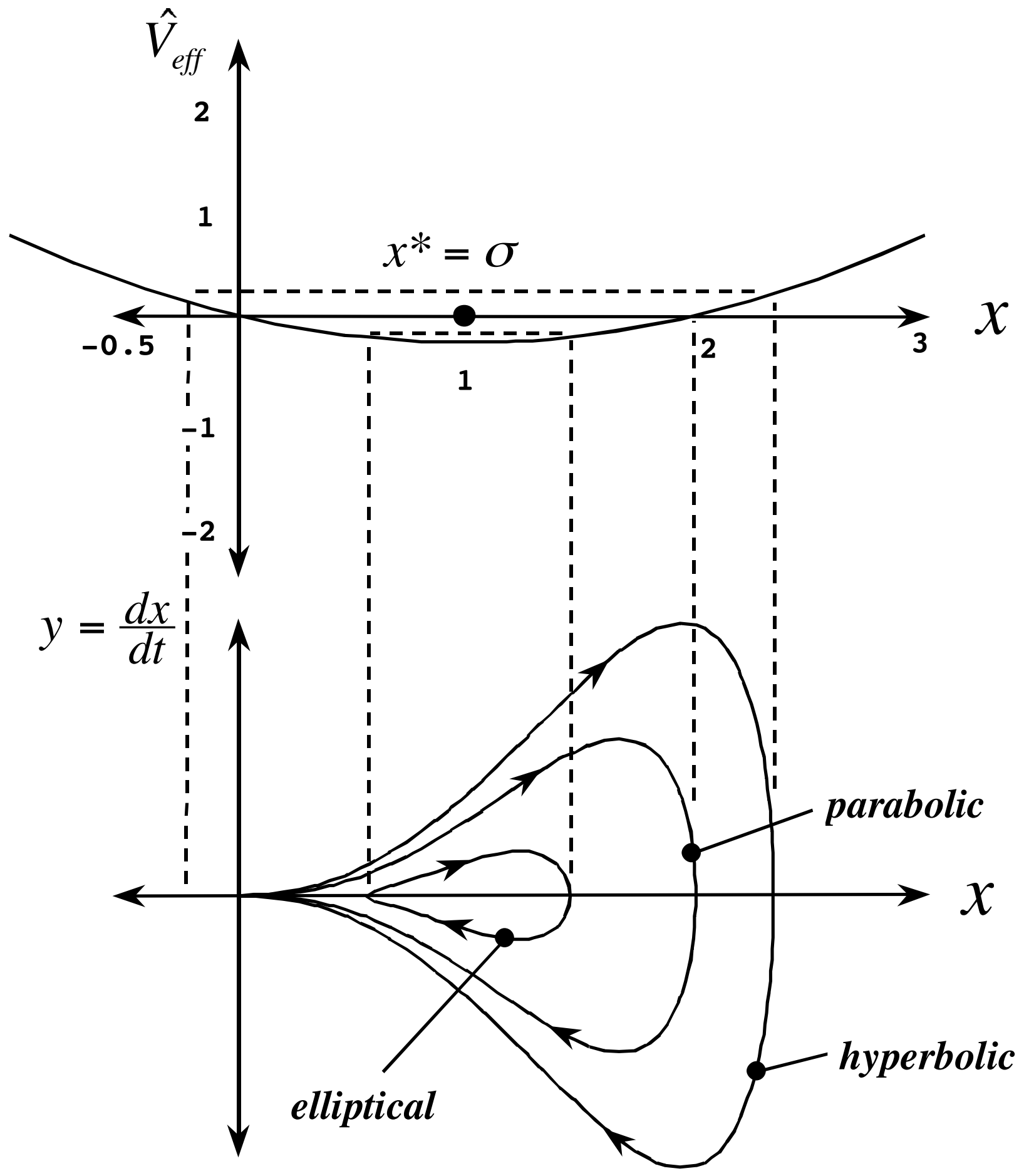

Figure 35. Newtonian Phase-Plane Diagram with $t$ as the Independent Variable ( $\sigma = 1$ )

# Appendix G: Miscellaneous Calculations – Chapter's 5 and 6

An important analysis tool will be *Mathematica's* Implicit Plotting routines (*ImplicitPlot*). There are several ways to specify a curve in the plane using Mathematica. A different plotting function is used for each of these methods. The most basic graphs of functions are plotted using the *Plot* routine. Curves given parametrically are plotted using the *ParametricPlot* routine. However, *ImplicitPlot* plots curves that are given implicitly as the solutions to a specified set of equations.

The function, *ImplicitPlot*, can use two methods to plot the solution to the given equations. The method that is used is determined by the form of the variable ranges given. One method uses the *Solve* routine to find solutions to the equation at each point in the $x$ range. It avoids singular points, plotting to within machine precision of those points, to generate an apparently smooth graph. This is the method used if only the range for $x$ is specified. The second method treats the equation as a function in 3-d space and generates a contour of the equation cutting through the plane wherever $z$ equals zero. This method is faster than the *Solve* method and handles a greater variety of cases, but may generate rougher graphs, especially around singularities or intersections of the curve. This method is used if the ranges for both $x$ and $y$ are specified.

Load the Libraries:

```
<< Graphics`ImplicitPlot`
<< Graphics`Graphics`
Off[General::spell1]
```

## Schwarzschild Calculations and Graphical Analysis

*Schwarzschild Equatorial Algebra:*

```
Simplify[Solve[
  {Sqrt[2 σ Ê - (2 σ + x^2) (1 - x)] == 0, (3/2 x^2 - x + σ) == 0
  {x, Ê}]]
```

$$\left\{\left\{\hat{E} \to \frac{1+\sqrt{1-6\sigma}-6\left(-3+\sqrt{1-6\sigma}\right)\sigma}{27\sigma},\ x \to \frac{1}{3}\left(1+\sqrt{1-6\right.\right.\right.$$

$$\left\{\hat{E} \to \frac{1-\sqrt{1-6\sigma}+6\left(3+\sqrt{1-6\sigma}\right)\sigma}{27\sigma},\ x \to \frac{1}{3}\left(1-\sqrt{1-6\sigma}\right.\right.$$

*Schwarzschild Proper Time Algebra:*

```
Simplify[Solve[{ x ((-1 + x) (x^2 + 2 σ)/(2 σ) + Ê) == 0,
   (x (-6 x^2 + 7 x^3 + 10 x σ + 8 σ (-1 + Ê))/(4 σ)) == 0},
  {x, Ê}]]
```

$$\left\{\left\{\hat{E} \to \frac{1+\sqrt{1-6\sigma}-6\left(-3+\sqrt{1-6\sigma}\right)\sigma}{27\sigma},\ x \to \frac{1}{3}\left(1+\sqrt{1-6\right.\right.\right.$$

$$\left\{\hat{E} \to \frac{1-\sqrt{1-6\sigma}+6\left(3+\sqrt{1-6\sigma}\right)\sigma}{27\sigma},\ x \to \frac{1}{3}\left(1-\sqrt{1-6\sigma}\right.\right.$$

$$\{x \to 0\}\}$$

*Schwarzschild Coordinate Time Algebra:*

```
Simplify[Solve[{ (-1 + x)^2 x ((-1 + x) (x^2 + 2 σ)/(2 σ) + Ê) == 0,
   ((-1 + x) x (-15 x^3 + 9 x^4 + 2 x^2 (3 + 7 σ) -
        8 σ (-1 + Ê) + 2 x σ (-11 + 6 Ê)) ) == 0},
  {x, Ê}]]
```

$$\left\{\left\{\hat{E} \to \frac{1+\sqrt{1-6\sigma}-6\left(-3+\sqrt{1-6\sigma}\right)\sigma}{27\sigma},\ x \to \frac{1}{3}\left(1+\sqrt{1-6\right.\right.\right.$$

$$\left\{\hat{E} \to \frac{1-\sqrt{1-6\sigma}+6\left(3+\sqrt{1-6\sigma}\right)\sigma}{27\sigma},\ x \to \frac{1}{3}\left(1-\sqrt{1-6\sigma}\right.\right.$$

$$\{x \to 0\},\ \{x \to 1\},\ \{x \to 1\}\}$$

```
Simplify[Solve[(-1 + x)^2 x ((-1 + x) (x^2 + 2 σ)/(2 σ) + Ê) == 0, Ê
```

$$\left\{\left\{\hat{E} \to -\frac{(-1+x)\ (x^2+2\sigma)}{2\sigma}\right\}\right\}$$

$0 < \sigma < \frac{1}{8}$ :

```
σn = .1;
Plot[effroots /. {σ -> σn}, {x, -2, 12}]
```

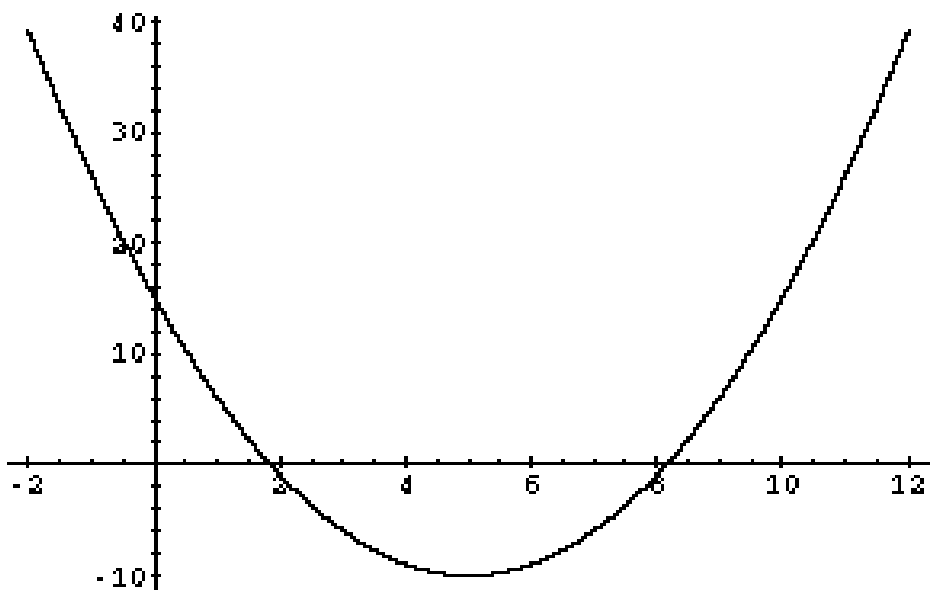

- Graphics -

```
Plot[V /. {σ -> σn}, {x, .7, 20}]
```

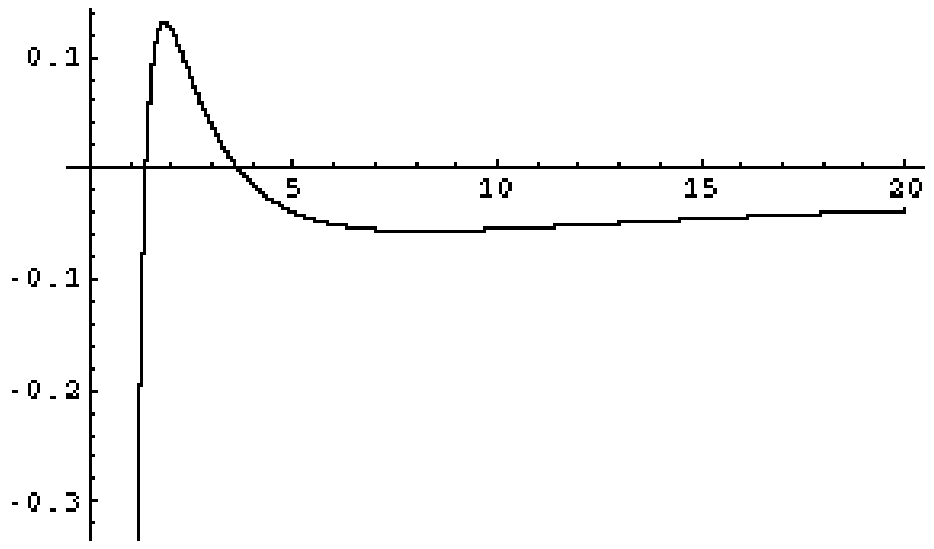

- Graphics -

```
Plot[V2 /. {λ -> λn, σ -> σn}, {x, -.5, 1}]
```

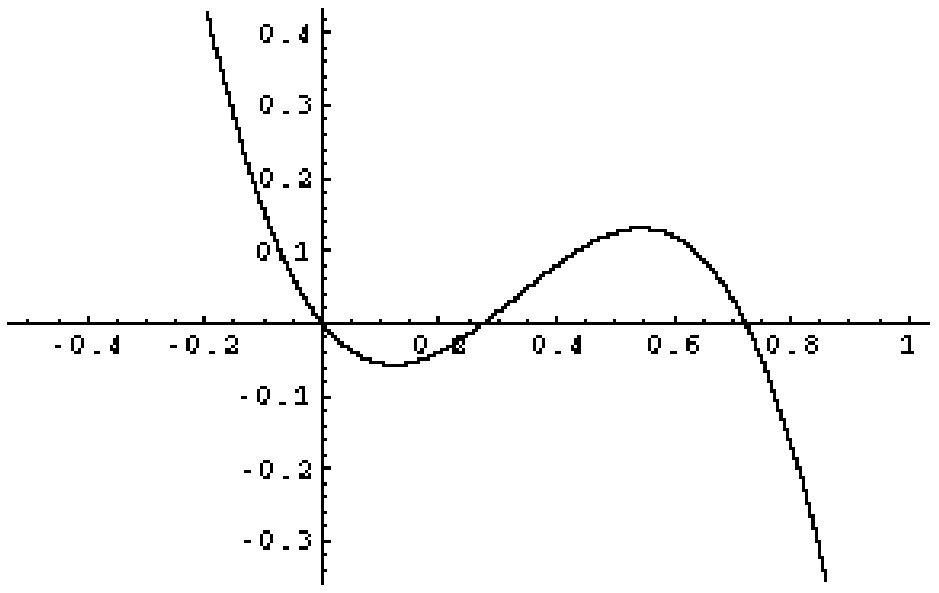

- Graphics -

```
e1 = V /. {σ -> σn, x -> 1.83772233983162109`}
0.130734
e2 = V /. {σ -> σn, x -> 8.16227766016837819`}
-0.0566601
xl = 1 / 2; xr = 1.5; yr = 1;
p1 = Table[ImplicitPlot[
    (β == (y^2 - x^3 + x^2)/(2 σ) - x) /. {λ -> λn, σ -> σn},
    {x, -xl , xr}, {y, -yr, yr}, PlotPoints -> 50],
  {β, -1, 1, .5}]
{- ContourGraphics -, - ContourGraphics -,
 - ContourGraphics -, - ContourGraphics -,
 - ContourGraphics -}
sep1 = ImplicitPlot[(e1 == (y^2 - x^3 + x^2)/(2 σ) - x) /. {σ -> σn
  {x, -xl , xr}, {y, -yr, yr}, PlotPoints -> 80]
- ContourGraphics -
sep2 = ImplicitPlot[
  (0 == (y^2 - x^3 + x^2)/(2 σ) - x + (x^4/(2 σ) + x^2)/(2 λ)) /. {λ -> λn, σ ->
  {x, -xl , xr}, {y, -yr, yr}, PlotPoints -> 80]
```

```
- ContourGraphics -
Show[{p1, sep1}, PlotRange -> {{-xl, xr}, {-yr, yr}},
 AspectRatio -> .8]
```

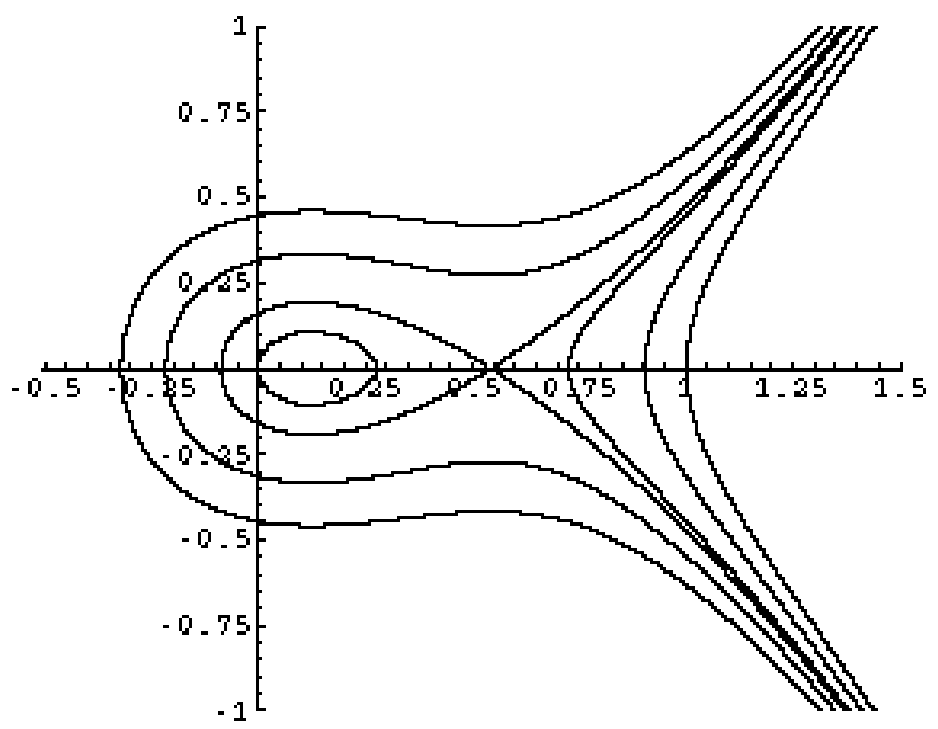

```
- Graphics -
```

The same calculations and numerical work were used in each of the following cases :

$\sigma = \frac{1}{8}$ :

$\frac{1}{8} < \sigma < \frac{1}{6}$ :

$\sigma = \frac{1}{6}$ :

$\sigma > \frac{1}{6}$ :

$\sigma \to -\infty$ :

$\sigma < 0$ :

Schwarzschild Null Geodesics :

```
Plot[-x^3 + x^2, {x, -1/2, 2}]
- Graphics -
Solve[D[-x^3 + x^2, x] == 0, x]
{{x -> 0}, {x -> 2/3}}
e1 = -x^3 + x^2 /. {x -> 2/3}
4/27
xl = 1; xr = 3; yr = 2;
p1 = Table[ImplicitPlot[
    (β == y^2 - x^3 + x^2 ), {x, -xl , xr}, {y, -yr, yr}],
   {β, -2, 2, .5}]
{- ContourGraphics -, - ContourGraphics -,
 - ContourGraphics -, - ContourGraphics -,
 - ContourGraphics -, - ContourGraphics -,
 - ContourGraphics -, - ContourGraphics -,
 - ContourGraphics -}
```

```
sep1 = ImplicitPlot[(e1 == y^2 - x^3 + x^2 ), {x, -xl , xr},
  {y, -yr, yr}, PlotPoints -> 80]
- ContourGraphics -
sep2 = ImplicitPlot[(.05 == y^2 - x^3 + x^2 ), {x, -xl , xr},
  {y, -yr, yr}, PlotPoints -> 50]
- ContourGraphics -
Show[{p1, sep1, sep2}, PlotRange -> {{-xl, xr}, {-yr, y
 AspectRatio -> .8]
```

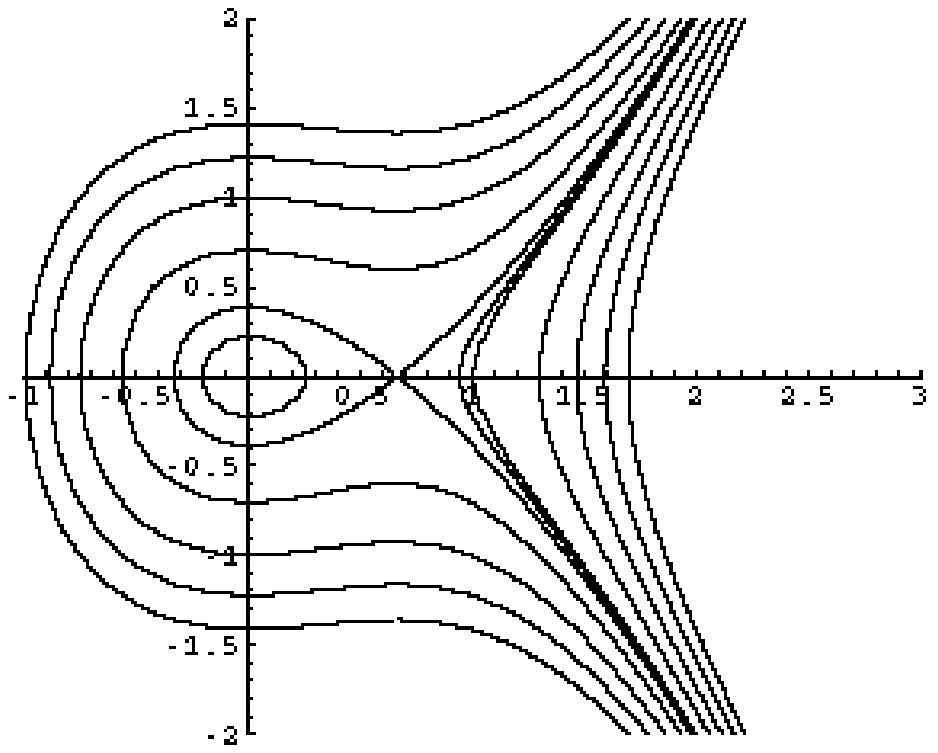

```
- Graphics -
```

Schwarzschild Phase – Plane :

```
V2Plot =
 Plot[V2 /. σ -> 1/9, {x, -1/2, 1.25}, PlotPoints -> 50
- Graphics -
Simplify[Solve[D[V2, x] == 0, x]]
```

$$\left\{\left\{x \to \frac{1}{3}\left(1-\sqrt{1-6\sigma}\right)\right\}, \left\{x \to \frac{1}{3}\left(1+\sqrt{1-6\sigma}\right)\right\}\right\}$$

```
e1 = Simplify[(V2 /. %%[[2]]) /. σ -> 1/9]
```

$$\frac{1}{9\sqrt{3}}$$

```
xl = 1/4; xr = 1.2; yr = .5;
p1 =
 Table[ImplicitPlot[(β == ((y^2 + x^2 - x^3)/(2 σ) - x) /. σ -> 1/9)
   {x, -xl , xr}, {y, -yr, yr}, PlotPoints -> 50],
  {β, -1.2, .2, .1}]
{- ContourGraphics -, - ContourGraphics -,
 - ContourGraphics -, - ContourGraphics -,
 - ContourGraphics -, - ContourGraphics -,
 - ContourGraphics -, - ContourGraphics -,
 - ContourGraphics -, - ContourGraphics -,
 - ContourGraphics -, - ContourGraphics -,
 - ContourGraphics -, - ContourGraphics -,
 - ContourGraphics -}
sep1 = ImplicitPlot[(e1 == ((y^2 + x^2 - x^3)/(2 σ) - x) /. σ -> 1/9
  {x, -xl , xr}, {y, -yr, yr}, PlotPoints -> 80]
- ContourGraphics -
sep2 = ImplicitPlot[(-.05 == ((y^2 + x^2 - x^3)/(2 σ) - x) /. σ -> 1/
  {x, -xl , xr}, {y, -yr, yr}, PlotPoints -> 50]
```

```
- ContourGraphics -
PhasePlane = Show[{p1, sep1, sep2},
  PlotRange -> {{-xl, xr}, {-.4, .4}}]
- Graphics -
Show[PhasePlane, V2Plot, AspectRatio -> .7]
- Graphics -
```

Coordinate Time Phase – Plane :

$$e = \text{Simplify}\left[\frac{x^4\left(1+\frac{x^2}{2\sigma}\right)(1-x)^3}{x^4(1-x)^2-y^2}-1\right]$$

$$-1-\frac{(-1+x)^3x^4(x^2+2\sigma)}{2(x^4-2x^5+x^6-y^2)\sigma}$$

## *Coordinate Time Phase-Plane Equations:*

$$y[\tau]^2 == \frac{x[\tau]^4}{\dot{t}^2}\left(\hat{E}^2-\left(1+\frac{x[\tau]^2}{2\sigma}\right)(1-x[\tau])\right) /.$$
$$\dot{t} \to \hat{E}/(1-x[\tau])$$

$$y[\tau]^2 == \frac{(1-x[\tau])^2x[\tau]^4\left(\hat{E}^2-(1-x[\tau])\left(1+\frac{x[\tau]^2}{2\sigma}\right)\right)}{\hat{E}^2}$$

```
Simplify[D[%5, τ]] /. y[τ] -> x'[τ]
```

$$2x'[\tau]\,y'[\tau] == \frac{1}{2\sigma\hat{E}^2}\Big((-1+x[\tau])\,x[\tau]^3\left(-8\sigma\left(-1+\hat{E}^2\right)+2\sigma\left(-11+6\hat{E}^2\right)x[\tau]+2(3+7\sigma)\,x[\tau]^2-15x[\tau]^3+9x[\tau]^4\right)x'[\tau]\Big)$$

```
Simplify[Solve[%8, y'[τ]]] /. x[τ] -> x
```

$$\left\{\left\{y'[\tau]\to\frac{1}{4\sigma\hat{E}^2}\left((-1+x)\,x^3\left(-15x^3+9x^4+2x^2(3+7\sigma)-8\sigma\left(-1+\hat{E}^2\right)+2x\sigma\left(-11+6\hat{E}^2\right)\right)\right)\right\}\right\}$$

$$\text{Simplify}\Big[\left(-15x^3+9x^4+2x^2(3+7\sigma)-8\sigma\left(-1+\hat{E}^2\right)+2x\sigma\left(-11+6\hat{E}\right.\right.$$

$$-15x^3+9x^4+2x^2(3+7\sigma)-8\sigma\left(-1+\hat{E}^2\right)+2x\sigma\left(-11+6\hat{E}^2\right)$$

## *Proper Time Phase-Plane Equations:*

$$y[\tau]^2 == x[\tau]^4\left(\hat{E}^2-\left(1+\frac{x[\tau]^2}{2\sigma}\right)(1-x[\tau])\right)$$

$$y[\tau]^2 == x[\tau]^4\left(\hat{E}^2-(1-x[\tau])\left(1+\frac{x[\tau]^2}{2\sigma}\right)\right)$$

```
Simplify[D[%, τ]] /. y[τ] -> x'[τ]
```

$$2x'[\tau]\,y'[\tau] == \frac{x[\tau]^3\left(8\sigma\left(-1+\hat{E}^2\right)+10\sigma x[\tau]-6x[\tau]^2+7x[\tau]^3\right)x'[\tau]}{2\sigma}$$

```
Simplify[Solve[%, y'[τ]]] /. x[τ] -> x
```

$$\left\{\left\{y'[\tau]\to\frac{x^3\left(-6x^2+7x^3+10x\sigma+8\sigma\left(-1+\hat{E}^2\right)\right)}{4\sigma}\right\}\right\}$$

```
V2 = Expand[Simplify[% /. y -> 0]]
```

$$-x+\frac{x^2}{2\sigma}-\frac{x^3}{2\sigma}$$

```
V2Plot = Plot[V2 /. σ -> 1/9, {x, 0, .8}]
- Graphics -
critical = Simplify[Solve[D[V2, x] == 0, x]]
```

$$\left\{\left\{x\to\frac{1}{3}\left(1-\sqrt{1-6\sigma}\right)\right\},\left\{x\to\frac{1}{3}\left(1+\sqrt{1-6\sigma}\right)\right\}\right\}$$

```
xl = .1; xr = 1.1; yr = .2;
```

```
p1 = Table[ImplicitPlot[(β == e /. σ -> 1/9 ),
   {x, -xl , xr}, {y, -yr, yr}, PlotPoints -> 201],
  {β, -.2, .2, .05}]
{- ContourGraphics -, - ContourGraphics -,
 - ContourGraphics -, - ContourGraphics -,
 - ContourGraphics -, - ContourGraphics -,
 - ContourGraphics -, - ContourGraphics -,
 - ContourGraphics -}
PhasePlane = Show[{p1},
  PlotRange -> {{-xl, xr}, {-yr, yr}}, AspectRatio -> 1]
```

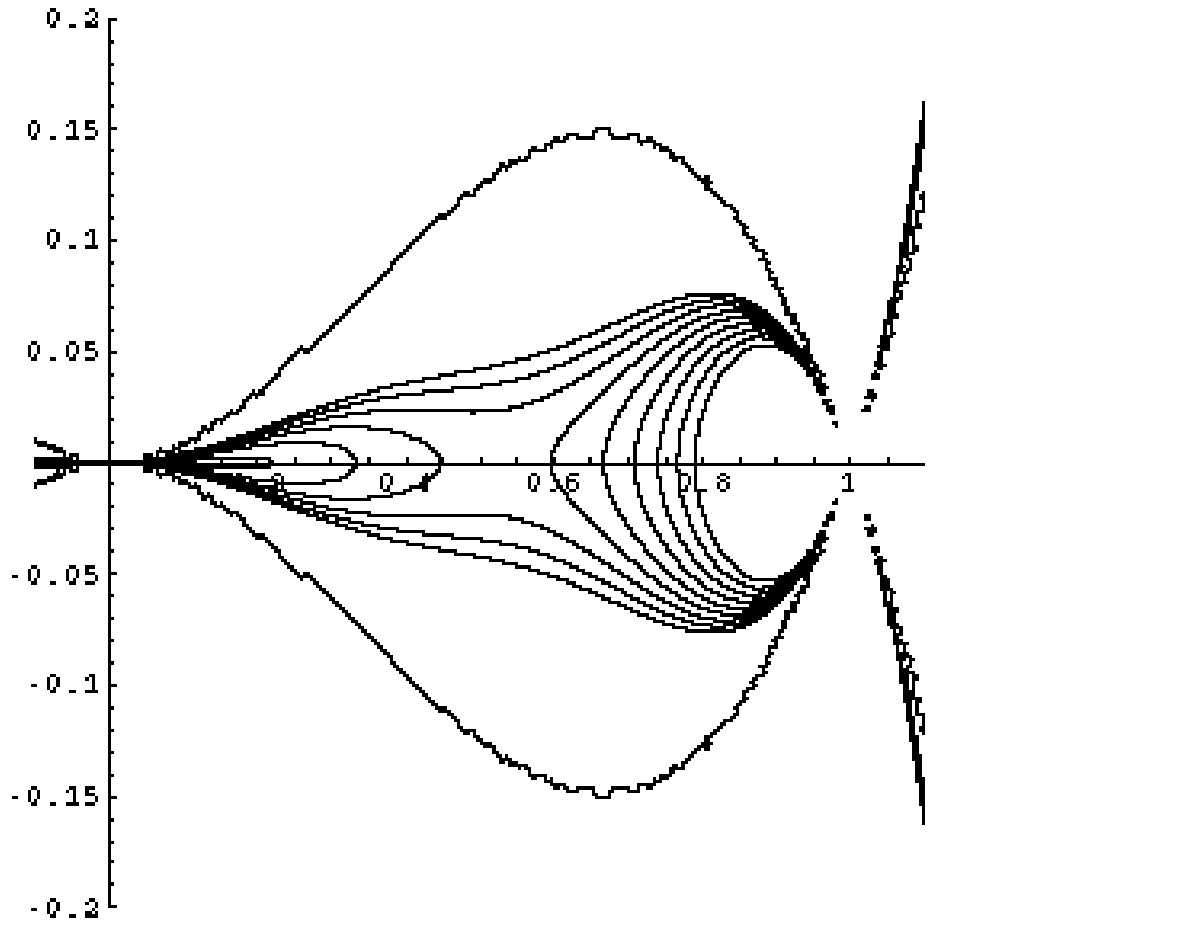

```
- Graphics -
```

# Reissner-Nordstrom Calculations and Graphical Analysis

*Effective Potential Equations: let* $V == V^2 - 1$

*Critical Points:*

$$V = -x + \frac{x^2 + \frac{x^4}{2\sigma}}{2\lambda} + \frac{x^2 - x^3}{2\sigma}$$

$$-x + \frac{x^2 + \frac{x^4}{2\sigma}}{2\lambda} + \frac{x^2 - x^3}{2\sigma}$$

```
Simplify[D[V, x]];
effroots =
```

$$\text{Collect}\left[\text{Expand}\left[\frac{2x^3 - 3x^2\lambda - 2\lambda\sigma + 2x(\lambda + \sigma)}{2\lambda\sigma}\right], x\right]$$

$$-1 + x\left(\frac{1}{\lambda} + \frac{1}{\sigma}\right) - \frac{3x^2}{2\sigma} + \frac{x^3}{\lambda\sigma}$$

```
Simplify[Solve[effroots == 0, x]]
```

```
x1 = 1/24 (12 λ + (6 I (I + √3) (-4 λ + 3 λ^2 - 4 σ)) /
     (-54 λ^2 + 27 λ^3 + 54 λ σ + 1/2
        √(2916 λ^2 (-2 λ + λ^2 + 2 σ)^2 + 4 (-9 λ^2 + 12 (λ + σ)
      (1/3) -
   2^(2/3) (1 + I √3) (-108 λ^2 + 54 λ^3 + 108 λ σ +
        √(2916 λ^2 (-2 λ + λ^2 + 2 σ)^2 + 4 (-9 λ^2 + 12 (λ + σ))
      (1/3));
x2 = 1/24 (12 λ + (2 (1 + I √3) (-9 λ^2 + 12 (λ + σ))) /
     (-54 λ^2 + 27 λ^3 + 54 λ σ + 1/2
        √(2916 λ^2 (-2 λ + λ^2 + 2 σ)^2 + 4 (-9 λ^2 + 12 (λ + σ)
      (1/3) +
   I 2^(2/3) (I + √3) (-108 λ^2 + 54 λ^3 + 108 λ σ +
        √(2916 λ^2 (-2 λ + λ^2 + 2 σ)^2 + 4 (-9 λ^2 + 12 (λ + σ))
      (1/3));
x3 =
  1/12 (6 λ + (6 (-4 λ + 3 λ^2 - 4 σ)) / (-54 λ^2 + 27 λ^3 + 54 λ σ +
        √(2916 λ^2 (-2 λ + λ^2 + 2 σ)^2 + 4 (-9 λ^2 + 12 (λ + σ)
      (1/3) +
   2^(2/3) (-108 λ^2 + 54 λ^3 + 108 λ σ +
        √(2916 λ^2 (-2 λ + λ^2 + 2 σ)^2 + 4 (-9 λ^2 + 12 (λ + σ))
      (1/3));
```

## *The second derivative - Inflection Points:*

```
Simplify[D[V, {x, 2}]]
```

$$\frac{3x^2 + \lambda - 3x\lambda + \sigma}{\lambda\sigma}$$

```
inflec = Collect[Expand[(3 x^2 + λ - 3 x λ + σ)/(λ σ)], x];
Simplify[Solve[inflec == 0, x]]
```

$$\left\{\left\{x \to \frac{1}{6}\left(3\lambda - \sqrt{9\lambda^2 - 12(\lambda+\sigma)}\right)\right\},\ \left\{x \to \frac{1}{6}\left(3\lambda + \sqrt{9\lambda^2 - 12(\lambda+\sigma)}\right)\right\}\right\}$$

## *Schwarzschild and RN Plots:*

```
VS = -x + (x^2 - x^3)/(2 σ);
λn = 2; σn = 1/9;
Plot[{V /. {λ -> λn, σ -> σn}, VS /. {λ -> λn, σ -> σn}},
 {x, -.2, 2.6}, PlotRange -> {Automatic, {-1, 1}}]
```

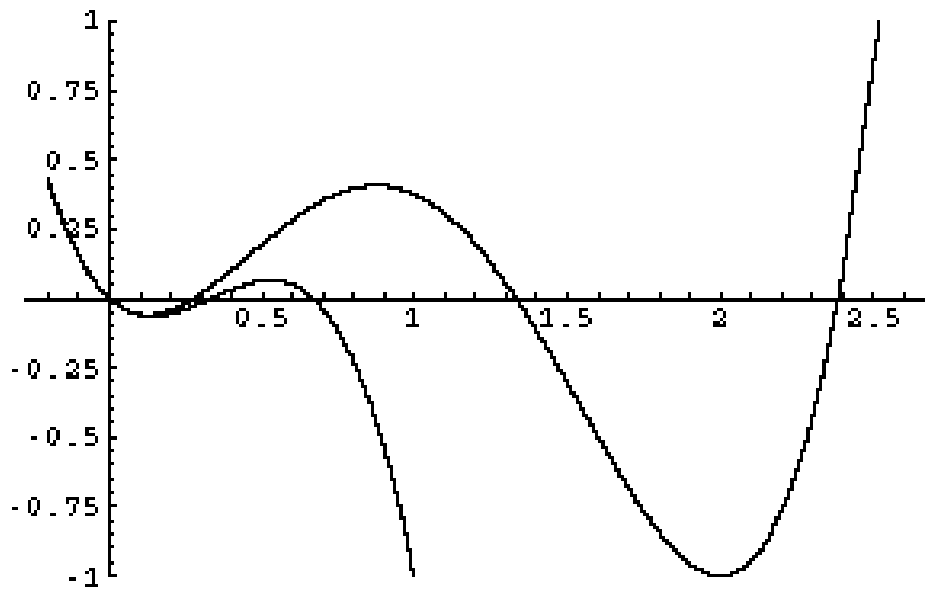

- Graphics -

*Effective Potential Descriminant:*

$$q = \frac{1}{3}a_1 - \frac{1}{9}(a_2)^2;$$
$$r = \frac{1}{6}(a_1 a_2 - 3a_0) - \frac{1}{27}(a_2)^3;$$
$$d = q^3 + r^2;$$

**Collect[Expand[λ σ effroots], x, Factor]**

$$x^3 - \frac{3x^2\lambda}{2} - \lambda\sigma + x(\lambda+\sigma)$$

**Simplify**$\left[d\ /.\ \left\{a_2 \to -\frac{3}{2}\lambda,\ a_1 \to (\lambda+\sigma),\ a_0 \to -\lambda\sigma\right\}\right]$

$$\frac{1}{64}\lambda^2(-2\lambda+\lambda^2+2\sigma)^2 + \left(-\frac{\lambda^2}{4}+\frac{\lambda+\sigma}{3}\right)^3$$

**desV = Collect**$\left[\text{Expand}\left[27\left(\frac{1}{64}\lambda^2(-2\lambda+\lambda^2+2\sigma)^2 + \left(-\frac{\lambda^2}{4}+\frac{\lambda+\sigma}{3}\right)^3\right)\right],\ \sigma,\ \text{Factor}\right]$

$$-\frac{1}{16}\lambda^3(-16+9\lambda) + \frac{3}{8}\lambda^2(8-21\lambda+9\lambda^2)\sigma - \frac{3}{16}\lambda(-16+3\lambda)\sigma^2 + \sigma^3$$

**σroots = Simplify[Solve[desV == 0, σ]]**

$$\sigma_1 = \frac{1}{16}\left(\lambda\,(-16+3\lambda) - \right.$$

$$\frac{3\lambda^3\,(-64+31\lambda)}{\left(\lambda^4\left(-512+352\lambda-47\lambda^2+16\sqrt{-2+\lambda}\,(-8+5\lambda)^{3/2}\right)\right)^{1}}$$

$$3\left(\lambda^4\left(-512+352\lambda-47\lambda^2+16\sqrt{-2+\lambda}\,(-8+5\lambda)^{3/2}\right)\right)$$

$$\sigma_2 = \frac{1}{32}\left(2\lambda\,(-16+3\lambda) + \right.$$

$$\frac{3\left(1+I\sqrt{3}\right)\lambda^3\,(-64+31\lambda)}{\left(\lambda^4\left(-512+352\lambda-47\lambda^2+16\sqrt{-2+\lambda}\,(-8+5\lambda)^{3/2}\right)\right)^{1}}$$

$$3\,I\left(I+\sqrt{3}\right)$$

$$\left(\lambda^4\left(-512+352\lambda-47\lambda^2+16\sqrt{-2+\lambda}\,(-8+5\lambda)^{3/2}\right)\right)^{1}$$

$$\sigma_3 = \frac{1}{32}\left(2\lambda\,(-16+3\lambda) + \right.$$

$$\frac{3\left(1-I\sqrt{3}\right)\lambda^3\,(-64+31\lambda)}{\left(\lambda^4\left(-512+352\lambda-47\lambda^2+16\sqrt{-2+\lambda}\,(-8+5\lambda)^{3/2}\right)\right)^{1}}$$

$$3\left(1+I\sqrt{3}\right)$$

$$\left(\lambda^4\left(-512+352\lambda-47\lambda^2+16\sqrt{-2+\lambda}\,(-8+5\lambda)^{3/2}\right)\right)^{1}$$

*Plot the $\sigma_i$ roots:*

```
Plot[{Re[N[σ1 /. {λ -> λn}]],
  Re[N[σ2 /. {λ -> λn}]], Re[N[σ3 /. {λ -> λn}]]},
 {λn, -1, 6}, PlotRange -> {{-1, 8}, {-6, 2}}]
```

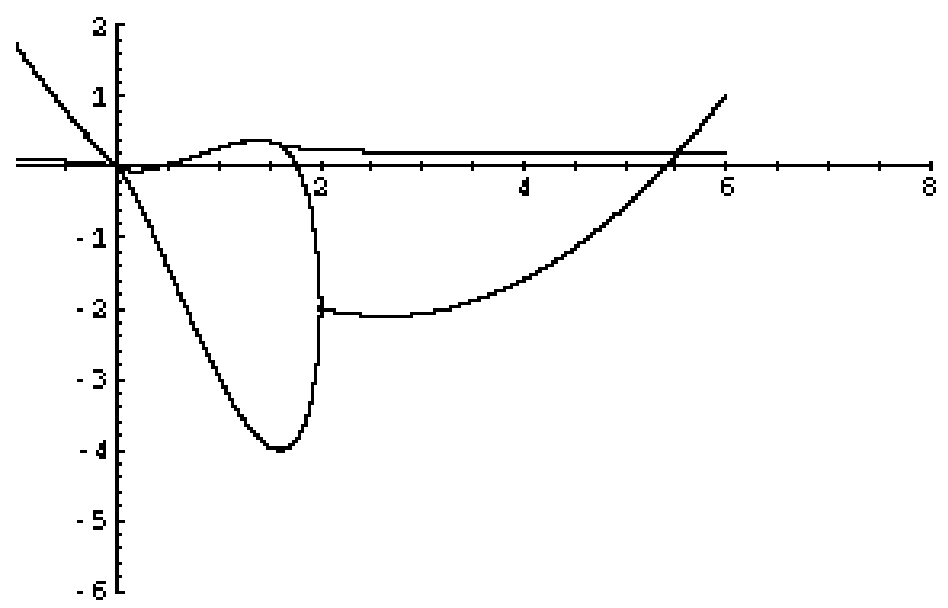

```
Plot[{Re[N[σ1 /. {λ -> λn}]], Re[N[σ2 /. {λ -> λn}]],
  Re[N[σ3 /. {λ -> λn}]], Im[N[σ1 /. {λ -> λn}]],
  Im[N[σ2 /. {λ -> λn}]], Im[N[σ3 /. {λ -> λn}]]},
 {λn, -2, 8}, PlotRange -> {{-1.5, 6}, {-4, 3}}]
```

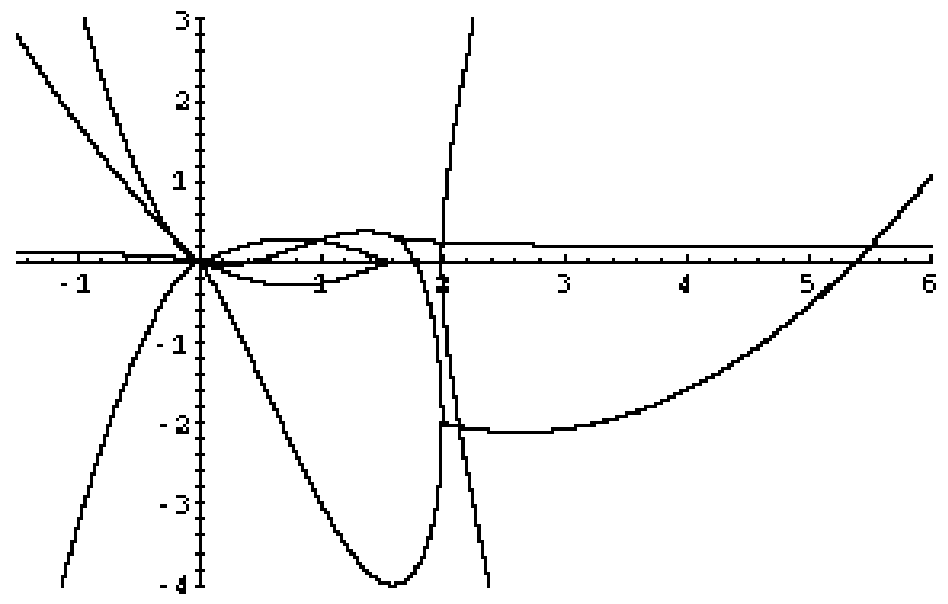

- Graphics -

*Linear Stability* $\lambda = -1$ *:*

```
λn = -1; TableForm[Table[
  {N[x1 /. {λ -> λn, σ -> σn}], N[x2 /. {λ -> λn, σ -> σn}
   N[x3 /. {λ -> λn, σ -> σn}], σn},
  {σn, -2, 2, .2}]]
Plot[{Re[N[x1 /. {λ -> λn, σ -> σn}]],
   Re[N[x2 /. {λ -> λn, σ -> σn}]],
   Re[N[x3 /. {λ -> λn, σ -> σn}]]}, {σn, -2, 5},
 PlotRange -> {Automatic, Automatic}, AspectRatio -> .8
```

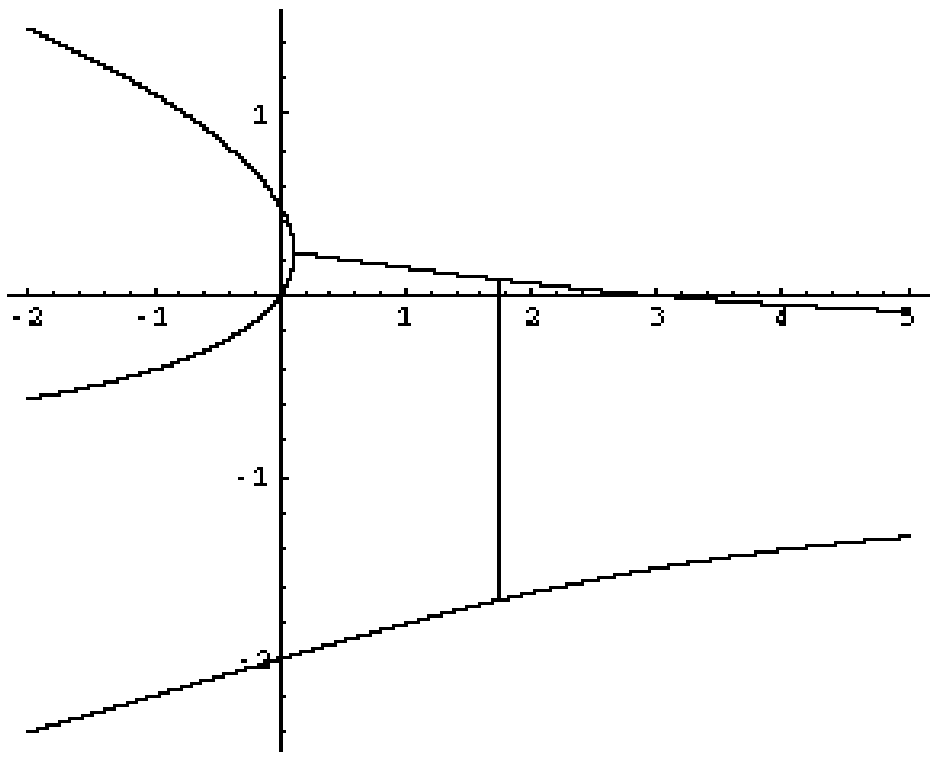

- Graphics -

$$V = -x + \frac{x^2 + \frac{x^4}{2\sigma}}{2\lambda} + \frac{x^2 - x^3}{2\sigma};$$

```
λn = -1; σn = -1;
Plot[V /. {λ -> λn, σ -> σn}, {x, -2, 2},
 PlotRange -> {Automatic, Automatic}]
```

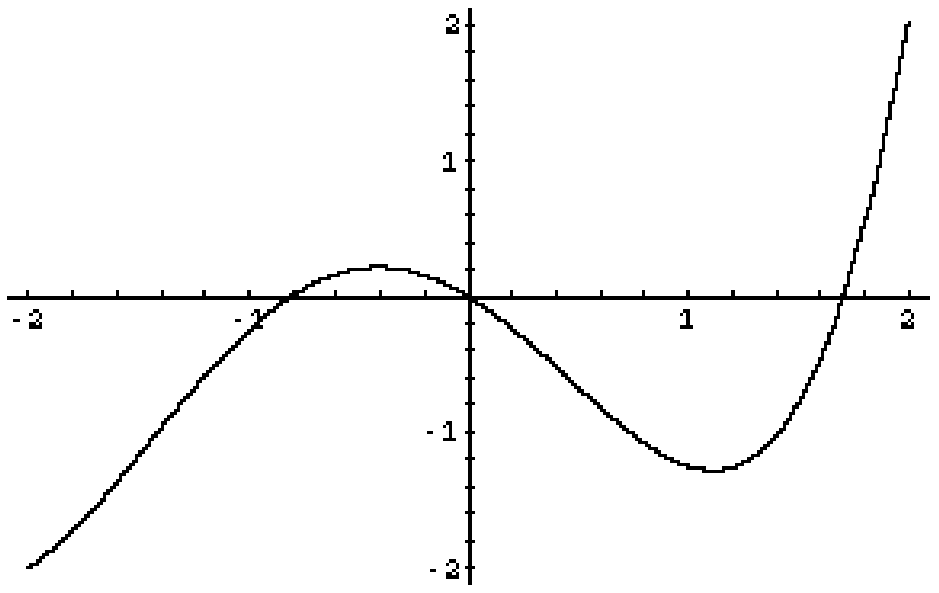

```
- Graphics -
λn = -100; σn = -1;
{N[x1 /. {λ -> λn, σ -> σn}], N[x2 /. {λ -> λn, σ -> σn}],
 N[x3 /. {λ -> λn, σ -> σn}]}
{-0.547134 + 1.89478 × 10^-14 I, -150.666 + 0. I,
 1.21309 - 1.89478 × 10^-14 I}
{N[σ1 /. {λ -> λn, σ -> σn}], N[σ2 /. {λ -> λn, σ -> σn}],
 N[σ3 /. {λ -> λn, σ -> σn}]}
{0.112725, 1.72489 + 3.29938 I, 1.72489 - 3.29938 I}
λn = -1; σn = 4000000;
{N[x1 /. {λ -> λn, σ -> σn}], N[x2 /. {λ -> λn, σ -> σn}],
 N[x3 /. {λ -> λn, σ -> σn}]}
{-0.25 - 2000. I, -0.25 + 2000. I, -1.}
```

*Find intersection point of the $\sigma_i$ :*

```
ExpandAll[σ3]
Re[%]
ExpandAll[σ1]
```

```
-λ + 3 λ^2/16 +
 (12 λ^3) / (-512 λ^4 + 352 λ^5 - 47 λ^6 - 128 √(-2 + λ) λ^4 √(-8 + 5
     80 √(-2 + λ) λ^5 √(-8 + 5 λ))^(1 / 3) -
 (93 λ^4) /
  (16 (-512 λ^4 + 352 λ^5 - 47 λ^6 - 128 √(-2 + λ) λ^4 √(-8 + 5 λ) +
        80 √(-2 + λ) λ^5 √(-8 + 5 λ))^(1 / 3)) +
 3/16 (-512 λ^4 + 352 λ^5 - 47 λ^6 -
      128 √(-2 + λ) λ^4 √(-8 + 5 λ) + 80 √(-2 + λ) λ^5 √(-8 + 5 λ))^
    (1 / 3)
(9 (64 λ^3 - 31 λ^4 +
       (λ^4 (-512 + 352 λ - 47 λ^2 + 16 √(-2 + λ) (-8 + 5 λ)^(3/2)))^2
   /
  (32 (λ^4 (-512 + 352 λ - 47 λ^2 + 16 √(-2 + λ) (-8 + 5 λ)^(3/2)))^1/
0
```

```
Simplify[Solve[9 (64 λ^3 - 31 λ^4 +
    (λ^4 (-512 + 352 λ - 47 λ^2 + 16 √(-2 + λ) (-8 + 5 λ)^(3/2))
  ==
 0,
 λ]]
```

$\left\{\{\lambda \to 0\}, \left\{\lambda \to \frac{16}{47}\left(11 + 3\sqrt{3}\right)\right\}\right\}$

```
N[%]
```

{{λ → 0}, {λ → 5.51358}}

*Linear Stability* $\lambda \to -\infty$:

*Series Expand* $\sigma_1$:

```
Normal[Series[σ1 /. λ -> 1/β, {β, 0, 1}]]
Expand[%]
% /. β -> 1/λ
Collect[%, λ, Factor]
N[%]
```

$0.166667 + \frac{0.0925926}{\lambda} + 0.\ \lambda - 1.30867 \times 10^{-16} \lambda^2$

```
10 + 8 / 10
```

$\frac{54}{5}$

*Series Expand* $\sigma_2$:

```
Normal[Series[σ2 /. λ -> 1/β, {β, 0, 1}]]
Expand[%]
% /. β -> 1/λ
Collect[%, λ, Factor]
N[%]
```

$(-0.0833333 - 0.839146\,I) - \frac{0.0462963 + 0.833409\,I}{\lambda} - (1.5 + 1.93649\,I)\,\lambda + (0.28125 + 1.81546\,I)\,\lambda^2$

*Series Expand* $\sigma_3$:

```
Normal[Series[σ3 /. λ -> 1/β, {β, 0, 1}]]
Expand[%]
% /. β -> 1/λ
Collect[%, λ, Factor]
N[%]
```

$(-0.0833333 + 0.839146\,I) - \frac{0.0462963 - 0.833409\,I}{\lambda} - (1.5 - 1.93649\,I)\,\lambda + (0.28125 - 1.81546\,I)\,\lambda^2$

*Series Expand* $x_1$:

```
Normal[Series[x1 /. {λ -> 1/β}, {β, 0, 1}, {σ, 0, 1}]]
```

$$\frac{1}{24}\left(\frac{1}{27}\,(1+I\sqrt{3})\,(108-36\,I\sqrt{3})+\right.$$
$$6\,I\,(I+\sqrt{3})\,\left(-\frac{4}{3}+\frac{1}{81}\,(54-18\,I\sqrt{3})\right)\Big)+$$
$$\frac{1}{24}\left(-3\,(1+I\sqrt{3})\,\left(-\frac{32\,I}{27\sqrt{3}}-\frac{(-108+36\,I\sqrt{3})^2}{13122}\right)+\right.$$
$$6\,I\,(I+\sqrt{3})\,\left(-\frac{4}{243}\,(54-18\,I\sqrt{3})+\right.$$
$$\frac{1}{2}\left(\frac{32\,I}{27\sqrt{3}}-\frac{4\,(54-18\,I\sqrt{3})\,(-54+18\,I\sqrt{3})}{6561}\right)\Big)\Big)$$
$$\beta+$$
$$\Big(\frac{1}{24}$$
$$(4\,I\sqrt{3}\,(1+I\sqrt{3})-4\sqrt{3}\,(I+\sqrt{3}))+\frac{1}{24}\Big(-3\,(1+I$$
$$\left(\frac{4\,I\,(-108+36\,I\sqrt{3})}{81\sqrt{3}}+\frac{1}{81}\,(108+156\,I\sqrt{3})\right)+$$
$$6\,I\,(I+\sqrt{3})\,\Big(-\frac{4}{3}-\frac{8\,I}{3\sqrt{3}}+\frac{1}{2}\,\Big(-\frac{2}{81}\,(54+78\,I\sqrt{3}$$
$$\frac{1}{6561}\,(4\,(-54\,I\sqrt{3}\,(54-18\,I\sqrt{3})+$$
$$54\,I\sqrt{3}\,(-54+18\,I\sqrt{3}))))\Big)\Big)$$
$$\beta\Big)$$
$$\sigma$$

**Expand[%] /. {β -> 1/λ}**

$$\frac{2}{3}+\frac{8}{27\,\lambda}-\sigma+\frac{2\,\sigma}{9\,\lambda}$$

## *Series Expand $x_2$ :*

**Normal[Series[x₂ /. {λ -> 1/β}, {β, 0, 2}, {σ, 0, 2}]]**

$$\frac{1}{24}\,(-4\,I\sqrt{3}\,(1+I\sqrt{3})+4\sqrt{3}\,(I+\sqrt{3}))\,\sigma+\Big(\frac{1}{24}$$
$$(-6\,I\sqrt{3}\,(1+I\sqrt{3})+6\sqrt{3}\,(I+\sqrt{3}))+\frac{1}{24}\Big(3\,I\,(I+$$
$$\left(4\,I\sqrt{3}+\frac{34992+324\,I\sqrt{3}\,(-108+36\,I\sqrt{3})}{13122}\right)+$$
$$2\,(1+I\sqrt{3})\,\Big(4\,I\sqrt{3}-\frac{3}{2}\,\Big(-4\,I\sqrt{3}-$$
$$\frac{1}{6561}\,(4\,(8748-81\,I\sqrt{3}\,(54-18\,I\sqrt{3})+$$
$$81\,I\sqrt{3}\,(-54+18\,I\sqrt{3})))))\Big)$$
$$\beta\Big)$$
$$\sigma^2$$

**Expand[%] /. {β -> 1/λ}**

$$\sigma+\frac{3\,\sigma^2}{2}-\frac{\sigma^2}{\lambda}$$

*Series Expand $x_3$ :*

```
Normal[Series[x3 /. {λ -> 1/β}, {β, 0, 1}, {σ, 0, 1}]]
```

$$\frac{1}{12}\left(\frac{1}{27}\left(-108+36 I\sqrt{3}\right)+6\left(-\frac{4}{3}+\frac{1}{81}\left(54-18 I\sqrt{3}\right)\right)\right)+$$

$$\frac{3}{2\beta}+\frac{1}{12}\left(3\left(-\frac{32 I}{27\sqrt{3}}-\frac{\left(-108+36 I\sqrt{3}\right)^2}{13122}\right)+6\left(-\frac{4}{243}\left(54-18 I\sqrt{3}\right)+\frac{1}{2}\left(\frac{32 I}{27\sqrt{3}}-\frac{4\left(54-18 I\sqrt{3}\right)\left(-54+18 I\sqrt{3}\right)}{6561}\right)\right)\right)\beta+$$

$$\frac{1}{12}\left(3\left(\frac{4 I\left(-108+36 I\sqrt{3}\right)}{81\sqrt{3}}+\frac{1}{81}\left(108+156 I\sqrt{3}\right)\right)+6\left(-\frac{4}{3}-\frac{8 I}{3\sqrt{3}}+\frac{1}{2}\left(-\frac{2}{81}\left(54+78 I\sqrt{3}\right)-\frac{1}{6561}\left(4\left(-54 I\sqrt{3}\left(54-18 I\sqrt{3}\right)+54 I\sqrt{3}\left(-54+18 I\sqrt{3}\right)\right)\right)\right)\right)\right)\beta\,\sigma$$

```
Expand[%] /. {β -> 1/λ}
```

$$-\frac{2}{3}-\frac{8}{27\lambda}+\frac{3\lambda}{2}-\frac{2\sigma}{9\lambda}$$

***A similar analysis has been considered for the other parameter values:***

*Linear Stability $\lambda = +1$ :*

*Linear Stability $\lambda = 3/2$ :*

*Linear Stability $\lambda = 8/5$ :*

*Linear Stability $\lambda = 16/9$ :*

*Linear Stability $\lambda = 1.99$ :*

*Linear Stability $\lambda = 2$ :*

*Linear Stability $\lambda = 2.1$:*

*This is Brigman's Center :*

```
f[1] = x[2]; f[2] = -1/λ x[1]^3 + 3/2 x[1]^2 - (1 + σ/λ) x[1] + σ;
A = Simplify[
  Table[D[f[i], x[j]], {i, 1, 2}, {j, 1, 2}] /. x[1] -> x
roots = Simplify[Solve[f[2] == 0, x[1]]] /. x[1] -> x;
terms = Expand[A /. roots[[1]]][[2, 1]];
Table[Normal[Series[
    Part[terms, i] /. {λ -> 1/β}, {β, 0, 1}]] /. {β -> 1
 {i, 1, Length[terms]}];
```

*This gives* $-\omega^2$ leading to equation (20) :

```
Expand[Sum[%[[i]], {i, 1, Length[terms]}]]
```

$$-1+\left(3-\frac{1}{\lambda}\right)\sigma$$

*Correspondingly, this gives* $-\omega^2$ *:*

$$3-\frac{9\lambda}{4}+\frac{4+3\sigma}{9\lambda}$$

*This is the usual Reissner-Nordström Center node* $x_2^*$ *:*

```
A /. {x -> x2}
Simplify[%]
Expand[%[[2, 1]]]
Length[%]
```

20

```
Table[Normal[Series[
    Part[%44, i] /. {λ -> 1/β}, {β, 0, 1}]] /. {β -> 1/λ
 {i, 1, 20}]
Sum[%[[i]], {i, 1, 20}]
Expand[%]
```

$$3+\frac{4}{9\lambda}-\frac{9\lambda}{4}+\frac{\sigma}{3\lambda}$$

```
Expand[%]
```

$$-\frac{2}{9\lambda}+\frac{4\sigma}{3\lambda}+\frac{I\sqrt{-2916\left(\frac{16}{3}-4\sigma\right)+2916\,(4+4\sigma)}}{36\sqrt{3}}+\frac{I\sqrt{-2916\left(\frac{16}{3}-4\sigma\right)+2916\,(4+4\sigma)}}{54\sqrt{3}\,\lambda}+\frac{32\,I\sqrt{3}\,\sqrt{-2916\left(\frac{16}{3}-4\sigma\right)+2916\,(4+4\sigma)}}{\lambda\,(-3888+23328\,\sigma)}-\frac{252\,I\sqrt{3}\,\sigma\sqrt{-2916\left(\frac{16}{3}-4\sigma\right)+2916\,(4+4\sigma)}}{\lambda\,(-3888+23328\,\sigma)}$$

```
Series[%49, {σ, 0, 1}]
```

$$-1+\left(3-\frac{1}{\lambda}\right)\sigma+O[\sigma]^2$$

```
Normal[%]
```

$$-1+\left(3-\frac{1}{\lambda}\right)\sigma$$

```
Expand[%]
```

$$-1+3\sigma-\frac{\sigma}{\lambda}$$

```
Collect[%37, λ]
```

$$3-\frac{9\lambda}{4}+\frac{\frac{4}{9}+\frac{\sigma}{3}}{\lambda}$$

# *Index*

## A

## B

## C

## D

## E

## F

## G

## H

## I

## J

## K

## L

## M

## N

## O

## P

## Q

## R

## S

## T

## U

## V

## W

## Y

## Z